\documentclass[12pt]{book}

\usepackage[utf8]{inputenc}
\usepackage{emptypage} 
\usepackage{setspace}
\usepackage{amssymb,latexsym,amsmath}
\usepackage{bbold} 
\usepackage[a4paper,top=3cm,bottom=2cm,left=3cm,right=3cm,marginparwidth=1.75cm]{geometry}
\usepackage{graphicx}
\usepackage{subcaption} 
\usepackage{rotating}
\usepackage[colorlinks=true, allcolors=blue]{hyperref}
\usepackage{nomencl}
\usepackage{booktabs} 
\usepackage[font=small]{caption} 
\usepackage{xcolor}  

\usepackage{booktabs}      
\usepackage{longtable}     
\usepackage{pdflscape}     
\usepackage{pdfpages}

\definecolor{LightGray}{gray}{0.9}
\usepackage{algorithm}
\usepackage{algpseudocode}
\usepackage{hyperref}
\usepackage[style=authoryear, maxcitenames=2, mincitenames=1, maxbibnames=5, minbibnames=5, backend=biber, sorting=nyt]{biblatex}
\usepackage{proof}
\usepackage{wrapfig}
\usepackage{listings}
\usepackage{comment}

\newtheorem{theorem}{Theorem}[section]

\makenomenclature 

\title{Goodness-of-Fit Tests and Calibration Machine-Learning Algorithms\\ for Logistic Regression with Sparse Data}
\author{Ebrahim Khaled Ebrahim\thanks{Department of Applied Statistics, Faculty of Business, Alexandria University, Egypt. \texttt{ebrahimkhaled@alexu.edu.eg}. M.Sc. thesis; supervisors: Prof.\ Osama Abd El-Aziz Hussein and Dr.\ Ahmed El-Kotory.}}
\date{Alexandria University \\ \today}

\begin{document}

\maketitle













\frontmatter 
.
\vspace{5em}
\begin{figure}[H]
    \centering
    \includegraphics[width=1\linewidth]{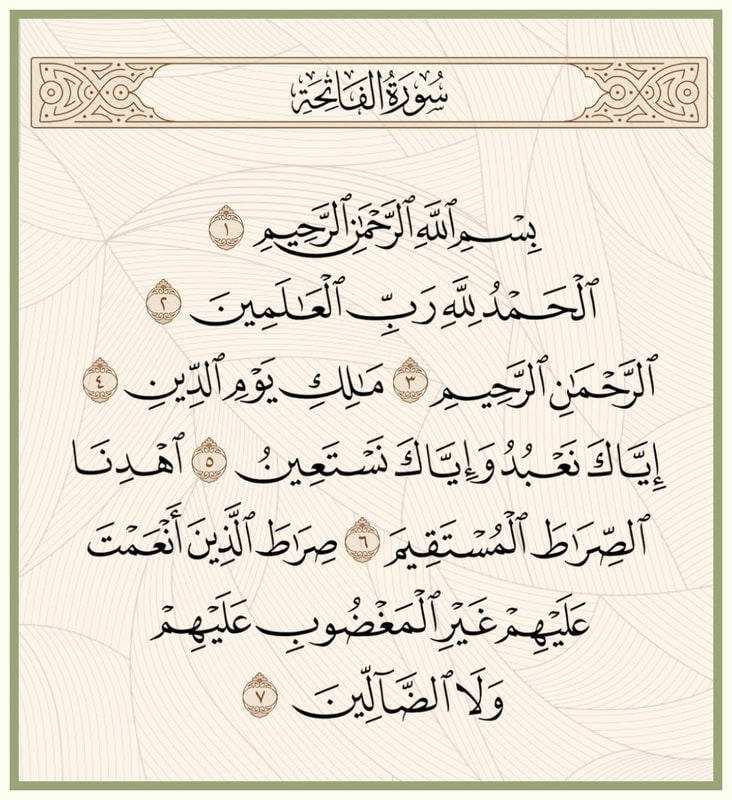}
\end{figure}

\begin{figure}[H]
    \centering
    \includegraphics[width=1\linewidth]{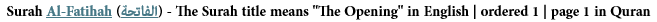}
\end{figure}

\chapter*{\Large \begin{center} Acknowledgments
  \end{center}}

First and foremost, all praise and gratitude are due to \textbf{Allah}, the Most Gracious, the Most Merciful. I thank Him for granting me the strength, patience, and perseverance to complete this work. Every success I have achieved is by His will and blessing alone. \textit{Alhamdulillah}.

\vspace{1.5em}
I would like to express my deepest appreciation to my supervisors, \textbf{Prof. Osama Abd El-Aziz Hussein} and \textbf{Dr. Ahmed El-Kotory}, for their invaluable guidance, continuous support, and profound expertise throughout this research journey. Their insightful feedback, constructive criticism, and unwavering encouragement have been instrumental in shaping this thesis. I am truly honored to have worked under their supervision.

\vspace{1.5em}
My heartfelt gratitude goes to my beloved \textbf{family}---my parents, whose endless love, sacrifices, and prayers have been my constant source of strength. To my father and mother, thank you for believing in me and supporting my academic pursuits unconditionally. Your encouragement has carried me through every challenge. I am forever indebted to you.

\vspace{1.5em}
Finally, I extend my sincere thanks to everyone who has contributed, directly or indirectly, to the completion of this work, especially the faculty members, and my colleagues. May Allah bless you all.

\chapter*{\Large \begin{center} Abstract
  \end{center}}

Assessing the goodness-of-fit of a logistic regression model is a critical prerequisite that must be satisfied before the model is used for inference. However, goodness-of-fit (GOF) tests, like the Chi-square and deviance, often give invalid results when the data is ``sparse"—a common issue when using continuous predictors like age or weight. In these scenarios, the asymptotic distribution assumptions required for binary logistic regression goodness-of-fit tests are often not satisfied. This research study classical GOF tests for binary logistic regression under both grouped and sparse data conditions, comparing the performance of approximately 30 statistical tests and machine learning calibration testing algorithms (as alternative for GOF tests). The tests included a different set of types, such as classic Chi-Square and Hosmer-Lemeshow variants, standardized Pearson statistics, covariate-space partitioning techniques, smoothing-based methods, as well as contemporary calibration machine learning and bootstrap procedures. Of these tests, At a fixed size, the GiViTi Calibration (2016), McCullagh(1989), Osius-Rojek(1992), le Cessie(1995) and Stute-Zhu tests(2002)—proved to be empirically high powerful, demonstrating a strong balance between correctly identifying bad models (high emperical power) and not liberally raising false alarms on good models (correct emperical Type I error).  Relying solely on these formal methods is insufficient and highlights visual diagnostics, such as calibration plots, as a vital exploratory step for detecting model deficiencies that formal tests often overlook. This was further discussed by the application to real-world data \textit{(Low Birth Weight Dataset)}, which showed that many of thoes statistical tests fail to provide valid conclusions when exposed to the complexities of actual datasets. The main conclusion from this study is that a model access requires a combination of several powerful statistical tests alongside a careful visual inspection of model calibration.

\newpage
\tableofcontents
\listoffigures
\listoftables
\listofalgorithms

\addcontentsline{toc}{chapter}{Table of Contents}

\renewcommand{\nomname}{List of Symbols}
\renewcommand{\nompreamble}{The next list describes several symbols that will be later used within the body of the document.}

\nomenclature{\(g\)}{The $g$-th group or covariate pattern.}
\nomenclature{\(\boldsymbol{\beta}\)}{Vector of regression coefficients, $(\beta_0, \beta_1, \ldots, \beta_p)'$.}
\nomenclature{\(\gamma_0, \gamma_1\)}{The calibration intercept and slope parameters, respectively.}
\nomenclature{\(\eta_i\)}{The linear predictor for the $i$-th observation, $\mathbf{x}_i^T\boldsymbol{\beta}$.}
\nomenclature{\(\pi_i\)}{The true probability of success for the $i$-th observation, $P(Y_i=1|\mathbf{x}_i)$.}
\nomenclature{\(\hat{\pi}_i\)}{The estimated probability of success for the $i$-th observation from a fitted model.}
\nomenclature{\(\overline{\hat{\pi}}_g\)}{The mean of the estimated probabilities within covariate pattern  (or Group) $g$.}
\nomenclature{\(\mathcal{L}(\boldsymbol{\beta})\)}{The log-likelihood function.}
\nomenclature{\(\chi^2_{df}\)}{The chi-square distribution with $df$ degrees of freedom.}

\nomenclature{\(B\)}{The number of bootstrap replications.}
\nomenclature{\(e_g\)}{The expected number of successes in group $g$, calculated as $n_g \overline{\hat{\pi}}_g$.}
\nomenclature{\(E_i\)}{The e-variable (likelihood ratio) for observation $i$ in the eHL test.}
\nomenclature{\(eHL\)}{The final e-value test statistic from the eHL test.}
\nomenclature{\(G\)}{The number of groups in a partitioning scheme (e.g., Hosmer-Lemeshow) or the number of unique covariate patterns in grouped data.}
\nomenclature{\(k_2, k_3, k_4\)}{Terms related to the second, third, and fourth moments of the binomial distribution used in McCullagh's test.}
\nomenclature{\(m_g\)}{The number of trials (observations) for the $g$-th unique covariate pattern.}
\nomenclature{\(\mathbb{n}\)}{The constant number of observations in each group for the Ebrahim-Farrington test, $\mathbb{n} = N/G$.}
\nomenclature{\(o_g\)}{The observed number of successes in group $g$.}
\nomenclature{\(p\)}{The number of covariates (independent variables) in the model.}
\nomenclature{\(p^*\)}{The total number of parameters in the model, including the intercept ($p^*=p+1$).}
\nomenclature{\(q_i\)}{The recalibrated probability for observation $i$ from an isotonic regression fit in the eHL test.}
\nomenclature{\(r_{i,pearson}\)}{The Pearson residual for the $i$-th observation.}
\nomenclature{\(r_{i,deviance}\)}{The Deviance residual for the $i$-th observation.}
\nomenclature{\(U(\boldsymbol{\beta})\)}{The score function, $\partial \mathcal{L}(\boldsymbol{\beta}) / \partial \boldsymbol{\beta}$.}
\nomenclature{\(\mathbf{W}\)}{The diagonal weight matrix with elements $W_{ii} = m_i\hat{\pi}_i(1-\hat{\pi}_i)$.}
\nomenclature{\(\mathbf{X}\)}{The $n \times p^*$ design matrix, with observations in rows and predictors (including an intercept column) in columns.}
\nomenclature{\(\mathbf{x}_i\)}{The vector of covariates for the $i$-th observation, $(1, x_{i1}, \ldots, x_{ip})'$.}
\nomenclature{\(X^2\)}{The Pearson chi-square statistic.}
\nomenclature{\(Y_i\)}{The observed binary outcome (0 or 1) for the $i$-th observation.}
\nomenclature{\(Z_{EF}, Z_{McC}\)}{The final standardized, normally-distributed test statistics for the Ebrahim-Farrington and McCullagh tests, respectively.}
\nomenclature{$M = J$}{Number of unique covariate patterns.}
\nomenclature{\(n_g\)}{The sample size in group $g$.}
\nomenclature{\(\pi(\cdot)\)}{The true probability function, mapping a covariate value (or vector) to the probability of success, i.e., $\pi(x) = P(Y=1|x)$.}
\nomenclature{\(\sigma(z)\)}{The sigmoid (logistic) function, defined as $\sigma(z) = \frac{1}{1 + e^{-z}}$. It maps any real-valued input $z$ to the interval $(0,1)$ and is used to convert the linear predictor in logistic regression to a probability.}

\nomenclature{\(I(\cdot)\)}{The indicator function, which is 1 if its argument is true and 0 otherwise.}
\nomenclature{\(\text{logit}(p)\)}{The logit function, $\ln(p/(1-p))$.}
\nomenclature{\(\text{diag}(\mathbf{v})\)}{A diagonal matrix with the elements of vector $\mathbf{v}$ on its diagonal.}

\nomenclature{GOF}{Goodness-of-Fit.}
\nomenclature{\(\hat{C}\)}{Hosmer-Lemeshow test statistic using groups of equal size (e.g., deciles of risk).}
\nomenclature{\(\hat{H}\)}{Hosmer-Lemeshow test statistic using groups of equal distances (equal-width intervals of predicted probabilities).}


\nomenclature{\(\mathbf{x}_g\)}{Vector of covariates including the intercept, $(1, x_{g1}, \ldots, x_{gp})'$, for the $g$-th covariate pattern.}
\nomenclature{\(Y_g = o_g\)}{The count of successes $0,1,\ldots,m_g$ in the $g$-th covariate pattern, $\sim \text{Binomial}(m_g, \pi_g)$, in the context of binary logistic regression.}
\nomenclature{\(\pi_g\)}{The probability of success for the $g$-th covariate pattern (or group).}
\nomenclature{\(\boldsymbol{\pi}\)}{The true probability vector, $(\pi_1, \pi_2, \ldots, \pi_N)'$.}
\nomenclature{\(\mu_g\)}{The quantity $m_g \pi_g$; in the sparse data case, equals $\pi_i$. Estimated by $E_g = m_g \hat{\pi}_g$.}
\nomenclature{\(\theta_i\)}{The $i$-th logit, $\text{logit}(\pi_i)$, in the context of logistic regression.}
\nomenclature{\(\mathcal{I}(\boldsymbol{\theta_0})\)}{The expected Fisher information matrix: $-E\left[\frac{\partial^2 \ell}{\partial \boldsymbol{\theta} \partial \boldsymbol{\theta}^\top}\right]$ evaluated at $\theta_0$.}
\nomenclature{\(\mathbf{e}\)}{The Pearson residual vector, $\mathbf{e} = (r_{1,\text{pearson}}, \ldots, r_{n,\text{pearson}})'$.}


\nomenclature{\(L(\theta; \mathbf{x})\)}{The likelihood function, where $\theta$ is the parameter vector and $\mathbf{x}$ is the data.}



\printnomenclature

\newpage
\mainmatter 
\chapter{Introduction}

\subsection*{Binary Logistic Regression: Purpose and Assumptions}

Binary logistic regression is a statistical method used to model the relationship between a set of predictor variables (covariates) and a binary outcome variable. The outcome variable $Y$ takes only two values: 0 (failure) or 1 (success). The goal of the model is to estimate the probability of success, $\pi$, given the observed values of the predictors \parencite{hosmer2013applied}.

The model relates the probability $\pi_i$ to the covariates through the logit function:
\begin{equation}
   \eta(\pi_i) = \log\left(\frac{\pi_i}{1 - \pi_i}\right) = \beta_0 + \beta_1 x_{i1} + \cdots + \beta_p x_{ip} = \mathbf{x}_i'\boldsymbol{\beta}
\end{equation}

This transformation ensures that the predicted probabilities lie between 0 and 1, the inverse of the logit gives the probability of success:
\begin{equation}
    \pi_i = \frac{\exp(\mathbf{x}_i'\boldsymbol{\beta})}{1 + \exp(\mathbf{x}_i'\boldsymbol{\beta})}
\end{equation}

The model makes several assumptions. First, the outcome variable is binary. Second, observations are independent of each other. Third, the relationship between the logit of the probability and the covariates is linear. Fourth, there is no multicollinearity among the predictors. Fifth, the model is correctly specified---meaning all relevant variables are included, no irrelevant variables are present, and the functional form is appropriate \parencite{hosmer2013applied, agresti2013categorical}.

\subsection*{Goodness-of-Fit: What It Means and Why It Matters}

After fitting a logistic regression model, it is not enough to simply report the parameter estimates. The model must be checked to see whether it fits the data well. This process is called \textit{goodness-of-fit}(GOF) assessment. \parencite{gof2002}

Goodness-of-fit testing addresses the following hypothesis:
\begin{align}
    H_0 & : \mathbb{E}[Y | \mathbf{x}] = \pi(\mathbf{x}, \boldsymbol{\beta}) \\
    H_1 & : \mathbb{E}[Y | \mathbf{x}] \neq \pi(\mathbf{x}, \boldsymbol{\beta})
\end{align}
where $\pi(\mathbf{x}, \boldsymbol{\beta})$ is the hypothesized functional form of the model. If we reject $H_0$, we conclude that the model is misspecified in some way. This could mean that an important variable is missing, that the wrong functional form was used, or that the assumed relationship does not hold in

Hosmer and Lemeshow (2000) stated that ``\textit{an assessment of the adequacy of the fitted model is an essential part of the model building process}''. Similarly, Nikulin et al. (2002) emphasized that ``\textit{statistical models that cannot be validated through goodness-of-fit measures, cannot be reliably employed}''. These statements reflect a fundamental principle: a model that does not fit the data cannot be trusted for inference or prediction \parencite{hosmer2000applied, gof2002}.

The standard approach to goodness-of-fit testing in logistic regression uses chi-square-based statistics, such as the Pearson chi-square statistic:
\begin{equation}
    X^2 = \sum_{g=1}^{G} \frac{(o_g - e_g)^2}{e_g(1 - e_g/n_g)}
\end{equation}
where $o_g$ is the observed number of successes in group $g$, $e_g = n_g \bar{\hat{\pi}}_g$ is the expected number, and $n_g$ is the sample size in group $g$.

Under certain conditions, this statistic follows a chi-square distribution with $(G - p - 1)$ degrees of freedom under $H_0$, where $G$ is the number of groups and $p$ is the number of predictors. However, these conditions are not always satisfied, which leads to the problems discussed below.

\subsection*{GOF Fail: The Problem with Continuous Covariates}

The chi-square approximation for goodness-of-fit statistics relies on asymptotic theory. For the approximation to be valid, each group must contain enough observations. More precisely, the expected number of successes and failures in each group should be large enough that the normal approximation to the binomial distribution holds \parencite{agresti2013categorical}. When the model includes continuous covariates (such as age, weight, or blood pressure), a different situation arises. Each observation tends to have a unique combination of predictor values. In the terminology of logistic regression, each \textit{covariate pattern} contains only one observation. This situation is called \textit{sparse data} or \textit{ungrouped data}.

In sparse data, the number of groups $G$ equals the sample size $n$. The classical chi-square and deviance statistics no longer follow a chi-square distribution, and the standard goodness-of-fit tests become not valid. The tests may reject the null hypothesis too often (liberal behavior) or too rarely (conservative behavior), depending on the situation \parencite{hosmer2000applied, farrington1996}.

Farrington (1996) showed that when $n/G$ tends to a finite constant (rather than infinity), the asymptotic chi-square distribution does not hold. Kuss (2002) confirmed that in practical settings with continuous covariates, standard tests such as the Pearson and deviance tests cannot be used directly.

\subsection*{Sparse Data}

The term ``sparse data'' refers to a situation where each observation has a unique covariate pattern, or where most covariate patterns contain very few observations. This is the opposite of ``grouped data'', where many observations share the same covariate pattern. In grouped data, we can write:
\begin{equation}
    Y_g \sim \text{Binomial}(m_g, \pi_g), \quad g = 1, 2, \ldots, G
\end{equation}
where $Y_g$ is the number of successes in group $g$, $m_g$ is the number of trials, and $\pi_g$ is the probability of success. The sum $\sum_{g=1}^G m_g = n$ gives the total sample size. When $m_g$ is reasonably large for each group, the binomial counts can be approximated by normal distributions, and chi-square tests work well.

In sparse data, each group contains at most one observation ($m_g = 1$ for all $g$), so $G = n$. Each observation is either a success ($Y_g = 1$) or a failure ($Y_g = 0$). There is no replication within covariate patterns. The consequence is that the expected counts are too small for the normal approximation to hold, and the chi-square distribution no longer applies \parencite{kuss2002, farrington1996, paul2013testing}.

\vspace{0.5cm}
\noindent Table~\ref{tab:grouped_vs_sparse} summarizes the key differences:

\begin{table}[ht]
    \centering
    \begin{tabular}{lcc}
        \toprule
        \textbf{Property}              & \textbf{Grouped Data}       & \textbf{Sparse Data} \\
        \midrule
        Number of groups ($G$)         & $G \ll n$                   & $G = n$              \\
        Observations per group ($m_g$) & $m_g > 1$                   & $m_g = 1$            \\
        Outcome per group ($Y_g$)      & Count ($0, 1, \ldots, m_g$) & Binary ($0$ or $1$)  \\
        Chi-square approximation       & Valid                       & Not valid            \\
        \bottomrule
    \end{tabular}
    \caption{Comparison of grouped and sparse data structures.}
    \label{tab:grouped_vs_sparse}
\end{table}

\paragraph{Numerical Example:}
Consider a logistic regression with two predictors, X1 (continuous) and X2 (binary), The first two rows share identical patterns (X1=1.5, X2=0), forming the $1^{st}$ group with $m_g = 2$ observations. The next four rows share pattern (X1=2.3, X2=1), forming the $2^{nd}$ group with $m_g = 4$ observations. 
\
\begin{table}[h]
    \centering
    \begin{tabular}{ccccccc}
        \hline
        Scenario & X1            & X2         & Y & Covariate Pattern & $m_g$ & $Y_g$ \\
        \hline
        Grouped  & \textbf{1.5 } & \textbf{0} & 1 & $1^{st}$          & 2     & 1     \\
                 & \textbf{1.5}  & \textbf{0} & 0 &                                   \\
        \\
                 & \textbf{2.3}  & \textbf{1} & 1 & $2^{nd}$          & 4     & 3     \\
                 & \textbf{2.3 } & \textbf{1} & 1 &                                   \\
                 & \textbf{2.3 } & \textbf{1} & 1 &                                   \\
                 & \textbf{2.3 } & \textbf{1} & 0 &                                   \\
        \hline
        Sparse   & 1.5           & 0          & 1 & $1^{st}$          & 1     & -     \\
                 & 1.6           & 0          & 0 & $2^{nd}$          & 1     & -     \\
                 & 2.3           & 1          & 1 & $3^{rd}$          & 1     & -     \\
                 & 2.4           & 1          & 0 & $4^{th}$          & 1     & -     \\
        \hline
    \end{tabular}
    \caption{Grouped vs. Sparse Data Example}
\end{table}

The presence of continuous covariates almost always leads to sparse data. Since continuous variables take many distinct values, it is rare for two observations to have exactly the same predictor values. This is why goodness-of-fit testing in logistic regression with continuous covariates requires special attention.

\section{GLM and Binary Logistic Regression}

Binary logistic regression is a specific case of the Generalized Linear Models (GLMs), a class of models introduced by Nelder and Wedderburn.  Generalized Linear Models (GLMs) extend linear models to encompass a broader range of probability distributions. The key to this extension is the use of the exponential family of distributions. (\cite{McCullaghNelder1989GLM})

\subsection*{Canonical Form}

For binary logistic regression with the binomial distribution, we can express the probability mass function in the canonical exponential family form. This formulation explicitly accounts for the grouped nature of the data, where each observation ($y_i$) represents a count of successes out of a number of trials (0,1,2,..,$m_j$), rather than individual binary outcomes (0,1) :


\begin{equation}
    f(y_j; \theta_j, \phi) = \exp\left\{\frac{y_j\theta_j - m_jb(\theta_j)}{a(\phi)} + c(y_j, \phi)\right\}
\end{equation}

where:
\begin{itemize}
    \item $j$ indexes the unique covariate patterns (groups), $j = 1, ..., J$
    \item $y_j$ is the number of successes in the $j$-th group
    \item $m_j$ is the number of trials in the $j$-th group
    \item $\theta_j = \log[\pi_j/(1-\pi_j)]$ is the natural parameter (logit)
    \item $\pi_j$ is the probability of success for the $j$-th group
    \item $\phi$ is the dispersion parameter
    \item The functions $a(\phi)$, $b(\theta_j)$, and $c(y_j, \phi)$ are defined as:
          \begin{itemize}
              \item $a(\phi) = \phi = 1/m_j$ (the dispersion parameter is the reciprocal of the number of trials)
              \item $b(\theta_j) = \log[1 + \exp(\theta_j)]$
              \item $c(y_j, \phi) = \log\binom{m_j}{y_j}$
          \end{itemize}
\end{itemize}
 
\begin{itemize}
    \item The expected value is:
          \begin{equation}
              E(Y_j) = \mu_j = m_jb'(\theta_j) = m_j\frac{\exp(\theta_j)}{1 + \exp(\theta_j)} = m_j\pi_j
          \end{equation}
    \item The variance is:
          \begin{equation}
              \text{Var}(Y_j) = m_jb''(\theta_j)a(\phi) = m_j\pi_j(1-\pi_j)
          \end{equation}
    \item The logit is:
          \begin{equation}
              g(\mu_j/m_j) = \log\left(\frac{\mu_j/m_j}{1-\mu_j/m_j}\right) = \log\left(\frac{\pi_j}{1-\pi_j}\right)= \theta_j = \mathbf{x}_j'\boldsymbol{\beta}
          \end{equation}
          where $\mathbf{x}_j$ is the vector of covariates for the $j$-th group and $\boldsymbol{\beta}$ is the vector of regression coefficients.
\end{itemize}

\section{Asymptotic Theory in Hypothesis Testing}

This section presents the asymptotic theory that forms the basis of most goodness-of-fit tests in logistic regression. Understanding this theory is important for two reasons. First, it explains why goodness-of-fit tests produce chi-square statistics. Second, it shows why these tests can fail when the data is sparse.

\subsection*{Asymptotic Theory}

Asymptotic theory studies the behavior of statistical procedures when the sample size $n$ becomes large. The word ``asymptotic'' means ``approaching a limit.'' In this context, it refers to what happens as $n \to \infty$.

The main idea is simple: many test statistics do not have exact, known distributions for finite samples. However, as the sample size grows, these statistics approach standard distributions (such as the normal or chi-square distribution). We can then use these limiting distributions to compute p-values and make decisions about hypotheses.

For goodness-of-fit testing, the chi-square distribution plays a central role. Most classical GOF tests---including the Pearson chi-square, deviance, Hosmer-Lemeshow, and many others---rely on the assumption that their test statistic follows a chi-square distribution when the sample is large enough. If this assumption does not hold (as is the case with sparse data), the tests become unreliable.\parencite{lehmann_testing_2005,rossi_mathematical_2018}

\subsection*{MLE and Its Properties}

In logistic regression, we estimate the parameters $\boldsymbol{\beta}$ by maximizing the likelihood function. The resulting estimates are called Maximum Likelihood Estimators (MLEs). Under certain regularity conditions, the MLE has two important properties that are used to construct hypothesis tests \parencite{lehmann_testing_2005, casella2002statistical}.

\begin{theorem}[Consistency of the MLE]
    As the sample size increases, the MLE converges to the true parameter value. Formally, if $\hat{\boldsymbol{\beta}}_n$ is the MLE based on $n$ observations and $\boldsymbol{\beta}_0$ is the true parameter, then:
    \begin{equation}
        \hat{\boldsymbol{\beta}}_n \xrightarrow{P} \boldsymbol{\beta}_0 \quad \text{as } n \to \infty
    \end{equation}
    where $\xrightarrow{P}$ denotes convergence in probability.
\end{theorem}

\begin{theorem}[Asymptotic Normality of the MLE]
    For large samples, the MLE is approximately normally distributed around the true parameter. Let $\mathcal{I}(\boldsymbol{\beta}_0)$ be the Fisher Information matrix. Then:
    \begin{equation}
        \sqrt{n}(\hat{\boldsymbol{\beta}}_n - \boldsymbol{\beta}_0) \xrightarrow{d} N\left(\mathbf{0}, \mathcal{I}(\boldsymbol{\beta}_0)^{-1}\right)
    \end{equation}
    where $\xrightarrow{d}$ denotes convergence in distribution. This means that for large $n$:
    \begin{equation}
        \hat{\boldsymbol{\beta}}_n \approx N\left(\boldsymbol{\beta}_0, \frac{1}{n}\mathcal{I}(\boldsymbol{\beta}_0)^{-1}\right)
    \end{equation}
\end{theorem}

These two properties form the foundation for the three classical hypothesis tests: the Likelihood Ratio Test (LRT), the Score Test, and the Wald Test.

\subsection{The Likelihood Ratio Test and Deviance} \label{LRT_and_Deviance}

The Likelihood Ratio Test (LRT) is one of the most commonly used methods for testing hypotheses about statistical models. It compares the fit of two models: a restricted model (under the null hypothesis) and an unrestricted model (the full model). The idea is straightforward: if the null hypothesis is true, the two models should fit the data equally well \parencite{lehmann_testing_2005, agresti2013categorical}.

\subsubsection*{The Likelihood Ratio}

Let $L(\boldsymbol{\beta})$ denote the likelihood function. For testing the null hypothesis $H_0$ against the alternative $H_1$, the likelihood ratio is:
\begin{equation}
    \Lambda = \frac{L(\tilde{\boldsymbol{\beta}})}{L(\hat{\boldsymbol{\beta}})}
\end{equation}
where $\tilde{\boldsymbol{\beta}}$ is the MLE under the null hypothesis (restricted model) and $\hat{\boldsymbol{\beta}}$ is the MLE under the full model (unrestricted model). Since the full model always fits at least as well as the restricted model, we have $0 < \Lambda \leq 1$.

\subsubsection*{The LRT Statistic}

The LRT statistic is defined as:
\begin{equation}
    G^2 = -2 \log \Lambda = -2 \left[ \ell(\tilde{\boldsymbol{\beta}}) - \ell(\hat{\boldsymbol{\beta}}) \right] = 2 \left[ \ell(\hat{\boldsymbol{\beta}}) - \ell(\tilde{\boldsymbol{\beta}}) \right]
\end{equation}
where $\ell(\cdot) = \log L(\cdot)$ is the log-likelihood function.

\begin{theorem}[Wilks' Theorem]
    Under the null hypothesis, and assuming standard regularity conditions, the LRT statistic follows a chi-square distribution asymptotically:
    \begin{equation}
        G^2 \xrightarrow{d} \chi^2_r \quad \text{as } n \to \infty
    \end{equation}
    where $r$ is the number of restrictions imposed by the null hypothesis (the difference in the number of parameters between the full and restricted models).
\end{theorem}

\subsubsection*{Derivation of the Deviance Statistic}

The deviance statistic is a specific application of the LRT. It compares the fitted model to the saturated model (a model with one parameter per observation, which fits the data perfectly).

For binary logistic regression with $n$ observations, the log-likelihood of any model is:
\begin{equation}
    \ell(\boldsymbol{\beta}) = \sum_{i=1}^{n} \left[ y_i \log(\pi_i) + (1 - y_i) \log(1 - \pi_i) \right]
\end{equation}
where $y_i \in \{0, 1\}$ is the observed outcome and $\pi_i = P(Y_i = 1 | \mathbf{x}_i)$ is the probability of success.

\paragraph{Step 1: Log-likelihood of the fitted model.}
For the fitted model with estimated probabilities $\hat{\pi}_i$:
\begin{equation}
    \ell(\hat{\boldsymbol{\beta}}) = \sum_{i=1}^{n} \left[ y_i \log(\hat{\pi}_i) + (1 - y_i) \log(1 - \hat{\pi}_i) \right]
\end{equation}

\paragraph{Step 2: Log-likelihood of the saturated model.}
The saturated model sets $\pi_i = y_i$ for each observation. Since $y_i \in \{0, 1\}$:
\begin{itemize}
    \item If $y_i = 1$: the contribution is $1 \cdot \log(1) + 0 \cdot \log(0) = 0$
    \item If $y_i = 0$: the contribution is $0 \cdot \log(0) + 1 \cdot \log(1) = 0$
\end{itemize}
Therefore, $\ell(\boldsymbol{\beta}_{sat}) = 0$.

\paragraph{Step 3: The deviance.}
Applying the LRT formula:
\begin{align}
    D &= -2 \left[ \ell(\hat{\boldsymbol{\beta}}) - \ell(\boldsymbol{\beta}_{sat}) \right] \\
      &= -2 \left[ \ell(\hat{\boldsymbol{\beta}}) - 0 \right] \\
      &= -2 \sum_{i=1}^{n} \left[ y_i \log(\hat{\pi}_i) + (1 - y_i) \log(1 - \hat{\pi}_i) \right]
\end{align}

This can be rewritten as:
\begin{equation}
    D = -2 \sum_{i=1}^{n} \left[ y_i \log\left(\frac{\hat{\pi}_i}{y_i}\right) + (1 - y_i) \log\left(\frac{1 - \hat{\pi}_i}{1 - y_i}\right) \right]
\end{equation}
where we use the convention that $0 \cdot \log(0/0) = 0$.

Under the null hypothesis (the model is correctly specified), the deviance follows a chi-square distribution:
\begin{equation}
    D \xrightarrow{d} \chi^2_{n - p^*}
\end{equation}
where $p^*$ is the number of parameters in the model.

\subsection{The Score Test}

The Score test (also called the Lagrange Multiplier test or Rao test) takes a different approach. Instead of comparing likelihoods, it looks at the gradient (slope) of the log-likelihood function at the null hypothesis \parencite{lehmann_testing_2005, rao1973}.

The score function is defined as:
\begin{equation}
    U(\boldsymbol{\beta}) = \frac{\partial \ell(\boldsymbol{\beta})}{\partial \boldsymbol{\beta}}
\end{equation}

At the MLE, the score equals zero (since that is where the likelihood is maximized). The Score test asks: if we evaluate the score at the restricted MLE $\tilde{\boldsymbol{\beta}}$, how far is it from zero? If the null hypothesis is true, the score should be close to zero.

\begin{theorem}[The Score Test]
    Let $\tilde{\boldsymbol{\beta}}$ be the MLE under the null hypothesis. The Score test statistic is:
    \begin{equation}
        S = U(\tilde{\boldsymbol{\beta}})^T \mathcal{I}(\tilde{\boldsymbol{\beta}})^{-1} U(\tilde{\boldsymbol{\beta}})
    \end{equation}
    Under the null hypothesis:
    \begin{equation}
        S \xrightarrow{d} \chi^2_r
    \end{equation}
    where $r$ is the number of restrictions.
\end{theorem}

The Score test has a practical advantage: it only requires fitting the model under the null hypothesis. This is useful when the full model is difficult to estimate.

\paragraph{Connection to Goodness-of-Fit Tests.}
Many goodness-of-fit tests for logistic regression are based on the Score test principle. Tests such as the Osius-Rojek test and the Tsiatis test check whether adding certain terms to the model would improve the fit. They do this by examining the score at the fitted model, without actually fitting the augmented model \parencite{osius1992normal}.

\subsection{The Wald Test}

The Wald test is perhaps the most intuitive of the three tests. It measures how far the parameter estimate is from the null hypothesis value, standardized by its standard error \parencite{lehmann_testing_2005}.

\begin{theorem}[The Wald Test]
    Let $\hat{\boldsymbol{\beta}}$ be the MLE from the full model. For testing $H_0: \boldsymbol{\beta} = \boldsymbol{\beta}_0$, the Wald test statistic is:
    \begin{equation}
        W = (\hat{\boldsymbol{\beta}} - \boldsymbol{\beta}_0)^T \mathcal{I}(\hat{\boldsymbol{\beta}}) (\hat{\boldsymbol{\beta}} - \boldsymbol{\beta}_0)
    \end{equation}
    Under the null hypothesis:
    \begin{equation}
        W \xrightarrow{d} \chi^2_r
    \end{equation}
    where $r$ is the number of parameters being tested.
\end{theorem}

The Wald test only requires fitting the unrestricted model. However, it can be unreliable when the sample size is small or when the parameter estimate is far from the null value.

\subsection{Why This Matters for Goodness-of-Fit Testing}

The goodness-of-fit tests discussed in this thesis are built on the asymptotic theory presented above. The Pearson chi-square and deviance statistics are special cases of the LRT. The Osius-Rojek and related tests are based on the Score test. Understanding this theory helps explain:

\begin{enumerate}
    \item The distribution of GOF test statistics as chi-square variables, based on the asymptotic properties of the MLE.
    \item The relationship between degrees of freedom and the number of parameters, as established by Wilks' theorem.
    \item The failure of these tests in sparse data settings, where the asymptotic approximation requires large counts per group rather than just a large total sample size.
\end{enumerate}

\section{Classical Goodness-of-Fit Tests}

Within the context of goodness-of-fit assessment for logistic regression, the \textbf{Pearson Chi-square and Deviance statistics} remain the most extensively documented and utilized methods. Derived from classical statistical principles, these tests have provided the standard framework for measuring model calibration and assessing the adequacy of the model \parencite{hosmer2013applied, agresti2015foundations}. Their asymptotic behavior and reliability across different data configurations have been the subject of exhaustive theoretical and empirical scrutiny.

\subsection{Pearson Chi-Square Statistic} \label{sec: pearson_chi_square}

\paragraph{Definition and Formulation}  \parencite{hosmer2013applied}
Let $Y_1, \ldots, Y_n$ be independent Bernoulli random variables with $E(Y_i) = \pi_i$, where $\pi_i$ is the probability of success for the $i$-th observation . The Pearson chi-square statistic is defined as:

\begin{equation}
    X^2 = \sum_{i=1}^n \frac{(Y_i - \hat{\pi}_i)^2}{\hat{\pi}_i(1-\hat{\pi}_i)}
\end{equation}\\
where $\hat{\pi}_i$ is the estimated probability from the fitted logistic regression model.\\
First, we can rewrite this as:
\begin{equation}
    X^2 = \sum_{i=1}^n \left(\frac{Y_i - \hat{\pi}_i}{\sqrt{\hat{\pi}_i(1-\hat{\pi}_i)}}\right)^2 \label{eq:pearson_ind}
\end{equation}
\paragraph{Individual Contributions and Residuals} 
Second, the Pearson residual for the $i$-th observation is defined as:
\begin{equation}
    r_{i,pearson} = \frac{Y_i - \hat{\pi}_i}{\sqrt{\hat{\pi}_i(1-\hat{\pi}_i)}}\label{eq:pearson_ind_res}
\end{equation}
Under the null hypothesis of a correctly specified model, these residuals possess an expected value of approximately zero and a variance near unity in large-sample contexts. The Central Limit Theorem implies that $r_{i,pearson}$ asymptotically follows a standard normal distribution. Since the sum of squares of independent standard normal variables follows a chi-square distribution, the statistic $X^2 = \sum_{i=1}^n r_{i,pearson}^2$ is asymptotically chi-square distributed.

\paragraph{Asymptotic Distribution} Third, The degrees of freedom are determined by the number of independent terms in the sum. This is equal to the number of sample size(n) minus the number of parameters estimated in the model ($p^*$), resulting in n-$p^*$ degrees of freedom. Thus, under the null hypothesis and for large samples, $X^2 \sim \chi^2_{n-p^*}$ asymptotically. \parencite{agresti2013categorical}
\newline \newline
In this context, it's important to note that the number of degrees of freedom should originally be the number of unique covariate patterns (G) minus the number of parameters in the model ($p^*$). However, when dealing with sparse data, we often find that (G) is approximately equal to the sample size (n).

The Pearson chi-square statistic for logistic regression with grouped data is defined as:

\begin{equation}
    X^2 = \sum_{g=1}^G \frac{(y_g - m_g\hat{\pi}_g)^2}{m_g\hat{\pi}_g(1-\hat{\pi}_g)}   \sim \chi^2_{g-p^*} \label{eq:pearson_group}
\end{equation}

where $G$ is the number of covariate patterns, $y_g$ is the observed number of successes within $m_g$ is the number of observations, and $\hat{\pi}_g$ is the estimated probability of success for pattern $g$.

\subsection{Deviance Statistic } \label{sec:deviance_statistic_introduction}
The deviance statistic is a fundamental measure for assessing the goodness-of-fit in logistic regression models. It quantifies the discrepancy between the fitted model and a saturated model, providing insight into how well the model explains the observed data. (recall subsection \ref{LRT_and_Deviance} )

\paragraph{Individual Contributions and Residuals}
The contribution of each observation to the overall deviance, known as the deviance component, is given by:
\begin{equation}
    d_i = -2[y_i\log(\hat{\pi}_i) + (1 - y_i)\log(1 - \hat{\pi}_i)]
    \label{eq:deviance_component}
\end{equation}
From this, we define the deviance residual for the $i$-th observation as:
\begin{equation}
    r_{i,deviance} = \text{sign}(y_i - \hat{\pi}_i)\sqrt{d_i}
    \label{eq:deviance_residual}
\end{equation}
These residuals provide a measure of the discrepancy between observed and fitted values on the scale of the linear predictor. $\sum{r_{i,deviance}}=D$.
\paragraph{Asymptotic Distribution}
Under specific conditions, notably the null hypothesis of model adequacy, the deviance statistic follows an asymptotic distribution:
\begin{equation}
    D \sim \chi^2_{n-p^*}
    \label{eq:deviance_distribution}
\end{equation}
where $p^*$ is the number of parameters in the model.

Agresti and Cox \parencite{Cox1989, agresti2013categorical} demonstrated that when the model holds true (i.e., $\pi_i = \hat{\pi}_i$ for all $i$), the difference $|X^2 - D|$ approaches zero as $n \to \infty$. This convergence is particularly reliable when the number of observations per covariate pattern $m_g$ is sufficiently large—a condition typically violated in ungrouped data where $m_g = 1$—and when $\hat{\pi}_i$ is near $0.5$. Conversely, if the model is misspecified, both $X^2$ and $D$ grow unboundedly as $n$ increases, following a non-central chi-squared distribution with non-centrality parameter $\lambda$. This behavior arises because the expected values of the residuals, such as $E(r_{i, \text{pearson}})$, no longer vanish \parencite{hosmer2000applied}. Furthermore, Marriott \parencite{Geo2015} notes that the deviance statistic often exhibits a more stable and tractable distribution in high-dimensional contexts, making it preferable when the asymptotic approximations for the Pearson statistic lack uniformity.

\subsection{Conducting Initial Investigations on Classical GOF Tests}

To evaluate the performance of various goodness-of-fit tests in logistic regression, an initial simulation study was conducted. This study compared the power of traditional tests—specifically the Pearson chi-square and deviance—against the Hosmer-Lemeshow test, a widely used method that involves \textbf{partitioning observations} based on their predicted probabilities to compare observed and expected frequencies.
The simulation generated data from a logistic regression model with a quadratic term, $\text{logit}(p) = \beta_0 + \beta_1x + \beta_2x^2$ ($\beta_0 = 0$, $\beta_1 = 1$, $\beta_2 = 0.5$), where $x$ was uniformly distributed on $[-3, 3]$ (Sparse Data). A misspecified linear model was then fitted to this data, omitting the quadratic term $(x^2)$. The study examined sample sizes ranging from 50 to 500, with 500 simulations per sample size. The power of each test was assessed as the proportion of simulations rejecting the null hypothesis at $\alpha = 0.05$.  

\begin{figure}[h]
    \centering
    \includegraphics[width=0.75\linewidth]{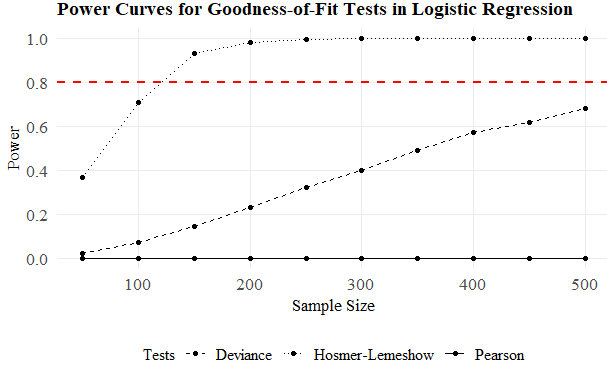}
    \caption[Power of Traditional Goodness-of-Fit Tests (Pilot Simulation)]{Power of Traditional Goodness-of-Fit Tests (Pilot Simulation)}
    \label{fig:power_start_simulation}

\end{figure}

The figure \ref{fig:power_start_simulation} illustrates the effectiveness of traditional goodness-of-fit tests (Pearson and Deviance) with a modified method (Hosmer-Lemeshow) in logistic regression (R Code in Appendix. \ref{3-sim}). It displays the power curves for these tests, illustrating their ability to detect model misspecification based on sample size and specific conditions. Findings are as follows: 1) The Pearson Chi-Square test shows consistently (Zero) power across all sample sizes, suggesting an inability to detect any thing about the simulated misspecification. 2) The Deviance test exhibits a gradual increase in power but remains below the 0.8 threshold (as example) even at 500 observations. 3) In contrast,the Hosmer-Lemeshow test demonstrates superior performance, reaching the conventional 0.8 power threshold at approximately 150 observations and approaching a power of 1 for larger sample sizes, but failed to have a good power in the low samples.  Notably, all tests show inadequate power for sample sizes below 100, highlighting a common limitation in small-sample scenarios.

\newpage
\section{Calibration vs. Discrimination in (Machine Learning)} \label{sec:calibration_vs_discrimination}

In the context of predictive modeling, two critical concepts are commonly used to evaluate model performance: \textbf{calibration} and \textbf{discrimination}. Both are essential for determining how well a model can predict outcomes. \parencite{huang2020tutorial, walsh2017beyond}

\subsection{Discrimination}

Discrimination refers to the model's ability to correctly differentiate between individuals who will experience the event of interest and those who will not. Calibration and discrimination assess different aspects of model performance. A model can have high discrimination (e.g., high AUROC) but still be poorly calibrated if its predicted probabilities do not match observed outcomes. For example, a model may rank patients correctly by risk but consistently overestimate the actual probability of disease. Both metrics should be reported, yet calibration remains under-reported: one review found $63\%$ of models reported discrimination, but only $36\%$ reported calibration \parencite{huang2020tutorial, wessler2015clinical}.



\noindent
\textbf{Test Statistic:}
\begin{itemize}
    \item \textbf{AUROC:} The AUROC is computed by plotting the true positive rate against the false positive rate at different classification thresholds. The area under this curve serves as a summary of the model's discrimination ability.
    \item \textbf{C-Statistic:} For binary outcomes, the C-statistic is equivalent to the AUROC and is frequently reported in clinical studies to assess discrimination\parencite{huang2020tutorial}.
\end{itemize}

\subsection{Calibration}

Calibration refers to how close the predicted probabilities of an event are to the actual observed probabilities. A well-calibrated model produces predictions where the predicted probability closely matches the actual probability of the outcome across different risk levels. \parencite{serrano2012calibration,hosmer1995confidence, Kueppers_2020_CVPR_Workshops}

To assess calibration, several statistical tools and graphical methods are employed. The most common method is the \textbf{Calibration plot, Reliability Index and GIVITI Test} (discussed in details in chapter 5). And now we need illustrate the difference between four \textbf{important concepts}: Calibration Methods, Calibration Tests, Calibration Measurements, and Calibration Plots.

\begin{itemize}
    \item \textbf{Calibration Plot:} A calibration plot is a graphical tool used to visually assess the agreement between predicted probabilities and observed outcomes. The most common is the reliability diagram, where predicted probabilities are grouped (e.g., into deciles), and the mean predicted probability is plotted against the observed event frequency in each group. A perfectly calibrated model will have points lying on the 45-degree line (where predicted probability = observed proportion of success, in each bin).
          \begin{figure}[h]
              \centering
              \includegraphics[width=0.4\linewidth]{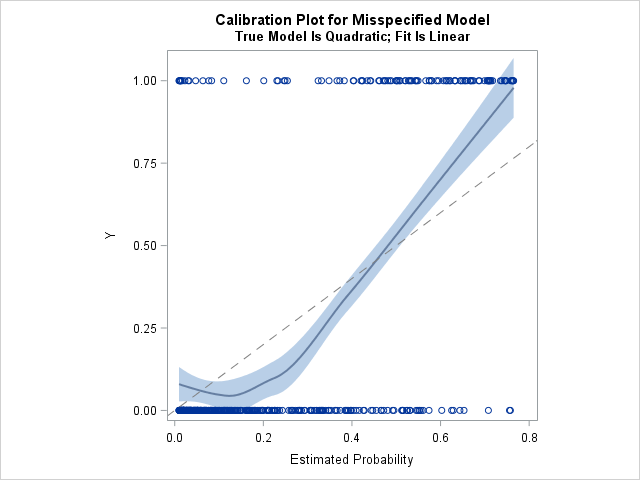}
              \caption[Calibration Plot (Reliability Diagram)]{Calibration Plot (Reliability Diagram) \cite{hastie2009elements}}
              \label{fig:calibrationplot2}
          \end{figure}
    \item \textbf{Calibration Measurement:} This refers to a quantitative metric that summarizes the degree of calibration of a model, often as a single number. Unlike tests, these do not provide a p-value but rather a score or error. Examples include the Brier score, Expected Calibration Error (ECE) \parencite{kull2017beta}, and calibration slope/intercept.

    \item \textbf{Calibration Method:} This refers to a procedure or algorithm used to adjust or improve the predicted probabilities from a model so that they better reflect the true likelihood of outcomes. Examples include Platt scaling \parencite{niculescu2005predicting}, isotonic regression \parencite{niculescu2005predicting, barlow1972isotonic}, and beta calibration \parencite{kull2017beta}. These methods are applied to the model's output to enhance its calibration (goodness of fit). they are ussually seek to make the slope = 1 and intercept = 0 of the relation between observed outcome (y) and logit of predicted probability (logit($\hat p$)) at this time its called \textbf{perfect calibrated} model.

    \item \textbf{Calibration Test:} A calibration test is a formal statistical procedure used to assess whether the predicted probabilities from a model are well-calibrated. These tests typically yield a p-value to help determine if there is a significant deviation from perfect calibration. Examples include the Hosmer-Lemeshow test, the GiViTI calibration test \parencite{nattino2014new}, and Spiegelhalter's z-test \parencite{spiegelhalter1986calibration}.

\end{itemize}

It is important to distinguish \textbf{calibration} from \textbf{discrimination}, as they represent two complementary yet distinct dimensions of model performance \parencite{steyerberg2010assessing}. \textbf{Discrimination} refers to a model's ability to rank or separate individuals who experience the event from those who do not, and is commonly measured by the Area Under the ROC Curve (AUC), also known as the c-statistic \parencite{hosmer2013applied}. \textbf{Calibration}, by contrast, evaluates the accuracy of the predicted probabilities themselves—whether, for example, among observations predicted to have a 70\% probability, approximately 70\% actually experience the event. A model can have good discrimination but poor calibration (correctly ranking individuals while systematically over- or under-estimating probabilities), or good calibration but poor discrimination (accurate probabilities but inability to separate outcome groups) \parencite{van2019calibration}. Since this thesis focuses on \textbf{goodness-of-fit testing}, we are primarily concerned with the calibration aspect of model evaluation; however, both dimensions should ideally be assessed together for a comprehensive evaluation.

\section{Research Problem}

Logistic regression is one of the most commonly used statistical methods for modeling binary outcomes. Before using a fitted model for inference or prediction, it is essential to verify that the model adequately represents the data. This verification process is known as goodness-of-fit (GOF) assessment \parencite{hosmer2013applied}.

The classical approach to GOF testing relies on the Pearson chi-square and deviance statistics. These statistics compare observed outcomes to expected values under the fitted model. Under suitable conditions, both statistics follow a chi-square distribution asymptotically, which allows researchers to compute p-values and make formal decisions about model adequacy \parencite{agresti2013categorical}.

However, these classical tests require a critical assumption: each covariate pattern must contain multiple observations so that expected cell counts are sufficiently large for the chi-square approximation to hold. This assumption is satisfied when all predictors are categorical with few levels, creating a limited number of distinct covariate patterns. In such grouped data settings, the classical tests perform well.

The problem arises when the model includes continuous covariates. In this case, each observation tends to have a unique combination of predictor values. The number of covariate patterns $G$ approaches the sample size $n$, and each pattern contains only one observation ($m_g = 1$). This situation is known as sparse data or ungrouped data. Under these conditions, the expected counts are too small, and the chi-square approximation fails \parencite{farrington1996, kuss2002}.

The consequence is that the Pearson and deviance statistics no longer follow a chi-square distribution when data are sparse. Researchers who apply these tests to sparse data may obtain misleading results---either rejecting good models or accepting bad ones. Since continuous covariates are common in practice (e.g., age, blood pressure, income), this limitation affects a large proportion of applied logistic regression analyses. Over the past four decades, statisticians have developed numerous alternative tests to address this problem. These include:
\begin{itemize}
    \small
    \item Partitioning-based tests, such as the Hosmer-Lemeshow test, which group observations by predicted probability \parencite{hosmer1980goodness};
    \item Normal approximation tests, such as the Osius-Rojek and Farrington tests, which adjust the Pearson statistic for sparse data \parencite{osius1992normal, farrington1996};
    \item Score-based tests, such as the Tsiatis test and le Cessie-van Houwelingen test, which detect specific forms of misspecification \parencite{Tsiatis1980, le1991goodness};
    \item Resampling-based tests, such as bootstrap and simulation approaches \parencite{Hosmer1997}.
\end{itemize}

Despite this variety of available methods, there is no clear consensus on which test performs best under different conditions. \parencite{Hosmer1997, paul2013testing}. This research addresses the following question: \textbf{Which goodness-of-fit tests perform best for binary logistic regression with continuous covariates?} To answer this question, we conduct a systematic comparison of available tests using simulation studies under various conditions of sample size and model misspecification. The goal is to evaluate the emperical power and Type I error rate of each test, and to provide practical recommendations for applied researchers.

\chapter{GOODNESS-OF-FIT (GOF)}

\section*{Introduction}
Many approaches have been developed for assessing GOF in logistic regression. A model can be inadequate in several ways: the linear predictor may be incorrectly specified, a covariate may appear in the wrong functional form, important covariates may be omitted, or the link function may be misspecified. Each of these can lead to biased coefficient estimates \parencite{Hauck1991, xie2008increasing}. 

This chapter reviews the main approaches to goodness-of-fit (GOF) testing in binary logistic regression, with emphasis on the challenges that arise under sparse data. The discussion is organized by mathematical principle: grouping-based, score-based, information-theoretic, and smoothing-based methods. The chapter also addresses specification error and residual diagnostics, providing the theoretical foundation for the test-specific details and simulation studies in subsequent chapters.

\newpage

\section{Specification Error}

\cite{GAM2021} stated that, when a model exhibits a specification error, the inference conducted on it can be not correct. This inadequacy may arise from various issues, such as incorrect specification of the linear predictors, omission of important covariates, or misspecification of the link function, all of which can lead to inconsistent coefficient estimation and biased treatment effect estimates. In this subsection we will dig into the sources of specification error and the implications it has on the usage of the model.

\subsubsection{Sources of Specification Error } \label{sec:sources_of_specification_error}

Several primary sources of specification error \parencite{gof2002,agresti2013categorical} in logistic regression are:

\begin{itemize}
    \item \textbf{Omitted Variable Bias:} This occurs when relevant predictors are excluded from the model, leading to biased estimates of the included variables and poor model fit.
    \item \textbf{Omitted Interaction Term:} This occurs when an interaction term is omitted from the model, leading to biased estimates of the included variables and poor model fit.
    \item \textbf{Incorrect Functional Form:} This arises when the relationship between predictors and the outcome is misspecified. Common issues include assuming linearity when the true relationship is non-linear, omitting necessary polynomial terms or time factors in time-series data, and neglecting interaction effects or multiplicative components.

    \item \textbf{Inclusion of Irrelevant Variables:} This occurs when extraneous variables are included, potentially leading to overfitting and reduced model efficiency.
    \item \textbf{Incorrect Link Function:} While the logit link is standard for binary logistic regression, other link functions (e.g., probit, complementary log-log) may be more appropriate in certain scenarios. An incorrect link function can lead to poor model fit and biased predictions.

\end{itemize}

\subsubsection{Consequences of Uncaptured Specification Error} \label{sec:consequences_of_uncaptured_specification_error}

\textit{Menard} mentioned that, \textit{"misspecification may result in biased logistic regression coefficients, coefficients that are systematically overestimated or underestimated"} \parencite{menard2002}. Consequently, the model's predictive performance can deteriorate, leading to incorrect conclusions and decisions based on the incorrect predictions and classification \parencite{myers2010, GAM2021} and certainly poor inferences.  Thus, ensuring a proper goodness of fit test or tool weather Formal test or Informal graph (initialized by Landwehr in JASA \parencite{GraphicalMethods1984}) is essential for the model usability. Moreover, selecting the appropriate tool for the right data scenario becomes even more critical, especially, when dealing with sparse data.

\newpage
\section{Methodological Approaches to GOF Testing}

Several classes of GOF tests have been developed for logistic regression. The following subsections describe each class organized by mathematical principle \parencite{Hosmer1997, xie2008increasing}.

\subsection{Grouping-Based Methods} \label{sec: Partitioning}

Grouping-based methods address sparsity by partitioning observations into groups, restoring the conditions under which chi-square statistics are valid \parencite{hosmer2013applied, myers2010}. Two main partitioning strategies exist:

\subsubsection{Deciles of Predicted Risk (DOR)}

DOR methods order observations by $\hat{\pi}_i$ and divide them into $G$ groups (typically $G = 10$). Within each group, observed and expected counts are compared. The Hosmer--Lemeshow test is the standard implementation \parencite{hosmer1980goodness}. Binning can be \textbf{equi-width} (equal intervals) or \textbf{equi-depth} (equal group sizes) \parencite{canary2015comparison}.

\cite{pigeon1999b} follows a similar DOR approach but introduces a within-group correction factor:
\begin{equation}
    \phi_g = \frac{\sum_{\{i: \hat{\pi}_i \in \text{group}_g\}} \hat{\pi}_i(1 - \hat{\pi}_i)}{n_g \bar{\hat{\pi}}_g(1 - \bar{\hat{\pi}}_g)}
\end{equation}
which accounts for heterogeneity of predicted probabilities within groups.

\subsubsection{Partition Covariate Space (PCS)} \label{sec:clustering_in_goodness_of_fit_testing}

PCS methods partition the covariate space directly rather than the predicted probabilities. Common approaches include \parencite{canary2015comparison}:
\begin{itemize}
    \item \textbf{User-specified:} Manual partitioning based on domain knowledge. \cite{Tsiatis1980} requires defining $G$ mutually exclusive regions of $\mathbb{R}^p$ and uses a score test to evaluate fit within each region.
    \item \textbf{Cluster-based:} Algorithmic grouping using K-means or hierarchical clustering. \cite{xie2008increasing} groups observations by covariate similarity, with $G = 10$ if $p < 5$ and $G = p + 5$ if $p \geq 5$.
    \item \textbf{Combo:} Combining categorical covariate patterns with median splits on $\hat{\pi}_i$.
\end{itemize}

Grouping restores the validity of chi-square approximations when each group has sufficient observations, enables localized assessment of fit across risk strata, and increases sensitivity to region-specific misspecification \parencite{hosmer2013applied, agresti2013categorical}.

\subsection{Score-Based Methods} \label{sec:score_test_in_logistic_regression}

Score tests evaluate model adequacy by examining the score function at the null hypothesis, without fitting the alternative model \parencite{Tsiatis1980, le1991goodness}. The score function for the $j$-th parameter is:
\begin{equation}
    U_j(\boldsymbol{\beta}) = \sum_{i=1}^n x_{ij}(y_i - \pi_i)
\end{equation}

The score test statistic for testing $H_0: \boldsymbol{\beta} = \boldsymbol{\beta}_0$ is:
\begin{equation}
    T = \mathbf{U}(\boldsymbol{\beta}_0)' \mathcal{I}(\boldsymbol{\beta}_0)^{-1} \mathbf{U}(\boldsymbol{\beta}_0) \sim \chi^2_q
\end{equation}
where $\mathcal{I}(\boldsymbol{\beta}_0)$ is the Fisher information matrix and $q$ is the number of restrictions tested.

\subsubsection{Null and Alternative Hypotheses for Score-Based Tests} \label{sec:null_and_alternative_hypotheses_for_score_based_tests}

Score-based GOF tests can target specific forms of misspecification:
\begin{itemize}
    \item \textbf{Omitted quadratic term:}
          \begin{align*}
              H_0: & \; \text{logit}(P) = \beta_0 + \beta_1 x \\
              H_A: & \; \text{logit}(P) = \beta_0 + \beta_1 x + \beta_2 x^2
          \end{align*}

    \item \textbf{Omitted interaction:}
          \begin{align*}
              H_0: & \; \text{logit}(P) = \beta_0 + \beta_1 x_1 + \beta_2 x_2 \\
              H_A: & \; \text{logit}(P) = \beta_0 + \beta_1 x_1 + \beta_2 x_2 + \beta_3 x_1 x_2
          \end{align*}

    \item \textbf{Misspecified link function (Stukel):}
          $H_0: g = \text{logit}$ \; vs.\ $H_A: g \neq \text{logit}$ \parencite{stukel1988generalized}

    \item \textbf{Overdispersion (le Cessie--van Houwelingen):}
          $H_0: \sigma^2 = 0$ \; vs.\ $H_A: \sigma^2 > 0$ \parencite{le1995goodness}

    \item \textbf{Mixed covariate patterns (Pulkstenis--Robinson):}
          Tests for systematic miscalibration across covariate subgroups \parencite{pulkstenis2002two}
\end{itemize}

\subsubsection{Comparison with LRT and Wald Tests}

The likelihood ratio test (LRT) compares nested models via $-2\Delta \log L$, and the Wald test uses the ratio $(\hat{\beta}_j / \text{SE}(\hat{\beta}_j))^2$ from the full model. All three tests are asymptotically equivalent under $H_0$ but may differ in finite samples. The score test is preferable when the alternative model is difficult to specify or fit, since it requires fitting only the restricted model \parencite{hosmer2000applied}.

\subsection{Information Matrix Test} \label{sec:information_theory}

White's information matrix (IM) test detects model misspecification by comparing two estimators of the Fisher information \parencite{White1982}. Under correct specification, the expected information matrix and the outer product of gradients are equal:
\begin{equation}
    H_0: \quad E\!\left[\frac{\partial \ell}{\partial \boldsymbol{\beta}} \frac{\partial \ell}{\partial \boldsymbol{\beta}^T}\right] = -E\!\left[\frac{\partial^2 \ell}{\partial \boldsymbol{\beta} \partial \boldsymbol{\beta}^T}\right]
\end{equation}

Orme \parencite{Orme1988} simplified the computation for binary data models. The IM test can detect omitted variables, nonlinearity, incorrect link functions, and heterogeneity without specifying a particular alternative.

\subsection{Smoothing-Based Methods} \label{sec:smoothing_methods}

Smoothing-based GOF tests replace each residual with a locally weighted average of nearby residuals \parencite{le1991goodness}:
\begin{equation}
    \text{Smoothed Residual}_i = \sum_{j=1}^n w_{ij}(Y_j - \hat{\pi}_j)
\end{equation}
where $w_{ij}$ depends on the distance $\|\mathbf{x}_i - \mathbf{x}_j\|$ relative to a bandwidth $h$. Smoothing-based GOF tests replace each residual with a locally weighted average of nearby residuals. The hypotheses are:

    \begin{align*}
        H_0: & \quad E[\text{Smoothed Residual}_i] = 0 \\
        H_A: & \quad E[\text{Smoothed Residual}_i] \neq 0
    \end{align*}

This approach avoids the arbitrary groupings of partition-based methods and can detect localized lack of fit. Key contributions include the kernel-based approach of Copas \parencite{copas1989unweighted} and the le Cessie--van Houwelingen test based on an unbiased estimator of the smoothed residual quadratic form \parencite{le1991goodness, le1995goodness}.

\subsubsection{Generalized Additive Models (GAM)}

The residual smoothing approach described above uses a fixed kernel to average nearby residuals. GAMs generalize this idea by replacing the linear predictor of the logistic model with a sum of smooth, data-driven functions \parencite{hastie2009elements, wood2017generalized}:
\begin{equation}
    g(E[Y|\mathbf{X}]) = \beta_0 + f_1(X_1) + f_2(X_2) + \cdots + f_p(X_p)
\end{equation}
where $g(\cdot)$ is the link function and each $f_j(\cdot)$ is a smooth function (typically spline-based) estimated from the data. In the context of GOF testing, a GAM serves as a flexible reference model: by fitting an over-specified GAM that includes nonlinear terms and interactions, one obtains estimated probabilities that can reveal departures from the simpler logistic model. These GAM-based probabilities are then used to construct groupings or as inputs for GOF statistics (such as modified Hosmer--Lemeshow or Xie tests), resulting in substantially increased power for detecting omitted interactions or nonlinear terms \parencite{GAM2021}.

\subsubsection{LOESS (Locally Estimated Scatterplot Smoothing)}

LOESS takes a different approach to smoothing: rather than fitting a single global function, it fits low-degree polynomials to localized subsets of the data \parencite{cleveland1979, cleveland1988}. For each point $x$, the LOESS estimate is:
\[
    \hat{m}(x) = \sum_{i=1}^n w_i(x)\, y_i
\]
where the weights $w_i(x)$ are determined by a tricube kernel, giving more influence to observations near $x$ and negligible weight to distant observations. Eubank and Spiegelman \parencite{eubank1990} derived GOF test statistics from nonparametric regression fits (including LOESS) applied to model residuals. If the logistic model is correctly specified, the smoothed residuals should show no systematic pattern across the covariate space; systematic deviations indicate lack of fit.

Both GAM and LOESS connect to the smoothing-based GOF framework by providing flexible, nonparametric alternatives against which the fitted logistic model can be compared. Their ability to capture nonlinear relationships without pre-specifying functional forms makes them useful tools for constructing GOF statistics with greater sensitivity to misspecification \parencite{GAM2021, le1991goodness}.

\subsection{Calibration Methods}

Calibration methods assess whether predicted probabilities $\hat{\pi}_i$ correspond to observed outcome rates. The underlying principle is to regress observed outcomes on predicted probabilities; departures of the slope from 1 or intercept from 0 indicate miscalibration \parencite{nattino2015new, harrell2015regression}. Calibration methods are discussed in detail in Chapter~4.

\subsection{Bootstrap Methods} \label{sec:bootstrap_in_gof}

The bootstrap, introduced by Efron \parencite{efron1979bootstrap}, is a resampling technique that estimates sampling distributions by repeatedly drawing samples with replacement from the observed data. In GOF testing, bootstrap methods are valuable when the theoretical null distribution of a test statistic is complex, unknown, or relies on asymptotic approximations that may not hold \parencite{efron1993bootstrap}.

\subsubsection{Basic Algorithm}

The bootstrap procedure consists of the following steps:
\begin{enumerate}
    \item Start with an observed sample $X = (x_1, \ldots, x_n)$ and compute the observed test statistic $T_{\text{obs}}$.
    \item Generate $B$ bootstrap samples $X_b^* = (x_{b,1}^*, \ldots, x_{b,n}^*)$, each of size $n$, by sampling with replacement from $X$.
    \item For each bootstrap sample, compute the test statistic $T_b^*$.
    \item Estimate the empirical p-value:
\end{enumerate}
\begin{equation}
    p\text{-value} = \frac{1}{B} \sum_{b=1}^{B} I(T_b^* \geq T_{\text{obs}})
\end{equation}
where $I(\cdot)$ is the indicator function \parencite{efron1993bootstrap}. This avoids the need for a closed-form null distribution.

\subsubsection{Model-Based Bootstrap for GOF}

In GOF testing, a model-based bootstrap is used to generate data under the null hypothesis. The procedure fits the logistic model to the original data, then generates new binary responses from $\text{Bernoulli}(\hat{\pi}_i)$ for each observation while keeping the covariates fixed. The model is refit to each bootstrap dataset and the GOF statistic is recomputed, building an empirical null distribution against which the observed statistic is compared.

This approach is employed in several GOF tests studied in this thesis:
\begin{itemize}
    \item The \textbf{Stute--Zhu test} \parencite{stute2002model}, whose asymptotic null distribution is complex and data-dependent, making theoretical p-value calculation impractical.
    \item The \textbf{projection-based test} \parencite{liu2024comprehensive}, which integrates over all projection directions, yielding an intractable theoretical distribution.
    \item The \textbf{Lai \& Liu standardized Hosmer--Lemeshow test}, which uses bootstrap for power estimation rather than p-value calculation, addressing the problem of statistical overpowering in large samples.
\end{itemize}

The main limitation of bootstrap methods is computational cost: stable p-value estimates typically require $B \geq 1000$ replications, each involving model fitting \parencite{efron1993bootstrap}. Nevertheless, bootstrap approaches enable the practical application of GOF tests that would otherwise be analytically intractable.

\subsection{Summary Measures}

In addition to formal test statistics, summary measures such as the Brier score, log-loss, and pseudo-$R^2$ statistics (Cox \& Snell, Nagelkerke, McFadden) provide complementary assessments of model performance \parencite{agresti2013categorical}. This study focuses primarily on test statistics rather than summary measures.

\section{Residuals in Logistic Regression} \label{sec:logistic_residuals}

Residual analysis complements formal GOF tests by providing observation-level diagnostics. The main residual types for binary logistic regression are \parencite{agresti2013categorical, hosmer2000applied}:
\begin{itemize}
    \item \textbf{Raw residuals:} $r_i = y_i - \hat{\pi}_i$
    \item \textbf{Pearson residuals:} $r_{i,P} = (y_i - \hat{\pi}_i) / \sqrt{\hat{\pi}_i(1-\hat{\pi}_i)}$
    \item \textbf{Deviance residuals:} $r_{i,D} = \text{sign}(y_i - \hat{\pi}_i)\sqrt{d_i}$, where $d_i$ is the $i$-th deviance contribution
    \item \textbf{Standardized residuals:} Adjusted by $\sqrt{1 - h_{ii}}$ to account for leverage \parencite{hosmer2013applied}
\end{itemize}

\noindent The distributions of these residuals depend on $\hat{\pi}_i$, which complicates their direct use as GOF diagnostics in sparse data settings. Graphical methods based on residuals, introduced by Landwehr et al.\ \parencite{GraphicalMethods1984}, remain useful for exploratory assessment.

\section*{Conclusion}

This chapter reviewed the statistical foundations of GOF testing in binary logistic regression. The primary challenge is that classical $X^2$ and $D$ statistics require grouped data with sufficient cell counts; with continuous covariates, these conditions are not met. Several methodological approaches address this limitation: grouping-based methods restore the chi-square approximation through partitioning, score-based methods test specific forms of misspecification, information matrix tests detect general model inadequacy, and smoothing-based methods provide continuous assessment across the covariate space. The tests introduced in this chapter are examined individually in the next chapter.

\chapter{Selected Goodness-of-fit Tests}

\section*{Introduction}

\paragraph{Overview and Classification of Goodness-of-Fit Tests for Logistic Regression}

\textbf{Goodness-of-fit testing} is essential for evaluating whether a logistic regression model adequately fits observed data. Numerous tests exist, each with unique theoretical bases and practical considerations. This chapter summarizes the main types of goodness-of-fit tests and their key features.

\vspace{0.5em}
\noindent We focus on four main categories:

\begin{enumerate}
    \item \textbf{Classical Chi-Square Type Tests:} These compare observed and expected frequencies, often using the chi-square ($\chi^2$) distribution. Examples include the \textit{Pearson Chi-Square} and \textit{Hosmer-Lemeshow} tests. Variants include:
          \begin{itemize}
              \item \textbf{Partitioning-Based:} Group data by predicted probabilities or covariate patterns (e.g., Hosmer-Lemeshow, Xie, Pulkstenis-Robinson).
              \item \textbf{Standardization-Based:} Standardize residuals or statistics for comparison to the normal distribution (e.g., Farrington, McCullagh, Windmeijer, Osius and Rojek).
          \end{itemize}

    \item \textbf{Score and Likelihood-Based Tests:} Derived from the likelihood or its derivatives, such as the \textit{Deviance test}, \textit{Tsiatis test}, and \textit{Stukel test}. These are useful for continuous covariates and detecting specific model misspecification.

    \item \textbf{Information Matrix Tests:} Compare observed and expected information matrices to detect model misspecification (e.g., \textit{White Information Matrix test}, IM test).

    \item \textbf{Smoothing-Based Methods:} Use nonparametric techniques to assess fit by examining residual patterns (e.g., \textit{Le Cessie and van Houwelingen kernel-based test}, GAM-based methods).
\end{enumerate}

\vspace{0.5em}
\noindent The chapter details the theory, algorithms, and practical use of each test type, helping readers select appropriate methods for their data and research goals.

\newpage
\section{Pearson Chi-Square Test}

The Pearson Chi-Square test \parencite{pearson1900criterion} is a classic goodness-of-fit test that can be applied to logistic regression models. For logistic regression, it compares the observed frequencies to the expected frequencies under the fitted model \parencite{agresti2013categorical}. for each observation $i$ ($i=1,\ldots,n$), the Pearson Chi-Square statistic is defined as:

\begin{align}
    X^2_{pearson} = \sum_{i=1}^n \sum_{k=0}^1 \frac{(o_{ki} - e_{ki})^2}{e_{ki}}
    = \sum_{i=1}^n \frac{(y_i - \hat{\pi}_i)^2}{\hat{\pi}_i(1-\hat{\pi}_i)}
\end{align}
Here, for each observation $i$ ($i=1,\ldots,n$):
\begin{itemize}
    \item $o_{1i}$ is 1 if $Y_i=1$ (event), 0 otherwise;
    \item $e_{1i} = \hat{\pi}_i$ is the model-predicted probability of event for observation $i$;
    \item $o_{0i}$ is 1 if $Y_i=0$ (non-event), 0 otherwise;
    \item $e_{0i} = 1 - \hat{\pi}_i$ is the model-predicted probability of non-event for observation $i$.
    \item $y_i$ is the observed outcome for the $i$-th observation
    \item $\hat{\pi}_i$ is the predicted probability for the $i$-th observation
    \item $n$ is the number of observations
\end{itemize}

\paragraph{Asymptotic Distribution} Under the null hypothesis of good fit and certain regularity conditions, $X^2$ follows a chi-square distribution asymptotically at fixed sample size \parencite{agresti2013categorical}, with $(n-p^*)$ degrees of freedom, where $p^*$ is the number of parameters in the model. In Sparse Data Scenarios, the $X^2$ chi-square distribution is not appropriate.

\paragraph{Limitations of the Pearson Chi-Square and Deviance tests (Section \ref{sec:deviance_test})}
Several limitations of the Pearson Chi-Square and Deviance tests have been noted by various authors. ~\cite{pulkstenis2002two} point out that for continuous covariates or when the number of covariate patterns is close to the sample size, the asymptotic chi-square distribution may not be appropriate. \cite{agresti2013categorical} and SAS documentation~\parencite{sas2020} highlight that when there are as many covariate patterns as observations ($J = n$), the statistics become uninformative, depending only on the fitted probabilities and sample size, and the resulting $p$-values can be \textbf{erroneous}. \cite{hosmer2000applied} note that Pearson and Deviance residuals can be exceptionally large when fitted probabilities are close to 0 or 1, potentially leading to misleading assessments of model fit. \cite{farrington1996} further observes that the $X^2$ statistic may provide no evidence of serious misfit in the presence of sparse data.

\section{Deviance Test (Likelihood Ratio Test)} \label{sec:deviance_test}

The deviance test \parencite{nelder1972} is another commonly used goodness-of-fit test for logistic regression models. It is based on the likelihood ratio principle and compares the fitted model to a saturated model \parencite{mccullagh1989generalized}. The deviance statistic is defined as:

\begin{equation}
    D = -2 \sum_{i=1}^n \left[y_i \log\left(\frac{\hat{\pi}_i}{y_i}\right) + (1-y_i) \log\left(\frac{1-\hat{\pi}_i}{1-y_i}\right)\right]
\end{equation}

Here, $y_i$ denotes the observed response for the $i$-th subject, $\hat{\pi}_i$ is the model-predicted probability for that subject, and $n$ is the total sample size.\textbf{The deviance statistic $D$ is obtained from the likelihood ratio test} (see section \ref{sec:score_test_in_logistic_regression}); a detailed derivation and discussion of the hypotheses can be found in Appendix \ref{sec:deviance_test_appendix}. \\

Like the Pearson Chi-Square test, under the null hypothesis and certain conditions, the deviance statistic follows a chi-square distribution with $(n-p^*)$ degrees of freedom. However, the deviance test also faces challenges when dealing with continuous covariates or sparse data \parencite{pulkstenis2002two}. One of the limitations is that assypmtotically, the deviance statistic will approach the pearson chi-square statistic \parencite{Cox1989,hosmer2000applied}. The computational procedure for calculating both statistics is detailed in Algorithm \ref{alg:pearson_and_deviance_statistics} (see Appendix).

\section{Hosmer-Lemeshow Test}
\label{sec:hosmer_lemeshow_test}
The Hosmer-Lemeshow test, introduced by Hosmer and Lemeshow, is a widely adopted goodness-of-fit test for logistic regression models. Its utility is particularly evident when dealing with continuous covariates or when the number of covariate patterns approaches the number of observations. \parencite{hosmer1980goodness, lemeshow1982review}

This test employs two main statistics: $\hat{C}$ and $\hat{H}$. The $\hat{C}$ statistic, based on groups of equal size binning (typically deciles of risk), is commonly implemented in statistical software. The $\hat{H}$ statistic, which uses fixed cutpoints, is less frequently available in software packages (recall section \ref{sec: Partitioning}). Hosmer and Lemeshow demonstrated through simulations that when the logistic regression model is correctly specified , and all cells have sufficiently large estimated expected values, both $\hat{C}$ and $\hat{H}$ for $G$ groups closely follow a chi-square distribution with $G -2$ degrees of freedom, $\chi^2(G-2)$.

The test's null hypothesis assumes that the model fits the data well, while the alternative hypothesis suggests a poor fit. A small p-value (typically $< 0.05$) indicates evidence against the null hypothesis, suggesting a poor fit.
The Hosmer-Lemeshow test statistic is calculated as follows:

\begin{equation}
    \hat{C} = \hat{H} = \sum_{g=1}^{G} \frac{(o_g - e_g)^2}{n_g \overline{\hat{\pi}}_g (1 - \overline{\hat{\pi}}_g)}
\end{equation}

where:
\begin{itemize}
    \item $G$ is the number of groups (commonly 10).
    \item $O_g$ is the observed number of events in group $g$: $O_g = \sum_{i \in \text{group } g} y_i$
    \item $e_g$ is the expected number of events in group $g$: $e_g = \sum_{i \in \text{group } g} \hat{\pi}_i = n_g \overline{\hat{\pi}}_g$
    \item $n_g$ is the number of observations in group $g$.
    \item $\overline{\hat{\pi}}_g$ is the mean predicted probability in group $g$: $\overline{\hat{\pi}}_g = \frac{1}{n_g} \sum_{i \in \text{group } g} \hat{\pi}_i$
\end{itemize}

The test statistic $\hat{C}$ (or $\hat{H}$, depending on the grouping method) approximately follows a chi-square distribution with $G-2$ degrees of freedom under the null hypothesis of good model fit. The $\hat{C}$ statistic uses equal-sized groups $n_g = n/G$ , typically deciles of risk, while the $\hat{H}$ statistic uses fixed cutpoints to create the groups (i.e. $[0,0.1) ,[0.1,0.2) \dots , [0.9,1]$ for g = 10) recall section \ref{sec: Partitioning}.

\paragraph{Algorithm for Hosmer-Lemeshow Test}
The computational procedure for both the $\hat{C}$ and $\hat{H}$ versions of the test is presented in Algorithm \ref{alg:hosmer_lemeshow_test} (see Appendix).

\vspace{0.5em}

\paragraph{Considerations:}
\begin{itemize}
    \item \cite{MHL2021} comment on Hosmer-Lemeshow test that it often rejects the null hypothesis of good fit for large samples. Thats why he introduced the modified Hosmer-Lemeshow test for large samples (Section \ref{sec:mHL_large}) not only but also \cite{Lai2018} porposed a simple test procedure in standardizing the power of Hosmer--Lemeshow test in large data sets.
    \item The test results can be sensitive to the choice of the number of groups and its the grouping techniques ($\hat{C}$ or $\hat{H}$), However, \cite{hosmer2013applied} proved that $\hat C$ (decile of risk grouping) is better and more stable.
    \item $\hat{H}$ has a non stable number of groups, if some groups (intervals) have no observations, the actual number of groups will be less than spacified $g$.
    \item May lack power to detect certain types of lack of fit \parencite{Hosmer1997}
    \item One frequently cited disadvantage of the decile of risk grouping is that subjects within each decile may have quite different values for the covariates.\parencite{Hosmer1997, Geo2015, GraphicalMethods1984} (recall section \ref{sec: Partitioning})
\end{itemize}

\section{Pigeon-Heyse ($J^2$) Test} \label{sec:pigeon_heyse_test}

The Pigeon-Heyse test, introduced by Pigeon and Heyse, extends the classic Pearson chi-squared test for goodness-of-fit but modifies it to account for continuous covariates, addressing situations where the standard logistic model fit tests are limited. Like the Hosmer-Lemeshow test, it provides a method to assess the alignment between observed outcomes and the model's predicted probabilities, but it includes an additional correction factor to handle variability within groups of observations. \parencite{pigeon1999b}

The J² statistic is based on grouping observations, similar to the Hosmer-Lemeshow test, but the key difference is the inclusion of a correction factor that adjusts for the variability in predicted probabilities within each group. The procedure can be applied using the deciles of risk (DOR) grouping or partitioning the covariate space (PCS), the latter allowing for more flexible grouping based on the distribution of covariates.

The Pigeon-Heyse test statistic ($J^2$) is defined as:
\begin{equation}
    J^2 = \sum_{g=1}^{G} \frac{(o_g - e_g)^2}{n_g \overline{\hat{\pi}}_g (1 - \overline{\hat{\pi}}_g)} \cdot \phi_g^{-1}
\end{equation}
where:
\begin{itemize}
    \item $o_g$ is the observed number of events in group $g$
    \item $e_g = n_g \overline{\hat{\pi}}_g$ is the expected number of events in group $g$
    \item $n_g$ is the number of observations in group $g$
    \item $\overline{\hat{\pi}}_g$ is the mean predicted probability in group $g$
    \item $\phi_g$ is the correction factor for group $g$, given by:
          \begin{equation}
              \phi_g = \frac{1}{n_g \overline{\hat{\pi}}_g (1 - \overline{\hat{\pi}}_g)} \sum_{i \in \text{group }g} \hat{\pi}_i (1 - \hat{\pi}_i)
          \end{equation}
\end{itemize}
Under the null hypothesis, $J^2$ is approximately distributed as a chi-squared distribution with $G-1$ degrees of freedom. Under certain conditions (when predicted probabilities are equal within each group), $J^2$ becomes equivalent to Hosmers's $HL_{stat}$ test ($ \forall g \: : \phi_g = 1 $). \\

\noindent The null hypothesis for the Pigeon-Heyse test assumes that the model fits the data adequately, with deviations between observed and expected outcomes being random. A small p-value ($< 0.05$) suggests that the model does not provide a good fit to the data.

\paragraph{Algorithm}
The implementation details are provided in Algorithm \ref{alg:pigeon_heyse_test} (see Appendix).

Where as we can notice, $O_g$, $E_g$ and $\overline{\hat{\pi}}_g$ are calculated as Hosmers algorithm (Sec. \ref{sec:hosmer_lemeshow_test}). The test proceeds by forming groups based on predicted probabilities, similarly to the Hosmer-Lemeshow approach. However, it \textbf{only} \textbf{introduces the correction factor $\phi_g$ to account for variability in predicted probabilities within each group}. This correction allows the Pigeon-Heyse test to remain robust even when predicted probabilities are heterogeneous within a group.

\paragraph{Advantages and Considerations:} \parencite{pigeon1999b}

\begin{itemize}
    \item[--] Strong Theoretical Foundation: Based on solid statistical theory relating to quadratic forms.
    \item[--] The inclusion of the correction factor makes the Pigeon-Heyse test more robust to group variability than the Hosmer-Lemeshow test \parencite{pigeon1999b}.
    \item[--] It is less sensitive to how groups are defined, but still depends on the choice of grouping strategy (DOR vs. PCS).
    \item[--] The Pigeon-Heyse test may be conservative, leading to fewer model rejections compared to other tests, potentially missing models with minor lack of fit.
    \item[--] The paper notes that the distribution of $J^2$ may differ from the reported $\chi^2_{G-1}$ in some cases, in Canary's simulation  found that $j^2$ often followed a distribution closer to $\chi^2_{G-2}$ in their simulations.  \parencite{canary2015comparison}

\end{itemize}

\section{Pulkstenis and Robinson (PR) Test} \label{sec:pr_test}

Pulkstenis and Robinson proposed a modification to traditional goodness-of-fit tests for logistic regression models, addressing limitations of the Hosmer-Lemeshow test when both categorical and continuous covariates are present (for mixed models only). \parencite{pulkstenis2002two}

\subsubsection{Key Differences from Hosmer-Lemeshow Test}
\begin{itemize}
    \item \textbf{Grouping Strategy}: Creates groups based on categorical covariate patterns first, then subdivides based on fitted probabilities.
    \item \textbf{Number of Groups}: Creates $2C$ groups, where $C$ is the number of \textbf{unique categorical covariate patterns}.
    \item \textbf{Preservation of Covariate Structure}: Maintains the original covariate pattern structure, potentially improving detection of interaction effects.
\end{itemize}

\subsubsection{Test Statistics}
PR test proposes two statistics, analogous to Pearson chi-square and deviance:

\begin{equation}
    \chi^{*2} = \sum_{g=1}^{2C} \frac{(o_g - e_g)^2}{n_g\overline{\hat{\pi}}_g(1-\overline{\hat{\pi}}_g)}
\end{equation}

\begin{equation}
    D^{*2} = - 2 \sum_{g=1}^{2C} \left[o_g \log\left(\frac{o_g}{e_g}\right) + (n_g - o_g) \log\left(\frac{n_g - o_g}{n_g - e_g}\right)\right]
\end{equation}

Where: $g$ indexes the $2C$ groups formed by \textbf{unique categorical covariate patterns} and \textbf{probability splits}, the rest of symbols are defined as in Hosmer-Lemeshow test (Sec. \ref{sec:hosmer_lemeshow_test}).

\paragraph{Degrees of Freedom} PR test uses $df = (2C - 1)(K' - 1) - k - 1$, where $C$ is the unique number of categorical covariate patterns, $K'$ is the total number of categories across all categorical variables, and $k$ is the number of categorical variables.

\subsubsection{Limitations}

\begin{itemize}
    \item The test relies on creating covariate patterns based on categorical variables as its first step. Without categorical variables, there is only one covariate pattern for the entire dataset. In this case, the degrees of freedom become $df = (2C - 1)(K' - 1) - k - 1$. If there are no categorical variables, then $C = K' = 1$, $k =  0$, so $df = 0$, which is problematic for the chi-square approximation. However, if there is only one binary variable (e.g., $C = 2$, $K' = 2$, $k = 1$), then $df = (2 \times 2 - 1)(2 - 1) - 1 - 1 = 1$, and the test is valid.
    \item Another criticism by Hosmer holds unless all the continuous covariates have either positive or negative coefﬁcients. When the signs of the coefﬁcients are different one can have similar estimated probabilities, but widely different values for the continuous covariates. \parencite{hosmer2013applied}
\end{itemize}

\subsubsection{Algorithm}

The computational steps for the PR test are detailed in Algorithm \ref{alg:pr_test} (see Appendix).

\section{Xie Test} \label{sec:xie_test}

Building on the foundation of the Hosmer-Lemeshow test, Xie et al. \parencite{xie2008increasing} proposed a modified approach that aims to improve the power of goodness-of-fit tests for logistic regression models, particularly when continuous covariates are involved. Unlike the Hosmer-Lemeshow test, which relies on partitioning based on the predicted probabilities, Xie's method employs a direct partitioning of the covariate space itself.

The key innovation in Xie's approach is the use of cluster analysis to divide the multidimensional covariate space into distinct regions (PCS). This partitioning strategy groups observations based on the similarity of their covariate values rather than their predicted risk scores, enabling a more robust assessment of model fit across different regions of the covariate space. The test statistic is then calculated using a formula similar to that in the Hosmer-Lemeshow test, which measures the deviation of observed outcomes from expected outcomes within each partitioned group.

\subsubsection*{Algorithm} Xie's test is presented in Algorithm \ref{alg:xie} (see Appendix).

{
Because the distribution of the test statistic is stochastically bounded by known chi-square distributions, $\chi_{g-1}<= ChiSquare Test stat <= \chi_{g-p-1} $ as noted by Molinari \parencite{molinari1977distribution}, it is conservative to consider the test statistic having an asymptotic chi-square distribution with $G-1$ degrees of freedom. Although this approximation might lead to low power in detecting certain alternative hypotheses, Hosmer and Lemeshow suggest using $G-2$ as the degrees of freedom for the Pearson-type statistic in their simulations.}
\paragraph{Xie et al. propose} choosing $G=10$ if $p < 5$ and $G=p+5$ if $p \geq 5$. Under the null hypothesis, they suggest comparing the test statistic to a chi-square distribution with degrees of freedom $df = G - (p/2) - 1$.

\paragraph{Considerations:}
\begin{itemize}
    \item[--] The test has claimed a higher power in detecting nonlinear effects or interactions compared to the Hosmer-Lemeshow test. Because The clustering-based partitioning may offer a more detailed diagnostic of model fit, highlighting specific regions in the covariate space where improvements are needed. \parencite{xie2008increasing}
    \item[--] Xie's method depends on the choice of clustering method and number of groups, which affects test performance.Unlike the Hosmer-Lemeshow test, which
          groups based on predicted probabilities.

\end{itemize}

\section{Nattino $\text{mHL}_{\text{large }} $} \label{sec:mHL_large}

Nattino, who is also the primary author of the GiViTI calibration test and the Belt test (see section \ref{sec:giviti_calibration_test_and_belt_framework}), introduced a modification to the Hosmer-Lemeshow (HL) test to overcome its shortcomings in large sample  \parencite{MHL2021}. In addition, Lai  proposed a straightforward procedure to standardize the power of the Hosmer-Lemeshow test for large datasets \parencite{Lai2018} (see Algorithm \ref{alg:lailiu_bootstrap}). \\

The conventional HL test tends to \textbf{reject the null hypothesis of adequate fit as sample size increases}, even when the model fits well in practice, due to its rising statistical power. The modified version, denoted as $\text{mHL}_{\text{large}}$, incorporates a new parameter $\epsilon$ that quantifies goodness of fit in a way that is not influenced by sample size. \parencite{MHL2021} \\

The key idea behind $\text{mHL}_{\text{large}}$ is to standardize the noncentrality parameter $\lambda$ of the noncentral chi-squared distribution that characterizes the HL statistic under the alternative hypothesis. This standardization results in the parameter $\epsilon$:

\begin{equation}
    \epsilon = \frac{\sqrt{\lambda}}{\sqrt{n}}
\end{equation}

where $n$ is the sample size. The parameter $\epsilon$ is independent of the sample size and can be used to measure the goodness of fit of a model. The $\text{mHL}_{\text{large}}$ hypotheses:

\begin{equation}
    H_0: \epsilon \leq \epsilon_0 \quad \text{vs} \quad H_1: \epsilon > \epsilon_0
\end{equation}

where $\epsilon_0$ is a threshold that defines acceptable fit. By default, $\epsilon_0$ is set to the value of $\epsilon$ expected from a model attaining a p-value of 0.05 in the traditional HL test with a sample size of one million. \\

\paragraph{Algorithm}
The procedure for the $\text{mHL}_{\text{large}}$ test is outlined in Algorithm \ref{alg:mHL_large} (see Appendix).

\paragraph{Advantages:} \parencite{MHL2021}
\begin{itemize}
    \item[--] The test uses the same HL statistic as the traditional test but compares it to a noncentral chi-squared distribution
    \item[--] It provides a measure of goodness of fit ($\epsilon$) critical's value, that is independent of sample size. The approach provides both a formal statistical procedure and a measure of effect size (the standardized parameter itself)
    \item[--] It maintains the intuitive interpretation of the traditional HL test while addressing its limitations in large samples.
    \item[--] Maintained power to detect truly poorly fitting models
\end{itemize}
\paragraph{Considerations:} \parencite{MHL2021}
\begin{itemize}
    \item[--] The test importance significantly appears in the very huge data, the author arbitrarily used $n_0=10^6$.
    \item[--] For moderate sample sizes (e.g., 10,000), the two tests perform similarly
    \item[--] It's a conservative test, as its proportion of the rejection is less that $ 5\% $ for the correct model.
\end{itemize}

By using this modified approach, researchers can more appropriately assess the goodness of fit of logistic regression models in large samples, avoiding the issue of rejecting models with practically acceptable fit due to large sample sizes.

\section{McCullagh's Test (Pearson Standardization)}
McCullagh's test \parencite{mccullagh1985, mccullagh1989generalized}, introduced in 1985, provides a modification of the Pearson chi-square statistic for goodness-of-fit assessment in logistic regression models. This test is particularly valuable when dealing with sparse data, where traditional tests may fail to provide reliable results. Unlike traditional approaches, McCullagh's method uses conditional asymptotic moments of the Pearson statistic given the parameter estimates.
The test statistic is based on the standardized Pearson chi-square statistic ($Z_{McC}$) and is computed as:
\begin{equation} \label{eq:mccullagh}
    Z_{McC} = \frac{X^2 - E(X^2|\hat{\beta})}{\sqrt{Var(X^2|\hat{\beta})}}
\end{equation}

Under the null hypothesis of a good fit, $Z_{McC}$ follows a standard normal distribution. The test rejects the null hypothesis at significance level $\alpha$ if $|Z_{McC}| > z_{1-\alpha/2}$, where $z_{1-\alpha/2}$ is the $(1-\alpha/2)$ quantile of the standard normal distribution. \\

\paragraph{Advantages and Considerations:}
\begin{itemize}
    \item The calculation of conditional moments for McCullagh's test is mathematically involved and can be computationally intensive.
    \item For sparse data, McCullagh's test claims to provide \textbf{better performance} than the Osius and Rojek test (see section \ref{sec:osius_rojek_test}); however, for grouped data, Farrington's test is generally superior \parencite{kuss2002}.
    \item McCullagh's test is more robust to the grouping strategy than the Hosmer–Lemeshow test. These tests may perform poorly in the following situations:
          \begin{itemize}
              \item When covariates are continuous, grouping can lead to a loss of information.
              \item When the sample size is moderate, group counts may be sparse or unreliable.
          \end{itemize}
    \item McCullagh’s approach addresses these issues by:
          \begin{itemize}
              \item Utilizing the likelihood at the individual observation level.
              \item Conditioning directly on the predicted probabilities, thus avoiding the drawbacks of grouping.
          \end{itemize}
\end{itemize}

\paragraph{The Algorithm}
This algorithm is captured from the CODE \parencite{kuss2002goflogit} of the macro \texttt{GOFLOGIT} written by Kuss \parencite{kuss2002paper} on SAS \parencite{sas2020}, On github documentation. The detailed implementation is provided in Algorithm \ref{alg:mccullagh} (see Appendix).

\section{Windmeijer Test (Pearson Standardization)}
Weesie  has written a STATA program implementing a method proposed by Windmeijer for computing the signiﬁcance of the Pearson chi-square statistic using the standard normal distribution \parencite{weesie1998, windmeijer1990}. The approach is similar to Osius and Rojek Test, but is \textbf{only} appropriate in settings when there are \textbf{n covariate patterns (not applicable for Sparse Data)}. Thus it is less general than the above method.\parencite{osius1992normal}
Windmeijer points out that both the Pearson chi-square and the estimator of its variance used to form $Z_{O\&R}$ (section \ref{sec:osius_rojek_test})  are quite sensitive, as noted above, to large or small estimated probabilities. Both values are inﬂated. He suggests that subjects with very small or large ﬁtted values, near 0 or 1, be excluded when using the Pearson chi-square statistic. The default exclusion criteria in Weesie's STATA program are $\hat{\pi} <1.0\times10^{-5}$  or $\hat{\pi}>(1-1.0\times10^{-5})$.

\section{Osius and Rojek's Test  (Pearson Standardization)}  \label{sec:osius_rojek_test}

Osius and Rojek developed a test that provides a normal approximation to the Pearson Chi-Square statistic for evaluating goodness-of-fit in logistic regression models \parencite{osius1992normal}. This test is particularly valuable when analyzing sparse data or models with continuous covariates.
The standardized test statistic is given by:
\[
    Z_{O\&R} = \frac{X^2 - (n - p^*)}{\sqrt{A + RSS}} \sim N(0,1) \text{ under } H_0
\]

\paragraph{Algorithm} Where, all its component are mentioned in Algorithm \ref{alg:osius_rojek_test} (see Appendix). However, this test main Algorithm is conducted on grouped data was well presented in \parencite{liu2024comprehensive} by Liu et al. and we hense adjust it for Individual Data ($n_g=1$).

\paragraph{The Osius and Rojek test addresses Sparse Data issue by:}
\begin{enumerate}

    \item Using a normal approximation instead of a chi-square approximation, which can be more accurate for sparse data.
    \item Incorporating a variance correction through the weighted linear regression, which accounts for the variability in the estimated probabilities.
    \item Providing a test statistic that is less sensitive to small expected frequencies in individual cells.

\end{enumerate}

\section{Farrington's Test (Pearson Standardization)} \label{sec:farrington_test}

Farrington proposed a modification to the Pearson chi-square statistic to enhance its performance for generalized linear models, especially when dealing with sparse data \parencite{farrington1996}. This test aims to address the limitations of the traditional Pearson chi-square statistic by incorporating a first-order correction term, which induces local orthogonality with the regression parameters. This modification reduces the dependency of the statistic on the regression parameter estimates, leading to increased power and more accurate goodness-of-fit assessments in sparse data scenarios.

\subsubsection{Test Statistic}

The Farrington test statistic, denoted as $X^2_F$, is defined by adding an adjustment to the traditional Pearson chi-square, with the following form by \cite{kuss2002}:
\begin{equation} \label{eq:farrington}
    X^2_F = \sum_{i=1}^{N} \frac{(y_i - m_i\hat{\pi}_i)^2}{m_i\hat{\pi}_i(1 - \hat{\pi}_i)} + \sum_{i=1}^{N} \frac{-(1 - 2\hat{\pi}_i)}{m_i\hat{\pi}_i(1 - \hat{\pi}_i)}(y_i - m_i\hat{\pi}_i)
\end{equation}

where $y_i$ represents the observed outcomes, $m_i$ is the total number of trials, and $\hat{\pi}_i$ is the predicted probability for the $i$th observation. This test statistic has minimal variance and removes the dependence of the distribution on the bias of parameter estimates, making it a more robust measure compared to the standard Pearson chi-square statistic.

\subsubsection{Algorithm for Farrington's Test}

The computational steps for Farrington's test statistic are outlined in Algorithm \ref{alg:farrington_test} (see Appendix).

\subsubsection{Properties and Limitations}

One of the notable properties of the Farrington test is that it is asymptotically independent of the regression parameters \parencite{farrington1996}. This independence implies that the test remains stable even with the inclusion of more covariates in the model, making it suitable for complex datasets with sparse data points. However, despite its theoretical advantages, the Farrington test has some practical limitations:
\begin{itemize}
    \item When all individual trials have exactly one observation ($m_i = 1$), the value of the Farrington statistic equals the number of observations, $X^2_F = N$, which implies that the test will never reject the null hypothesis of a good fit in this scenario, power expected to be Zero in this case.
    \item The calculation of the test statistic is not straightforward and requires detailed knowledge of its moments, which are not easily computed or available in standard statistical software packages \parencite{hosmer2013applied}.
    \item Kuss simulated the performance of the Farrington test and found that it often outperforms the traditional tests when dealing with simi-sparse data (where $M \neq N$) \parencite{kuss2002}. However, its practical implementation remains challenging due to the complexity of its moment calculations.
\end{itemize}

\subsubsection{Comparison with Other Goodness-of-Fit Tests}

Farrington's test provides a significant improvement over traditional tests like the Hosmer-Lemeshow test, particularly for datasets where not full sparse data is a concern. The simulation studies conducted by Kuss demonstrated that the Farrington test outperformed other goodness-of-fit tests in terms of size and power properties, especially in scenarios with low-frequency counts \parencite{kuss2002}. Despite these advantages, the complexity and lack of software support limit its widespread adoption in practical applications.

\subsubsection{Conclusion}

While Farrington's modification to the Pearson statistic holds theoretical promise for handling not full sparse data, its practical limitations cannot be overlooked. The value of the test statistic is identical to the number of observations when each observation is unique, and its implementation is not straightforward due to the complexity of its moments. Therefore, although it has been shown to outperform other tests in specific scenarios, it is not widely used in practice, and simpler alternatives like the Hosmer-Lemeshow test are often preferred for their ease of use.

\newpage
\section{Tsiatis Score Test} \label{section:tsiatis}

The Tsiatis test, introduced by Tsiatis, is a goodness-of-fit test for logistic regression models, particularly useful for models with continuous covariates. \parencite{Tsiatis1980}

\subsubsection{Theoretical Foundation}

Consider a logistic regression model extended with a new categorical variable:

\begin{equation}
    \log\left(\frac{\pi(\mathbf{x},\mathbf{I})}{1-\pi(\mathbf{x},\mathbf{I})}\right) = \mathbf{x}'\boldsymbol{\beta} + \mathbf{I}'\boldsymbol{\gamma}
\end{equation}

where:
\begin{itemize}
    \item $\mathbf{x}$ is the vector of original covariates
    \item $\boldsymbol{\beta}$ is the vector of regression coefficients
    \item $\mathbf{I} = (I^{(1)}, ..., I^{(G-1)})'$ is a vector of indicator variables for $G-1$ mutually exclusive regions of the covariate space (PCS)  ($G$ = number of groups)
    \item $\boldsymbol{\gamma} = (\gamma_1, ..., \gamma_{G-1})'$ are additional coefficients for these regions
\end{itemize}

\subsubsection{Hypothesis Testing}

The null hypothesis is $H_0: \gamma_1 = ... = \gamma_g = 0$, tested using the Score test statistic:

\begin{equation}
    T = \mathbf{S}'V^{-}\mathbf{S}
\end{equation}

\noindent where $\mathbf{S}$ is the score vector and $V^{-}$ is a generalized inverse of its covariance matrix. The distribution of $T$ is $\chi^2(R_\nu)$, where $R_\nu$ denotes the rank of $V$.

\noindent \textbf{Note on Score Test:} The Score test evaluates the slope of the likelihood function at the null hypothesis value. It's computationally efficient as it doesn't require estimating the model under the alternative hypothesis, i.e. Liklyhood ratio test (Full and reduced model are needed) (see section \ref{sec:score_test_in_logistic_regression}).

\paragraph{Algorithms}
The implementation of the Tsiatis test is provided in Algorithm \ref{alg:tsiatis_test} (see Appendix). The matrix form implementation is detailed in Algorithm \ref{alg:tsiatis_score_test_matrix_form} (see Appendix).

\paragraph{Considerations and Limitations of the Tsiatis Score Test:}
\begin{itemize}
    {\small
    \item \textbf{Partitioning is subjective:} There is no universally accepted rule for how to partition the covariate space, especially with continuous or high-dimensional covariates. Both the number and boundaries of groups are somewhat arbitrary and can substantially affect test results. this test is very sensitive to the choice of the number of groups and the partitioning method.

    \item \textbf{Challenges with high-dimensional or sparse data:} Partitioning becomes more difficult as dimensionality increases or data become sparse, often resulting in small or unbalanced group sizes. This can undermine the validity of asymptotic approximations and reduce test reliability.
    \item \textbf{Detection scope:} The Tsiatis test is most sensitive to lack-of-fit that appears as group-specific effects, but may miss model misspecification that does not align with the chosen partitions.
    \item \textbf{Sample size and group balance:} Small sample sizes or highly unbalanced groups can reduce the power and validity of the test, as with most goodness-of-fit tests for logistic regression with continuous covariates.
          }
\end{itemize}

\section{Stukel's Score Test} \label{sec:stukel_test}

Stukel's test is designed to assess the adequacy of the logistic link function in logistic regression models, particularly for detecting asymmetry and heavy tails in the data \parencite{stukel1988generalized}. The test is based on a generalized logistic model that introduces two shape parameters, $\alpha_1$ and $\alpha_2$, to allow for deviations from the standard logistic link. Generalized Logistic Model:
\begin{equation}
    \text{logit}(\pi_i) = \eta_i + \alpha_1 \eta_i^2 I(\eta_i > 0) + \alpha_2 \eta_i^2 I(\eta_i \leq 0)
\end{equation}
where $\eta_i = x_i'\beta$ is the linear predictor, $I(.)$ is the indicator function, and $\alpha_1$ and $\alpha_2$ are shape parameters. The standard logistic model is recovered when $\alpha_1 = \alpha_2 = 0$. Hypothesis:
\begin{itemize}
    \item[--] Null hypothesis ($H_0$): $\alpha_1 = \alpha_2 = 0$ (the logistic link is correctly specified)
    \item[--] Alternative: At least one of $\alpha_1$, $\alpha_2$ is nonzero (evidence of link misspecification)
\end{itemize}

\paragraph{Test Statistic:}
The score test statistic for $H_0$ is:
\begin{equation}
    T_S = U'(\hat{\beta})\mathcal{I}^{-1}(\hat{\beta})U(\hat{\beta})
\end{equation}
where $U(\hat{\beta})$ is the score vector and $\mathcal{I}(\hat{\beta})$ is the information matrix, both evaluated at the maximum likelihood estimate $\hat{\beta}$ under the standard logistic model. Under $H_0$, $T_S$ follows a chi-square distribution with 2 degrees of freedom. A significant result indicates that the logistic link may not be appropriate for the data.

\paragraph{Variants and Implementation:}
Stukel's test can be implemented in several ways, depending on which quadratic terms are included in the model: one can include only the $z_1$ term (testing $\alpha_1$), only the $z_2$ term (testing $\alpha_2$), or both $z_1$ and $z_2$ terms (testing both $\alpha_1$ and $\alpha_2$ jointly, which is the most comprehensive version). Each of these model augmentations can be evaluated using either the \textbf{Score Test} or the \textbf{Likelihood Ratio Test}, resulting in \textbf{SIX possible versions}.
\begin{itemize}
    \item[--] $z_1$ only: Score test and Likelihood ratio test
    \item[--] $z_2$ only: Score test and Likelihood ratio test
    \item[--] Both $z_1$ and $z_2$: Score test and Likelihood ratio test
\end{itemize}
This structure allows for flexible detection of different types of link function misspecification, as each variant may be sensitive to different forms of asymmetry or tail behavior in the data.

\paragraph{Practical Considerations:}
Hosmer  reports that Stukel's test is especially useful for identifying misspecification of the link function, such as asymmetry or unusual tail behavior. However, it is not designed to detect all forms of model inadequacy. We will check the claimed performance of this test in the simulation \parencite{Hosmer1997}.

\paragraph{Algorithm} The detailed implementation of Stukel's test is provided in Algorithm \ref{alg:stukel_test} (see Appendix).

\section{The Information Matrix (IM) White Test} \label{sec:information_matrix}
The Information Matrix (IM) test, originally proposed by White  and later refined for binary data models by Orme, is a powerful goodness-of-fit procedure for detecting model misspecification \parencite{White1982, Orme1988}. The practical implementation used in this research follows the algorithm developed by Kuss, who provided a comprehensive evaluation of global goodness-of-fit tests for logistic regression with sparse data \parencite{kuss2002}. Unlike tests that focus only on the residuals, the IM test strikes at the heart of maximum likelihood (ML) estimation theory. The detailed algorithm is presented in Algorithm \ref{alg:im_test} (see Appendix).

\paragraph{Theoretical Foundation and Kuss's Insights}
A fundamental property of a correctly specified likelihood model is the \textbf{information matrix equality} (recall section \ref{sec:information_theory}).
\begin{equation}
    E\left[ \left( \frac{\partial \mathcal{L}}{\partial \boldsymbol{\beta}} \right) \left( \frac{\partial \mathcal{L}}{\partial \boldsymbol{\beta}} \right)^T \right] = - E\left[ \frac{\partial^2 \mathcal{L}}{\partial \boldsymbol{\beta} \partial \boldsymbol{\beta}^T} \right]
\end{equation}

\noindent Kuss notes that the Information Matrix test relies on the idea of comparing two different estimators of the information matrix which should give comparable results under a satisfactory model \parencite{kuss2002}. As Hosmer and Lemeshow termed the IM-test 'elegant, but difficult to compute in practice' \parencite{hosmer2013applied}, Kuss showed how to calculate this test for logistic regression models efficiently \parencite{kuss2002paper,kuss2002goflogit}. Evaluating the difference of the diagonal elements of the two estimators results in the $((p^*) \times 1)$-vector:
\begin{equation}
    \hat{d} = \frac{1}{M} \sum_{i=1}^{M} (y_i - \hat{\pi}_i)(1 - 2\hat{\pi}_i)z_i
\end{equation}

where $z_i = (1, x_{i1}^2, \ldots, x_{ip}^2)^T$ and the components of $\hat{d}$ sum to 0 in the case of a good model fit. After standardization of $\hat{d}$ with an appropriate variance, the test statistic ($IM_{DIAG}$) can be compared to a $\chi^2_{p^*}$-distribution. \\

\noindent Importantly, Kuss emphasizes that the IM-test is calculated for the individual and not for the grouped observations, making it particularly suitable for sparse data scenarios where traditional goodness-of-fit tests may fail. \\

\noindent Chesher explained that the IM-test can be seen as a score test for detecting whether regression coefficients vary randomly across observations in logistic regression \parencite{Chesher1989}. In simple terms, it checks if the effect of each predictor is truly constant for all data points or not. This interpretation blurs the line between "specific" and "global" goodness-of-fit tests, since both can use the same test statistic and provide insight into overall model fit. \\

\noindent If the model is misspecified (e.g., an incorrect link function or omitted non-linear terms), this equality no longer holds. The IM test is designed to detect a statistically significant difference between these two estimators of the information matrix, thereby signaling a misspecification.

\paragraph{Implementation via Auxiliary Regression}
The practical implementation of the IM test, as described by Orme and implemented in the \texttt{goflogit} SAS macro by Kuss, is based on a clever auxiliary OLS regression \parencite{kuss2002}. The test statistic is constructed from the explained sum of squares of this regression. The logic is to regress the Pearson residuals on a set of constructed variables that capture the components of the information matrix estimators. A significant relationship in this regression implies that the information matrix equality is violated. The test statistic, often referred to as a score or Lagrange Multiplier (LM) statistic, is then compared to a $\chi^2$ distribution to assess its significance.

\section{Anova F-Test for Deviances } \label{sec:anova_f_test_for_deviances}

Also known in some literature as Modified Hosmer-Lemeshow Test  ($\boldsymbol {mHL}$), we call \textbf{Anova(Deviance) Test}, This test is commonly used in the statistical Packages, mentioned  by   \cite{liu2024comprehensive} and found in famous R packages like \texttt{LogisticDx} \parencite{LogisticDx} and \texttt{LDdiag}\parencite{LDdiag}, and used by \cite{rady2021comparison} in his comparison study  (Cairo University).
The Anova(Deviance) Test assesses the change in model deviance when the covariate group $G$ is added as a predictor. That is, fit a linear model $r_{i,deviance}\sim G$, where $r_{i,deviance}$ is the deviance residual, and the effect of adding $G$ can be assessed using the ANOVA analysis on the linear regression of $r_{i,deviance}$ against $G$.

\paragraph{Algorithm}The ANOVA F-test procedure is outlined in Algorithm \ref{alg:anova_f_test_for_deviances} (see Appendix).

\newpage
\section{Smoothed Residual Tests } \label{sec:smoothed_residual_tests}
The Hosmer-Lemeshow test, while widely used for assessing goodness-of-fit in logistic regression models, has been criticized for several limitations, particularly its sensitivity to the number of groups and sample size. To address these issues, le Cessie, van Houwelingen, Copas, and Hosmer proposed alternative approaches based on smoothing residuals. The key idea behind smoothing is to overcome the discrete nature of the Hosmer-Lemeshow test by providing a continuous assessment of model fit across the range of predicted probabilities. \parencite{Hosmer1997,le1991goodness,le1995goodness}

Smoothing techniques (recall section \ref{sec:smoothing_methods}) allow for a more nuanced evaluation of model fit by considering local patterns in the residuals, rather than relying on arbitrary groupings. This approach can potentially detect subtle departures from the model that might be missed by the Hosmer-Lemeshow test, especially in cases with continuous covariates or when the sample size is large.

Two notable tests were proposed based on this smoothing concept:

\subsection{Le Cessie and van Houwelingen (1991) Smoothed Residual Tests}

Le Cessie and van Houwelingen (1991) \parencite{le1991goodness} proposed a goodness-of-fit test for logistic regression based on smoothing standardized residuals using kernel methods. The core idea is to assess model fit by examining whether the smoothed residuals, when appropriately weighted, deviate significantly from zero.

\paragraph{Test Statistic (Scalar Form):}
let $r_s(X_i)$ be the smoothed residual at $X_i$, typically obtained via kernel smoothing of the standardized (Pearson) residuals and $v(X_i)$ be a weighting factor, specifically the inverse of the variance of the smoothed residual at $X_i$ (crucial for achieving desirable asymptotic properties).
\begin{equation}
    T = \frac{1}{n} \sum_{i=1}^n r_s(X_i)^2 v(X_i)
    \label{eq:cessie1991_scalar}
\end{equation}
The multiplication by $v(X_i)$ is crucial for achieving desirable asymptotic properties.

\paragraph{Test Statistic (Matrix Form):}

Let $\mathbf{r}$ be the vector of Pearson residuals, and let $\mathbf{W}$ be the $n \times n$ kernel smoothing matrix (with $W_{ij}$ representing the kernel weight for $X_j$ when smoothing at $X_i$). Let $\mathbf{D}_v$ be the diagonal matrix with entries $v(X_i)$. Then, the test statistic can be written in matrix notation as:
\begin{equation}
    T = \frac{1}{n} \mathbf{r}^\top \mathbf{W}^\top \mathbf{D}_v \mathbf{W} \mathbf{r}
    \label{eq:cessie1991_matrix}
\end{equation}
This form highlights the quadratic nature of the test and its dependence on the smoothing and weighting structure.

\paragraph{Standardization and Inference:}
The test statistic $T$ can be standardized to facilitate inference. Under the null hypothesis, a standardized version of $T$ can be compared asymptotically to a standard normal distribution, or, alternatively, a scaled version can be compared to a chi-square distribution with degrees of freedom determined by the estimated mean and variance of $T$.

\paragraph{Critical Considerations and Limitations:}
\begin{itemize}
    \item \textbf{Computational Complexity:} The calculation of the smoothed residuals and their variances (especially using the Nadaraya-Watson estimator or similar methods) can be computationally intensive, particularly for large sample sizes.
    \item \textbf{Bandwidth Sensitivity:} The choice of kernel bandwidth ($h$) significantly affects the test's performance and its asymptotic properties.
    \item \textbf{Applicability:} The method is primarily suited for continuous covariates; its extension to categorical variables is limited.
\end{itemize}

\textbf{Summary:} While the Le Cessie and van Houwelingen (1991) test provides a flexible, continuous assessment of model fit, it is computationally demanding and sensitive to smoothing parameters. These factors should be carefully considered in practical applications.

\subsubsection*{Algorithm} The kernel-based smoothed residual test algorithm is detailed in Algorithm \ref{alg:kernel_based_smoothed_residual_test} (see Appendix).

\subsection{Le Cessie and van Houwelingen (1995) Score Tests}

Le Cessie and van Houwelingen (1995) proposed an alternative approach based on score tests in random effects models. This method aimed to overcome limitations of the previous kernel-based approach. The definition of the  goodness-of-fit test in le Cessie and van Houwelingen (1991) is somewhat ad hoc (very specific), and results were  only obtained for a logistic model with continuous covariates. This paper shows that their type of  goodness-of-fit test is a score test in a random effect model. Because it is a score test, it is an efficient  test if the alternatives specified are valid. We can use the statistic for continuous covariates as well  as for categorical covariates. \parencite{le1995goodness}

\paragraph{Algorithm} The score test implementation is provided in Algorithm \ref{alg:le_cessie_van_houwelingen_1995_score_test} (see Appendix).

The goodness-of-fit methodology developed by Le Cessie and van Houwelingen (1995) employs a nested modeling approach that embeds the candidate model within an expanded family incorporating random effects. This framework establishes two competing model specifications:
\begin{itemize} { \small
    \item[--] \textbf{Baseline Model (without random effects):} $E(Y_i) = h^{-1}(\mathbf{x}_i\boldsymbol{\beta})$
    \item[--] \textbf{Extended Model (with random effects):} $E(Y_i) = h^{-1}(\mathbf{x}_i\boldsymbol{\beta} + r_i)$ }
\end{itemize}

where $\mathbf{r} = (r_1, \ldots, r_n)'$ represents a random vector characterized by zero mean and covariance structure $\sigma^2\mathbf{R}$, with $\mathbf{R} = \mathbf{I}_n$ (identity matrix) in most practical applications assuming independence among random effects.

\begin{itemize}
    { \small
    \item[--] \textbf{Null Hypothesis} $(H_0)$: The baseline model provides adequate representation of the data structure, formally expressed as testing $\sigma^2 = 0$.
              [$H_0: E(Y_i) = h^{-1}(\mathbf{x}_i\boldsymbol{\beta})$ for all $i$]

    \item[--] \textbf{Alternative Hypothesis} $(H_1)$: The baseline model exhibits inadequate fit, necessitating additional stochastic variation to capture unexplained systematic patterns, formally expressed as testing $\sigma^2 > 0$.
              [$H_1: E(Y_i) \neq h^{-1}(\mathbf{x}_i\boldsymbol{\beta})$ for some $i$]
          }
\end{itemize}

The implemented score test methodology evaluates the statistical evidence for rejecting the null hypothesis in favor of the alternative through a test statistic $T = U_{\sigma^2}^2/\text{Var}(U_{\sigma^2})$ that follows an asymptotic $\chi^2_1$ distribution under $H_0$. This approach determines whether the parsimonious model specification (excluding random effects) provides sufficient explanatory power for the observed data. The analytical framework offers considerable flexibility for goodness-of-fit assessment across generalized linear modeling contexts, with particular utility in logistic regression applications involving mixed covariate structures. The method is especially sensitive to overdispersion and unmodelled heterogeneity, complementing traditional diagnostic approaches such as residual plots and Hosmer-Lemeshow-type tests.

\paragraph{Considerations:}
\begin{itemize}
    \item Required specification of the correlation structure $R$, which can be challenging in practice, because it is very sensitive depends on the bandwidths parameter.
    \item Computational complexity may increase with large datasets or complex correlation structures
\end{itemize}

\subsection{Copas Unweighted Sum of Squares Test with Osius and Rojek Normal Approximation}

\paragraph{Definition}
The Copas Unweighted Sum of Squares (USS) Test, introduced by Copas   (Some authors call it RSS Test \parencite{kuss2002}), is a goodness-of-fit test for proportions \parencite{copas1989unweighted}. This test has been adapted for use in logistic regression, with a normal approximation proposed by Osius and Rojek. The test statistic is based on the unweighted sum of squared differences between observed outcomes and fitted probabilities.

\paragraph{Historical Context}
Hosmer et al. (1997) \parencite{Hosmer1997} noted that the Copas test can be considered a special case of the le Cessie and van Houwelingen (1995) test. This connection highlights the evolving nature of goodness-of-fit tests in logistic regression and the relationships between different approaches. Hosmer (2013) mentioned it in his book \parencite{hosmer2013applied} as $Z_s$.

\paragraph{Implementation in R}
In various R packages, this test is often referred to as the "le Cessie-van Houwelingen-Copas-Hosmer unweighted sum of squares test" or the "le Cessie-van Houwelingen normal test statistic for the unweighted sum of squared errors." These names reflect the collaborative development and refinement of the test over time.

\subsubsection*{Algorithm} The Copas USS test algorithm is presented in Algorithm \ref{alg:copas_uss_test} (see Appendix).

\paragraph{Note on Implementation}
The algorithm presented here is based on the implementation in the \texttt{RMS} package, which adapts the Copas test using the Osius and Rojek normal approximation. This approach allows for application to logistic regression models with continuous covariates, extending beyond the original grouped data context of Copas' test.

\section{GAM-Based Algorithm (GAM Probability Estimates for Grouping)} \label{sec:gam_test}
The GAM-based modified tests, introduced by Shang, propose a novel approach to improve the power of traditional goodness-of-fit tests in logistic regression by utilizing Generalized Additive Models (GAM) for grouping observations \parencite{GAM2021}. This approach specifically addresses the limitation of existing tests where model misspecification can lead to poor grouping and consequently low power in detecting lack of fit.
The modified test statistic is based on groupings derived from probabilities estimated using an over-fitted GAM model:
\begin{equation}
    \mathrm{logit}(\pi(x_i)) = s_0 + \sum_{j=1}^p\beta_jx_{i,j} + \sum_{j=p^*}^qs_j(x_{i,j})
\end{equation}
where $s_0$ is the intercept, $x_{i,j}, j=1,...,p$ are categorical variables, $x_{i,j}, j=p^*,...,q$ are continuous variables, and $s_j$ are smooth functions estimated using local smoothing algorithms.
The GAM-based grouping is then used to compute traditional test statistics (e.g., Hosmer-Lemeshow, Pulkstenis-Robinson, or Xie tests), resulting in three modified tests:

\textbf{(HL+GAM)}: Modified Hosmer-Lemeshow test,
\textbf{(PR+GAM)}: Modified Pulkstenis-Robinson test
\textbf{(XIE+GAM)}: Modified Xie test \\

\paragraph{Algorithm} The GAM-based test procedures are detailed in Algorithm \ref{alg:gam_gof_test} (see Appendix). a very important note is that the algorithm here used GAM not for smoothing the probabilities, instead it is used to produce an over-fitted probabilities.

\paragraph{Implementation Steps:}
\begin{enumerate}
    \item Fit an over-specified GAM model including:
          - All available main effects
          - Two-way interactions
          - Smooth functions for continuous variables (up to 3rd order polynomials).
    \item Use estimated probabilities from GAM to form groups according to the original test's grouping algorithm.
    \item Calculate the test statistic using the GAM-based groupings.
    \item Compare to the appropriate chi-square distribution for hypothesis testing.
\end{enumerate}

\newpage
\textbf{Key Features:}
\begin{itemize}
    \item Generally provides higher power than traditional tests for detecting (Shang simulation):
          - Omitted interaction terms
          - Missing quadratic terms
          - Missing first-order terms approach
    \item Sometimes an over-fitted GAM model to better approximate true underlying probabilities.
    \item Incorporates both categorical and continuous predictors through a mixture model.
    \item Can incorporate interaction terms and higher-order polynomials automatically.
    \item Maintains appropriate Type I error rates under the null hypothesis.
\end{itemize}

\paragraph{Considerations:}
\begin{itemize}
    \item The Hosmer test employs predicted probabilities for partitioning, while the Pulkstenis and Robinson tests also split each categorical level based on the median of the predicted data.The Xie test differs from traditional clustering methods as it does not rely on probabilities. Instead, it emphasizes clustering within covariate space. The modification proposed by the Shang model aims to enhance the Xie test by clustering in covariate space while incorporating smoothed terms and interactions.
    \item Shang et al. mentioned that XIE+GAM, showed no significant power gain in simulation studies comparing to XIE without GAM, but could be subject to type II error inflation. \parencite{GAM2021}
    \item The method for calculating probability to make partitions may have variables and interactions that mislead, resulting in biased probabilities.
    \item Power can be affected by the distribution of covariates.
    \item Requires more computational resources than traditional tests.
\end{itemize}

\section*{Conclusion}

\noindent
This chapter surveyed goodness-of-fit tests for logistic regression, from classic Pearson Chi-Square and Deviance tests—best suited for grouped data—to more modern approaches that address challenges like continuous predictors and sparse data. there is no universal goodness-of-fit test; the best choice depends on data characteristics and the type of model inadequacy of concern. Understanding the strengths and limitations of each method enables practitioners to select appropriate tools and critically assess model fit.
\vspace{0.5em}
\noindent \paragraph{In the following chapter,} we will explore advanced methods, including bootstrap-based and machine learning calibration tests, which enhance model assessment in complex and high-dimensional settings.

\chapter{Machine Learning and Bootstrap Algorithms}

\section*{Introduction}

In statistics, the question of whether a model adequately describes the data is addressed through \textbf{goodness-of-fit (GOF)} testing \parencite{gof2002}. In machine learning, the same question is framed as \textbf{calibration assessment}  \parencite{Bohdal2023}. This chapter treats these as equivalent perspectives: a model that fails a goodness-of-fit test is, in effect, miscalibrated. The chapter presents the definitions and diagnostic tools for calibration, followed by the formal hypothesis tests that are the main subject of this thesis.

\newpage

\section{Calibration in Machine Learning}

\textbf{Calibration} refers to the agreement between predicted probabilities and observed event frequencies \parencite{dawid1982well}. Many modern models, particularly deep neural networks and gradient-boosted trees, tend to be miscalibrated \parencite{guo2017calibration, niculescu2005predicting}. In applied many domains, miscalibrated probabilities can lead to poor decisions \parencite{denaxas2019}.

\subsection{Perfect Calibration}

Let $Y \in \{0, 1\}$ be the binary outcome and $X$ the covariate vector. A model produces probability estimates $\hat{\pi}(X) = P(Y=1 \mid X=x)$. The model is \textbf{perfectly calibrated} if, for every $\pi \in [0,1]$ \parencite{mayr2012}:
\begin{equation}
    \label{eq:perfect_calibration}
    P(Y=1 \mid \hat{\pi}(X) = \pi) = \pi, \quad \forall \pi \in [0, 1]
\end{equation}
That is, among all instances assigned a predicted probability of $\pi$, the proportion with $Y=1$ should equal $\pi$. \parencite{nattino2014new}

\subsection{Types of Calibration}
The condition of perfect calibration is often relaxed into weaker forms. The literature distinguishes two main levels \parencite{harrell2015regression, van2019calibration}.

\paragraph{(I) Average Calibration (Calibration-in-the-Large)}
This requires only that the mean predicted probability equals the overall event rate:
\begin{equation}
    E[\hat{\pi}(X)] = P(Y=1)
\end{equation}
A model satisfying this condition estimates the marginal event rate correctly but may still be miscalibrated within subgroups. This property is assessed by the \textbf{calibration intercept $\hat{\gamma}_0$}, where $\hat{\gamma}_0 = 0$ indicates good average calibration.

\paragraph{(II) Strong Calibration (Pointwise Calibration)}
This corresponds to Equation \ref{eq:perfect_calibration} and requires calibration across the full range of predicted probabilities. It is assessed by the \textbf{calibration slope $\hat{\gamma}_1$}, where $\hat{\gamma}_1 = 1$ indicates that the spread of predictions matches the spread of true probabilities. A slope below one suggests over-confidence; a slope above one suggests under-confidence.

\begin{equation}
    \text{logit}(\text{Pr}(Y_i=1)) = \gamma_0 + \gamma_1 \cdot \text{logit}(\hat{\pi}_i)
\end{equation}

\subsection{Miscalibration: Over- and Under-Confidence}
Miscalibration occurs when Equation \ref{eq:perfect_calibration} is violated. this can be due to overfitting or underfitting. \parencite{harrell2015regression}

\paragraph{(I) Over-Confidence}
The model produces probabilities that are too extreme (too close to 0 or 1). This is common in high-capacity models such as deep neural networks and gradient-boosted trees. On a calibration plot, this appears as an S-shaped deviation from the diagonal.

\paragraph{(II) Under-Confidence}
The model produces probabilities that are too close to the overall event rate. This can occur in heavily regularized models .

\subsection{Calibration Tests (Hypothesis Testing)} \label{sec:calibration_tests_formally}
Calibration tests are formal statistical procedures that test the null hypothesis of perfect calibration. Most tests are based on one of two principles.

\paragraph{(I) The Grouping Principle}
Data are partitioned into bins. Within each bin, observed events ($o_k$) are compared to expected events ($e_k = \sum \hat{\pi}_i$). The differences are aggregated into a test statistic, typically of the form $\sum (o_k - e_k)^2 / v_k$. Tests differ in how bins are formed: the Hosmer--Lemeshow test groups by predicted probability \parencite{hosmer1980goodness}, while the Xie test groups by covariate clustering \parencite{xie2008increasing}.

\paragraph{(II) The Regression Principle}
A logistic regression is fitted to the model's own predictions \parencite{Cox1958, Miller1991,nattino2014new,harrell2015regression}:
\begin{equation}
    \text{logit}(\text{Pr}(Y_i=1)) = \gamma_0 + \gamma_1 \cdot \text{logit}(\hat{\pi}_i)
\end{equation}
Under perfect calibration, $\gamma_0 = 0$ and $\gamma_1 = 1$. A calibration test can be constructed as a joint test of $H_0: \gamma_0 = 0,\; \gamma_1 = 1$.

\begin{figure}[H]
    \centering
    \includegraphics[width=0.9\linewidth]{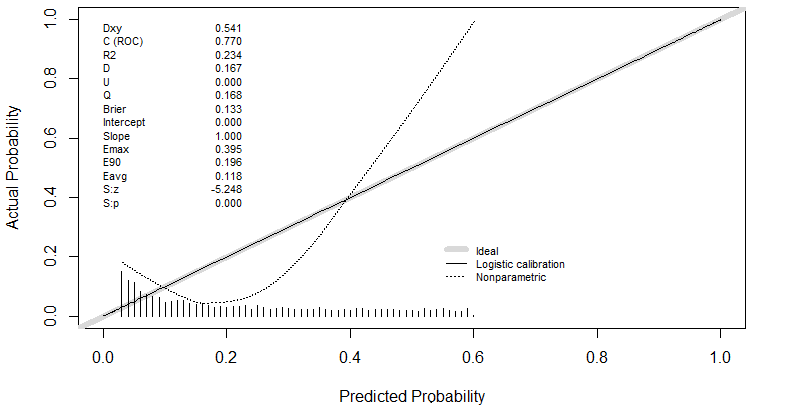}
    \caption{Calibration plot for a model with \textbf{neglected quadratic effects}. \cite{harrell2015regression}}
    \label{fig:calibration_plot_quadratic}
\end{figure}

\begin{figure}[H]
    \centering
    \includegraphics[width=0.9\linewidth]{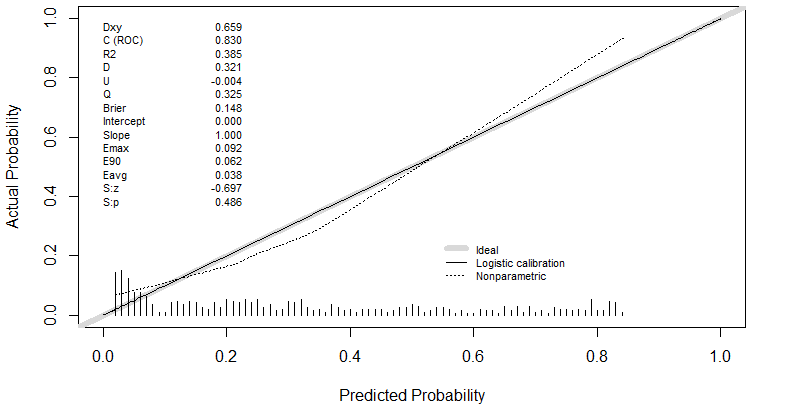}
    \caption{Calibration plot for a model with \textbf{neglected interaction effects}.\cite{harrell2015regression}}
    \label{fig:calibration_plot_interaction}
\end{figure}

However, the regression principle has limitations. As shown in Figures \ref{fig:calibration_plot_quadratic} and \ref{fig:calibration_plot_interaction}, the slope and intercept can equal their ideal values ($\hat{\gamma}_1 = 1$, $\hat{\gamma}_0 = 0$) even when the model is miscalibrated due to omitted terms (see Section \ref{sec:omitted_quadratic_term}). This motivates the use of more flexible calibration tests \parencite{nattino2014new}.

\subsection{The Bridge to Hypothesis Testing}
The visualization techniques and metrics described above are invaluable for diagnosing and quantifying miscalibration. However, they do not provide a formal statistical framework for decision-making. Metrics like ECE tell us the \textbf{magnitude} of the calibration error, but they do not tell us whether this error is \textbf{statistically significant} or could have arisen simply due to random chance in the finite sample.

This is where formal hypothesis tests for calibration become indispensable. These tests, which form the core of the remainder of this chapter, provide a formal test statistic that allows us to make a formal decision about the adequacy of the model's calibration under the null hypothesis of perfect calibration. They bridge the gap between the descriptive, diagnostic world of machine learning calibration metrics and the inferential, decision-oriented world of statistical goodness-of-fit testing. The following sections will detail the algorithms and theoretical underpinnings of several such tests, including both classical and cutting-edge approaches.



\newpage
\section{Selected some Calibration Tests}

in this section we will cover some of the most important and widely used calibration tests, including the Unreliability (U) test, the GiViTI test and the eHL test, Spiegelhalter z-test.

\subsection{The Unreliability (U) Test (Perfect Calibrationed Model LHR test)} \label{sec:unreliability_test}
A comprehensive approach to assessing and quantifying model calibration is provided by the validation framework developed by Frank Harrell Jr., wrote on his book \textit{Regression Modeling Strategies} and implemented in his R package \texttt{rms} \parencite{harrell2015regression}. This framework extends the foundational ideas of Cox \parencite{Cox1958} and the diagnostic strategy of Miller, Hui, and Tierney \parencite{Miller1991} to produce a suite of indices that measure different aspects of predictive accuracy. Central to this framework is the Unreliability (U) test, which specifically quantifies a model's lack of calibration.

\paragraph{The Logistic Calibration Model}
The entire framework is built upon fitting a simple logistic ``calibration model" to a validation sample. This model uses the observed binary outcome ($Y_i$) as the response and the linear predictor (or logit of the predicted probability, denoted as $X\hat{\beta}$) from the original model as the sole covariate.
\begin{equation}
    \label{eq:harrell_cal_model}
    \text{logit}(\text{Pr}(Y_i=1)) = \gamma_0 + \gamma_1 (X\hat{\beta})
\end{equation}
The estimated coefficients from this model, the calibration intercept ($\hat{\gamma}_0$) and calibration slope ($\hat{\gamma}_1$), diagnose miscalibration. For a perfectly calibrated model, the theoretical values are $\gamma_0=0$ and $\gamma_1=1$.

\paragraph{The U, D, and Q Indices}
From this calibration model, Harrell derives a set of key performance indices:
\begin{itemize}
    \item \textbf{Discrimination (D):} A measure of how well the model separates patients with and without the outcome. It is a scaled version of the original model's likelihood-ratio $\chi^2$ statistic\footnote{The likelihood-ratio $\chi^2$ statistic is a measure of how well the model fits the data. It is calculated as the difference between the log-likelihood of the model and the log-likelihood of a null model. A higher likelihood-ratio $\chi^2$ statistic is better. discussed in details in Frank's book \parencite{harrell2015regression}.}. A higher D is better.

    \item \textbf{Unreliability (U):} A measure of the model's lack of calibration. It is a scaled version of the likelihood-ratio test for the hypothesis that calibration is perfect ($H_0: \gamma_0=0, \gamma_1=1$). A lower U is better.

    \item \textbf{Overall Quality (Q):} A composite index that combines discrimination and calibration, defined as $Q = D - U$. It represents the model's discriminatory power, penalized by its lack of calibration. (introduced by Hosmer et. al. \parencite{hosmer1995confidence})
\end{itemize}

\paragraph{The Unreliability Test Statistic and Hypothesis Test}
The Unreliability index (U) and its associated hypothesis test are distinct but related. The hypothesis test evaluates whether there is statistically significant evidence of miscalibration, while the index `U` quantifies the magnitude of this miscalibration on a scaled metric.

The core of the test is the likelihood-ratio (LR) statistic for the joint null hypothesis $H_0: \gamma_0=0, \gamma_1=1$. Let:
\begin{itemize}
    \item $\mathcal{L}(\hat{\gamma}_0, \hat{\gamma}_1)$ be the maximized log-likelihood from the fitted calibration model (Equation \ref{eq:harrell_cal_model}).
    \item $\mathcal{L}(0, 1)$ be the log-likelihood calculated by forcing the intercept to 0 and slope to 1, effectively using the original model's linear predictor directly.
\end{itemize}
The unscaled LR statistic, which is used for the hypothesis test, is:
\begin{equation}
    \label{eq:u_lr_statistic}
    \text{LR}_{U}= 2 \left( \mathcal{L}_{cal} - \mathcal{L}_{null} \right) = 2 \left( \mathcal{L}(\hat{\gamma}_0, \hat{\gamma}_1) - \mathcal{L}(0, 1) \right)
\end{equation}
This $\text{LR}_{U}$ statistic follows a $\chi^2$ distribution with 2 degrees of freedom under the null hypothesis. The p-value from this test indicates the significance of the miscalibration.

\paragraph{The Unreliability Index U} as reported by \texttt{rms}, is this test statistic scaled by the sample size $n$:
\begin{equation}
    \label{eq:u_index}
    U = \frac{\text{LR}_{U}}{n} = \frac{2 \left( \mathcal{L}(\hat{\gamma}_0, \hat{\gamma}_1) - \mathcal{L}(0, 1) \right)}{n}
\end{equation}
This scaling allows U to be directly compared with the D index and makes it interpretable as the lack of quality due to poor calibration.

\paragraph{The Unreliability (U) Test Statistic and Algorithm}
The Unreliability (U) statistic provides a single summary measure for the joint hypothesis test of perfect calibration ($H_0: \alpha = 0, \beta = 1$). It quantifies the total amount of miscalibration by capturing deviations in both the average prediction (intercept) and the predictive spread (slope).

\begin{enumerate}
    {\small
    \item \textbf{The Calibration Model (Alternative Hypothesis):} This is the Cox validation model \parencite{Cox1958} ($\text{logit}(p) = \hat{\gamma}_0 + \hat{\gamma}_1 \cdot X\hat{\beta}$) which is allowed to have an intercept $\hat{\gamma}_0 \neq 0$ and slope $\hat{\gamma}_1 \neq 1$. Its maximized log-likelihood, $\mathcal{L}_{cal}$, represents the best possible fit achievable by a linear recalibration of the original model's linear predictor.

    \item \textbf{The Ideal/Null Model (Null Hypothesis):} This model assumes the original predictions are perfect. Its log-likelihood, $\mathcal{L}_{null}$, is calculated using the original model's linear predictor directly, which is equivalent to forcing $\gamma_0=0$ and $\gamma_1=1$. in this case, $\text{logit}(\Pr(Y_i = 1)) = 0 + 1 \cdot (X\hat{\beta}) = X\hat{\beta}$. This is equivalent to saying that the predicted probabilities from the original model ($\hat{\pi}_i = \frac{1}{1 + e^{-X_i\hat{\beta}}}$) are the true probabilities for all $i$. Therefore, $\mathcal{L}_{null}$ evaluates how well this assumption holds.
          }
\end{enumerate}

\paragraph{Algorithm} The complete algorithmic procedure for the Unreliability test is provided in Algorithm \ref{alg:unreliability_test} (see Appendix).

\subsection{The GiViTI Calibration Test and Belt Framework} \label{sec:giviti_calibration_test_and_belt_framework}
The GiViTI calibration framework, developed by Nattino, Finazzi, and Bertolini, represents a significant advancement in the assessment of goodness-of-fit for logistic regression models. It moves beyond the limitations of traditional grouping-based methods (like the Hosmer-Lemeshow test) by providing a more flexible, powerful, and diagnostically rich approach. The framework consists of two main components: \textbf{a formal statistical test of calibration} and an intuitive \textbf{graphical tool known as the calibration belt}.

The core innovation is the use of a polynomial logistic regression to model the relationship between a model's predicted probabilities and the observed event rates. This approach has been rigorously developed for two distinct but related scenarios: the \textbf{external validation} of a pre-existing model on new data \parencite{nattino2014new}, and the \textbf{internal validation} (or goodness-of-fit) of a model on its development dataset \parencite{nattino2015new}.

\subsubsection{The Core Concept: A Polynomial Calibration Curve}
The fundamental idea is to move beyond a simple linear check of calibration (like the one used in the Unreliability test) and allow for more complex, non-linear forms of miscalibration to be detected. This is achieved by fitting a polynomial logistic model to the validation data.

Let $\hat{\pi}_i$ be the predicted probability for observation $i$ from the model being tested, and let $Y_i$ be the observed binary outcome. The method first transforms the predictions to the logit scale, $g_i = \text{logit}(\hat{\pi}_i)$. Then, it fits the following calibration model:
\begin{equation}
    \label{eq:giviti_poly_model}
    \text{logit}(\text{Pr}(Y_i=1)) = \gamma_0 + \gamma_1 g_i + \gamma_2 g_i^2 + \dots + \gamma_m g_i^m
\end{equation}
A key feature of the GiViTI method is that the polynomial degree, $m$, is not fixed in advance. Instead, it is determined by a data-driven \textbf{forward selection procedure}. Starting with a base model (e.g., linear, used in the Unreliability test Equation \ref{eq:harrell_cal_model}), higher-order terms (e.g., quadratic, cubic) are added sequentially. At each step, a likelihood-ratio (LR) test is performed to see if the new, more complex model offers a statistically significant improvement over the previous one. The process stops when adding the next term does not yield a significant improvement, resulting in the most parsimonious yet adequate polynomial curve.

\subsubsection{The Calibration Belt (A Graphical Diagnostic Tool)[Informal]}
The calibration belt is the graphical component of the framework, originaly proposed by Finazzi et al. \parencite{Finazzi2011}. It is a \textbf{confidence band} plotted around the selected polynomial calibration curve. This provides a visual representation of the model's calibration performance across the entire range of predicted probabilities.

The belt is interpreted by comparing it to the line of perfect calibration (the 45-degree bisector):
\begin{itemize}
    \item \textbf{If the belt completely contains the the 45-degree bisector}, there is no significant evidence of miscalibration at the chosen confidence level.
    \item \textbf{If the belt deviates from the bisector}, it indicates statistically significant miscalibration in those regions of predicted probability. For example, if the belt is entirely above the bisector for low predicted probabilities, it means the model is systematically under-predicting the risk for low-risk individuals.
\end{itemize}
This graphical tool is highly diagnostic, as it allows researchers to pinpoint exactly \textbf{where} and \textbf{in what direction} the model is failing, a level of detail not available from a single test statistic. the following figure captured from Nattino et al. \parencite{Nattino2018}.

\begin{figure}[H]
    \centering
    \includegraphics[width=0.6\linewidth]{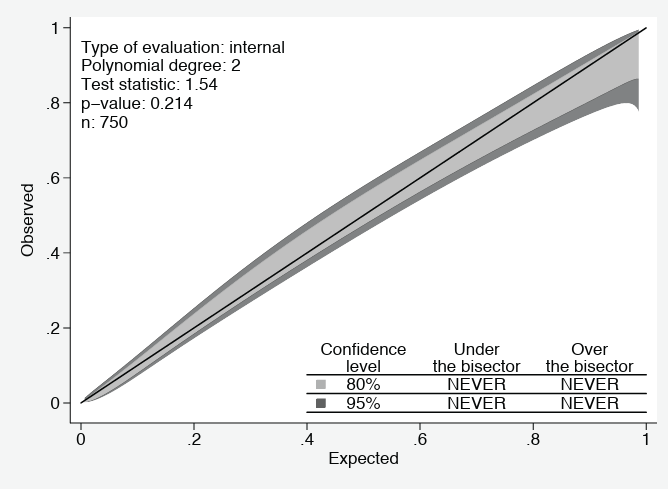}
    \caption[Calibration Belt Plot on The Developmental (Internal) Sample.]{Calibration belt plot on the developmental (Internal) sample. We can conclude that both the 80\% and 95\% calibration belts encompass the bisector over the whole range of the predicted probabilities. This suggests that the predictions of the model do not significantly deviate from the observed rate. \textbf{This suggests that the model is well-calibrated.} P-values and Sample Type discussed next page.}
    \label{fig:calibration_belt_fit}
\end{figure}

\begin{figure}[H]
    \centering
    \includegraphics[width=0.6\linewidth]{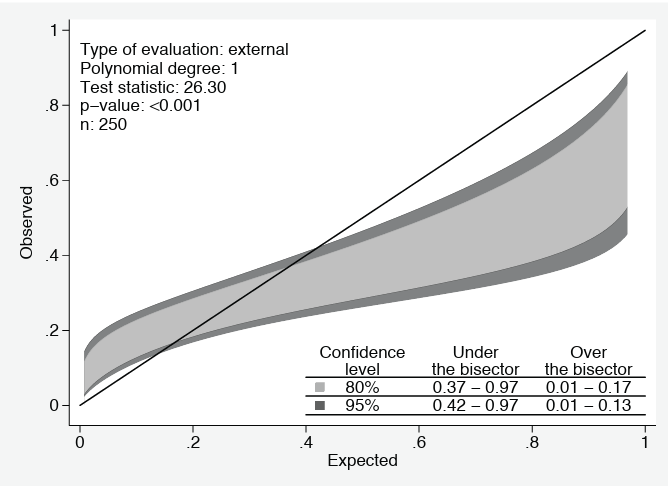}
    \caption[Calibration Belt Plot on (External) Sample.]{Calibration belt plot on  (External) sample. }
    \label{fig:calibration_belt_misfit}
\end{figure}
Because belt lie above and do not include the bisector for small predictions, the predictions of the model significantly underestimate the actual risk in the low range of probabilities. With 95\% and 80\% confidence, the estimated probabilities are underestimated for estimated probabilities smaller than 0.13 and 0.17, respectively (see the table in the bottom-right corner of the plot). However, the model also overestimates the mortality rates for high predicted probabilities. Indeed, the calibration belts are below the bisector for probabilities higher than 0.42 and 0.37 with 95\% and 80\% confidence, respectively. \textbf{This suggests that the model is miscalibrated in the low and high range of predicted probabilities.}

\subsubsection{The GiViTI Calibration Test: Formal Hypothesis Testing}
The formal statistical test is a likelihood-ratio test based on the selected polynomial model. The test statistic, $T_m$, measures the distance between the final $m$-degree polynomial model and the "ideal" model of perfect calibration (where $\gamma_0=0, \gamma_1=1$, and all higher-order $\gamma_i=0$). It is defined as:
\begin{equation}
    T_m = 2 \left( \mathcal{L}_{m} - \mathcal{L}_{\text{null}} \right)
\end{equation}
where $\mathcal{L}_{m}$ is the log-likelihood of the selected $m$-degree model, and $\mathcal{L}_{\text{null}}$ is the log-likelihood of the ideal model.

The main contribution of Nattino et al. was deriving the correct null distribution for $T_m$. Because $m$ is itself a random variable determined from the data, the distribution of $T_m$ is not a standard $\chi^2$ distribution. The correct distribution depends critically on the validation context.

\subsubsection{The 2014 Test for External Validation (Cross-Validation)}
In the external validation setting \parencite{nattino2014new}, a pre-specified model is \textbf{tested on a new, independent dataset}, to check if it generalizes well to new data. for instance, if we have a model that was built somewhere else (e.g., from a publication, or from a different dataset) predicting the risk of a certain outcome. I now have a new, independent set of patients. How well does my old model predict outcomes for these new patients?
\begin{itemize}
    \item \textbf{Logic and Starting Point:} The predicted probabilities, $\hat{\pi}_i$, are treated as fixed, known values. A simple linear recalibration ($\text{logit}(p) = \gamma_0 + \gamma_1 g_i$) may be necessary. Therefore, the forward selection procedure starts by fitting a linear model ($m=1$) and tests whether higher-order terms are needed.
    \item \textbf{Null Distribution:} The 2014 paper derives the complex analytical formula for the cumulative density function (CDF) of the $T_m$ statistic under this external validation scheme.
\end{itemize}

\subsubsection{The 2015 Test for Internal Validation (Goodness-of-Fit)}
In the internal validation setting, a model is tested on the same data used to fit it. This changes the problem fundamentally.
\begin{itemize}
    \item \textbf{Logic and Starting Point:} Nattino et al. \parencite{nattino2015new} showed that when we fit a logistic regression model via maximum likelihood, the resulting estimates are optimal for the linear combination of the predictors. A direct consequence of this optimality is that a linear calibration fit on the same data will \textbf{always} yield $\hat{\gamma}_0=0$ and $\hat{\gamma}_1=1$ \parencite{nattino2015new}. This means the test statistic for a linear fit, $T_1$, is always zero and provides no information about model fit. The first possible sign of miscalibration must come from non-linearities.
    \item \textbf{Methodology Adaptation:} To account for this, the forward selection procedure \textbf{must} start by testing a quadratic model ($m=2$) against the (perfectly-fitting) linear one.
    \item \textbf{Null Distribution:} This change in the starting point alters the underlying probability space. The 2015 paper derives the correct, and different, null distribution for $T_m$ that is conditional on the fact that the procedure begins at $m=2$.
\end{itemize}

\paragraph{Algorithm} The detailed algorithm for the GiViTI calibration test framework is presented in Algorithm \ref{alg:giviti} (see Appendix).

It worth to say that the GiViTI test offers a more robust and flexible assessment of model calibration than the unreliability index. While the unreliability index only detects linear miscalibration, GiViTI uses data-driven polynomial modeling to capture both linear and non-linear calibration issues, making it more sensitive to a wider range of problems in both internal and external validation.

\subsection{Spiegelhalter's z-test (Calibration Test)} \label{sec:spiegelhalter_z_test}

Spiegelhalter's z-test is a statistical test used to assess the calibration of binary logistic regression models. One of the ways to assess the calibration is to evaluate how well the predicted probabilities from a model agree with the observed outcomes (recall section \ref{sec:calibration_tests_formally}). In cases where continuous covariates are present, calibration becomes particularly important, as poor calibration can lead to incorrect conclusions from a model that may otherwise seem well-fitted based on traditional goodness-of-fit measures \parencite{spiegelhalter1986calibration}.

Although this test was originally introduced in 1986, it continues to be widely used and has demonstrated strong performance in evaluating calibration in many recent studies \parencite{walsh2017beyond,huang2020tutorial,harrell2015regression,lindhiem2018importance,cabanillas2024longitudinal}. In the literature, it is sometimes referred to as the Z-test for calibration. the main reason for using this test is that it is built on the Brier score, which is a proper scoring rule, and it is a good and very famous measure of calibration.

\subsubsection{The Mathematical Relationship Between Spiegelhalter's Z-Test and the Brier Score}

\paragraph{The Brier Score Foundation}

The Brier score, originally proposed by Brier (1950) \parencite{brier1950}, measures the mean squared error between predicted probabilities and binary outcomes:

\begin{equation}
    \text{BS} = \frac{1}{n}\sum_{i=1}^{n}(\hat\pi_i - y_i)^2
\end{equation}

where $\hat\pi_i$ is the predicted probability for observation $i$, $y_i$ is the actual binary outcome (0 or 1), and $n$ is the total number of observations. Spiegelhalter’s z-test is derived by isolating the calibration component of the Brier score defined as
\begin{equation}
    \text{BS} = \frac{1}{n}\sum_{i=1}^n (\hat\pi_i - y_i)^2,
\end{equation}
where \(\hat\pi_i\) is the predicted probability and \(y_i\in\{0,1\}\) the outcome. Expanding and rearranging yields:
\begin{equation}
    \text{BS}
    = \frac{1}{n}\sum_{i=1}^n (y_i - \hat\pi_i)(1 - 2\hat\pi_i)
    + \frac{1}{n}\sum_{i=1}^n \hat\pi_i(1 - \hat\pi_i).
\end{equation}
Under perfect calibration \((E[y_i|\hat\pi_i]=\hat\pi_i)\), the first term has mean zero and variance
\begin{align}
    \text{E(BS)}   & = \frac{1}{n}\sum_{i=1}^n \hat\pi_i(1 - \hat\pi_i)                      \\
    \text{Var(BS)} & = \frac{1}{n^2}\sum_{i=1}^n \hat\pi_i(1 - \hat\pi_i)(1 - 2\hat\pi_i)^2.
\end{align}
Thus the z-statistic is
\begin{equation}
    Z =\frac{\text{BS - E(BS)}}{\sqrt{\text{Var(BS)}}} = \frac{\sum_{i=1}^n (y_i - \hat\pi_i)(1 - 2\hat\pi_i)}{\sqrt{\sum_{i=1}^n \hat\pi_i(1 - \hat\pi_i)(1 - 2\hat\pi_i)^2}},
\end{equation}
which follows \(N(0,1)\) under the null, providing a formal test of calibration.\parencite{spiegelhalter1986calibration, brier1950}

\paragraph{what is the purpose of the term $1 - 2E_i$ in the Z-test formula:}
The term $1 - 2E_i$ in the Spiegelhalter's z-test formula serves to account for the symmetry of binary outcomes around $E_i = 0.5$. Specifically, the binary outcome $O_i$ is either 0 or 1, while the predicted probability $E_i$ lies between 0 and 1. When $E_i$ is close to 0 or 1, deviations from the observed outcome $O_i$ are weighted more heavily to reflect the greater certainty implied by the extreme probabilities. In contrast, when $E_i$ is near 0.5, the model is more uncertain, and hence deviations contribute less to the z-statistic.

The factor $1 - 2E_i$ effectively adjusts the contribution of each observation to the test statistic based on the level of uncertainty associated with the predicted probability $E_i$. This makes the test more sensitive to discrepancies in model calibration, especially when predicted probabilities approach the extremes of 0 or 1, where calibration issues are most impactful.

\paragraph{Comparison with Pearson and Hosmer-Lemeshow Tests:}
In other goodness-of-fit tests, such as the Pearson chi-square test or the Hosmer-Lemeshow test, the test statistic involves summing the squared differences between observed and expected outcomes $(O_i - E_i)^2$. This squaring of residuals can increase the test statistic either due to many small residuals or a few large residuals.

However, squaring has a disproportionate effect: it reduces smaller residuals less dramatically compared to larger residuals, particularly when the residuals fall between 0 and 1. As a result, in these tests, even a few large deviations can have an outsized influence on the overall test statistic, possibly leading to rejection of the null hypothesis even when many observations show small discrepancies. Spiegelhalter's z-test, with its adjustment via $1 - 2E_i$, addresses this issue by appropriately weighting deviations based on the uncertainty in the predicted probabilities, thus offering a more nuanced evaluation of calibration.

\paragraph{Algorithm} The algorithmic implementation of Spiegelhalter's z-test for calibration assessment is presented in Algorithm \ref{alg:spiegelhalter_z_test} (see Appendix).

Spiegelhalter's z-test provides a robust measure of calibration, particularly in scenarios involving continuous covariates. Its main advantage lies in its focus on assessing whether predicted probabilities reflect observed outcomes, which is crucial for predictive accuracy in logistic regression models with sparse or ungrouped data.

\newpage

\subsection{eHL Test: E-value Based Goodness-of-Fit} \label{sec:ehl_test}

The eHL test, proposed by Henzi et al. (2024), evaluates the calibration of predicted probabilities using the framework of e-values. This approach avoids the fixed binning strategies used in the classical Hosmer-Lemeshow test by employing a data-adaptive strategy based on sample splitting and isotonic regression.

\subsubsection{Definition of E-values}
An e-variable is a non-negative random variable $E$ such that its expected value under the null hypothesis $H_0$ is bounded by one \parencite{ Shafer2021}. Formally, if $P$ denotes the probability distribution under $H_0$:
\begin{equation}
    E_P[E] \le 1, \quad \forall P \in H_0
\end{equation}
The realized value of an e-variable is called an e-value. In the context of hypothesis testing, a large e-value provides evidence against the null hypothesis. This can be interpreted in terms of betting: if a skeptic bets against $H_0$ starting with a capital of 1, the e-value represents the resulting capital. A high final capital indicates that the null hypothesis is likely false \parencite{Shafer2021, Grunwald2019}.

\subsubsection{The eHL Test Statistic}
The null hypothesis for the eHL test assumes the model is perfectly calibrated. The test constructs an e-variable by comparing the original model predictions ($\hat{\pi}_i$) against recalibrated probabilities ($q_i$) obtained via isotonic regression.\parencite{Henzi2024SafeHL}

The test procedure involves splitting the data into a training set and a test set. The training set is used to estimate a calibration function using isotonic regression (a non-parametric method that fits a non-decreasing function to the data) \parencite{barlow1972isotonic}. This function maps the original predictions $\hat{\pi}$ to recalibrated values $q$. 

On the independent test set, an e-variable $E_i$ is calculated for each observation $i$ as a likelihood ratio:
\begin{equation}
    E_i = \frac{q_i^{y_i}(1-q_i)^{1-y_i}}{\hat{\pi}_i^{y_i}(1-\hat{\pi}_i)^{1-y_i}}
\end{equation}
where $y_i \in \{0,1\}$ is the observed outcome. If the recalibrated probability $q_i$ describes the outcome better than the original prediction $\hat{\pi}_i$, the ratio $E_i$ will be greater than 1, accumulating evidence against the null hypothesis \cite{Henzi2024SafeHL}.

\subsubsection{Aggregation and Decision Rule}
Because the result depends on the random split of the data, the procedure is repeated $B$ times (e.g., $B=100$) to stabilize the result. The final test statistic $eHL$ is the arithmetic average of the e-values obtained from these splits:
\begin{equation}
    eHL = \frac{1}{B} \sum_{b=1}^{B} \left( \prod_{j \in D_{test}^{(b)}} E_j \right)
\end{equation}
To maintain compatibility with standard statistical reporting, the e-value can be converted into a p-value. Based on Markov's inequality, the probability that an e-variable $E$ exceeds a threshold $1/\alpha$ under the null hypothesis is at most $\alpha$ \cite{Shafer2021}. Therefore, a valid p-value is obtained by:
\begin{equation}
    p\text{-value} = \min\left(1, \frac{1}{eHL}\right)
\end{equation}
For a significance level of $\alpha = 0.05$, the null hypothesis is rejected if the aggregate $eHL$ statistic exceeds 20. (Check algorithm \ref{alg:ehl} for more details)

\section{The Bootstrap Methods : Approaching statistical inference} \label{sec:bootstrap_methods}

The bootstrap method, introduced by Bradley Efron in 1979, is a powerful resampling technique that has revolutionized statistical inference by providing a computationally intensive but conceptually simple approach to estimating sampling distributions and calculating p-values. This method has become particularly valuable in goodness-of-fit testing, where theoretical distributions may be complex, unknown, or inappropriate for the data at hand. \parencite{efron1979bootstrap}

\paragraph{The Bootstrap Method} is based on the principle of resampling with replacement from the observed data to estimate the sampling distribution of a statistic. The core idea is to treat the empirical distribution of the observed data as an approximation to the true underlying population distribution. By repeatedly sampling from this empirical distribution, we can generate multiple "bootstrap samples" and compute the statistic of interest for each sample, thereby approximating its sampling distribution.\parencite{efron1979bootstrap}

\paragraph{The Basic Bootstrap Algorithm}
The fundamental bootstrap procedure consists of the following steps \parencite{efron1993bootstrap} :

\begin{enumerate}
    \item \textbf{Original Sample}: Start with an observed sample $X = (x_1, x_2, \ldots, x_n)$ of size $n$.

    \item \textbf{Bootstrap Sampling}: Generate $B$ bootstrap samples ($b=1, \ldots, B$), each of size $n$, by sampling with replacement from the original data. Each bootstrap sample $X^*_b = (x^*_{b,1}, x^*_{b,2}, \ldots, x^*_{b,n})$ is created by randomly selecting $n$ observations from $X$, allowing for repeated selections.

    \item \textbf{Statistic Computation}: For each bootstrap sample $X^*_b$, compute the statistic of interest $\hat{\theta}^*_b = T(X^*_b)$.

    \item \textbf{Distribution Estimation}: Use the collection of bootstrap statistics $\{\hat{\theta}^*_1, \hat{\theta}^*_2, \ldots, \hat{\theta}^*_B\}$ to estimate the sampling distribution of $\hat{\theta}$.
\end{enumerate}

\subsection{Bootstrap as a Replacement for Theoretical Distributions}

Traditional statistical inference relies heavily on theoretical distributions (such as the normal, chi-square, or t-distributions) to calculate p-values and construct confidence intervals. However, these theoretical distributions often require strong assumptions about the underlying data-generating process, which may not hold in practice. The bootstrap method offers several advantages as a replacement:

\paragraph{Advantages of Bootstrap over Theoretical Distributions}

\begin{itemize}
    \renewcommand\labelitemi{--}
    \item \textbf{No Distributional Assumptions}: Bootstrap methods do not require assumptions about the underlying population distribution, making them more robust and widely applicable.

    \item \textbf{Handles Complex Statistics}: Bootstrap can be applied to any statistic, regardless of how complex or non-standard it may be, without needing to derive its theoretical distribution.

    \item \textbf{Small Sample Performance}: Bootstrap methods often perform better than asymptotic approximations in small samples, where theoretical distributions may not be accurate.

\end{itemize}

\paragraph{Empirical P-value Calculation}
One of the most important applications of bootstrap in hypothesis testing is the calculation of empirical p-values. Instead of relying on theoretical distributions, the bootstrap p-value is computed as:\parencite{efron1993bootstrap}

\begin{equation}
    p\text{-value} = \frac{1}{B} \sum_{b=1}^{B} I(T_b^* \geq T_{obs})
\end{equation}

where $T_{obs}$ is the observed test statistic, $T_b^*$ is the bootstrap test statistic for the $b$-th bootstrap sample, and $I(\cdot)$ is the indicator function. This empirical approach provides a data-driven method for assessing statistical significance without requiring knowledge of the theoretical null distribution.

\subsection{Model-Based Bootstrap in Goodness-of-Fit Testing}

In the context of goodness-of-fit testing, the model-based bootstrap is particularly valuable because it generates data under the null hypothesis (that the model fits well). This approach ensures that the bootstrap samples reflect the structure of the fitted model while preserving the observed covariate patterns. (i.e. BAGof Test, has approach to calculate p-value using bootstrap (Section \ref{sec:bagoft_test}))

\paragraph{Model-Based Bootstrap Procedure for Goodness-of-Fit Tests}

\begin{enumerate}
    \item \textbf{Fit the Model}: Estimate the parameters of the model under the null hypothesis using the original data.

    \item \textbf{Generate Bootstrap Responses}: For each observation, generate a new response from the fitted model's distribution (e.g., Bernoulli distribution for logistic regression with fitted probabilities $\hat{\pi}_i$).

    \item \textbf{Compute Bootstrap Statistics}: Fit the model to each bootstrap dataset and compute the test statistic of interest.

    \item \textbf{Calculate Empirical P-value}: Compare the observed test statistic to the distribution of bootstrap statistics to obtain the empirical p-value.
\end{enumerate}

\subsection{Bootstrap Applications in Goodness-of-Fit Tests}

The thesis presents several examples where bootstrap methods are employed to overcome the limitations of theoretical distributions in goodness-of-fit testing. These examples demonstrate the practical utility and flexibility of bootstrap approaches.

\subsubsection{(I)Bootstrap in the Stute-Zhu Test}

The Stute-Zhu test, as described in (Section \ref{sec:stute_zhu_test}), employs a model-based bootstrap procedure to determine the significance of its test statistic. The asymptotic null distribution of the Stute-Zhu statistic is complex and data-dependent, making theoretical p-value calculation impractical. The bootstrap approach provides a computationally feasible solution in this case (Algorithm \ref{alg:tsz_bootstrap}). \parencite{stute2002model}
This bootstrap approach allows the Stute-Zhu test to be applied without requiring knowledge of the complex asymptotic distribution of its test statistic.

\subsubsection{(II)Bootstrap in the Projection-Based Test}

The projection-based test, an extension of the Stute-Zhu test, also relies heavily on bootstrap methods. As described in (Section \ref{sec:projection_test}), this test integrates checks over all possible projection directions, resulting in a test statistic with an intractable theoretical distribution (Algorithm \ref{alg:projection_bootstrap}).

The computational complexity of the projection-based test (O($n^3$)) makes the bootstrap approach particularly valuable, as it provides a practical method for significance testing without requiring theoretical distributional results.

\subsubsection{(III)Bootstrap in the Lai \& Liu Standardized H-L Test}

The Lai \& Liu (Section \ref{sec:lai_liu_test}) procedure for standardizing the Hosmer-Lemeshow test power also employs bootstrap methods, but in a different context. Here, bootstrap is used not for p-value calculation, but for power estimation in this case (Algorithm \ref{alg:lailiu_bootstrap}).This example demonstrates how bootstrap can be used creatively to address specific challenges in goodness-of-fit testing, such as the problem of statistical overpowering in large samples.

\subsection{Practical Considerations and Limitations}

While bootstrap methods offer significant advantages, they also present several practical considerations \parencite{efron1993bootstrap}:

\paragraph{Computational Cost}
Bootstrap methods require substantial computational resources, especially for large datasets or complex statistics. The number of bootstrap replications $B$ should be large enough to provide stable estimates (typically 1000 or more for p-value calculation).

\paragraph{Consistency and Convergence}
Bootstrap methods work especially well with large datasets, where they tend to give results that are close to the true values (this is called asymptotic consistency). However, when the sample size is small, the accuracy of bootstrap results can vary. In these cases, how well the bootstrap works depends on how representative the sample is of the whole population and how complicated the statistic being estimated is. In short, bootstrap methods are most reliable with large samples, but with smaller samples, their effectiveness can be affected by the quality of the data and the complexity of the analysis.

\paragraph{Implementation Challenges}
Proper implementation of bootstrap methods requires careful attention to details such as random number generation, parallel processing for efficiency, and appropriate handling of edge cases. \newline

\paragraph{Conclusion} \textit{The bootstrap method represents a fundamental shift in statistical inference, moving from reliance on theoretical distributions to data-driven approaches. In the context of goodness-of-fit testing, bootstrap methods have enabled the development and practical application of tests that would otherwise be computationally intractable or theoretically complex. The examples from this thesis demonstrate the versatility and power of bootstrap approaches in addressing real-world challenges in model assessment and validation. As computational resources continue to improve, bootstrap methods are likely to become even more prevalent in statistical practice, providing researchers with flexible and robust tools for statistical inference without the constraints of traditional distributional assumptions. \textbf{Now let's dive into the algorithms.}}

\subsection{A Binary Regression Adaptive Goodness-of-fit Test (BAGofT) } \label{sec:bagoft_test}

The Binary Regression Adaptive Goodness-of-fit Test (BAGofT), introduced by Zhang et al., addresses the limitations of traditional goodness-of-fit tests for binary regression models, particularly in scenarios involving continuous covariates. BAGofT employs a two-stage approach that combines adaptive data partitioning with chi-square testing to enhance power while maintaining proper size control. \parencite{BAGofT2019}

\subsubsection{Methodology}

BAGofT operates through the following key steps:

\begin{enumerate}
    {\small
    \item \textbf{Data Splitting}: The dataset $D$ is randomly divided into a training set $D_{n_1}$ of size $n_1$ and a test set $D_{n_2}$ of size $n_2$.

    \item \textbf{Model Fitting}: The model to assess (MTA) is fitted on the training set, yielding an estimated regression function $\hat{\pi}_{D_{n_1}}(\cdot)$.

    \item \textbf{Adaptive Partitioning}: Using the training data $D_{n_1}$, an adaptive partition \\
          $\{\hat{G}_{D_{n_1},1}, \ldots, \hat{G}_{D_{n_1},K}\}$ of the covariate space $\mathbb{S}$ is generated, for $K$ partitions.

    \item \textbf{Test Statistic Computation}: The BAGofT statistic is calculated on the test set:

          \begin{equation}
              \mathrm{BAG} = \sum_{k=1}^K \left(\frac{\sum_{\{i: \boldsymbol{x}_{e,i} \in \hat{G}_{D_{n_1},k}\}} R_i}{\sqrt{\sum_{\{i: \boldsymbol{x}_{e,i} \in \hat{G}_{D_{n_1},k}\}} \sigma_i^2}}\right)^2
          \end{equation}

          where $R_i = y_{e,i} - \hat{\pi}_{D_{n_1}}(\boldsymbol{x}_{e,i})$ and $\sigma_i^2 = \hat{\pi}_{D_{n_1}}(\boldsymbol{x}_{e,i})\{1-\hat{\pi}_{D_{n_1}}(\boldsymbol{x}_{e,i})\}$.
          }
\end{enumerate}

Here, $\boldsymbol{x}_{e,i}$ represents the covariate vector for the $i$-th observation in the test set. The test statistic is computed using these test set observations, grouped according to the partition derived from the training data.

\subsubsection{Adaptive Partitioning Schemes on Trained Data}

BAGofT's power stems from its flexible partitioning approach. The authors propose several schemes, including:

\begin{enumerate}
    \item \textbf{Grid-based partitions}: The covariate space is divided using a grid formed by quantiles of each covariate. This method is particularly useful for continuous covariates and can capture non-linear relationships.

    \item \textbf{Tree-based greedy adaptive partition selection}: This approach uses a decision tree algorithm to recursively partition the covariate space. It can efficiently handle interactions between covariates and automatically select relevant features for partitioning.

    \item \textbf{Partitions based on fitted probabilities}: This method uses quantiles of fitted probabilities from non-parametric models (e.g., random forests or neural networks) to create partitions. It can capture complex relationships that might be missed by simpler partitioning schemes.
\end{enumerate}

The optimal partition is selected using a criterion that maximizes the discrepancy between observed and fitted values:

\begin{equation}
    \mathcal{B}_q = \sum_{k=1}^K \left(\frac{\sum_{\{i: \boldsymbol{x}_{t,i} \in G_k^q\}} \{y_{t,i} - \hat{\pi}_{D_{n_1}}(\boldsymbol{x}_{t,i})\}}{\sqrt{\sum_{\{i: \boldsymbol{x}_{t,i} \in G_k^q\}} \hat{\pi}_{D_{n_1}}(\boldsymbol{x}_{t,i})\{1-\hat{\pi}_{D_{n_1}}(\boldsymbol{x}_{t,i})\}}}\right)^2
\end{equation}

In this equation, $\mathcal{B}_q$ represents the sum of squared standardized differences between observed and fitted values for a given partition $q$. The numerator sums the differences between observed outcomes $y_{t,i}$ and fitted probabilities $\hat{\pi}_{D_{n_1}}(\boldsymbol{x}_{t,i})$ for all observations in group $k$. The denominator standardizes this sum by the estimated variance of the Bernoulli outcomes in that group. The partition that maximizes $\mathcal{B}_q$ is chosen as it is most likely to reveal discrepancies between the model and the data.

\subsubsection{Practical Considerations (Splitting)}

To mitigate the randomness introduced by data splitting, BAGofT employs multiple splits and combines the results. The authors suggest using the median of p-values from multiple splits and rejecting the null hypothesis when this median is smaller than the 0.05 quantile of $\mathcal{N}(0.5, 1/(12S))$, where $S$ is the number of splits.

BAGofT offers a powerful and flexible approach to goodness-of-fit testing for binary regression models, particularly in scenarios involving continuous covariates where traditional methods may fail. Its adaptive nature allows for the detection of various types of model misspecification, providing researchers with a robust tool for model assessment in complex data environments.

\paragraph{Algorithm} The algorithmic implementation of the BAGofT test with adaptive partitioning is detailed in Algorithm \ref{alg:bagoft_test} (see Appendix).

\subsubsection{Bootstrap p-value (\texttt{Simulation})}
While the previous approach—taking the median of the p-values and making a decision directly or by comparing it to the 0.05 quantile of a normal distribution—can be used, it is not the most effective implementation of the BAGofT test. The authors recommend a more robust method: rather than using the median p-value as the final decision criterion, it should be used to compute an empirical p-value through a bootstrap procedure. This approach, as illustrated in Algorithm~\ref{alg:bagoft_bootstrap_pvalue}, aligns with the methodology described earlier (Algorithm~\ref{alg:bagoft_test}) and provides a more accurate assessment of model fit.

\paragraph{Algorithm} The bootstrap-based implementation of BAGofT for empirical p-value calculation is provided in Algorithm \ref{alg:bagoft_bootstrap_pvalue} (see Appendix).

\paragraph{Limitations and Practical Considerations}

While BAGofT offers significant advantages in detecting model misspecification, it has important limitations that researchers must consider when applying it in practice. The most critical limitation concerns the requirement for knowledge of potentially omitted variables.

\subparagraph{The Omitted Variable Problem}
A fundamental constraint of BAGofT is that the test can only detect misspecification related to variables that are included in the partitioning algorithm. As noted by the original authors, "For the covariates we choose to generate the partitions, we should not only consider the covariates in the model, but also the outside covariates, in order to find out whether there can be a further improvement of the model." This requirement presents a significant practical challenge because:

\begin{itemize}
    \item \textbf{Unknown Omitted Variables}: In real-world applications, researchers typically do not know which variables have been omitted from the model. The very nature of omitted variable bias means that these variables are unobserved or unmeasured, making it impossible to include them in the partitioning strategy. However, this omitted variable is only known if the deleted variable is polynomial or interaction of the included variables.

    \item \textbf{Detection Dependency}: BAGofT's ability to detect misspecification is entirely dependent on the researcher's ability to identify and include the relevant variables in the partitioning algorithm. If important omitted variables are not included in the partitioning, the test may fail to detect the resulting misspecification.

    \item \textbf{Theoretical vs. Practical Gap}: While the test is theoretically powerful when the correct variables are used for partitioning, this theoretical advantage is often unattainable in practice where the true data-generating process is unknown.
\end{itemize}

\subparagraph{BAGofT and the Single Predictor Limitation}
In the course of implementing the Binary Regression Adaptive Goodness-of-fit Test (BAGofT) for logistic regression models, an important practical limitation was encountered: the BAGofT R package raises an error when the dataset includes only a single predictor variable. Specifically, even if the covariate is present in the dataset and correctly specified in the model formula, the function will return an "object not found" error at prediction or data manipulation stages. This behavior appears to arise from the way the package handles variable environments during internal data partitioning and the scoping of variable names, which does not gracefully handle single-variable inputs. Notably, this issue is not mentioned in the original BAGofT methodological publication, which focuses exclusively on models with multiple covariates and provides no implementation warnings or examples for the one-predictor case.

A pragmatic workaround is to artificially introduce an additional "dummy" variable—such as a column full of ones or zeros—to the dataset. As long as this variable is not included in the model formula, it has no statistical impact on parameter estimation or model fit, but it ensures the package functions correctly. While this solution maintains computational integrity, it highlights an unaddressed edge case in the original algorithm’s implementation and its documentation. \newline

This fundamental limitation highlights the importance of understanding that no single goodness-of-fit test can provide a complete assessment of model adequacy. The effectiveness of BAGofT, like many statistical methods, depends critically on the quality of the input information and the researcher's understanding of the underlying data-generating process.

\subsection{The Stute-Zhu Test} \label{sec:stute_zhu_test}
The test proposed by Stute and Zhu (2002) is a powerful goodness-of-fit (GoF) procedure for generalized linear models that avoids the arbitrary binning required by traditional tests like the Hosmer-Lemeshow statistic. Its core idea is to check for systematic patterns in the model's residuals by ordering them along a single, intuitive direction: the model's own linear predictor.\parencite{stute2002model}

\paragraph{The Test Statistic}
Let $\hat{\epsilon}_i = Y_i - \hat{\pi}_i$ be the residual for observation $i$, and let $\hat{\eta}_i = X_i^T\hat{\beta}$ be the estimated linear predictor from the fitted logistic regression model. The test first constructs an empirical process, $R_n(u)$, which represents the cumulative sum of residuals for all observations whose linear predictor value is less than or equal to $u$. The test statistic, $T_{SZ}$, is a Cramér-von Mises type statistic which measures the average squared deviation of this cumulative residual process from zero. It is formally defined as:
\begin{equation}
    T_{SZ} = \int_{-\infty}^{\infty} \left[ R_n(u) \right]^2 dF_n(u)
\end{equation}
where $F_n(u)$ is the empirical distribution function of the linear predictors. In practice, this simplifies to a computationally efficient formula based on the sorted residuals.

\paragraph{Algorithm and Bootstrap Procedure}
The asymptotic null distribution of $T_{SZ}$ is complex and data-dependent. Therefore, as recommended by Liu et al. (2024), its significance is determined using a Model-Based Bootstrap (MBB) procedure.

\paragraph{Algorithm} The complete Stute-Zhu test procedure with model-based bootstrap for p-value calculation is outlined in Algorithm \ref{alg:tsz_bootstrap} (see Appendix).

The primary limitation of this test is its reliance on a single projection direction. While powerful, it may fail to detect complex forms of model misspecification that are not apparent along the axis of the linear predictor.

\subsection{The Projection-Based Test} \label{sec:projection_test}
The projection-based test, proposed by Liu et al. (2024), is a direct and powerful extension of the Stute-Zhu test. It is specifically designed to overcome the single-direction limitation by checking for systematic residual patterns along \textit{all possible directions} in the covariate space. This makes the test consistent, meaning it is guaranteed to detect any form of misspecification given a sufficiently large sample size. \parencite{liu2024comprehensive}

\paragraph{Algorithm and Bootstrap Procedure}
The calculation of the matrix $\mathbf{A}$ is computationally intensive, with a complexity of $O(n^3)$. As with the Stute-Zhu test, the null distribution of $T_{proj}$ is intractable, necessitating a bootstrap procedure for p-value calculation.

\paragraph{Algorithm} The projection-based test algorithm with bootstrap procedure for comprehensive model assessment is detailed in Algorithm \ref{alg:projection_bootstrap} (see Appendix).

\paragraph{The Test Statistic}
Instead of checking for patterns along a single direction, this test integrates the checks over all possible projection directions. A projection direction is defined by a unit vector $W$ on the $p$-dimensional unit sphere. The final test statistic, $T_{proj}$, can be expressed in a convenient computational form as a quadratic form:
\begin{equation} \label{eq:projection_test}
    T_{proj} = \frac{1}{n^2} \hat{\boldsymbol{\epsilon}}^T \mathbf{A} \hat{\boldsymbol{\epsilon}}
\end{equation}
where $\hat{\boldsymbol{\epsilon}}$ is the vector of residuals and $\mathbf{A}$ is a non-negative definite $n \times n$ matrix. Each element $A_{ij}$ of this matrix represents an aggregation over all projection directions. Following Equation (\ref{eq:projection_test}) from Liu et al. (2024), the components can be calculated as:
\begin{equation} \label{eq:A_ij}
    A_{ij} = \sum_{l=1}^{n} A_{ijl} \quad \text{where} \quad A_{ijl} = C_p \left( \pi - \arccos\left(\frac{(X_i-X_l)^T(X_j-X_l)}{\|X_i-X_l\| \|X_j-X_l\|}\right) \right)
\end{equation}
and $C_p = \pi^{(p/2-1)} / \Gamma(p/2+1)$ is a constant dependent on the number of model parameters, $p$. The key advantage of the projection-based test is its consistency and superior power against complex alternatives, as demonstrated in the simulation studies by Liu et al. (2024). Its primary drawback is the high computational cost, which makes efficient implementation and powerful hardware (e.g., GPUs) important for practical application in simulation studies.

\subsection{The Lai \& Liu Standardized Power Procedure for the H-L Test} \label{sec:lai_liu_test}
The Hosmer-Lemeshow (H-L) test is one of the most widely used methods for assessing the goodness-of-fit of logistic regression models due to its straightforward implementation. However, it suffers from a significant drawback: its statistical power is highly sensitive to sample size. In very large datasets ($n \gg 1000$), the H-L test can detect even trivial deviations between predicted probabilities and observed outcomes, leading to a statistically significant result (i.e., a small p-value) and the conclusion of "poor fit," even when the model is practically useful. \parencite{Lai2018}

To address this issue of statistical overpowering, Lai and Liu (2018) proposed a simple but clever procedure that does not modify the H-L test itself, but instead standardizes its power. The core idea is to reframe the research question from "Is this model a perfect fit for the data?" to a more relevant one: \textbf{"Does this model fit the large dataset as well as a good model is expected to fit a smaller, 'standard-sized' dataset?"}

\paragraph{The Core Idea: Power Standardization via Resampling}
The procedure reframes the goodness-of-fit assessment by comparing the model's performance to a target scenario. The core idea is to use resampling to simulate the power that the H-L test would have on a smaller, pre-specified "standard" sample size, $n_0$ (e.g., $n_0=500$ or $n_0=1000$). The final decision about the model's fit is then made probabilistically, based on this standardized power.

\paragraph{Interpretation of the Result}
The key output of this procedure is the \textbf{standardized power}, $p_k$. This value is not a p-value. It is an estimate of statistical power, and for this test, a low value is desirable.
\begin{itemize}
    \item \textbf{A low standardized power} (e.g., $p_k < 0.20$) suggests that while the H-L test might have been significant on the original large dataset, the actual magnitude of the misfit is so small that it would likely go undetected in a standard-sized study. In this case, the researcher can conclude that the model's fit is acceptable for practical purposes.
    \item \textbf{A high standardized power} (e.g., $p_k > 0.80$) is a strong indictment of the model. It implies that the lack of fit is so severe that it would be easily detected even in a small sample. This provides strong evidence that the model is poorly calibrated and should be rejected or revised.
\end{itemize}

\paragraph{The Proposed Test Procedure and Algorithm}
The test uses a bootstrap method to estimate the power of the H-L test under the target sample size $n_0$. This estimated power then becomes the probability of rejecting the null hypothesis for the original large dataset.

\paragraph{Algorithm} The detailed procedure for the Lai and Liu standardized power approach to address overpowering in large datasets is presented in Algorithm \ref{alg:lailiu_bootstrap} (see Appendix).

By standardizing the power, this procedure effectively recalibrates the H-L test, making it a viable and interpretable tool for assessing goodness-of-fit in the era of big data. \parencite{Lai2018}

\paragraph{Choosing $n_0$:}
The parameter $n_0$ acts as a tuning knob that controls the strictness of the test, much like adjusting the focus of a magnifying glass. If $n_0$ is set to a small value (for example, 100), the test becomes very lenient: only models with extremely poor fit are likely to be rejected, since even a small, underpowered study would struggle to detect subtle misfit. This leniency, however, comes with the risk of overlooking practically important deficiencies in model fit. On the other hand, choosing a large $n_0$ (such as 5000) makes the test highly stringent, closely resembling the original problem of excessive power in large samples—here, even minor and clinically unimportant deviations from the model may lead to rejection. To strike a balance between these extremes, it is generally recommended to select $n_0$ within a "Goldilocks" range, typically between 500 and 1000, where the test maintains both robustness and practical relevance.

\section*{Conclusion}
\vspace{0.5em}
\noindent
This chapter has highlighted the central role of \textbf{model calibration} in modern machine learning, emphasizing that trustworthy probability estimates are essential for responsible decision-making. Sometimes, while classical goodness-of-fit tests remain valuable, their practical utility can be limited in complex, high-dimensional, or non-standard data environments—precisely the settings where machine learning models are most often deployed. As a result, there is a growing need for advanced, flexible, and computationally robust calibration assessment methods that can adapt to the challenges posed by contemporary data science.

\vspace{0.5em}
\noindent
A key insight developed throughout this chapter is the \textbf{unification the concepts} of statistical goodness-of-fit testing and machine learning calibration assessment. Despite differences in terminology and tradition, both frameworks fundamentally address the same question: \emph{Do the model's predicted probabilities faithfully represent the true frequencies of observed outcomes?} By bridging these perspectives, we have established a comprehensive foundation for evaluating model adequacy that leverages the strengths of both classical statistics and modern machine learning.

\vspace{0.5em}
\noindent
\paragraph{In the next chapter,} we will discuss the detailed methodology and simulation study that underpin the comparative evaluation of goodness-of-fit tests and machine learning calibration methods for logistic regression. This includes a thorough description of the simulation design, the scenarios considered (such as varying sample sizes, covariate distributions, and types of model misspecification), and the rationale for adopting the framework established by Hosmer et al.~(1997). By systematically outlining the experimental setup, parameter choices, and evaluation criteria, the forthcoming chapter provides the foundation for the empirical results and insights presented later in the thesis.

\chapter{Simulation Study and Real-Data-Application}

This chapter details the simulation study designed to evaluate and compare various goodness-of-fit tests and machine learning calibration methods for logistic regression models. The study systematically assesses test performance across different scenarios, including varying sample sizes, covariate distributions, and types of model misspecification. The design is heavily influenced by the foundational work of Hosmer and colleagues, ensuring a rigorous and relevant comparison.

\section{Simulation Design}

\subsection{Methodology Based on Hosmer et al. (1997)}
The framework for this simulation study is methodologically grounded in the seminal paper by D. W. Hosmer, T. Hosmer, S. Le Cessie, and S. Lemeshow, titled \textit{``A Comparison of Goodness-of-Fit Tests for the Logistic Regression Model,''} published in \textit{Statistics in Medicine} (1997). By adopting their rigorous and comprehensive approach, we ensure that our evaluation is built upon a well-established and respected standard for comparing goodness-of-fit tests.\parencite{Hosmer1997}
\vspace{1em}

\noindent This established framework provides a robust basis for assessing not only the classic tests but also for extending the comparison to more contemporary methods developed since 1997. It allows for a direct and meaningful evaluation of how newer tests perform against the benchmarks and conditions originally proposed, particularly in challenging scenarios involving continuous covariates and sparse data.

\vspace{1em}
\noindent Adopting this simulation framework provides multiple strategic advantages for this thesis. Most importantly, it grounds the comparative evaluation of goodness-of-fit tests in a methodology that is widely recognized as a benchmark within the statistical and applied literature~\parencite{Hosmer1997}. By leveraging their well-designed scenarios—including investigating the null distribution of the test statistics, the omission of quadratic terms, interaction effects — this thesis ensures that new results are directly \textbf{standardized to a respected and reproducible reference}. This allows for transparent and meaningful comparisons both across tests and in relation to much of the published work, thereby maximizing the academic impact of this study.

\vspace{1em}
\noindent A key benefit of this approach is the use of simulation settings that reliably represent real-world challenges encountered when fitting logistic regression models. These settings have been proven to stress-test the sensitivity, size, and power properties of different goodness-of-fit tests under realistic conditions. By employing the same scenarios, this research provides clear and interpretable evidence on how novel or existing GOF methods perform relative to established standards. This in turn enhances the generalizability and credibility of the conclusions. \parencite{Hosmer1997}

\vspace{1em}
\noindent Nonetheless, there are inherent limitations in strictly following the Hosmer et al.~(1997) design. While it enables valuable comparability, it also means that the thesis inherits some of the constraints of the original paper—such as the \textbf{reliance on independent and identically distributed (IID) data} \parencite{Archer2007}, focus on \textbf{low-dimensional classical logistic regression}, and \textbf{acknowledged limitations in detecting certain types of model misspecification, especially in small   samples} \parencite{MHL2021}. (Hosmer use 200 and 500 only, however this thesis use 200, 500, 1000, 2000, 5000). As a result, the thesis may not directly address more modern issues arising in high-dimensional, penalized, or clustered data settings.

\vspace{1em}
\noindent In summary, by adopting the well-tested simulation scenarios of Hosmer et al.~(1997), this thesis achieves methodological rigor, facilitates direct comparisons with foundational research, and produces findings that are both robust and relevant for current statistical practice. At the same time, this methodology has been used for a lot of PhDs (e.g.\ \cite{nygaard2019simulation})  and academic papers (e.g.\ \cite{liu2024comprehensive}, \cite{canary2015comparison}, and \cite{hosmer2002goodness}), therefore it is a good start for this thesis.

\subsection{Study Parameters}
The simulation study examines the following key parameters:
\begin{itemize}
    \item \textbf{Sample sizes}: $n \in \{200, 500, 1000, 2000, 5000\}$
    \item \textbf{Number of replications}: 10,000 per scenario
    \item \textbf{Significance level}: $\alpha = 0.05$
\end{itemize}

\subsection{Simulation Structure: Two Experimental criteria}
To provide a comprehensive evaluation, the simulation is structured around two distinct experimental goals, each with its own set of models and objectives. This distinction is critical for understanding the source and purpose of the logistic coefficients used in the following sections.

\begin{enumerate}
    \item \textbf{Type I Error Analysis:} The first goal is to assess the ``size'' of each test, which is its probability of making a Type I error. This involves generating data from a \textbf{correctly specified model} and checking if the tests maintain the nominal significance level (e.g., rejecting the null hypothesis of good fit approximately 5\% of the time). For this experiment, the model coefficients are \textbf{pre-defined} as shown in Section \ref{sec:type1_scenarios}.

    \item \textbf{Power Analysis:} The second goal is to assess the statistical power of each test, which is its ability to correctly detect a model misspecification. This involves generating data from a known, more complex ``true'' model and fitting an \textbf{incorrectly specified, simpler model}. To precisely control the degree of misspecification, the coefficients for the true model in these scenarios are not pre-defined; instead, they are \textbf{calculated} to satisfy specific conditions, as detailed in Section \ref{sec:power_scenarios}.
\end{enumerate}

\section{Scenarios for Empirical Type I Error Analysis (Null Distribution)}
\label{sec:type1_scenarios}
This experiment evaluates test performance when the fitted model is the correct model. A key aspect of this is to generate data under various distributional conditions. The choice of covariate distribution and the pre-defined logistic model coefficients directly influences the distribution of the estimated probabilities, $\hat{\pi}(x)$, which can range from being symmetric to highly skewed. Table \ref{tab:sim_situations}, adapted from prior research, outlines these scenarios.

\begin{table}[htbp]
    \centering
    \caption{Simulation: Scenarios Used to Examine the Null Distribution of Test Statistics}
    \label{tab:sim_situations}
    \begin{tabular}{@{}llccc@{}}
        \toprule
        \textbf{Covariate Distribution}                                                  & \textbf{Logistic Coefficients}                                                                    & \multicolumn{3}{c}{\textbf{Probabilities (n=100)}}                                      \\
        \cmidrule(l){3-5}
                                                                                         &                                                                                                   & \textbf{Q1}                                        & \textbf{Q2 (Median)} & \textbf{Q3} \\
        \midrule
        $U(-6, 6)$                                                                       & $\beta_0=0, \beta_1=0.8$                                                                          & 0.087                                              & 0.5                  & 0.913       \\
        $U(-3, 3)$                                                                       & $\beta_0=0, \beta_1=0.8$                                                                          & 0.231                                              & 0.5                  & 0.769       \\
        $N(0, 1.5)$                                                                      & $\beta_0=0, \beta_1=0.8$                                                                          & 0.304                                              & 0.5                  & 0.696       \\
        $\chi^2(4)$                                                                      & $\beta_0=-4.9, \beta_1=0.65$                                                                      & 0.025                                              & 0.062                & 0.202       \\
        \begin{tabular}[c]{@{}l@{}}Indep. U(-6,6), \\ N(0,1.5), $\chi^2(4)$\end{tabular} & \begin{tabular}[c]{@{}l@{}}$\beta_0=-1.3, \beta_1=\beta_2=0.8/3,$\\ $\beta_3=0.65/3$\end{tabular} & 0.204                                              & 0.386                & 0.608       \\
        \bottomrule
    \end{tabular}
\end{table}

These scenarios create distinct patterns in the resulting probabilities:
\begin{itemize}
    \item \textbf{Symmetric Probabilities (Median = 0.5)}: Generated using Uniform or Normal covariates with $\beta_0=0$.
    \item \textbf{Highly Skewed Probabilities (Median = 0.062)}: Using a Chi-squared covariate ($\chi^2(4)$), leading to inflated zeros and rare ones of $\hat{y}$. Probably, this setting may give rise to outliers in the tail of the distribution. This is an important scenario for the simulation study.
    \item \textbf{Symmetric, Non-centered Probabilities (Median = 0.386)}: Using a combination of multiple independent covariates. Some tests make good use of the covariate space for partitioning data, and some tests use predicted probabilities; this scenario is important for those who only use predicted probabilities to check their performance.
\end{itemize}

\newpage
\section{Scenarios for Empirical Power Analysis}
\label{sec:power_scenarios}
To evaluate the power of the tests, we introduce three types of model misspecification. For these scenarios, the true data-generating model is more complex, and its coefficients are calculated to create specific, controlled degrees of error when a simpler model is fitted.

\subsubsection{5.3.1 Omitted Quadratic Term} \label{sec:omitted_quadratic_term}
This scenario evaluates the power of tests to detect non-linearity. Data is generated from a model containing a quadratic term, but the fitted model incorrectly omits it.
\begin{itemize}
    {
    \item \textbf{True Model}: $g(x) = \beta_0 + \beta_1x + \beta_2x^2$
    \item \textbf{Fitted Model}: $g(x) = \beta_0 + \beta_1x$
          }
\end{itemize}
Following Hosmer et al. (1997), where $x \sim U(-3, 3)$, we control the degree of non-linearity by defining the true quadratic model based on three fixed points, which allows for the direct calculation of the coefficients $\beta_0, \beta_1,$ and $\beta_2$. The fixed points are:
\noindent
\begin{center}
    \begin{tabular}{l l l l l}

        \textbullet~$\pi(-1.5) = 0.05$ &  & \textbullet~$\pi(3.0) = 0.95$ &  & \textbullet~$\pi(-3.0) = J$ \\
    \end{tabular}
\end{center}

\noindent We examine two levels of misspecification by using extreme values for $J$:
\begin{itemize}
    {\small
    \item \textbf{Slight Non-linearity}: $J = 0.01$
    \item \textbf{Pronounced Non-linearity}: $J = 0.4$
          }
\end{itemize}

\paragraph{Calculating the Coefficients}
For each value of $J$, the coefficients are the unique solution to the system of three linear equations derived by applying the logit transformation, $\text{logit}(p) = \ln(p/(1-p))$,
to the fixed points\footnote{Calculated as $\text{logit}(0.05) = \ln(0.05 / (1-0.05)) = \ln(0.05/0.95) \approx -2.944$.}:

\begin{align*}
    \beta_0 + (-1.5)\beta_1 + (-1.5)^2 \beta_2 & = \text{logit}(0.05) \approx -2.944 \\
    \beta_0 + (3.00)\beta_1 + (3.00)^2 \beta_2 & = \text{logit}(0.95) \approx 2.944  \\
    \beta_0 + (-3.0)\beta_1 + (-3.0)^2 \beta_2 & = \text{logit}(J)
\end{align*}

Solving this system yields the coefficients for each scenario:
\begin{itemize}
    \item For \textbf{slight non-linearity ($J=0.01$)}, where $\text{logit}(0.01) \approx -4.595$, the solution is:
          \[ \beta_0 \approx -1.138, \quad \beta_1 \approx 1.257, \quad \beta_2 \approx 0.035 \]
          Here, the small quadratic coefficient ($\beta_2$) indicates a subtle departure from linearity.

    \item For \textbf{pronounced non-linearity ($J=0.4$)}, where $\text{logit}(0.4) \approx -0.405$, the solution is:
          \[ \beta_0 \approx -3.232, \quad \beta_1 \approx 0.558, \quad \beta_2 \approx 0.500 \]
          In this case, the much larger quadratic coefficient signifies a substantial and more easily detectable curvature.
\end{itemize}

\subsubsection{5.3.2 Omitted Interaction Term} \label{sec:omitted_interaction_term}
This scenario assesses the ability of goodness-of-fit tests to detect an omitted interaction term between a continuous and a dichotomous covariate.
\begin{itemize}
    \item \textbf{True Model}: $g(x,d) = \beta_0 + \beta_1x + \beta_2d + \beta_3xd$
    \item \textbf{Fitted Model}: $g(x,d) = \beta_0 + \beta_1x + \beta_2d$
\end{itemize}
The magnitude of the interaction is controlled by defining the true model via a set of four fixed points. The covariates are a continuous variable, $x \sim U(-3, 3)$, and a dichotomous variable, $d \sim \text{Bernoulli}(1/2)$. The fixed points are:
\begin{itemize}
    \item $\pi(-3, 0) = 0.1$
    \item $\pi(-3, 1) = 0.1$
    \item $\pi(3, 0) = 0.2$
    \item $\pi(3, 1) = 0.2 + I$
\end{itemize}
We examine two levels of misspecification using extreme values for $I$:
\begin{itemize}
    \item \textbf{Slight Interaction}: $I = 0.1$
    \item \textbf{Pronounced Interaction}: $I = 0.7$
\end{itemize}

\paragraph{Calculating the Coefficients}
For each value of $I$, the four coefficients are the unique solution to the system of four linear equations derived from the logit-transformed fixed points:
\begin{align*}
    \beta_0 - 3\beta_1 \qquad \qquad        & = \text{logit}(0.1) \approx -2.197 \\
    \beta_0 - 3\beta_1 + \beta_2 - 3\beta_3 & = \text{logit}(0.1) \approx -2.197 \\
    \beta_0 + 3\beta_1 \qquad \qquad        & = \text{logit}(0.2) \approx -1.386 \\
    \beta_0 + 3\beta_1 + \beta_2 + 3\beta_3 & = \text{logit}(0.2 + I)
\end{align*}
This system is solved to find the coefficients for each scenario:
\begin{itemize}
    \item For \textbf{slight interaction ($I=0.1$)}, the solution is:
          \[ \beta_0 \approx -1.792, \quad \beta_1 \approx 0.135, \quad \beta_2 \approx 0.270, \quad \beta_3 \approx 0.090 \]
          The small interaction coefficient ($\beta_3$) represents a weak effect.

    \item For \textbf{pronounced interaction ($I=0.7$)}, the solution is:
          \[ \beta_0 \approx -1.792, \quad \beta_1 \approx 0.135, \quad \beta_2 \approx 1.791, \quad \beta_3 \approx 0.597 \]
          The substantially larger interaction coefficient ($\beta_3$) signifies a strong effect.
\end{itemize}

\subsubsection{5.3.3 Link Function Misspecification} \label{sec:link_function_misspecification}
The scenario of link function misspecification, such as generating data using Stukel's generalized logistic model and fitting a standard logistic model, is commonly used to assess the sensitivity of goodness-of-fit tests to deviations from the canonical logit link. In this approach, the true model is given by:
\begin{itemize}
    \item \textbf{True Model}: $\text{logit}(\pi_i) = \eta_i + \alpha_1 \eta_i^2 I(\eta_i > 0) + \alpha_2 \eta_i^2 I(\eta_i \leq 0)$, with $\eta_i = 0.8x$.
    \item \textbf{Fitted Model}: Standard logistic model (where $\alpha_1 = \alpha_2 = 0$).
\end{itemize}
Five different link functions are typically examined by varying the $\alpha$ parameters to simulate departures from the logit link. However, a detailed investigation of this type of \textbf{link function misspecification} was mentioned by Hosmer et al. (1997) \textit{but is \textbf{beyond the scope} of the present research}.

\section{Data Generation and Analysis Framework}
For every scenario in both the Type I error and power experiments, the general process for each of the 10,000 replications is as follows:
\begin{enumerate}
    \item \textbf{Generate Covariates}: Draw covariate values $X$ from the specified distribution(s).

    \item \textbf{Calculate True Probabilities}: Use the logistic function $\pi(x) = e^{g(x)} / (1 + e^{g(x)})$, where $g(x)$ is the linear predictor for the relevant true model.

    \item \textbf{Generate Binary Outcomes}: For each observation $i$, generate a uniform random number $u_i \sim U(0,1)$ and set the binary outcome $y_i$ as:
          \begin{equation}
              y_i =
              \begin{cases}
                  1 & \text{if } u_i \leq \pi(x_i) \\
                  0 & \text{if } u_i > \pi(x_i)
              \end{cases}
          \end{equation}
          This inverse transform method is a standard and theoretically sound approach for simulating Bernoulli trials.

    \item \textbf{Fit Logistic Regression Model}: Fit the appropriate logistic regression model to the generated data (either the correct model for Type I error analysis or the misspecified model for power analysis).

    \item \textbf{Calculate and Record Test Results}: Compute all relevant goodness-of-fit test statistics and their p-values. Record whether each test rejects the null hypothesis at the $\alpha=0.05$ level.
\end{enumerate}
The final reported result for each scenario is the proportion of the 10,000 replications in which the null hypothesis was rejected. This proportion represents the Type I error rate or the statistical power, depending on the experiment.

\section{Statistical Packages different algorithms} \label{sec:statistical_packages}
When discussing statistical techniques or modeling, a crucial question is which statistical package or software to use, such as \textit{SAS, R, Minitab, SPSS, Stata, or Python }libraries like \textit{SciPy} or \textit{Statsmodels}. It's important to note that not all tests or required graphs are available in every package. Therefore, it's essential to inform the reader about the most suitable statistical package for each specific analysis.\\

\noindent Packages often include important scientific notes. For example, the \textit{SAS} manual warns that with few events per profile, the deviance may not follow a true chi-square distribution, leading to inaccurate p-values \parencite{sas2020}. Similarly, \textit{Minitab} support notes that the reliability of deviance goodness-of-fit tests decreases as the number of trials per row drops, especially in binary response/frequency data \parencite{minitab}. \\

\noindent The problem is that the packages use different algorithms for the same test (or estimation method), and the results may differ. for example, the \textit{R} package \textit{logistic} uses the \textit{Fisher scoring} algorithm, while the \textit{R} package \textit{glm} uses the \textit{Newton-Raphson} algorithm.
    { \small
        {\footnotesize
            \begin{table}[H]
                \centering
                \caption[Values of the Hosmer--Lemeshow $\hat{C}$ by six different packages.]{Values of the Hosmer--Lemeshow decile of risk statistic, $\hat{C}$, computed by six different packages. (table from Hosmer et al. 1997 \parencite{Hosmer1997})}
                \begin{tabular}{lccc}
                    \hline
                    \textbf{Statistic}    & \textbf{Value} & \textbf{D.F.} & \textbf{$p$-value} \\
                    \hline

                    BMDPLR's $\hat{C}$    & 18.11          & 8             & 0.020              \\
                    LOGXACT's $\hat{C}$   & 13.08          & 8             & 0.109              \\
                    SAS's $\hat{C}$       & 11.83          & 8             & 0.159              \\
                    STATA's $\hat{C}$     & 12.59          & 8             & 0.128              \\
                    STATISTIX's $\hat{C}$ & 12.11          & 8             & 0.147              \\
                    SYSTAT's $\hat{C}$    & 14.70          & 8             & 0.065              \\
                    \hline
                \end{tabular}
            \end{table}
        }
    }

\noindent Hosmer mentioned, \textit{`The problem is that the packages use diﬀerent algorithms to select cutpoints that deﬁne the deciles. It is disconcerting to note that the statistic seems sensitive to choice of groups. All packages may produce the same ﬁtted model. However, depending on our choice of level of signiﬁcance and particular package used, we might reach diﬀerent conclusions on overall model ﬁt.'} \parencite{Hosmer1997}.

\begin{landscape}
    \footnotesize 
    \section{Tests Involved in the Study} \label{sec:tests_involved}
    \begin{longtable}{@{}llp{11cm}@{}}

        \caption{Summary of All Goodness-of-Fit Tests Implemented in the Simulation Study (before excluding)}
        \label{tab:test_summary}                                                                                                                                                                                                                                                                                                                                                                                                                                                                                                                                                  \\

        \toprule
        \textbf{Test Name in Simulation} & \textbf{Authors \& Citation}                                 & \textbf{Implementation Notes \& Special Settings}                                                                                                                                                                                                                                                                                                                                                                                                                       \\
        \midrule
        \endfirsthead

        \caption[]{(Continued)}                                                                                                                                                                                                                                                                                                                                                                                                                                                                                                                                                   \\
        \toprule
        \textbf{Test Name in Simulation} & \textbf{Authors \& Citation}                                 & \textbf{Implementation Notes \& Special Settings}                                                                                                                                                                                                                                                                                                                                                                                                                       \\
        \midrule
        \endhead

        \midrule
        \multicolumn{3}{r}{\textit{Continued on next page}}                                                                                                                                                                                                                                                                                                                                                                                                                                                                                                                       \\
        \endfoot

        \bottomrule
        \endlastfoot

        \multicolumn{3}{l}{\textit{\textbf{--- Hosmer-Lemeshow (HL) Variants ---}}}                                                                                                                                                                                                                                                                                                                                                                                                                                                                                               \\
        traditional\_HL                  & \cite{hosmer1980goodness}                                    & Library: \texttt{ResourceSelection}, Function: \texttt{hoslem.test()}. Groups based on deciles of risk (G=10). \textbf{See Algorithm~\ref{alg:hosmer_lemeshow_test}}.                                                                                                                                                                                                                                                                                                   \\
        hosmer\_equal\_width             & \cite{hosmer1980goodness}                                    & Custom implementation based on equal-width intervals of predicted probabilities. Special Parameter: G=10.                                                                                                                                                                                                                                                                                                                                                               \\
        large\_sample\_HL                & \cite{MHL2021}                                               & Library: \texttt{largesamplehl}, Function: \texttt{hltest()}. A modification of the HL test designed to be robust in very large samples. \textbf{See Algorithm~\ref{alg:mHL_large}}.                                                                                                                                                                                                                                                                                    \\
        HL\_GAM                          & \cite{GAM2021}                                               & Custom implementation. Hosmer-Lemeshow test where groups are formed based on probabilities from an over-specified GAM. \textbf{See Algorithm~\ref{alg:gam_gof_test}}.                                                                                                                                                                                                                                                                                                   \\
        pigeon\_heyse                    & \cite{pigeon1999b}                                           & Custom implementation. A modification of the HL test with a correction factor, using \textit{DOR} partioning. \textbf{See Algorithm~\ref{alg:pigeon_heyse_test}}.                                                                                                                                                                                                                                                                                                       \\
        hosmer\_bootstrap                & \cite{Lai2018}                                               & Standardizes the power of the Hosmer-Lemeshow test \texttt{G=10} by estimating, via bootstrap \texttt{K=200}, the probability that the H-L statistic would reject the null in a "standard-sized" sample ($n_0=min(n, max(100, n * 0.2))$). \texttt{pvalue = 1} if the test accepts good fit, \texttt{pvalue = 0} if not (per Lai \& Liu test decision) . Library: custom implementation (\texttt{lai\_liu\_test()}). \textbf{See Algorithm~\ref{alg:lailiu_bootstrap}}. \\

        \midrule
        \multicolumn{3}{l}{\textit{\textbf{--- Classic and Standardized Pearson/Deviance Tests ---}}}                                                                                                                                                                                                                                                                                                                                                                                                                                                                             \\
        pearson                          & \cite{pearson1900criterion}                                  & Standard Pearson $\chi^2$ statistic from ungrouped data. \textbf{See Algorithm~\ref{alg:pearson_and_deviance_statistics}}.                                                                                                                                                                                                                                                                                                                                              \\
        deviance                         & \cite{nelder1972}                                            & Standard Deviance statistic from ungrouped data. \textbf{See Algorithm~\ref{alg:pearson_and_deviance_statistics}}.                                                                                                                                                                                                                                                                                                                                                      \\
        osius\_rojek                     & \cite{osius1992normal}                                       & Library: \texttt{LogisticDx}, Pearson statistic standardized to a normal distribution. Implemented via \texttt{gof.glm()}. \textbf{See Algorithm in \ref{alg:osius_rojek_test}}.                                                                                                                                                                                                                                                                                        \\
        mccullagh\_test                  & \cite{mccullagh1985}                                         & Custom implementation from SAS documentation \cite{kuss2002goflogit}. Pearson statistic standardized using conditional moments. GPU used for the efficient implementation. \textbf{See Algorithm~\ref{alg:mccullagh}}.                                                                                                                                                                                                                                                  \\
        farrington\_standalone           & Farrington \parencite{farrington1996}                        & Custom implementation of original Farrington test applied to grouped (non-sparse) data, based on Kuss SAS  \parencite{kuss2002goflogit}.                                                                                                                                                                                                                                                                                                                                \\

        \midrule
        \multicolumn{3}{l}{\textit{\textbf{--- Tests Based on Partitioning the Covariate Space ---}}}                                                                                                                                                                                                                                                                                                                                                                                                                                                                             \\
        tsiatis\_clustering              & \cite{Tsiatis1980}                                           & Custom implementation. Score test applied to groups formed by k-means clustering on covariates. \textbf{See Algorithm~\ref{alg:tsiatis_score_test_matrix_form}}.                                                                                                                                                                                                                                                                                                        \\
        xie\_test                        & \cite{xie2008increasing}                                     & Custom implementation. Pearson-type statistic applied to groups formed by clustering the covariate space. \textbf{See Algorithm~\ref{alg:xie}}.                                                                                                                                                                                                                                                                                                                         \\
        XIE\_GAM                         & \cite{GAM2021}                                               & Custom implementation. Xie test using probabilities from an over-specified GAM. \textbf{See Algorithm~\ref{alg:gam_gof_test}}.                                                                                                                                                                                                                                                                                                                                          \\
        PR\_test                         & \cite{pulkstenis2002two}                                     & Custom implementation. \texttt{Pearson-type statistic} applied to groups partitioned by categorical variables and median probability. \textbf{See Algorithm~\ref{alg:pr_test}}.                                                                                                                                                                                                                                                                                         \\
        PR\_GAM                          & \cite{GAM2021}                                               & Custom implementation. Pulkstenis-Robinson test using probabilities from an over-specified GAM. \textbf{See Algorithm~\ref{alg:gam_gof_test}}.                                                                                                                                                                                                                                                                                                                          \\
        bagoft\_split1\_sim0             & \cite{BAGofT2019}                                            & Binary Adaptive Goodness-of-fit Test. Library: \texttt{BAGofT}, Special Parameters: \texttt{nsplits=1, nsim=0, testModel = testGlmBi(formula = ..., link = "logit"),
            parFun = parRF()},
        If your dataset contains only a single predictor, you must add an additional constant variable (e.g., a column of ones) to the \texttt{Data} parameter, but not appear in the \texttt{formula}. This is necessary because the BaGofT function requires at least two predictors to operate correctly. \textbf{See Algorithm~\ref{alg:bagoft_test}}.                                                                                                                                                                                                                        \\
        bagoft\_split20\_sim0            & \cite{BAGofT2019}                                            & Library: \texttt{BAGofT}, Special Parameters: \texttt{nsplits=20, nsim=0}, the rest same as \texttt{bagoft\_split1\_sim0}. \textbf{See Algorithm~\ref{alg:bagoft_test}}.                                                                                                                                                                                                                                                                                                \\
        F\_test                          & Used by \parencite{liu2024comprehensive, rady2021comparison} & Library: \texttt{LogisticDx}, Anova F-test, for deviance residuals on grouping factor. Implemented via \texttt{gof.glm()}, using \texttt{groups = 10} . \textbf{See Algorithm in \ref{alg:anova_f_test_for_deviances}}.                                                                                                                                                                                                                                                 \\


        \midrule
        \multicolumn{3}{l}{\textit{\textbf{--- Machine Learning \& Calibration Tests ---}}}                                                                                                                                                                                                                                                                                                                                                                                                                                                                                       \\
        unreliability\_index             & \cite{harrell2015regression}                                 & Likelihood-ratio test of the calibration model ($H_0: \gamma_0=0, \gamma_1=1$). Library: \texttt{rms}, Function: \texttt{val.prob()}. \textbf{See Algorithm~\ref{alg:unreliability_test}}.                                                                                                                                                                                                                                                                              \\
        giviti\_external                 & \cite{nattino2014new}                                        & Flexible polynomial calibration test for external validation. Library: \texttt{givitiR}, Setting: \texttt{devel="external"}. \textbf{See Algorithm~\ref{alg:giviti}}.                                                                                                                                                                                                                                                                                                   \\
        giviti\_internal                 & \cite{nattino2015new}                                        & Flexible polynomial calibration test for internal validation (goodness-of-fit). Library: \texttt{givitiR}, Setting: \texttt{devel="internal"}. \textbf{See Algorithm~\ref{alg:giviti}}.                                                                                                                                                                                                                                                                                 \\
        eHL                              & \cite{Henzi2024SafeHL}                                       & Extracted from Simulation Code: e-value Hosmer-Lemeshow test in \cite{MariusCP2023eHL}, this test compares the Calibrated Model to the original Model to make the decision (\texttt{b=10,s=0.5}). \textbf{See Algorithm~\ref{alg:ehl}}. The null hypothesis is rejected if the \texttt{eHL} statistic exceeds \texttt{20} (and, in another test, \texttt{10}).                                                                                                              \\
        spiegelhalter                    & \cite{spiegelhalter1986calibration}                          & Z-test based on the Brier score. Library: \texttt{rms}, Function: \texttt{val.prob()}. \textbf{See Algorithm~\ref{alg:spiegelhalter_z_test}}.                                                                                                                                                                                                                                                                                                                           \\

        \midrule
        \multicolumn{3}{l}{\textit{\textbf{--- Bootstrap-Based Tests ---}}}                                                                                                                                                                                                                                                                                                                                                                                                                                                                                                       \\
        stute\_zhu                       & \cite{stute2002model}                                        & Cramér-von Mises test on cumulative residuals. P-value obtained via model-based bootstrap (B=200). \textbf{See Algorithm~\ref{alg:tsz_bootstrap}}.                                                                                                                                                                                                                                                                                                                      \\
        projection\_test                 & \cite{liu2024comprehensive}                                  & Extension of the Stute-Zhu test that checks all projection directions. P-value via bootstrap. \textbf{See Algorithm~\ref{alg:projection_bootstrap}}.                                                                                                                                                                                                                                                                                                                    \\
        bagoft\_split20\_sim100          & \cite{BAGofT2019}                                            & Library: \texttt{BAGofT}, Special Parameters: \texttt{nsplits=20, nsim=100}, the rest same as \texttt{bagoft\_split1\_sim0}. \textbf{See Algorithm~\ref{alg:bagoft_test}}.                                                                                                                                                                                                                                                                                              \\

        \midrule
        \multicolumn{3}{l}{\textit{\textbf{--- Other Specialized and Smoothing-Based Tests ---}}}                                                                                                                                                                                                                                                                                                                                                                                                                                                                                 \\
        stukel\_score\_test              & \cite{stukel1988generalized}                                 & Three Score tests for misspecification of the logit link function, from Library: \texttt{LogisticDx}, Function: \texttt{gof.glm()}. \textbf{See Algorithm~\ref{alg:stukel_test}}.                                                                                                                                                                                                                                                                                      \\
        IM\_efficient                    & \cite{White1982} /  \cite{Orme1988}                          & Custom implementation from SAS documentation \cite{kuss2002goflogit}. Information Matrix Test for general model misspecification. \textbf{See Algorithm~\ref{alg:im_test}}.                                                                                                                                                                                                                                                                                             \\
        copas\_unweighted                & \cite{copas1989unweighted}                                   & Unweighted sum of squares test. Library: \texttt{rms}, Function: \texttt{resid(..., 'gof')}. \textbf{See Algorithm~\ref{alg:copas_uss_test}}.                                                                                                                                                                                                                                                                                                                           \\
        lecessie\_test                   & le Cessie \& van Houwelingen \parencite{le1995goodness}      & Custom implementation, originally coded in Library: \texttt{smwrStats
        } function: \texttt{leCessie.test()}. Smoothed residual test using a GPU-accelerated Python backend. \textbf{See Algorithm~\ref{alg:le_cessie_van_houwelingen_1995_score_test}}.                                                                                                                                                                                                                                                                                                                                                                                          \\
    \end{longtable}
\end{landscape}

\section{Liberal and Zero Power Tests} \label{sec:liberal_and_zero_power_tests}

In simulation studies evaluating goodness-of-fit tests for logistic regression, it is crucial to recognize that not all available tests are appropriate or informative in every context. Some tests are \textbf{excluded from the main simulation results} due to methodological limitations, practical constraints, or undesirable statistical properties. A key concept in this context is \textbf{liberal tests}. A test is considered \textbf{liberal} if it rejects the null hypothesis more often than the nominal significance level would suggest, even when the null is true. In other words, liberal tests have inflated Type I error rates, leading to frequent false positives and unreliable conclusions about model fit. The following sections outline the main reasons for excluding certain tests from the primary simulation results, focusing on issues such as excessive conservatism (zero power), computational infeasibility, lack of applicability to certain data types, and liberal behavior.

\begin{itemize}
    \item \textbf{Tests that never reject under full sparsity:} Some tests, such as the \texttt{Farrington} test, are known to be extremely conservative in sparse data settings. In our simulations, these tests almost never reject the null hypothesis, even when the model is misspecified. This lack of sensitivity renders them uninformative for evaluating power in such scenarios.(recall section~\ref{sec:farrington_test})

    \item \textbf{Tests requiring excessive computational resources:} Certain methods, like the \texttt{Projection Bootstrap} test, are computationally intensive to the point of being impractical for large-scale simulation studies. Their runtime and memory requirements far exceed those of other tests, making them infeasible for inclusion in comprehensive simulation comparisons. (recall section~\ref{sec:projection_test})

    \item \textbf{Tests requiring categorical variables for partitioning:} Some tests, such as the \texttt{Pulkstenis-Robinson} (PR) test and its variants, are specifically designed for situations where the covariate space can be partitioned using categorical variables. In scenarios where only continuous predictors are available, these tests cannot be meaningfully applied. (recall section~\ref{sec:pr_test})

    \item \textbf{Liberal tests that frequently reject the true null:} Tests like the \texttt{Pearson}, \texttt{Deviance}, \texttt{F-test for deviances}, and \texttt{Hosmer (Bootstrap)} goodness-of-fit tests are known to be liberal, often rejecting the true null hypothesis with unacceptably high probability, especially in the presence of sparse data or small group sizes. This leads to inflated Type I error rates and unreliable conclusions. (recall section~\ref{sec: pearson_chi_square}, \ref{sec:deviance_test}, \ref{sec:anova_f_test_for_deviances}, \ref{sec:lai_liu_test}, \ref{sec:anova_f_test_for_deviances})

    \item \textbf{Tests targeting linear calibration:} The \texttt{Unreliability Index} ($U$ test) of Harrell Jr. \parencite{harrell2015regression} specifically assesses \textbf{linear calibration} by fitting a logistic model of observed outcomes on the original model's linear predictor and testing if the calibration intercept is 0 and slope is 1 (see section~\ref{sec:unreliability_test}). This test is sensitive to systematic miscalibration (e.g., average prediction too high/low or over/under-confidence), but is not designed to detect non-linear misspecification. Thus, it is not directly comparable to omnibus goodness-of-fit tests in our power analysis.

\end{itemize}

\begin{landscape}

    \begin{figure}[h]
        \centering
        \includegraphics[width=1.5\textwidth]{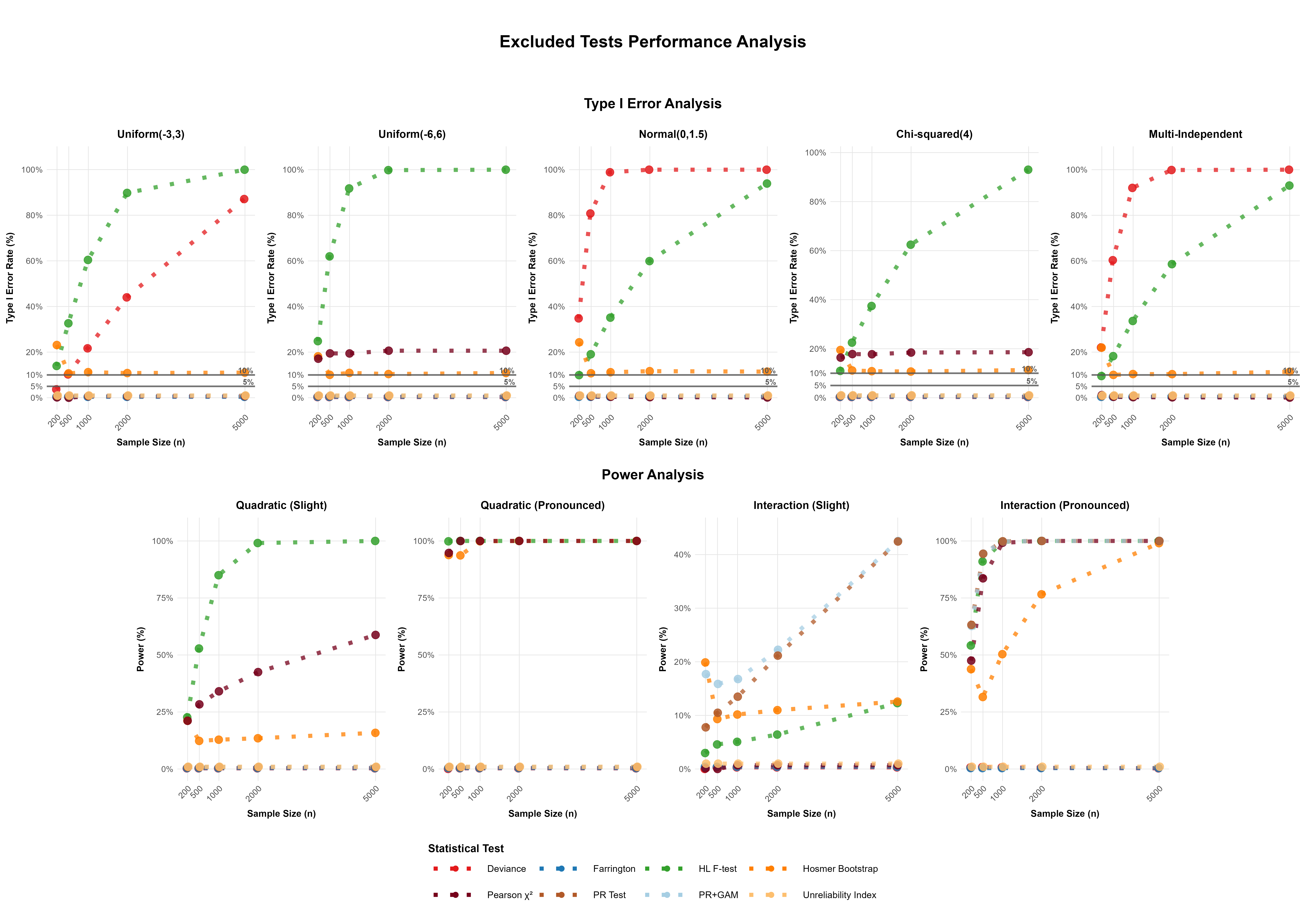}
        \caption{Performance of Excluded Tests: Type I Error and Power Analysis.}
        \label{fig:liberal_zero_power_infeasible_tests}
    \end{figure}

\end{landscape}

\subsection*{Results}
Figure~\ref{fig:liberal_zero_power_infeasible_tests} provides a detailed overview of the rejection rates (\%) for the excluded goodness-of-fit tests for logistic regression, evaluated across a range of sample sizes and simulation scenarios. In the top row (\emph{Type I Error Analysis}), results are shown under the null hypothesis for five different scenarios: Uniform($-3,3$), Uniform($-6,6$), Normal($0,1.5$), Chi-squared(4), and Multi-Independent (recall section~\ref{sec:type1_scenarios}). The bottom row (\emph{Power Analysis}) displays outcomes for four misspecified models: Quadratic (Slight), Quadratic (Pronounced), Interaction (Slight), and Interaction (Pronounced) (recall section~\ref{sec:power_scenarios}). Each colored line in the plots represents a specific test (refer to the legend for details). In the top row, the dotted horizontal lines at 5\% and 10\% mark the nominal significance thresholds.

\subsubsection*{Liberal Tests: Inflated Type I Error}

Several classical goodness-of-fit tests, including the \texttt{Pearson} chi-square, \texttt{Deviance}, \texttt{HL F-test}, and \texttt{Hosmer Bootstrap}, are observed to be \emph{liberal} in the simulation results. As shown in the top row of Figure~\ref{fig:liberal_zero_power_infeasible_tests}, these tests frequently reject the null hypothesis at rates far exceeding the nominal significance levels (5\% and 10\%), even when the model is correctly specified. This inflation of Type I error is especially pronounced in scenarios with sparse data or small group sizes (e.g., Uniform($-6,6$), Multi-Independent, and Chi-squared(4)), where the rejection rates can approach or exceed 100\% for moderate to large sample sizes. Such behavior renders these tests unreliable for assessing model fit, as they may lead to frequent false positives and unwarranted model criticism.

\subsubsection*{Zero Power and Extremely Conservative Tests}

In contrast, some tests demonstrate \emph{zero power} or extreme conservatism, failing to detect model misspecification even when it is present. The \texttt{Farrington} test is a prime example: across all power scenarios in the bottom row of Figure~\ref{fig:liberal_zero_power_infeasible_tests}, its rejection rate remains at or near zero, regardless of sample size or the degree of model misspecification. This indicates a complete lack of sensitivity to departures from the null hypothesis, making it uninformative for practical use in detecting lack of fit in case of Full Sparsity of data.

The \texttt{Unreliability Index} ($U$ test) also showed zero power in these simulations, consistently failing to reject the null in all misspecified scenarios (overviews of other tests in Figures~\ref{fig:power_omitted_variable_heatmap} and \ref{fig:power_omitted_interaction_heatmap}). This is expected, as the simulation design ensured the fitted and true models had nearly identical predicted probabilities, making the model appear well-calibrated \textbf{in a linear sense}. Since the $U$ test is only sensitive to linear calibration, it could not detect the non-linear misspecification present, and was therefore excluded from the power analysis.

\subsubsection*{Infeasible or Not Generally Applicable Tests}
Some tests, such as the \texttt{PR} and \texttt{PR+GAM} tests, are only applicable in scenarios with categorical predictors and thus lack general applicability in settings with only continuous variables. Additionally, the \texttt{Projection Test} requires extremely high computational resources, making it infeasible for large-scale simulation studies.

\section{Simulation Results} \label{sec:simulation_results}
This section presents the core findings from the simulation study, evaluating the performance of various goodness-of-fit (GOF) and calibration machine learning tests for logistic regression. The results are detailed through an analysis of empirical Type I error rates and statistical power under different model specifications and data conditions. For clarity and comprehensive comparison, the results are visualized in a series of heatmaps (Figures~\ref{fig:type1error_uniform_heatmap} through \ref{fig:power_omitted_interaction_heatmap}), which display rejection rates across a range of sample sizes from $n=200$ to $n=5000$.

\subsection*{Type I Error Performance}
The first objective of the simulation was to assess the ability of each GOF test to maintain the nominal significance level, set at $\alpha = 0.05$. Figures~\ref{fig:type1error_uniform_heatmap}, \ref{fig:type1error_chi2_multi_heatmap}, and \ref{fig:type1error_normal_heatmap} (to be shown in the following pages) display the empirical Type I error rates under the null hypothesis across five distinct covariate distributions.

In scenarios with less data sparsity, such as the Uniform(-3,3) and Normal(0,1.5) distributions, most of the evaluated tests perform well. Tests including the traditional Hosmer-Lemeshow (\texttt{traditional\_HL}), Osius-Rojek (\texttt{osius\_rojek}), Information Matrix (\texttt{IM\_efficient}), and McCullagh's conditional test (\texttt{McCullagh\_Test}) demonstrate excellent control of Type I error, with rejection rates hovering close to the 5\% nominal level, particularly for sample sizes of $n \geq 500$.

Performance diverges in the more challenging scenarios characterized by greater data sparsity (Uniform(-6,6), Chi-squared(4), and Multi-Independent). Even under these conditions, several tests remain robust. The \texttt{osius\_rojek}, \texttt{stute\_zhu\_test}, and \texttt{Copas\_unweighted\_S} tests, for example, continue to hold their size well across all sample sizes. In contrast, some tests become overly conservative in these settings. The \texttt{spiegelhalter} and \texttt{eHL} tests consistently show rejection rates at or near zero, indicating a very low probability of Type I error but suggesting a potential lack of power to detect true model misspecification. As expected, for most well-calibrated tests, the empirical error rates converge toward the nominal 5\% level as the sample size increases.

\subsection*{Power to Detect Model Misspecification}
The second objective was to evaluate the statistical power of the tests, defined as their ability to correctly reject the null hypothesis when the fitted model is misspecified. We examined two primary forms of misspecification: an omitted quadratic term and an omitted interaction term, each with a "slight" and a "pronounced" effect size (recall section~\ref{sec:power_scenarios}).

\subsubsection*{Power for Omitted Quadratic Term}
Figure~\ref{fig:power_omitted_variable_heatmap} summarizes the power of the tests to detect an omitted quadratic variable. When the misspecification is pronounced, most tests exhibit high power that rapidly approaches 100\% as the sample size grows. The \texttt{McCullagh\_Test}, \texttt{osius\_rojek}, and \texttt{leCessie\_Test} are particularly effective, achieving substantial power even at smaller sample sizes.

The "slight" misspecification scenario provides a more challenging benchmark. Here, power is considerably lower for all tests, but the relative performance differences are more distinct. The \texttt{McCullagh\_Test}, \texttt{leCessie\_Test}, and \texttt{stute\_zhu\_test} consistently emerge as the most powerful options, highlighting their superior sensitivity in detecting more subtle forms of model inadequacy.

\subsubsection*{Power for Omitted Interaction Term}
The results for detecting an omitted interaction term are presented in Figure~\ref{fig:power_omitted_interaction_heatmap}. In the \textbf{`pronounced'} interaction scenario, the findings mirror those from the omitted variable case. Power is generally very high, and many tests reach near-perfect detection rates for $n \geq 1000$. The \texttt{McCullagh\_Test} again stands out, achieving over 70\% power even with a small sample of $n=200$. Other strong performers in this scenario include the \texttt{osius\_rojek}, \texttt{stukel}, and \texttt{Copas\_unweighted\_S} tests.

When the interaction effect is \textbf{`slight'}, the detection task becomes more difficult, and the power of all tests is correspondingly reduced. Nonetheless, the tests that performed well in the pronounced scenario generally maintain their superior relative ranking. For all tests, the ability to detect this subtle misspecification steadily improves with increasing sample size.

\paragraph{In summary} the simulation results highlight a subset of tests that offer a strong balance of reliable Type I error control and high statistical power. Notably, the \texttt{McCullagh\_Test}, \texttt{osius\_rojek}, and \texttt{leCessie\_Test} consistently demonstrate robust performance across a wide range of conditions, making them strong candidates for practical applications.

\begin{landscape}

    \begin{figure}[H]
        \centering
        \includegraphics[width=1.4\textwidth]{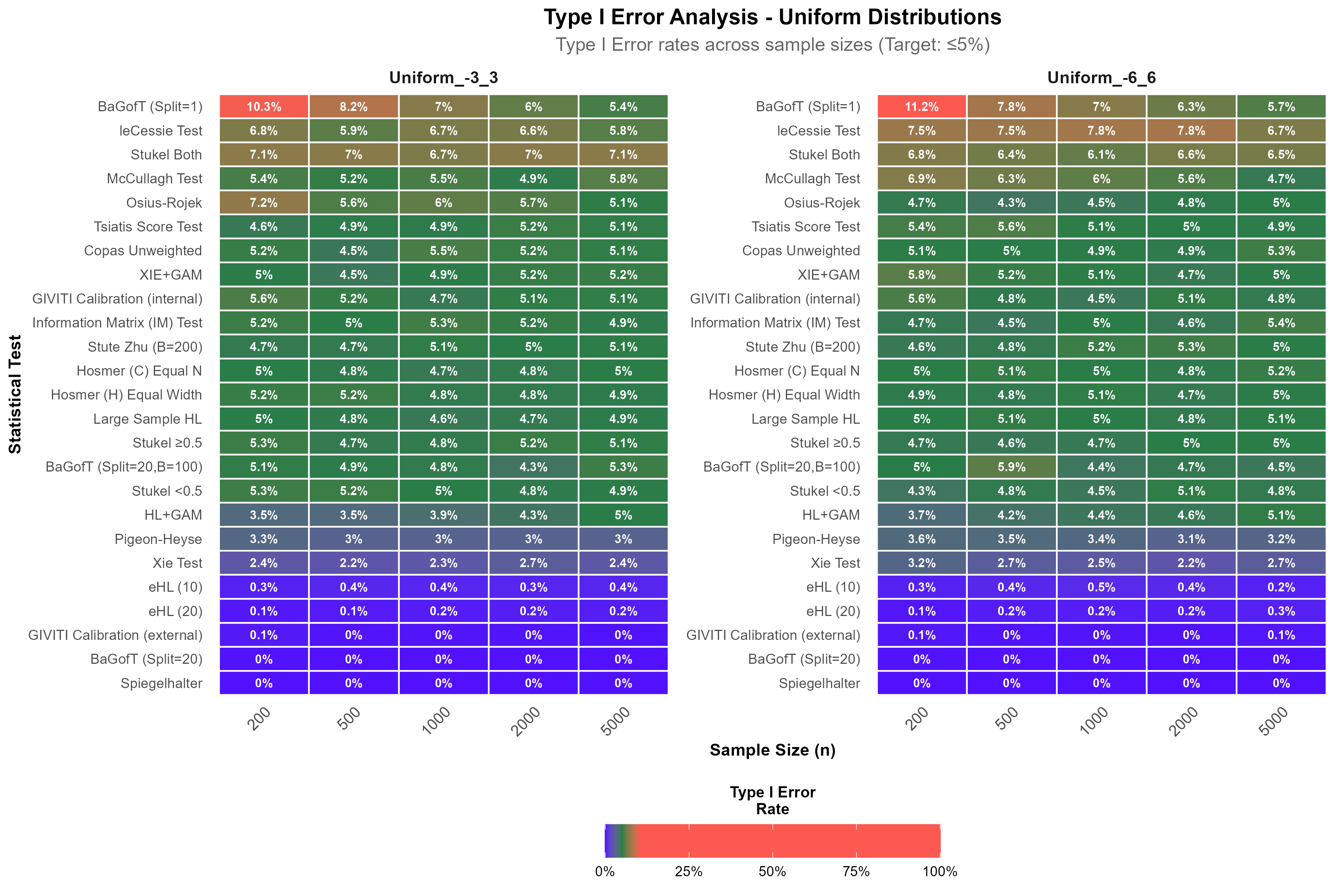}
        \caption[Type I Error: Uniform Scenarios]{Type I error heatmap for Uniform(-3,3) and Uniform(-6,6) scenarios.}
        \label{fig:type1error_uniform_heatmap}
    \end{figure}

    \newpage
    \begin{figure}[H]
        \centering
        \includegraphics[width=1.5\textwidth]{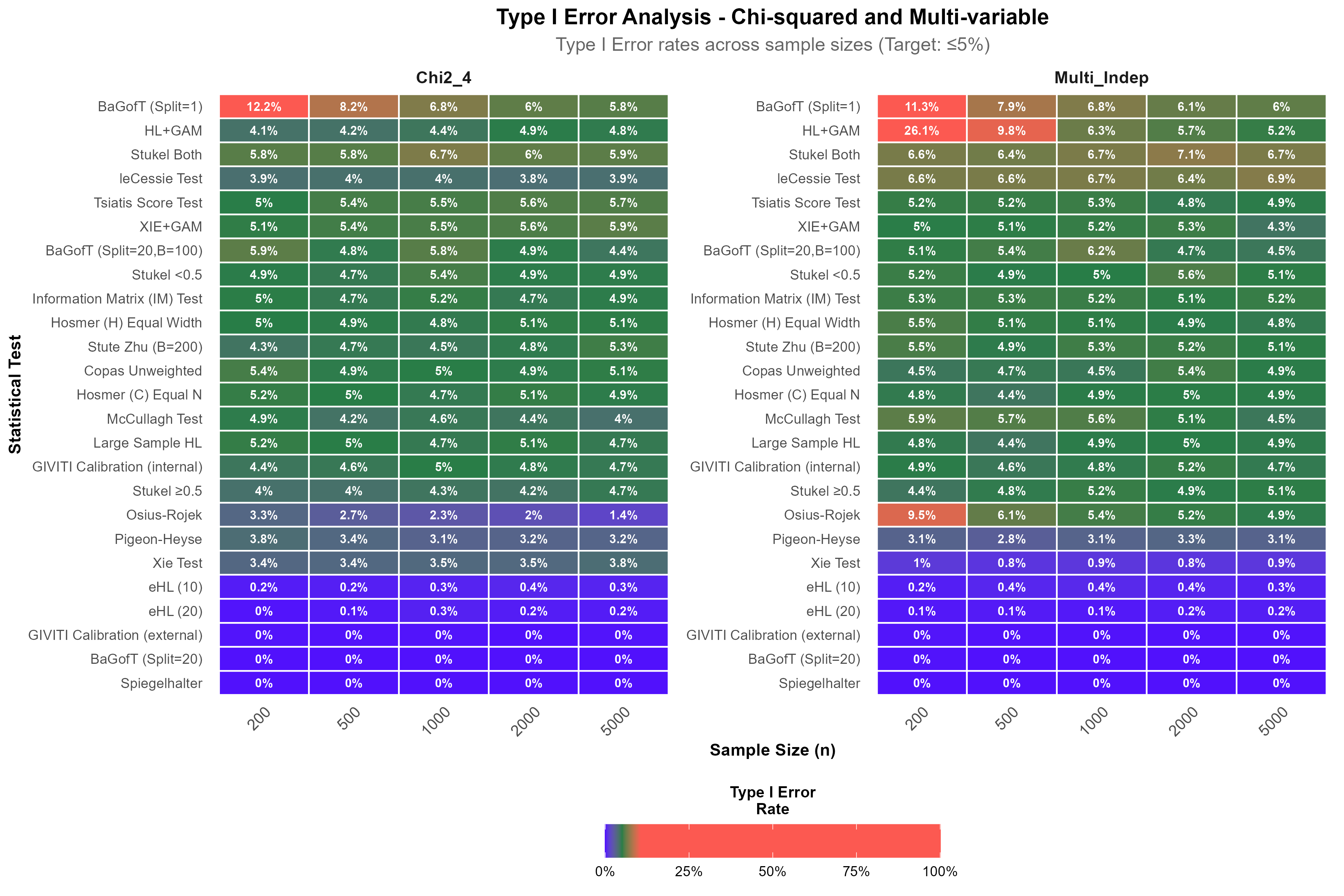}
        \caption[Type I Error: Chi-squared and Multi-Independent Scenarios]{Type I error heatmap for Chi-squared(4) and Multi-Independent scenarios.}
        \label{fig:type1error_chi2_multi_heatmap}
    \end{figure}

    \begin{figure}[H]
        \centering
        \includegraphics[width=1.5\textwidth]{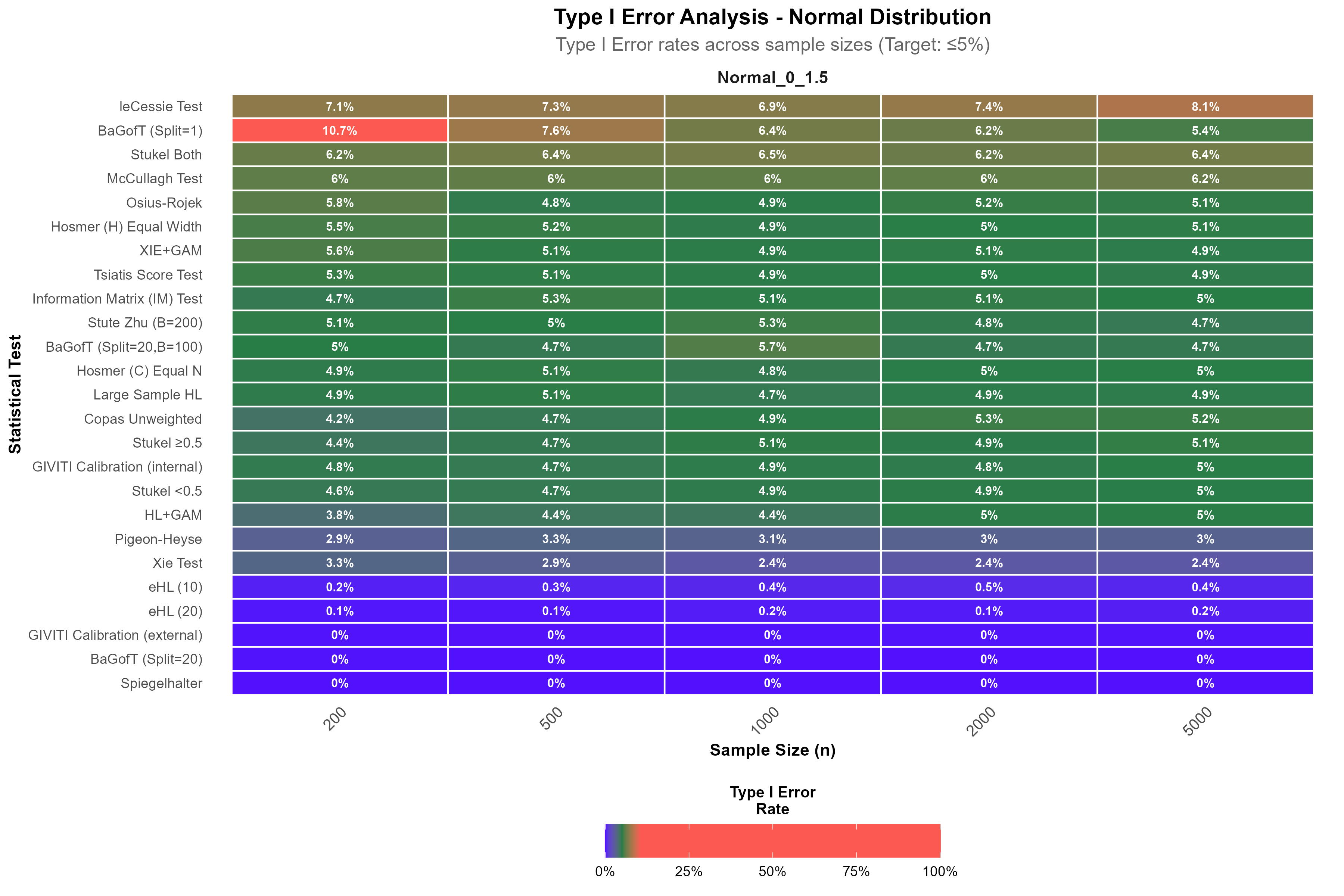}
        \caption[Type I Error: Normal Scenario]{Type I error heatmap for the Normal(0,1.5) scenario.}
        \label{fig:type1error_normal_heatmap}
    \end{figure}

    \begin{figure}[H]
        \centering
        \includegraphics[width=1.5\textwidth]{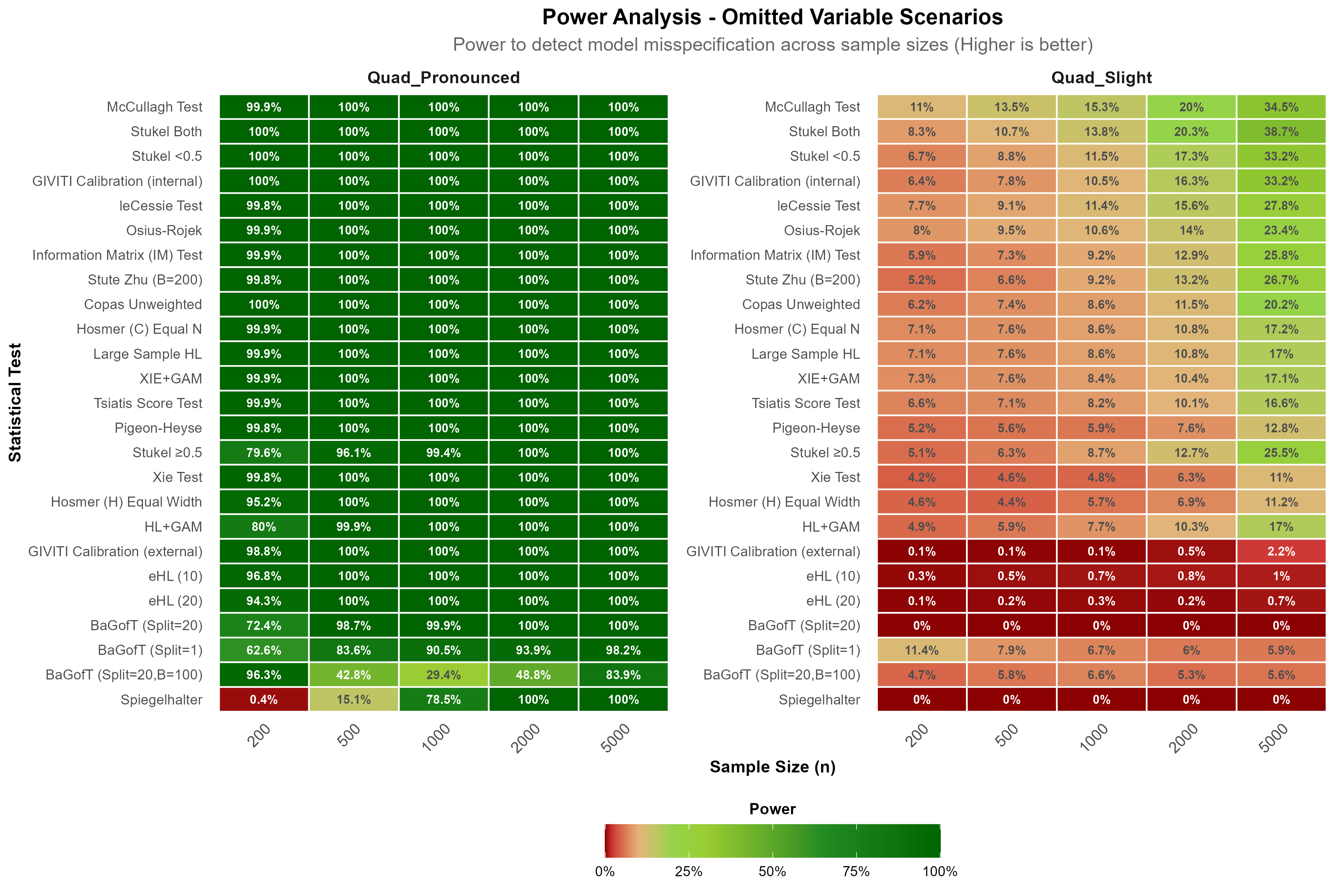}
        \caption[Power Analysis: Omitted Variable Scenarios]{Power analysis heatmap for omitted variable scenarios.}
        \label{fig:power_omitted_variable_heatmap}
    \end{figure}

    \begin{figure}[H]
        \centering
        \includegraphics[width=1.5\textwidth]{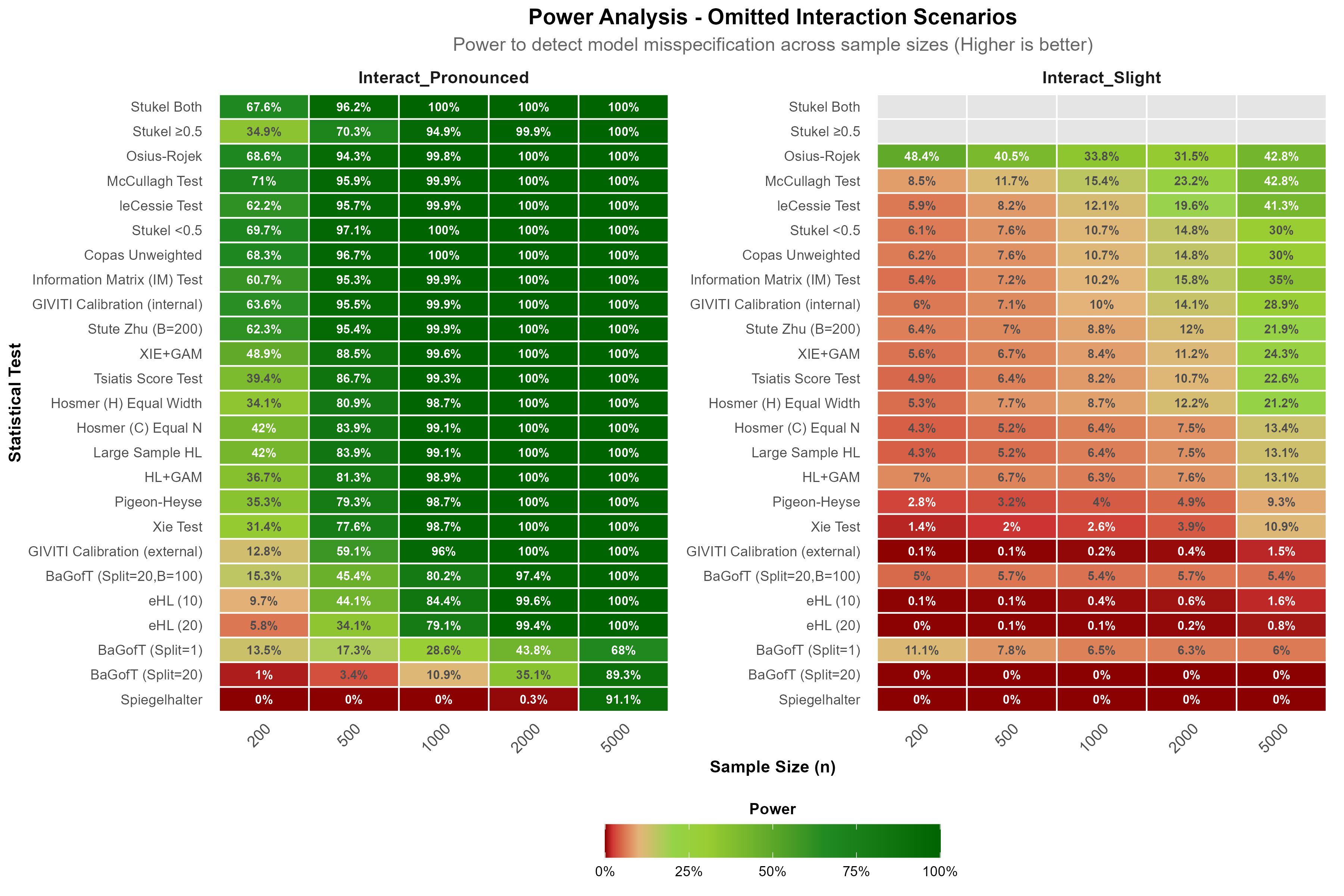}
        \caption[Power Analysis: Omitted Interaction Scenarios]{Power analysis heatmap for omitted interaction scenarios.}
        \label{fig:power_omitted_interaction_heatmap}
    \end{figure}

\end{landscape}

\newpage
\section{Real Application} \label{sec:real_application}
\subsection{Revisiting Hosmer's et al. Low Birth Weight Dataset}
To demonstrate some of the challenges associated with existing methods for assessing overall goodness-of-fit, we utilize the well-known low birth weight dataset originally presented in Hosmer and Lemeshow Paper \parencite{Hosmer1997}. This dataset was collected at Baystate Medical Center in Springfield, Massachusetts, in 1986, and comprises information on 188 births. The primary outcome variable is whether the newborn's birth weight was less than 2500 grams (classified as low birth weight). Of the 188 cases, 58 were low birth weight and 129 were normal birth weight.\footnote{\url{https://github.com/jcizel/WRDS-SAS-UTILITIES/blob/master/SAS-macro/goflogit.sas}}

The purpose of this example is not to provide a comprehensive analysis of these data, but rather to illustrate the practical issues that can arise when evaluating model fit. The independent variables included in the analysis are: the mother's age at last menstrual period \texttt{(AGE)}, her weight at last menstrual period \texttt{(LWT)}, race (coded as two indicator variables: \texttt{RACE\_1} and \texttt{RACE\_2}, with white as the reference group), and smoking status during pregnancy \texttt{(SMOKE; 1 = smoker, 0 = non-smoker)}.This real-data example, taken from Hosmer and Lemeshow \parencite{Hosmer1997}, serves as a benchmark for comparing the performance and limitations of various goodness-of-fit tests for logistic regression. However, this is the same dataset Kuss used in his work \parencite{kuss2002goflogit,kuss2002paper}. The dataset is jittered to avoid ties in the estimated probabilities.
    {\small
        \begin{verbatim}
Call:
glm(formula = birthwgt$y ~ birthwgt$age + birthwgt$lwt + birthwgt$race2 +
    birthwgt$race3 + birthwgt$smoke, family = binomial(), data = birthwgt)

Coefficients:
                Estimate Std. Error z value Pr(>|z|)
(Intercept)     0.290653   1.112732   0.261  0.79393
birthwgt$age   -0.020445   0.034291  -0.596  0.55104
birthwgt$lwt   -0.012749   0.006436  -1.981  0.04759 *
birthwgt$race2  1.273472   0.519216   2.453  0.01418 *
birthwgt$race3  0.971791   0.417956   2.325  0.02007 * 
birthwgt$smoke  1.024814   0.381765   2.684  0.00727 **
---
Signif. codes:  0 '***' 0.001 '**' 0.01 '*' 0.05 '.' 0.1 ' ' 1
\end{verbatim}
    }

The fitted model presented in the previous R output includes variables that are established risk factors for low birth weight \texttt{(Y)}. Although mother's age \texttt{(age)} was not statistically significant (using Wald test\footnote{Wald test is a statistical test used to test the significance of coefficients in a logistic regression model. It is named after Abraham Wald, who first proposed the test in 1943. The Wald test statistic \parencite{Wald1943} is a test statistic that is used to test the significance of a coefficient in a logistic regression model. The Wald statistic is calculated by dividing the coefficient by its standard error. The Wald test is a popular method for testing the significance of coefficients in logistic regression models.}), it was kept in the model due to its recognized biological relevance \parencite{Hosmer1997}. The coefficients obtained from our model match those reported by Hosmer and Lemeshow, providing a consistent starting point for reproducing their analysis, while Hosmers et al add to some predictors some noise to the data to make it less discrete (jittering), we still can apply all the goodness-of-fit tests discussed in our thesis, as well as our proposed method, to this benchmark example (Appendix \ref{ap:hosmer_lemeshow_1997}).  It also the same dataset that kuss used in his work (grouped) (Appendix \ref{ap:kuss_2002}). The fitted model used by both Hosmer and Kuss is as follows:

\begin{align}
    \text{logit}({\pi}_i) =\ \beta_0 + \beta_1\,\text{age} + \beta_2\,\text{lwt} + \beta_3\,\text{race2} + \beta_4\,\text{race3} + \beta_5\,\text{smoke}
\end{align}

After estimation of the parameters, the model becomes:
\begin{align}
    \text{logit}(\hat{\pi}_i) =\  & 0.290653
    - 0.020445 \times \text{age} \notag                                   \\
                                  & - 0.012749 \times \text{lwt} \notag   \\
                                  & + 1.273472 \times \text{race2} \notag \\
                                  & + 0.971791 \times \text{race3} \notag \\
                                  & + 1.024814 \times \text{smoke}
\end{align}

However, as noted by Hosmer et al. in their original paper (p. 976) \parencite{Hosmer1997}, \textbf{the TRUE underlying model}    should also account for interaction effects between \texttt{AGE} and \texttt{LWT}, as well as between \texttt{SMOKE} and \texttt{LWT}. Specifically, the more comprehensive model is:
\begin{align} \label{eq:true_model_real_data}
    \text{logit}(\pi_i) =\  & \beta_0
    + \beta_1 \times \text{age}
    + \beta_2 \times \text{lwt} \notag                                        \\
                            & + \beta_3 \times \text{race2}
    + \beta_4 \times \text{race3}
    + \beta_5 \times \text{smoke} \notag                                      \\
                            & + \beta_6 \times (\text{age} \times \text{lwt})
    + \beta_7 \times (\text{smoke} \times \text{lwt})
\end{align}

\paragraph{Why are these interaction terms important?}
Including interaction effects between \texttt{AGE} and \texttt{LWT}, as well as between \texttt{SMOKE} and \texttt{LWT}, is crucial for accurately modeling the risk of low birth weight. Logically, the effect of a mother's weight on the probability of low birth weight may not be constant across all ages; for example, low weight might have a more pronounced impact on younger mothers compared to older ones, or vice versa. Similarly, the adverse effect of smoking during pregnancy could be exacerbated or mitigated depending on the mother's weight. Ignoring these interactions assumes that the influence of each predictor is independent and additive, which may oversimplify the true biological relationships. By incorporating these interaction terms, the model can capture more complex, real-world dependencies among risk factors, leading to more accurate predictions and a better understanding of how combinations of maternal characteristics jointly affect birth outcomes. This is especially important in clinical and epidemiological research, where nuanced relationships between risk factors can have significant implications for intervention and prevention strategies.

\newpage
\subsection{EDA (Exploratory Data Analysis):}
before we start the analysis, we need to understand the data and the relationships between the predictors and the outcome. we will use the \texttt{givitiCalibrationBelt} plot for this purpose.
\begin{figure}[H]
    \centering
    \includegraphics[width=1\textwidth]{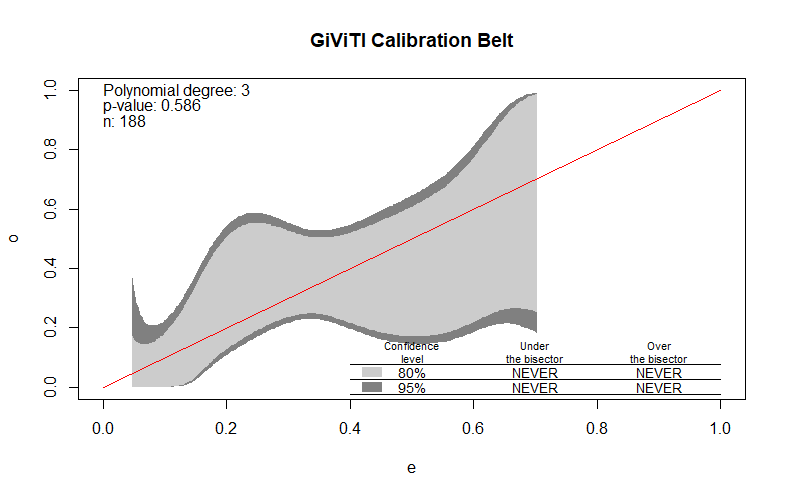}
    \caption[GIVITI Calibration Belt for the low birth weight dataset.]{GIVITI Calibration Belt for the low birth weight dataset. The plot visualizes the agreement between observed outcomes and predicted probabilities from the fitted logistic regression model.
    }
    \label{fig:calib_belt_1997}
\end{figure}

Figure~\ref{fig:calib_belt_1997} displays a lot of information and suggestions:
\begin{itemize}
    {\small
    \item \textbf{Polynomial degree:} The best-fitting calibration curve is of degree 3 (non-linear), indicating an evidence of nonlinearity in the calibration relationship.
    \item \textbf{p-value:} The p-value for the calibration test is 0.586. This means there is no evidence to reject the null hypothesis of good calibration.
    \item \textbf{Confidence belt:} The entire 95\% confidence belt contains the bisector (red line) across the range of predicted probabilities, further supporting the conclusion that the model is well-calibrated.
    \item \textbf{Summary table:} The table below the plot indicates that, at both the 80\% and 95\% confidence levels, the observed calibration curve is \emph{never} significantly under or over the bisector at any point.
          }
\end{itemize}

The GIVITI Calibration Belt shows that the fitted logistic regression model is well-calibrated for the low birth weight data, in line with Hosmer and Lemeshow's findings. However, this model still omits important interaction effects between AGE and LWT, and between SMOKE and LWT.

\begin{figure}[H]
    \centering
    \includegraphics[width=0.85\textwidth]{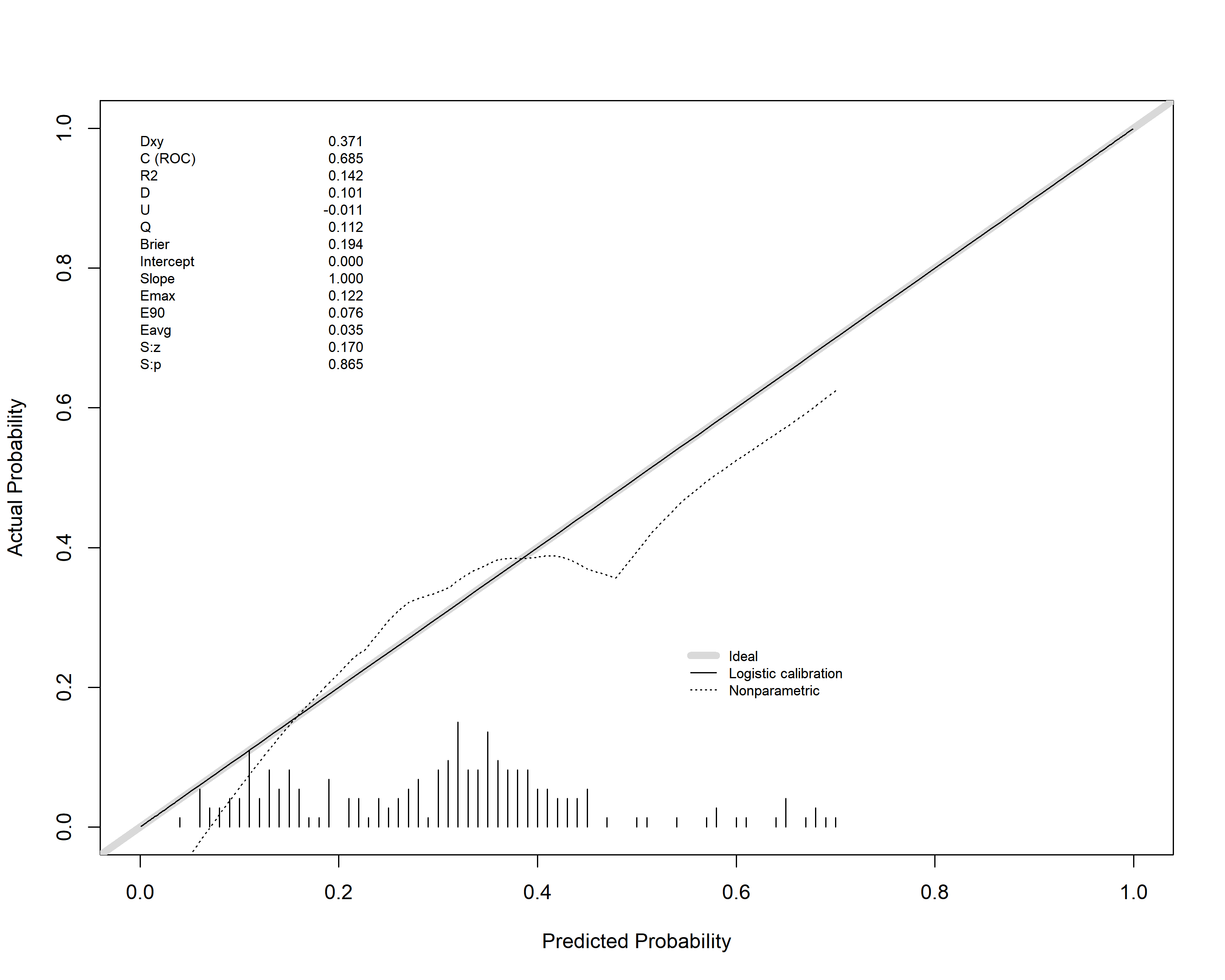}
    \caption[RMS Reliability Plot for the low birth weight dataset.]{RMS Reliability Plot for the low birth weight dataset. The plot compares predicted probabilities from the fitted logistic regression model to observed event rates, with the solid line representing perfect calibration, the dashed line showing logistic calibration, and the dotted line indicating a nonparametric (loess) fit. The histogram at the bottom displays the distribution of predicted probabilities. Key calibration metrics (e.g., $C$-statistic, $Brier$ score, $E_{\text{max}}$, $S:p$) are reported in the upper left.}
    \label{fig:rms_reliability}
\end{figure}

Figure~\ref{fig:rms_reliability} provides a complementary assessment of model calibration using the \texttt{rms} package's reliability plot. The plot visualizes the agreement between predicted and observed probabilities across the range of fitted values. The proximity of the logistic calibration curves to the 45-degree line indicates that the model's predicted probabilities are well-aligned with actual outcomes, with only minor deviations at higher risk levels. however nonparametric line show some quadratic or non-linearity component that maybe missing.

The $C$-statistic (0.685) and Brier score (0.191) further support the model's reasonable discrimination and calibration.\footnote{A $C$-statistic (AUC) above 0.7 is generally considered acceptable, while values between 0.6 and 0.7 indicate modest discrimination; values below 0.6 are considered poor\parencite{hosmer2013applied}. For the Brier score, values closer to 0 indicate better calibration, with scores below 0.25 typically considered acceptable for binary outcomes\parencite{brier1950}.} The $E_{\text{max}}$ and $E_{\text{avg}}$ values quantify the maximum and average absolute calibration error, respectively, both of which are low.\footnote{For $E_{\text{max}}$ and $E_{\text{avg}}$, lower values indicate better calibration. While there is no universal cutoff, $E_{\text{max}}$ values below 0.1 and $E_{\text{avg}}$ values below 0.05 are often interpreted as indicating good calibration in clinical prediction models\parencite{van2019calibration}.}
The histogram shows that most predictions are concentrated at lower probabilities, reflecting the prevalence of the outcome in the dataset. Overall, the RMS reliability plot corroborates the findings from the GIVITI Calibration Belt, confirming that the fitted model provides reliable probability estimates for this data.

\subsection{Goodness-of-Fit Tests (RESULTS):} \label{sec:gof_tests_results}
We now proceed to apply various goodness-of-fit tests to the low birth weight dataset, utilizing the \texttt{birthwgt} data. In this analysis, we  Ungrouped the data to make it Sparse, meaning that each observation in \texttt{birthwgt\_trials} is set to 1 (See Appendix \ref{sec:ungrouping_data}). For certain tests—such as those by McCullagh, Farrington, Pearson, and Deviance—which have specific formulations for grouped data, we also use the Grouped original data, treating \texttt{birthwgt\_trials} as the number of covariate patterns $n_g$ in the grouped data context. However, it is important to note that, out of the 188 total observations, only 17 unique covariate patterns of $m_g = 2 or 3$ exist; the remaining patterns each occur only once, resulting in a sparse data structure.

\begin{figure}[H]
    \centering
    \includegraphics[width=1\textwidth]{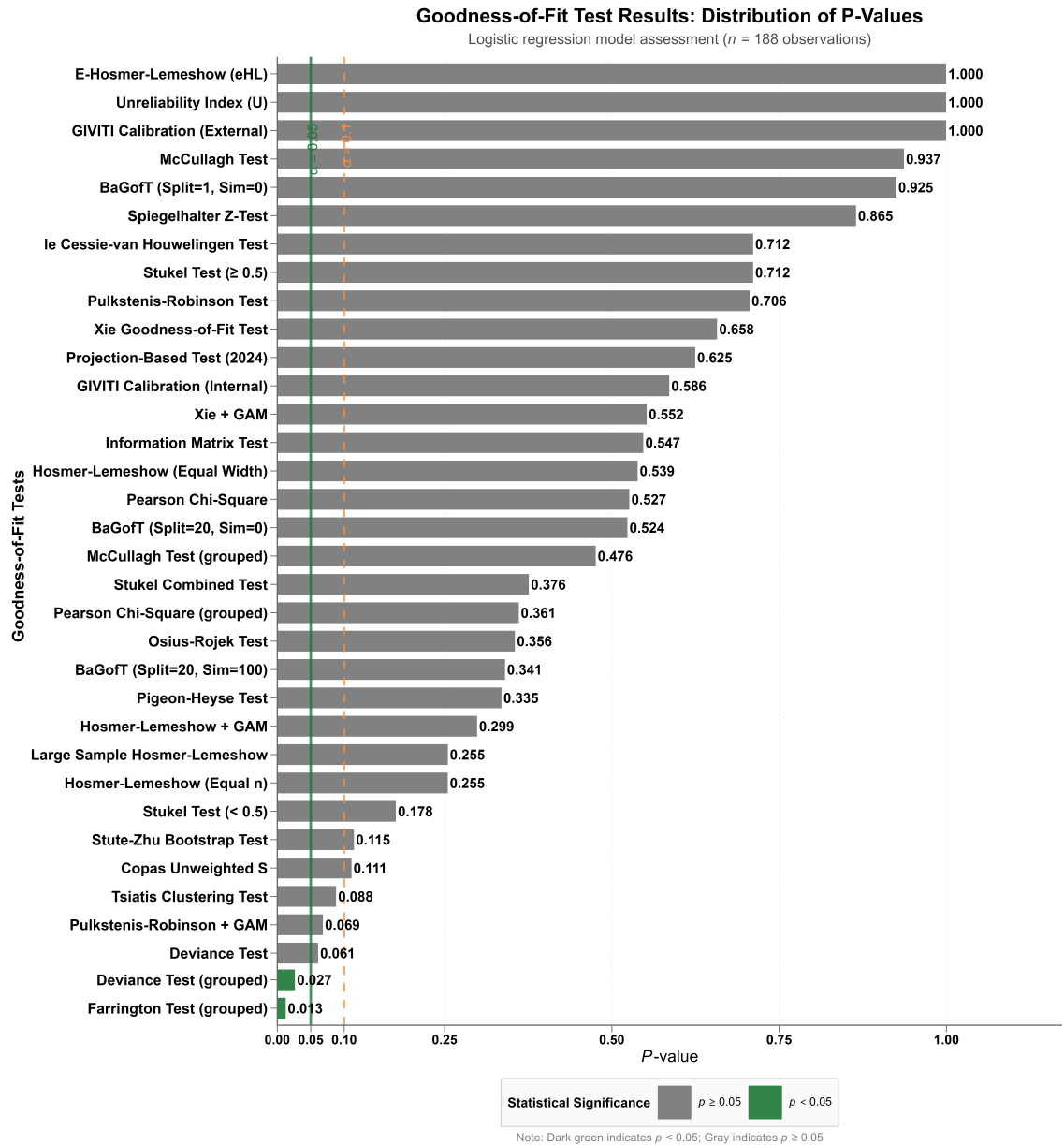}
    \caption[Goodness-of-Fit and Calibration Tests P-Values for the Low Birth Weight Dataset.]{Goodness-of-Fit and Calibration Tests P-Values for the LBW dataset.}
    \label{fig:gof_pvalues_lbw}
\end{figure}

\paragraph{Hosmer-Lemeshow and Related Tests:}
The results in (Figure~\ref{fig:gof_pvalues_lbw}) shows that, the traditional Hosmer-Lemeshow (equal sample size) test (\texttt{traditional\_HL}, $p=0.255$), the equal-width variant (\texttt{hosmer\_equal\_width}, $p=0.539$), and the large-sample HL (\texttt{large\_sample\_HL}, $p=0.255$) all yield p-values greater than 0.05, indicating no evidence against model fit. The \texttt{HL\_GAM} test ($p=0.299$), which uses probabilities from an over-specified GAM, also supports model adequacy. The Pigeon-Heyse modification ($p=0.335$) is consistent with these findings.

\paragraph{Classic and Standardized Pearson/Deviance Tests:}
The ungrouped \texttt{Pearson} ($p=0.527$) and \texttt{Deviance} ($p=0.061$) tests do not reject the null hypothesis, though the Deviance test is borderline. The grouped versions show more sensitivity: grouped \texttt{Pearson} ($p=0.361$) and especially grouped \texttt{Deviance} ($p=0.027$), the latter being significant and suggesting possible lack of fit. The \texttt{Osius-Rojek} ($p=0.356$) and \texttt{McCullagh's} test ($p=0.937$ ungrouped, $p=0.476$ grouped) also indicate no evidence of misfit.

\paragraph{Partitioning and Clustering-Based Tests:}
Tests that partition the covariate space, such as \texttt{Tsiatis clustering} ($p=0.088$), \texttt{Xie test} ($p=0.658$), and \texttt{PR\_test} ($p=0.706$), generally do not reject the null. The GAM-based extensions (\texttt{XIE\_GAM}, $p=0.552$; \texttt{PR\_GAM}, $p=0.069$) show slightly lower p-values for the \texttt{XIE\_GAM} test, with \texttt{PR\_GAM} being borderline. the \texttt{BaGofT} tests (e.g., \texttt{bagoft\_split1\_sim0}, $p=0.925$; \texttt{bagoft\_split20\_sim0}, $p=0.524$) also indicate no evidence of misfit. These results suggest that while most partitioning-based tests do not detect misfit, some GAM-based variants are more sensitive to subtle deviations.

\paragraph{Machine Learning and Calibration-Focused Tests:}
Modern calibration tests, including the \texttt{Unreliability Index} ($p=1.000$), \texttt{eHL} ($p=1.000$), and both \texttt{GiViTI} tests (external: $p=1.000$, internal: $p=0.586$), all strongly support the model's calibration. The \texttt{Spiegelhalter} test ($p=0.865$) also indicate no evidence of misfit.

\paragraph{Other Specialized and Smoothing-Based Tests:}
The \texttt{Stukel score} tests (geq0.5: $p=0.712$, l0.5: $p=0.178$, both: $p=0.376$), \texttt{IM-efficient} ($p=0.547$), \texttt{Copas unweighted} ($p=0.111$), and \texttt{le Cessie-van Houwelingen} test ($p=0.712$) all fail to reject the null hypothesis, further supporting the model's adequacy.

\paragraph{Bootstrap-Based Tests:}
The \texttt{Stute-Zhu} test ($p=0.115$) and the \texttt{Projection-Based} test ($p=0.625$) both yield non-significant p-values, supporting the adequacy of the model. The \texttt{bagoft\_split20\_sim100} test ($p=0.341$) also indicates no evidence of misfit.

\paragraph{Summary:}
Overall, the vast majority of tests—including classical, modern, calibration-focused, and machine learning-based approaches—do not reject the null hypothesis of good model fit. However, a few tests, particularly those based on grouped data (grouped Deviance, Farrington). Tests incorporating GAM-based grouping also tend to produce lower p-values, indicating increased sensitivity to nonlinearity or unmodeled effects.

\newpage
\subsection{Adding the interaction effects to the model:}
By incorporating the interaction effects into the model, equation~\ref{eq:true_model_real_data} represents a more complete specification. This revised model extends the approach of Hosmer et al.~\parencite{Hosmer1997} by explicitly including interaction terms between \texttt{AGE} and \texttt{LWT}, as well as between \texttt{SMOKE} and \texttt{LWT}.

\begin{figure}[H]
    \centering
    \includegraphics[width=0.9\textwidth]{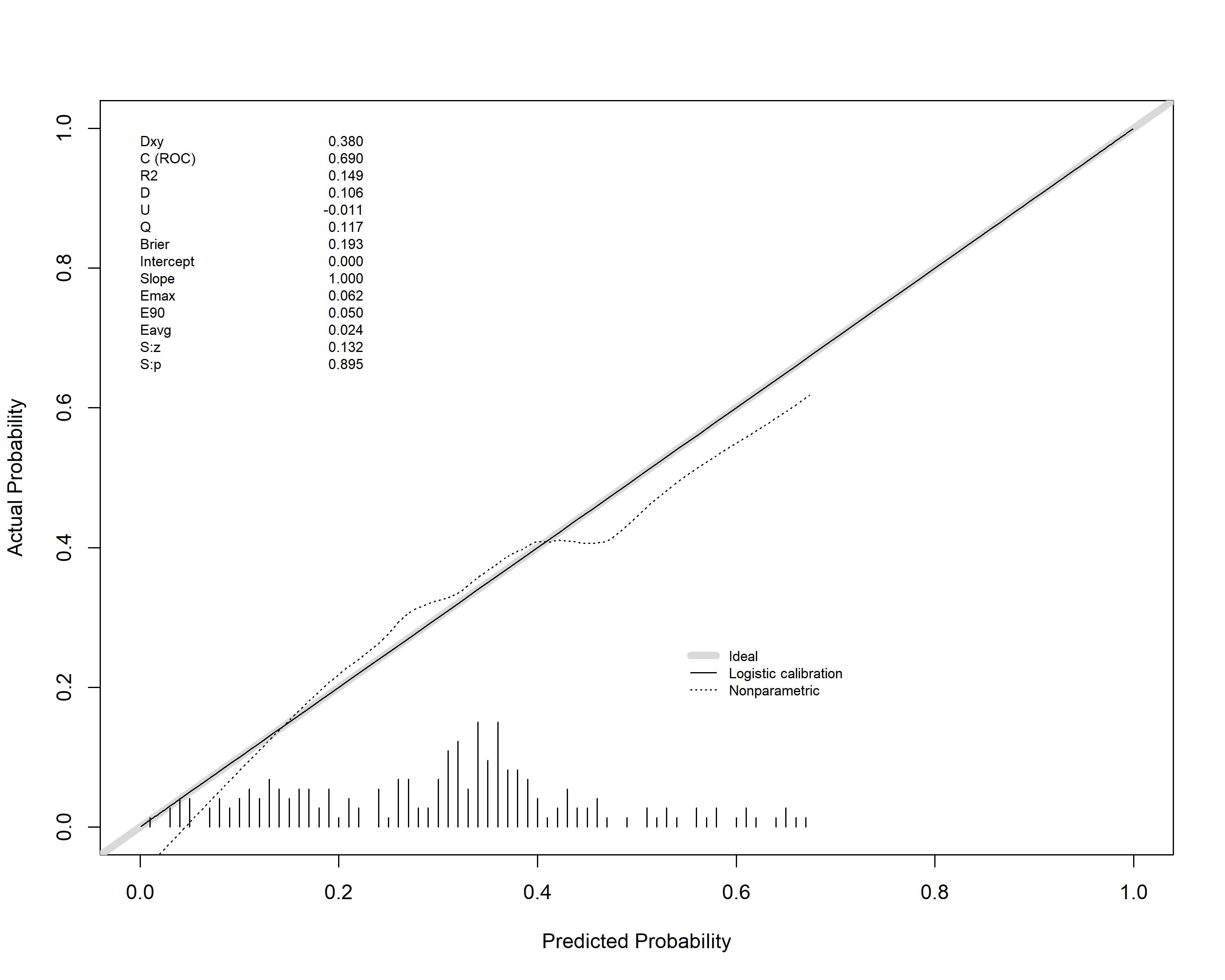}
    \caption[RMS Reliability Plot for the low birth weight dataset (with interaction effects).]{RMS Reliability Plot for the low birth weight dataset, after including interaction effects in the logistic regression model.}
    \label{fig:rms_reliability_interaction}
\end{figure}

With these interaction effects added to the logistic regression, we re-evaluate the model's calibration using the RMS reliability plot. Figure~\ref{fig:rms_reliability_interaction} presents the updated reliability plot for the improved model, now reflecting the inclusion of the \texttt{AGE:LWT} and \texttt{SMOKE:LWT} interaction terms.

Compared to the previous model (see Figure~\ref{fig:rms_reliability}), the nonparametric (loess) calibration curve in Figure~\ref{fig:rms_reliability_interaction} now shows much less deviation from the ideal 45-degree line. This indicates that the inclusion of interaction terms has improved the model's calibration, particularly in regions where the original model exhibited nonlinearity or systematic miscalibration. The logistic calibration curve and the nonparametric fit are now both closely aligned with the ideal line, suggesting that the updated model provides more reliable probability estimates across the range of predicted risks. The calibration metrics in the upper left of the plot (e.g., $C$-statistic, Brier score, $E_{\text{max}}$, $E_{\text{avg}}$) also reflect this improvement, with values indicating better discrimination and lower calibration error.

This result demonstrates the importance of considering interaction effects in logistic regression modeling, especially when initial calibration plots suggest potential model misfit or missing nonlinearities.

\begin{figure}[H]
    \centering
    \includegraphics[width=0.8\textwidth]{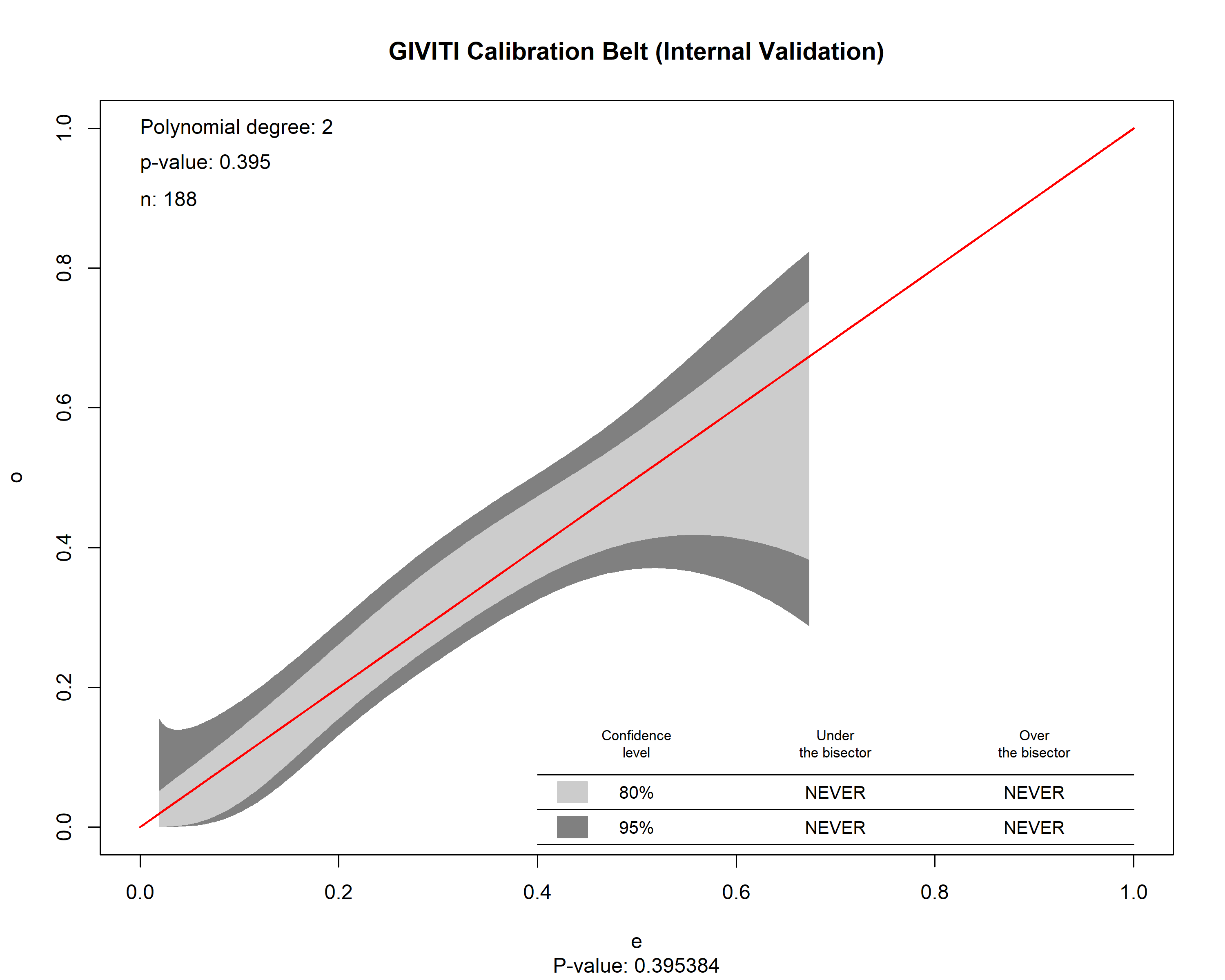}
    \caption[GIVITI Calibration Belt (Internal Validation) for the low birth weight dataset with interaction effects.]{GIVITI Calibration Belt (Internal Validation) for the low birth weight dataset after including interaction effects in the logistic regression model.}
    \label{fig:giviti_calibration_belt_internal_interaction}
\end{figure}

To illustrate the effect of including interaction effects, it is useful to compare the updated GIVITI Calibration Belt (Figure~\ref{fig:giviti_calibration_belt_internal_interaction}) to the original version (see Figure~\ref{fig:calib_belt_1997}). \textbf{In the original model (before adding interaction terms)}, the calibration curve required a higher polynomial degree (3), and the confidence bands (80\% and 95\%) were relatively wide, especially at the extremes of predicted risk. The curve also showed clear nonlinearity and cyclic deviations from the ideal 45-degree line, suggesting the presence of unmodeled effects or interactions. \textbf{After adding the interaction terms}, the updated calibration belt (Figure~\ref{fig:giviti_calibration_belt_internal_interaction}) demonstrates a notable improvement: the calibration curve is now well-approximated by a polynomial of degree 2, and the confidence bands are both narrower and more symmetric across the range of predicted probabilities. The reduction in cyclic or nonlinear patterns, along with the improved symmetry and tightness of the confidence bands, provides strong visual and statistical evidence that the inclusion of interaction effects has addressed much of the miscalibration observed in the original model. Interestingly, the calibration test p-value decreased after adding interaction terms (from 0.586 to 0.395), contrary to expectations. This shows that p-values can be sensitive to sample variation and model structure, so both statistical and graphical evidence should be used when assessing model fit.

\paragraph{GOF Test Statistics:} The inclusion of interaction terms in the logistic regression model is widely recognized in the literature as a means to improve model calibration and overall fit. This enhancement is often reflected in the p-values of goodness-of-fit (GOF) tests, where an increase in p-values typically indicates stronger evidence supporting model adequacy. In particular, tests that previously signaled poor fit—such as the \texttt{Deviance test (grouped)} and the \texttt{Farrington test (grouped)}—are expected to be less likely to reject the model once relevant interaction effects are incorporated. This expectation aligns with established research emphasizing the value of both graphical calibration diagnostics and the explicit modeling of interactions to achieve more accurate and reliable predictions.

\begin{figure}[H]
    \centering
    \includegraphics[width=1\textwidth]{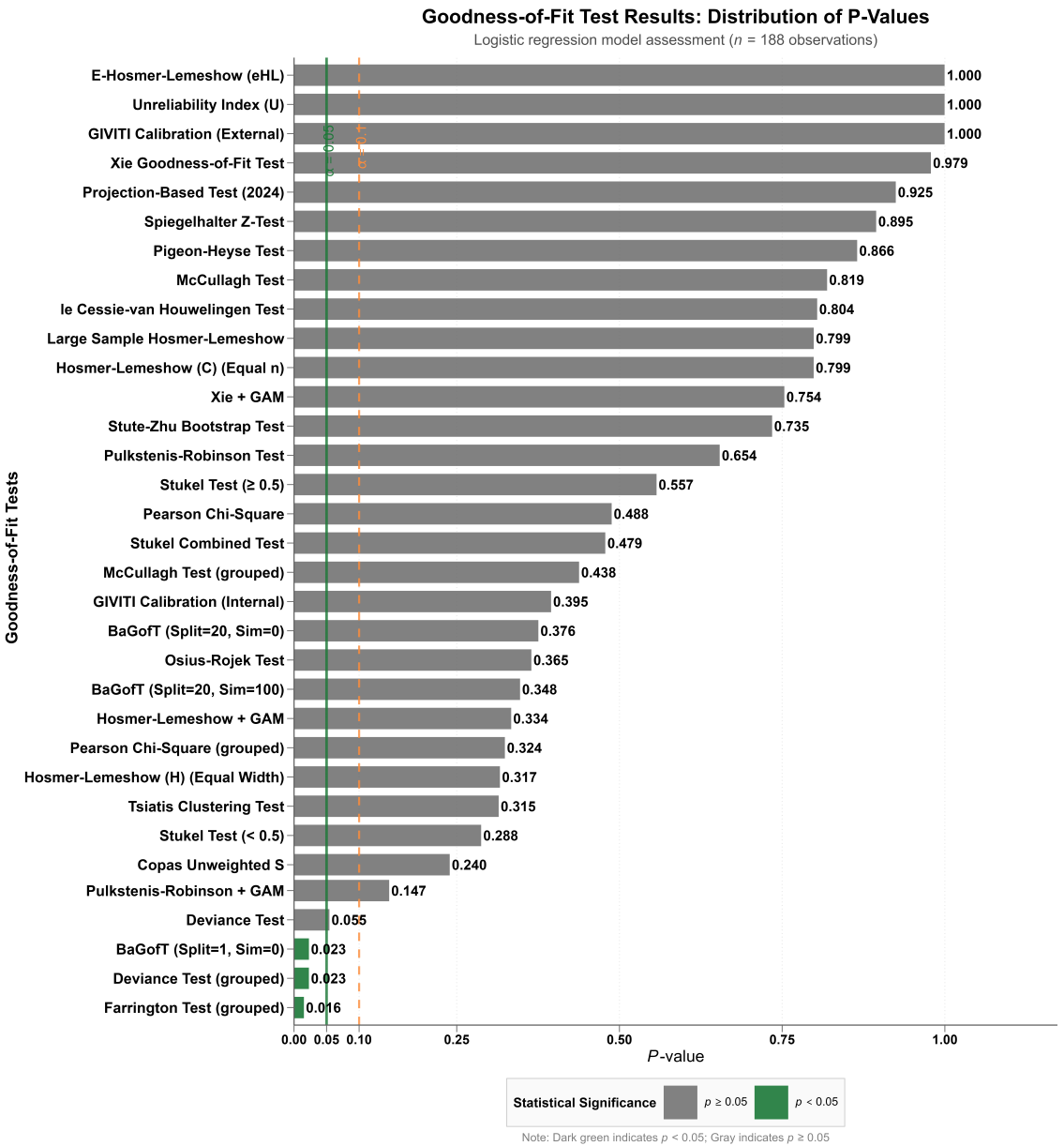}
    \caption[Goodness-of-Fit and Calibration Tests P-Values for the Low Birth Weight Dataset (with interaction effects).]{Goodness-of-Fit and Calibration Tests P-Values for the Low Birth Weight Dataset, after including interaction effects in the logistic regression model.}
    \label{fig:gof_pvalues_lbw_interaction}
\end{figure}

Figure~\ref{fig:gof_pvalues_lbw_interaction} displays the updated p-values for a comprehensive set of GOF and calibration tests following the addition of interaction terms to the logistic regression model. The visualization mirrors the format of Figure~\ref{fig:gof_pvalues_lbw}, allowing for direct comparison between the original and improved model specifications.

The revised p-values indicate that, for most tests, there is an upward shift, consistent with improved model fit after accounting for interaction effects. However, a few tests—such as certain configurations of the BAGoFT test (e.g., without simulation or with minimal splits)—show a decrease in p-value, underscoring that not all methods respond identically to model modifications. Notably, the p-values for the Deviance test (grouped) and the Farrington test (grouped) exhibit only marginal increases and continue to reject the model at the conventional $\alpha = 0.05$ threshold. The majority of other GOF and calibration tests, both classical and modern, maintain p-values above this threshold, supporting the adequacy of the updated model.

These findings illustrate that, while the addition of interaction terms generally enhances model calibration and fit—as reflected in most GOF test results—some tests remain sensitive to residual model misfit. This underscores the necessity of employing a combination of graphical and statistical GOF assessments, as well as careful model specification, to ensure robust evaluation in logistic regression analyses.

\section*{Conclusion}
This chapter provided a comprehensive framework for evaluating goodness-of-fit (GOF) tests in logistic regression, combining a large-scale simulation study with a practical application to a real-world dataset. The methodology was anchored in the established work of Hosmer et al. (1997) to ensure rigor and comparability, while extending the analysis to a broader range of sample sizes and a more diverse collection of modern and classical tests.
\vspace{0.5em}

\noindent The simulation study systematically assessed the performance of GOF tests in terms of their size (Type I error rate) and power under various scenarios of model misspecification. This controlled environment allowed for a nuanced understanding of how different tests respond to challenges such as non-linearity, incorrect link functions, and omitted interaction effects.
\vspace{0.5em}

\noindent The application to the low birth weight dataset served as a practical demonstration of the principles discussed. The initial analysis revealed a key insight: while most GOF tests suggested the model was adequate, graphical diagnostics (RMS plot, GIVITI belt) and a few specific tests (e.g., the grouped Deviance test) hinted at underlying miscalibration. However, these same tests continued to reject the null hypothesis even after the model was corrected, failing to reflect the improved fit. This discrepancy highlighted the danger of relying on a single test and underscored the value of a multi-ways assessment strategy. However, after incorporating interaction terms, the model's calibration was visibly improved, as confirmed by the updated graphical plots. The subsequent GOF analysis reflected this improvement, with most tests yielding higher p-values. However, the varied responses of different tests to this model refinement further illustrated their differential sensitivities.

\vspace{0.5em}
\noindent  Building on these results, the following chapter provides a broader discussion that synthesizes the key findings, explores their practical implications, and outlines avenues for future research.

\chapter{Discussion, Conclusion and Recommendation}

\section*{Introduction}

This chapter synthesizes and interprets the findings from the comprehensive simulation study and the real-world data application presented in the preceding chapters. The core objective of this thesis was to rigorously evaluate and compare a diverse array of goodness-of-fit (GOF) and calibration tests for binary logistic regression, with a particular focus on their performance in sparse data settings characterized by continuous covariates. The simulation study revealed significant heterogeneity in test performance, highlighting that no single test is uniformly superior across all conditions. Key findings indicated that while some tests exhibit strong power against specific forms of model misspecification, others suffer from inflated Type I error rates or fail to detect subtle yet important model deficiencies.
\vspace{0.5em}

\noindent The application to the low birth weight dataset served as a practical illustration of these empirical findings, revealing a critical insight: a disconnect often exists between the conclusions drawn from statistical tests and those from graphical diagnostics. This case study demonstrated that while most tests initially suggested an adequate model fit, a more nuanced assessment incorporating visual tools and a corrected model specification told a different story. These results challenge the common practice of relying on a single p-value and underscore the necessity of a holistic validation strategy that integrates multiple sources of evidence.
\vspace{0.5em}

\noindent This discussion situates these results within the broader academic discourse, connecting them to the foundational work of Hosmer et al. (1997) and the ongoing challenges in model assessment noted in recent literature. We argue that the findings advocate for a paradigm shift from isolated test application towards an integrated, multi-faceted validation framework. The remainder of this chapter will delve deeper into the implications of our results, discuss the rationale for excluding certain tests from the main analysis, acknowledge the study's limitations, and future research that can build upon the insights gained.

\newpage
\section{Justification for Excluded Tests}
In any large-scale simulation study, it is common to find that certain tests are unsuitable for the specific conditions being investigated. This section details the tests that were excluded from the main comparative analysis and provides both theoretical and empirical justifications for their exclusion. The primary reasons for exclusion were non-informative results (e.g., zero power), severe lack of control over the Type I error rate (liberal tests), or a lack of applicability to the simulation scenarios.

\subsection{Tests with Inflated Type I Error Rates (Liberal Tests)}
A test is considered liberal if its empirical Type I error rate significantly exceeds the nominal significance level ($\alpha=0.05$), leading to an unacceptably high rate of false positives.

\paragraph{Pearson and Deviance Tests (Ungrouped)}
The classic Pearson and Deviance tests are theoretically known to be invalid in sparse data settings, as the test statistics do not follow the assumed $\chi^2$ distribution. Our simulations confirmed this theoretical failure. As seen in the Type I error analysis (Figure~\ref{fig:liberal_zero_power_infeasible_tests}), the \texttt{pearson} test exhibited a highly inflated error rate, reaching 20.6\% in the Uniform(-6,6) scenario at $n=2000$. The \texttt{deviance} test was often overly conservative but became liberal in other scenarios, making it unreliable. This unreliability was less apparent in the real data application, where the \texttt{pearson} test yielded non-significant p-values of 0.527 and 0.488 for the initial and corrected models, respectively. The \texttt{deviance} test produced borderline p-values of 0.061 and 0.055. However, given their demonstrated lack of size control in the simulations, they were excluded from the primary power comparisons.

\paragraph{(Deviance)F-Test (\texttt{hl\_ftest})}
This test, which uses an F-distribution, proved to be catastrophically liberal in our simulations. Its empirical Type I error rate was consistently near 100\% across most scenarios and sample sizes. For example, in the Uniform(-6,6) scenario with $n=2000$, its error rate was 99.8\% (Figure~\ref{fig:liberal_zero_power_infeasible_tests}). Such behavior renders the test completely unreliable for practical use, as it rejects nearly every correctly specified model.

\paragraph{Hosmer-Lemeshow Bootstrap Test (Lai \& Liu)}
The method proposed by Lai \& Liu is not a conventional statistical test but a power-standardization technique. It utilizes bootstrap power estimation to derive a probabilistic rule, which results in a binary accept/reject (1/0) decision depending on the test confidence to accept the model rather than a continuous p-value. While this approach is innovative and potentially powerful, our simulations revealed it to be consistently liberal. Across the tested scenarios, its Type I error rate was approximately 11\% for most sample sizes, and in some cases, escalated to as high as 25\% at N=500. This inflated error rate suggests its decision framework is overly aggressive in sparse data settings. Consequently, it was excluded from the main comparative analysis, which was limited to tests that provide a standard p-value for direct comparison.

\subsection{Tests with Zero or Near-Zero Power}
Some tests were found to be excessively conservative, rarely or never rejecting the null hypothesis even when the model was severely misspecified.

\paragraph{Farrington Standalone Test}
The test labeled \texttt{farrington\_standalone} applies the original Farrington test formulas, which are designed for grouped data, to ungrouped binary data. As theoretically predicted in Section \ref{sec:farrington_test}, the variance term of the statistic collapses to zero when all trials are equal to one. The simulation results confirmed this perfectly: the test never rejected the null hypothesis under any condition, yielding a power of 0.0 across all power scenarios (see Figures~\ref{fig:liberal_zero_power_infeasible_tests}). Interestingly, in the real data application, the test produced significant p-values of 0.013 and 0.016 for the initial and corrected models, respectively, suggesting model rejection in both cases. This discrepancy is explained by the data's structure. The Farrington test is designed for grouped data, and the real-world dataset contained the necessary covariate patterns (multiple subjects with identical covariate values) for the test to function as intended. Our simulations, however, used ungrouped data to model sparse settings, a context in which the test is theoretically invalid and was empirically observed to fail. Given its failure in the controlled simulation environment, it was excluded from the main comparative analysis.

\paragraph{Unreliability Index (\texttt{unreliability\_index})}
This test is designed to assess the calibration model by testing the joint hypothesis that the calibration intercept is 0 and the slope is 1. It is a powerful tool for detecting linear miscalibration (recall \ref{sec:unreliability_test}) but is not designed to detect other forms of misspecification, such as omitted non-linear or interaction terms. This was confirmed in our power analysis, where the \texttt{unreliability\_index} had a power of 0.0 for all omitted variable and interaction scenarios. Its performance in the real data application further validated this: it yielded a p-value of 1.0 for both the initial and corrected models, failing to detect any misspecification. As it was not sensitive to the types of misspecification studied here, it was excluded from the power analysis comparisons.

\subsection{Tests Not Generally Applicable}
The \texttt{PR\_test} and its GAM-based variant are specifically designed for models that include at least one categorical covariate for the initial grouping. Since our primary simulation scenarios focused exclusively on continuous covariates to assess performance in the sparsest settings, these tests were not applicable and were thus excluded from the main simulation results tables. However, they were applicable to the low birth weight dataset, which includes categorical predictors. In that application, the \texttt{PR\_test} yielded p-values of 0.706 and 0.654 for the initial and corrected models, respectively, suggesting adequate fit for both the initial and corrected models (in correct model p-value falsely decreased). Its GAM-based variant, \texttt{PR\_GAM}, produced p-values of 0.069 (Very Close to Reject tje incomplete model) and 0.147, showing a notable improvement in fit after adding the interaction term. These results suggest the tests are potentially useful tools when their applicability conditions are met.

\newpage
\section{Performance of Key Goodness-of-Fit Tests}
This section provides a detailed discussion of the performance of the remaining tests, organized according to the methodological families outlined in Chapter 5. The analysis integrates theoretical expectations with the empirical findings from the simulation study (Figures~\ref{fig:type1error_uniform_heatmap} through \ref{fig:power_omitted_interaction_heatmap}) and the real-world application (Figure~\ref{fig:gof_pvalues_lbw} and \ref{fig:gof_pvalues_lbw_interaction}). \textbf{It is important to recall the note mentioned in previously, that tests classification are not mutually exclusive, a test can belong to multiple categories}, this classification is just for the sake of clarity and organization.

\subsection{Hosmer-Lemeshow Variants}
The Hosmer-Lemeshow (HL) test and its variants are the most common benchmarks for GOF testing in logistic regression. They all rely on partitioning the data by predicted probabilities.

\paragraph{Traditional HL Test (\texttt{traditional\_HL})}
The traditional HL test performed as expected, serving as a solid benchmark. It generally maintained good control over the Type I error rate, with empirical rates staying close to the nominal 0.05 level across all scenarios and sample sizes (e.g., 5.0\% in the Uniform(-6,6) scenario at $n=1000$). Its power was moderate; for instance, in the pronounced quadratic misspecification, it achieved a respectable power of 99.8\% at $n=2000$, but struggled with the slight quadratic misspecification, reaching only 10.8\% power at the same sample size. In the real application, it correctly failed to reject the initial misspecified model ($p=0.255$) and, as expected, its p-value increased substantially to $p=0.799$ after the model was corrected with interaction terms, correctly reflecting the improved fit. (check figure \ref{fig:gof_pvalue_change_plot_A4})

\paragraph{Large Sample HL Test (\texttt{large\_sample\_HL})}
The large sample modification by Nattino et al. is designed to be more robust for very large $n$. In our simulations, which went up to $n=5000$, its performance was almost identical to the traditional HL test. For example, in the slight quadratic scenario at $n=5000$, its power was 17.0\% compared to 17.2\% for the traditional test. This is expected, as the modification's primary benefit manifests at even larger sample sizes ($> 5000$). In the real application, its p-values ($0.255$ initial, $0.799$ final) were virtually identical to the traditional test, confirming their similarity at this sample size.

\paragraph{Pigeon-Heyse Test (\texttt{pigeon\_heyse})}
This test modifies the HL statistic with a correction factor for within-group variability. Theoretically, this should make it more robust. Empirically, it was found to be consistently more conservative than the traditional HL test, with Type I error rates often falling below 5\% (e.g., 3.4\% in the Uniform(-6,6) scenario at $n=1000$). This conservatism came at the cost of power; in the slight quadratic scenario at $n=2000$, its power was 7.6\%, noticeably lower than the traditional HL's 10.8\%. In the real application, it behaved as expected, with its p-value increasing from a non-significant $0.335$ to a very high $0.866$ after model correction.

\newpage
\paragraph{Hosmer-Lemeshow with Equal-Width Intervals (\texttt{hosmer\_equal\_width})}
This variant partitions the probability scale `[0, 1]` into equal-width intervals. It demonstrated good control of the Type I error rate, performing comparably to the traditional HL test. Its power profile showed a notable divergence: it was less powerful in detecting the slight quadratic misspecification (6.9\% power vs. 10.8\% for traditional HL at $n=2000$) but was surprisingly more sensitive to the slight omitted interaction term (12.2\% power vs. 7.5\% for traditional HL at $n=2000$). In the real application, its p-value in falsely decreased from $0.539$ to $0.317$ after model correction, though the final p-value was lower than for other HL variants. This suggests that the choice of grouping strategy can influence a test's sensitivity to different types of misfit.

\subsection{Classic and Standardized Tests}
This family of tests moves beyond simple grouping by using more sophisticated statistical properties to assess fit.

\paragraph{Osius-Rojek and McCullagh Tests}
The \texttt{osius\_rojek} and \texttt{mccullagh\_test}, both of which standardize the Pearson statistic, were standout performers. They consistently maintained excellent Type I error control, even in the most sparse scenarios (e.g., for Osius-Rojek, the error rate was 4.5\% in the Uniform(-6,6) scenario at $n=1000$). Their key advantage was their statistical power. The \texttt{mccullagh\_test} was consistently one of the most powerful tests, achieving 100\% power in the pronounced quadratic scenario at $n=1000$, and an impressive 20.0\% in the difficult slight quadratic case at $n=2000$. In the real application, the McCullagh test behaved differently, with its p-value decreasing from an already high $0.937$ to $0.819$ after correction (which means it did not detect any misspecification, not only but also, the p-value falsely decreased after adding the interactions term), the test was strongly supporting both models. The Osius-Rojek test performed well in the application, with its p-value increasing from $0.356$ to $0.365$ (both p-values in general less than thoes of McCullagh tests).

\paragraph{Spiegelhalter's Test}
This Brier-score based test is theoretically sound, as it directly assesses the calibration component of a proper scoring rule. However, in our simulations, it proved to be the most conservative of all the tests evaluated. Its empirical Type I error rate was consistently 0.0 across almost all scenarios (Figures \ref{fig:type1error_uniform_heatmap}, \ref{fig:type1error_chi2_multi_heatmap}, \ref{fig:type1error_normal_heatmap}), guaranteeing an extremely low risk of false positives. This strict control, however, came at a very high cost to its statistical power, rendering it insensitive to all but the most severe forms of model misspecification.

This lack of sensitivity was most apparent in the scenarios with subtle model failures. For both the slight quadratic and the slight interaction misspecifications, the power of Spiegelhalter's test was 0.0 across all sample sizes, from $n=200$ to $n=5000$. It completely failed to detect these more nuanced forms of lack of fit, whereas other top-tier tests like McCullagh's or Stute-Zhu's showed increasing power with sample size.

Even when the model misspecification was pronounced, the test required a very large sample size to respond. In the pronounced quadratic scenario (Figure~\ref{fig:power_omitted_variable_heatmap}), its power was negligible for smaller samples, only climbing to a useful level at $n=1000$ (power of 78.5\%) and becoming reliable at $n=2000$ (power of 99.9\%). In contrast, a test like McCullagh's already achieved 100\% power by $n=1000$ in the same scenario. The test's insensitivity was even more stark in the pronounced interaction scenario (Figure~\ref{fig:power_omitted_interaction_heatmap}), where its power remained effectively zero until the sample size reached $n=5000$, at which point it finally achieved a power of 91.1\%.

\subsection{Tests Based on Partitioning the Covariate Space}
These tests group observations based on their covariate values rather than their predicted probabilities.

\paragraph{Tsiatis and Xie Tests}

The Tsiatis test (\texttt{tsiatis\_clustering}), a score test that checks for the significance of dummy variables created by clustering the covariate space, and the Xie test (\texttt{xie\_test}), a Hosmer-Lemeshow-like test that partitions data based on the covariate space, both showed good control over the Type I error rate, though the Xie test was often conservative (e.g., 2.5\% in the Uniform(-6,6) scenario at $n=1000$). Their power was generally lower than the top-tier standardized tests but still respectable. In the pronounced interaction scenario at $n=1000$, the \texttt{tsiatis\_clustering} test achieved a power of 99.3\%. In the real application, both tests behaved as expected, with the Tsiatis p-value perfectly increasing from a borderline $0.088$ (more likely to reject the incomplete model) to $0.315$, however, both p-values in general less than thoes of Hosmer, McCullagh and Osius-Rojek tests, and the Xie p-value increasing from $0.658$ to $0.979$, correctly signaling the improved model fit.

\paragraph{BAGofT Test (NO Bootstraping)}
The BAGofT (\texttt{bagoft\_test}) uses a two-stage, data-splitting strategy to test for model misspecification. The dataset is first split into training and test sets. The training data is used to learn an "adaptive" partition of the covariate space, specifically designed to find regions where the model fits poorly. This partition is then used to group the observations in the independent test set, where a chi-square-like statistic is computed. This separation of the partition-finding and testing stages prevents overfitting and ensures valid statistical inference. The number of random data splits is a key tuning parameter; the procedure is typically repeated multiple times to produce a more stable, aggregated result.

The two main configurations tested were a single data split (`nsplits=1`, denoted \texttt{bagoft\_split1\_sim0}) and an aggregated version with 20 splits (`nsplits=20`, denoted \texttt{bagoft\_split20\_sim0}). The version with 20 splits proved to be extremely conservative across almost all scenarios. Its empirical Type I error rate was consistently 0.0, and as a direct consequence, its power to detect any form of misspecification was also effectively zero (Figures \ref{fig:power_omitted_variable_heatmap} and \ref{fig:power_omitted_interaction_heatmap}). This suggests that the aggregation method, while intended to improve stability, renders the test completely insensitive.

The single-split version was less conservative but demonstrated highly problematic and unstable behavior. In terms of size, it was found to be liberal, with its Type I error rate climbing as high as 11.2\% in the sparse Uniform(-6,6) scenario at n=200. More concerning was its power profile. In the slight interaction scenario, the test's power exhibited a counter-intuitive and statistically unsound trend: its power to detect the misspecification \textbf{decreased} as the sample size increased, dropping from 11.1\% at n=200 to 6.3\% at n=2000. This is a critical failure for a statistical test, as it implies that with more data, the test becomes more confident that a poorly specified model is actually a good fit. This paradoxical behavior suggests that the test's adaptive partitioning mechanism may be unstable, leading to unreliable conclusions.

In the real-world application, this instability was also evident. For the initial misspecified model, the single-split test produced a non-significant p-value of 0.925, failing to detect the misfit. However, for the *corrected* model, its p-value dropped to a highly significant 0.023, incorrectly signaling a lack of fit for the better model. Given its liberal tendencies, unstable power characteristics, and erratic performance in the real-world application, the \textbf{BAGofT test (with no bootstraping)}, in its current implementation and with the tested parameters, \textbf{cannot be recommended as a reliable tool for goodness-of-fit assessment.}

\paragraph{GAM-based Tests} The GAM-based tests were introduced with the hypothesis that forming groups based on probabilities from a more flexible, over-specified GAM would lead to more powerful goodness-of-fit assessments. The simulation results, however, reveal a complex and inconsistent picture, where the utility of the GAM preprocessing was highly dependent on the underlying base test.

A key finding is that the GAM-based grouping significantly improved the Type I error control of the \texttt{xie\_test}. As shown in the Type I error analysis (Figures \ref{fig:type1error_uniform_heatmap}, \ref{fig:type1error_chi2_multi_heatmap}, \ref{fig:type1error_normal_heatmap}), the standard Xie test was often overly conservative, with error rates falling as low as 2.2\% in the Uniform(-3,3) scenario and 2.5\% in the Uniform(-6,6) scenario at $n=1000$. After applying the GAM preprocessing, the resulting \texttt{XIE\_GAM} test was much better calibrated, with its Type I error rate consistently closer to the nominal 5\% level across all scenarios (e.g., 4.9\% and 5.1\% in the same Uniform scenarios, respectively). This improved size control directly translated into better power. While the standard Xie test struggled to detect the slight interaction (power of 3.9\% at $n=2000$), the \texttt{XIE\_GAM} test was nearly three times more powerful (11.2\%), demonstrating that the GAM-based grouping can be effective when the base test is too conservative. (check figure \ref{fig:power_omitted_interaction_heatmap})

In contrast, the GAM preprocessing had a detrimental effect on the Hosmer-Lemeshow test. The \texttt{traditional\_HL} test already had good Type I error control, but the \texttt{HL\_GAM} variant became noticeably liberal, particularly in the complex Multi-Independent scenario where its error rate reached 26.1\% at a small sample size of $n=200$. This inflated Type I error did not translate into a consistent power advantage. In fact, the \texttt{traditional\_HL} test was often more powerful than the \texttt{HL\_GAM} test. For instance, in the slight quadratic scenario at $n=2000$, the traditional HL had a power of 10.8\% compared to only 10.3\% for the GAM version. \\

In the real-world application, both GAM-based tests behaved as expected, with their p-values increasing after the model was corrected with interaction terms (\texttt{HL\_GAM}: $0.299 \to 0.334$; \texttt{XIE\_GAM}: $0.552 \to 0.754$). However, they did not offer a clear diagnostic advantage over their simpler counterparts. The results suggest that while GAM-based grouping can be a good tool for recalibrating an overly conservative test like Xie's, its utility is not universal and can even be harmful when applied to an already well-behaved test like Hosmer-Lemeshow, leading to a loss of size control without a consistent gain in power.

\subsection{Machine Learning and Advanced Methods}
This category includes modern, computationally intensive approaches that often originate from the machine learning literature but have strong parallels with statistical goodness-of-fit testing.

\paragraph{GiViTI Calibration Test (\texttt{giviti\_internal} and \texttt{giviti\_external})}
The GiViTI calibration framework represents a significant methodological advancement over the fixed-bin approach of the Hosmer-Lemeshow test. Instead of using rigid groups, it employs a data-driven polynomial regression to flexibly model the calibration curve, allowing it to detect more complex, non-linear patterns of miscalibration. The framework provides two distinct tests for two different validation scenarios: internal validation (goodness-of-fit) and external validation.

\subparagraph{The Internal (Goodness-of-Fit) Test}
The \texttt{giviti\_internal} test is the correct version for assessing the goodness-of-fit of a model on the same data used to train it. Theoretically, its flexible polynomial approach should give it more power to detect non-linear misfit compared to the linear bins of the Hosmer-Lemeshow test.

Our simulation results strongly support this theoretical advantage. First, the test demonstrated excellent control of the Type I error rate, maintaining its size close to the nominal 5\% level across all scenarios (e.g., 4.7\% in the Uniform(-3,3) scenario and 4.8\% in the complex Multi-Independent scenario at $n=1000$). More importantly, it showed a clear power advantage over the traditional HL test in detecting subtle misspecifications. For example, in the slight quadratic scenario at $n=2000$, the GiViTI test achieved a power of 16.3\%, substantially higher than the 10.8\% power of the traditional HL test. The advantage was even more pronounced in the slight interaction scenario at $n=2000$, where the GiViTI test's power was 14.1\%, nearly double that of the HL test (7.5\%). This confirms that its data-driven flexibility is highly effective. For low samples $n=200$, slight interaction scenario, power GiViTI test's power was approximately 64\%, while Hosmer was 42\%.

In the real-world application, the internal test produced a counter-intuitive result: its p-value decreased slightly from a non-significant $0.586$ for the initial misspecified model to $0.395$ for the corrected model. While a better model should ideally yield a higher p-value, this result highlights the primary strength of the GiViTI framework. The p-value alone was less informative than the accompanying calibration belt (Figures~\ref{fig:calib_belt_1997} and \ref{fig:giviti_calibration_belt_internal_interaction}), which visually confirmed a dramatic improvement in fit. The belt for the corrected model was smoother, required a lower polynomial degree, and showed less systematic deviation from the ideal line. This demonstrates that the GiViTI test's main value lies in its combination of a reasonably powerful formal test with an invaluable graphical diagnostic tool.

\subparagraph{The External Test and a Tale of Two Contexts}
The \texttt{giviti\_external} test is designed for a different purpose: to validate a pre-existing, "frozen" model on a new, independent dataset. Its statistical theory assumes that the model's parameters are fixed. However, in our simulation study, the model was refit for every single replication. This means we were applying the external test in a context that violated its core assumption, making it a test of its robustness outside its intended use case.

The results clearly show that the external test is not appropriate for this goodness-of-fit context. It was found to be extremely conservative, with empirical Type I error rates consistently near zero (e.g., 0.05\% in the Uniform(-6,6) scenario at $n=1000$, far below the nominal 5\%). This extreme conservatism led to a catastrophic loss of power. For the slight quadratic and slight interaction misspecifications, its power was less than 1\% across all sample sizes. Even for the pronounced interaction at $n=1000$, where the internal test achieved 99.9\% power, the external test's power was only 96.0\%.

This stark difference in performance is explained by the mismatch in statistical context. The `internal` test correctly accounts for the fact that the model parameters were estimated from the data, leading to a correctly calibrated null distribution. The `external` test's null distribution, which assumes fixed parameters, is incorrect for this setting, resulting in an overly conservative procedure. In the real application, the external test's p-value was 1.000 for both the initial and corrected models, confirming its inability to provide any diagnostic information when used for a goodness-of-fit assessment. This finding underscores the critical importance of selecting a statistical test that is theoretically matched to the experimental design.

\paragraph{eHL Test}
The e-value based Hosmer-Lemeshow test (\texttt{eHL}) is a modern, non-parametric approach that replaces the arbitrary, fixed bins of the traditional HL test with an adaptive partitioning scheme based on isotonic regression. The test uses sample-splitting to learn a calibration map on a training portion of the data and then calculates an e-value—a measure of evidence against the null hypothesis of good calibration—on the remaining test portion. The final test statistic, `eHL`, is the average of these e-values over many random splits, and a p-value is derived using the transformation $p = \min(1, 1/eHL)$.

Theoretically, an e-value greater than 1 provides evidence against the null hypothesis, suggesting the calibrated probabilities from the isotonic regression are a better fit to the data than the original model's probabilities. Conversely, an e-value less than 1 indicates that the original model's predictions were actually a better fit than the recalibrated ones. The test's authors note that this latter case should be rare, as isotonic regression is designed to improve calibration.

\begin{figure}[H]
    \centering
    \includegraphics[width=0.8\textwidth]{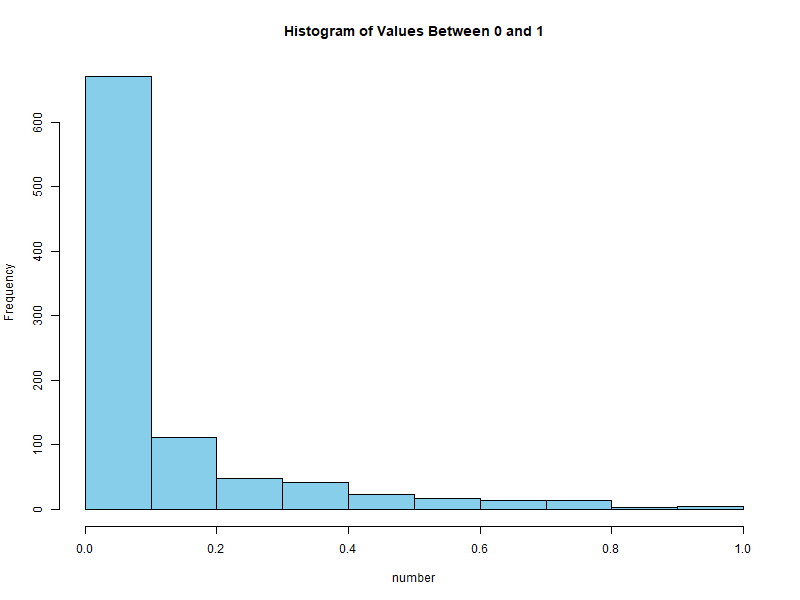}
    \caption{Distribution of the final eHL test statistic over 1000 simulation replications for the slight quadratic misspecification scenario ($n=1000$). The vast majority of the values are concentrated between 0 and 1.}
    \label{fig:ehl_distribution_slight_quad}
\end{figure}

However, our simulation study reveals a critical performance characteristic of the test in scenarios with only subtle model misspecification. Figure~\ref{fig:ehl_distribution_slight_quad} shows the distribution of the final `eHL` statistic from 1000 replications of the slight quadratic misspecification scenario. The histogram is heavily right-skewed, with the overwhelming majority of its mass—approximately 95\% of the 1000 replications—producing an eHL value less than 1. This empirical distribution demonstrates that in situations of very subtle misfit, the non-parametric isotonic regression, when fit on a small, random training split, consistently fails to find a better calibration map. The noise in the training split leads it to produce recalibrated probabilities that are a poorer fit to the test data than the original, slightly flawed model's probabilities.

This directly explains the test's extremely conservative behavior. Since the final `eHL` statistic is almost always less than 1, it is mathematically impossible for it to cross the conventional rejection thresholds of 10 (for $p < 0.10$) or 20 (for $p < 0.05$). Consequently, its empirical Type I error rate was consistently near zero across all scenarios (e.g., 0.2\% in the Uniform(-6,6) scenario at $n=1000$), and its power was less than 1\% in all slight misspecification scenarios. In the real application, it produced a p-value of 1.000 for both the misspecified and corrected models, failing to provide any diagnostic information. While the eHL test is theoretically robust against false positives, our findings suggest that it is \textbf{ not sensitive enough to be a practical tool for detecting subtle but potentially important forms of model inadequacy.}

\subsection{Bootstrap-Based Tests}
These tests use resampling to derive the null distribution, which is theoretically advantageous.

\paragraph{Stute-Zhu Test (\texttt{stute\_zhu})}
The Stute-Zhu test demonstrated an excellent balance of properties, positioning it as one of the most reliable tests in this study. The test operates on a powerful principle that avoids arbitrary grouping. It sorts the data based on the model's linear predictor (the logit of the predicted probabilities) and then calculates a cumulative sum of the residuals. A significant, systematic drift in this sum away from zero indicates model misfit. Because the theoretical distribution of this cumulative deviation statistic is complex and data-dependent, the test relies on a model-based bootstrap procedure to simulate the null distribution and derive a reliable p-value.

In our simulations, this methodology proved highly effective. The test maintained its nominal size almost perfectly across all scenarios (e.g., 5.2\% in the sparse Uniform(-6,6) scenario at $n=1000$), confirming that the bootstrap procedure provides excellent Type I error control. It also showed very strong power, achieving 99.9\% in the pronounced interaction case at $n=1000$. While its power is consistently high, it is noteworthy that in the most challenging scenario—the slight quadratic misspecification at $n=2000$—its power of 13.2\% was slightly lower than that of other top-tier standardized tests like McCullagh's (20.0\%). This positions it as a highly reliable and powerful test, though perhaps not the single most sensitive option for detecting very subtle non-linearity.

In the real application, it performed ideally. For the initial misspecified model, it produced a non-significant but borderline p-value of 0.115 (Very close to Reject), correctly suggesting potential room for improvement where many other tests failed to do so. After the model was corrected with interaction terms, the p-value increased substantially to 0.735, clearly and correctly reflecting the improved model fit. This combination of robust size control, high power, and sensible performance on real data makes the Stute-Zhu test a top recommendation.

\paragraph{BAGofT Test with Bootstrapping (nsim=100)}
The BAGofT authors recommend a bootstrap procedure to derive the null distribution of the test statistic, which is intended to provide better control over the Type I error rate. This version of the test (\texttt{bagoft\_split20\_sim100}) implements a model-based bootstrap where, for each of the 100 simulations (\texttt{nsim=100}), the entire 20-split aggregation procedure is performed to calculate a single statistic from the null distribution.

Our simulation results confirm that this bootstrap procedure is successful in its primary goal: it effectively controls the Type I error rate. Unlike the liberal single-split version, the bootstrapped test maintained its size very close to the nominal 5\% level across all scenarios (e.g., 4.7\% in the Normal scenario and 4.9\% in the challenging Multi-Independent scenario at $n=2000$). However, this rigorous control of false positives came at a severe and unacceptable cost to statistical power. The test's ability to detect misspecification was dramatically reduced, rendering it one of the least powerful tests in the entire study. For instance, in the slight quadratic scenario at $n=2000$, its power was only 5.3\%, significantly lower than even the traditional Hosmer-Lemeshow test (10.8\%) and far below top-tier performers like the McCullagh test (20.0\%). The performance was even worse in the pronounced quadratic case at $n=1000$, where its power was a dismal 29.4\% while most other reliable tests had already achieved 100\% power.

This lack of sensitivity was also apparent in the real-world application. The bootstrapped test failed to detect the misfit in the initial model, producing a non-significant p-value of 0.341. After the model was corrected with interaction terms, the p-value barely changed to 0.348, indicating the test was unresponsive to the clear improvement in model fit. Furthermore, the nested resampling structure—100 bootstrap simulations, each performing 20 data splits and fitting a Random Forest—makes this one of the most computationally expensive tests in the study. In conclusion, while the bootstrap successfully addresses the Type I error inflation of the simpler BAGofT variants, the resulting test has unacceptably low power and a prohibitive computational cost, \textbf{making it a poor choice for practical use compared to more efficient and powerful alternatives.}

\paragraph{Projection-Based Test (\texttt{projection\_test})}
As this test was too computationally intensive for the full simulation, its performance is best assessed from the real data application. In that application, it produced a non-significant p-value of 0.625 for the initial model and an even higher p-value of 0.925 for the corrected model, aligning with the conclusions of other reliable tests and correctly reflecting the improvement in model fit.

\subsection{Other Specialized and Smoothing-Based Tests}
This category includes a variety of tests that use unique statistical principles, ranging from unweighted error sums to properties of the likelihood function's information matrix.

\paragraph{le Cessie-van Houwelingen Test}
This smoothing-based test performed exceptionally well, confirming its theoretical advantages. It maintained good size control, though it could be slightly liberal in some sparse settings (e.g., 7.8\% in the Uniform(-6,6) scenario at $n=2000$). However, it demonstrated high power, often comparable to the McCullagh and Osius-Rojek tests. In the real application, its p-value correctly increased from a non-significant $0.712$ to $0.804$, reflecting the improved model specification.

\paragraph{Copas Unweighted Sum of Squares (USS) (or RSS) Test (\texttt{copas\_unweighted})}
The Copas test is based on the simple and intuitive idea of summing the squared raw residuals ($y_i - \hat{\pi}_i$). In our study, this straightforward test demonstrated a commendable balance of properties. It exhibited excellent control over the Type I error rate, maintaining its size very close to the nominal 5\% level across all scenarios, including the most challenging sparse ones (e.g., 4.9\% in the Uniform(-6,6) scenario at $n=1000$). Its statistical power was consistently respectable and often superior to the Hosmer-Lemeshow benchmark. For example, in the slight interaction scenario at $n=2000$, the Copas test achieved a power of 14.8\%, nearly double that of the traditional HL test (7.5\%). It was also highly effective in detecting pronounced misspecification, reaching 100\% power in the pronounced quadratic scenario at $n=1000$. In the real-world application, the test behaved exactly as expected for a reliable diagnostic: its p-value was a non-significant but suggestive 0.111 for the initial misspecified model, and it correctly increased to 0.240 after the model was improved with interaction terms, reflecting the better fit. Overall, the Copas test proved to be a reliable, well-calibrated, and reasonably powerful tool.

\paragraph{Stukel's Test (\texttt{stukel\_both})}
Stukel's test is a score-based test specifically designed to detect misspecification of the logit link function by checking for issues like asymmetry or non-standard tail behavior. While its Type I error rate was generally well-controlled, it could be slightly liberal in some sparse settings (e.g., 6.1\% in the Uniform(-6,6) scenario at $n=1000$). The most notable finding was its surprisingly strong power against omitted non-linearity, a type of misspecification it is not explicitly designed for. In the slight quadratic scenario at $n=2000$, it was one of the most powerful tests, achieving a power of 20.4\%, outperforming even the McCullagh (20.0\%) and Osius-Rojek (14.0\%) tests, while in $n=5000$, Stukel reached 38\% while  McCullagh (34\%) and Osius-Rojek (23\%) . This suggests that omitting a quadratic term can manifest in a way that mimics a misspecified link function, to which this test is highly sensitive. In the real application, its p-value correctly increased from 0.376 to 0.479 after the model was corrected.

A critical implementation detail to note is the occurrence of `NA` values for certain Stukel's test variants in the simulation results. Stukel's test works by creating two new auxiliary predictors, $z_1 = 0.5 \cdot \eta^2 \cdot I(\eta > 0)$ and $z_2 = 0.5 \cdot \eta^2 \cdot I(\eta \le 0)$, where $\eta$ is the linear predictor (logit). The test for positive deviations (\texttt{s\_st\_pgeq0\_5}) relies only on $z_1$. In scenarios that produce highly skewed probabilities, such as the Chi-squared(4) distribution, it is possible for all linear predictors ($\eta_i$) in a given simulated dataset to be negative. In such a case, the auxiliary predictor $z_1$ becomes a vector of all zeros. A variable with zero variance is perfectly collinear with the model's intercept, making it impossible to fit the auxiliary regression required by the test. This computational failure correctly results in an `NA`. This behavior highlights that while powerful, the test's applicability can be limited by the distribution of the predicted probabilities.

\paragraph{Information Matrix Test (\texttt{IM\_efficient})}
The Information Matrix (IM) test is a theoretically elegant global test for general model misspecification, based on the principle that two different estimators of the Fisher information matrix should be equal for a correctly specified model. In our simulations, it demonstrated excellent and reliable performance. Its Type I error rate was almost perfectly controlled across all scenarios and sample sizes (e.g., 5.0\% in the Uniform(-6,6) scenario at $n=5000$, and 4.7\% in the Chi2(4) scenario at $n=2000$). Its statistical power was also consistently strong, making it a very good all-around performer. For example, in the slight interaction scenario at $n=2000$, it achieved a power of 15.8\%, more than double that of the traditional HL test. In the real-world application, it  failed to reject the initial, misspecified model with a p-value of 0.547. However, the test failed to compute for the corrected model that included interaction terms, returning an `NA`. This suggests a practical limitation: the auxiliary regression used to calculate the test statistic can become numerically unstable or suffer from collinearity when the primary model itself becomes more complex (e.g., by including interaction terms that are correlated with the main effects and their squares). Therefore, while the IM test is a powerful and reliable tool for standard models, its practical utility may be limited when assessing more complex model specifications.

\newpage
\section{Real-World Application Synthesis}
The real-world application to the low birth weight dataset provided a crucial lesson that complements the simulation findings. The initial analysis of the misspecified model (without interaction terms) revealed a significant conflict. The majority of tests, including reliable ones like the Stute-Zhu and McCullagh tests, produced non-significant p-values (Figure~\ref{fig:gof_pvalues_lbw}), suggesting the model was adequate. However, the graphical diagnostics told a different story. Both the GIVITI belt (Figure~\ref{fig:calib_belt_1997}) and the RMS reliability plot (Figure~\ref{fig:rms_reliability}) showed clear signs of systematic miscalibration, hinting at unmodeled non-linearity.
\vspace{0.5em}

\noindent This discrepancy highlights the primary conclusion of this thesis: \textbf{no single goodness-of-fit test is sufficient for a comprehensive model evaluation.} A p-value, even from a powerful test, is a single-number summary that can miss important patterns of misfit that are immediately apparent in a graphical analysis.

\vspace{0.5em}

\noindent The conflict was resolved by fitting the correctly specified model, which included the interaction terms known to be important. As shown in the updated plots (Figures~\ref{fig:rms_reliability_interaction} and \ref{fig:giviti_calibration_belt_internal_interaction}), the calibration improved dramatically. This improvement was reflected in the p-values of most GOF tests, which generally increased (Figure~\ref{fig:gof_pvalues_lbw_interaction}), confirming that they were correctly responding to the improved model specification. This exercise demonstrates the power of a holistic approach that combines robust statistical tests with careful graphical diagnostics.

\vspace{0.5em}

\begin{landscape}

    \begin{figure}[t]
        \centering
        \includegraphics[width=1.3\textwidth]{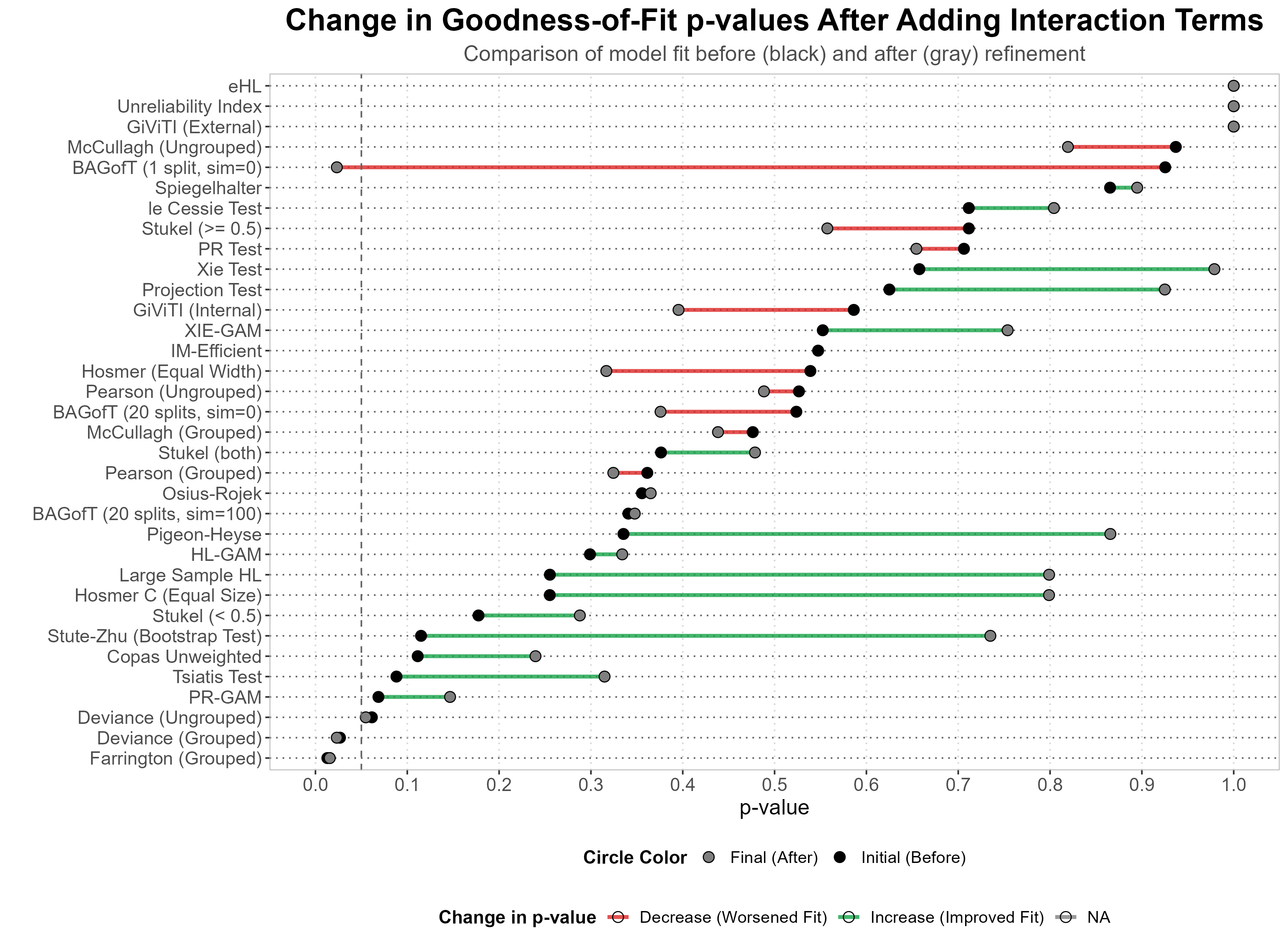}
        \caption[Change in goodness-of-fit p-values for a wide range of tests after adding interaction terms to the logistic regression model in the real data application]{Change in goodness-of-fit p-values for a wide range of tests after adding interaction terms to the logistic regression model in the real data application. Each line connects the p-value for the initial (misspecified) model (black) to the p-value for the refined (interaction) model (gray). Red lines indicate an increase in p-value (improved fit), while green lines indicate a decrease (worsened fit). The plot demonstrates that most reliable tests show a substantial increase in p-value after model refinement, confirming improved model adequacy.}
        \label{fig:gof_pvalue_change_plot_A4}
    \end{figure}
\end{landscape}

\noindent A detailed examination of the changes in p-values, as visualized in Figure \ref{fig:gof_pvalue_change_plot_A4}, reveals a clear hierarchy of test sensitivity. For the initial, misspecified model, a small group of tests correctly signaled a potential lack of fit by producing p-values that were either statistically significant ($\alpha < 0.05$) or borderline. The most sensitive tests were the \texttt{Farrington (Grouped)} test ($p=0.013$) and the \texttt{Deviance (Grouped)} test ($p=0.027$), both of which formally rejected the null hypothesis. Other tests that raised a warning flag with p-values below 0.15 included the \texttt{Deviance (Ungrouped)}, \texttt{PR-GAM}, \texttt{Tsiatis Clustering}, \texttt{Copas Unweighted}, and \texttt{Stute-Zhu} tests. This group of seven tests proved to be the most effective at detecting the subtle misspecification caused by the omitted interaction terms.

\vspace{0.5em}

\noindent Furthermore, the plot highlights which tests were most responsive to the model's correction. An ideal test should show a substantial increase in its p-value after the model is improved. The most dramatic and correct responses were seen from the bootstrap-based tests. The \texttt{Stute-Zhu} test's p-value increased more than sixfold, from a borderline 0.115 to a decisively non-significant 0.735. Similarly, the \texttt{Projection Test} p-value increased from 0.625 to 0.925. In stark contrast, several tests behaved erratically. The \texttt{BAGofT (1 split)} test produced a highly misleading result, with its p-value plummeting from a non-significant 0.925 for the poor model to a significant 0.023 for the superior model, incorrectly penalizing the improved specification. The \texttt{GiViTI (Internal)} test also showed a counter-intuitive decrease, though it remained non-significant. These results strongly suggest that tests like Stute-Zhu are not only sensitive to misfit but are also reliable indicators of model improvement, whereas other, more adaptive tests like BAGofT may exhibit unstable behavior in real-world applications.

\section{Conclusion and Recommendations}

This thesis has delivered an in-depth assessment of a broad spectrum of goodness-of-fit tests for binary logistic regression, with a particular focus on the challenges posed by sparse data. Through extensive simulation studies, we systematically evaluated the performance of nearly 35 different tests (including various versions and parameter settings), spanning Chi-Square-based, Partitioning-based, Score-based, Likelihood Ratio-based, Smoothing-based, Bootstrap-based, Information Matrix-based, and Machine Learning Calibration-based approaches. Our review encompassed both classic and contemporary methods, from the early 20th-century Pearson’s Chi-square test, through the widely adopted 1980 Hosmer-Lemeshow test, up to the latest 2024 Projection Bootstrap Test. The main findings of this work can be summarized as follows:

\begin{itemize}
    \item The Hosmer-Lemeshow test, while robust and widely used, is not the optimal choice at all for sparse data situations. No modified version of the Hosmer-Lemeshow test provided a substantial improvement for sample sizes of 5000 or fewer.
    \item Classic Pearson's Chi-square and Deviance tests are particularly ill-suited for sparse data and are generally not recommended in such contexts.
    \item Relying solely on hypothesis tests and their p-values is insufficient for assessing model fit. Visual diagnostics, such as the RMS reliability plot and the GIVITI calibration belt, are essential components of a thorough evaluation.
    \item Employing multiple goodness-of-fit tests is necessary to gain a comprehensive understanding of model adequacy.
    \item Simulation results indicate that the five best (non-liberal) options for detecting omitted polynomial terms or interaction misspecification are: the McCullagh and Osius-Rojek tests (standardized Chi-squared based), the GIVITI test (machine learning calibration-based), the le Cessie-van Houwelingen test (smoothing and score-based, though slightly liberal), and the Stute-Zhu test (bootstrap-based).
    \item Further analysis is required to evaluate those five tests across additional complex scenarios, more types of misspecification and larger sample sizes before drawing definitive conclusions. For example, Kuss \parencite{kuss2002} demonstrated that the Farrington test outperforms the McCullagh and Osius-Rojek tests in many scenarios where data sparsity is not extreme. Conversely, he found that the Copas and information matrix-based tests are much more effective at detecting link function misspecification and functional form misspecification.
    \item The Stukel test performed well in identifying the types of misspecification considered in this study, although it was originally designed to detect link function misspecification. More investigation is needed to clarify its effectiveness for omitted polynomial or interaction terms. Liu et al. (2024) \parencite{liu2024comprehensive} found that the Stukel test's performance closely matches that of the Stute-Zhu and Projection tests for sample sizes up to 500, except in two scenarios where the Projection test outperformed for those smaller samples. We anticipate that with larger sample sizes, the performance of the Projection, Stute-Zhu, and Stukel tests will converge.
    \item The GIVITI machine learning calibration test is a strong choice, particularly when using its 'Internal' version. The 'External' version is intended for situations where a global model has already been established and the goal is to assess its fit on a new dataset—a scenario that is relatively uncommon in practice.

    \item GAM-based versions of the Hosmer test are not recommended, particularly for small sample sizes. However, for the PR test, the GAM-based approach performs slightly better than the original, and for the XIE test, the GAM version is substantially superior to the original (though the original XIE test is slightly more conservative than its GAM counterpart).
    \item The Adaptive Partitioning test (BAGofT) is not advised. The non-bootstrap (sim=0) version lacks reliability, while the bootstrap version is both computationally expensive and exhibits low power. Additionally, it requires prior knowledge of the omitted variable for its partitioning, which is impractical in real-world applications. It is also unsuitable for simple logistic regression models with fewer than two predictors.
    \item The e-value version of the Hosmer test (eHL), which compares original predicted probabilities to calibrated probabilities, is not recommended. The calibration process often yields unreliable probabilities, and in some cases, the calibrated values perform worse than the original predictions.
    \item Assessing model fit by examining the linearity between logits and observed responses (i.e., checking if the reliability index has slope=1 and intercept=0) is not a reliable indicator of goodness-of-fit, especially when the model is misspecified due to omitted polynomial or interaction terms.

    \item It is generally not advisable to ungroup data that is already grouped (see section~\ref{sec:ungrouping_data}). For tests that accept the "number of trials" (i.e., the count for each covariate pattern) as input—such as Pearson's Chi-square test and its variants—performance is typically better when the data remain in grouped form rather than being expanded to individual observations.

    \item There are notable computational complexity considerations for these tests:
          \begin{itemize}
              \item The \textbf{McCullagh test} is computationally Moderate, with a complexity of approximately $O(n^2 p)$. It is well-suited for standard multi-core CPU execution (or better GPU preform matrix heavy calculations), However, it's not feasible for high sample sizes for large-scale simulation studies with high numbers of bootstrap replications without GPU.

              \item The \textbf{le Cessie-van Houwelingen test} is computationally demanding, with a complexity of $O(n^2)$ due to large matrix multiplications. A GPU is highly recommended to accelerate the core calculations and make the test practical for sample sizes beyond a few hundred.

              \item The \textbf{Projection-Based test} is extremely computationally intensive, with a complexity of $O(n^3 p)$. A powerful NVIDIA GPU is practically a requirement for this test, as its triple-nested loop structure is prohibitively slow on a CPU for all but the smallest sample sizes. not only but also it is a Bootstrap-based version, the matter which make it even more computationally intensive.
              \item The \textbf{BAGofT test} with simulation parameters (e.g., \texttt{nsplits=20}, \texttt{nsim=100}) is highly computationally intensive due to its nested resampling and adaptive partitioning structure. The primary bottleneck is the repeated fitting of the Random Forest partitioning model, a CPU-bound task that does not benefit from a GPU. A high-core-count CPU is strongly recommended to effectively parallelize the numerous independent simulation runs.
          \end{itemize}
\end{itemize}

In summary, while significant progress has been made in understanding and improving goodness-of-fit testing for logistic regression, important challenges and open questions remain. The findings and recommendations presented here provide a strong foundation for both applied practice and further methodological development.Looking ahead, several promising directions for future research and methodological innovation are outlined in the following section.

\section{Methodology Limitations}

\begin{itemize}
    \item \textbf{Simulation Framework Inheritance:} The simulation design closely follows the methodology of Hosmer et al. (1997), which, while providing a rigorous and widely accepted benchmark, also means that the study inherits certain limitations of the original framework. These include a focus on independent and identically distributed (IID) data, low-dimensional settings, and classical logistic regression, potentially limiting the generalizability to high-dimensional, penalized, or clustered data scenarios.

    \item \textbf{Limited Covariate Structures:} The scenarios primarily involve simple covariate distributions (e.g., univariate or low-dimensional multivariate settings with uniform, normal, or chi-squared distributions). The study does not address more complex or realistic data structures, such as correlated covariates, categorical predictors with many levels, or time-dependent covariates.

    \item \textbf{Pre-defined and Calculated Coefficients:} For Type I error analysis, model coefficients are pre-defined, while for power analysis, coefficients are calculated to satisfy specific fixed points. This approach, while systematic, may not capture the full range of parameter values encountered in real-world applications.

    \item \textbf{Misspecification Types Considered:} The power analysis is limited to specific forms of model misspecification: omitted quadratic terms, omitted interaction terms, and (in principle) link function misspecification. Other important sources of misspecification, such as omitted relevant variables, inclusion of irrelevant variables, or incorrect functional forms for categorical predictors, are not explored.

    \item \textbf{Link Function Misspecification Not Fully Explored:} Although the methodology acknowledges the importance of assessing sensitivity to link function misspecification (e.g., using Stukel's generalized logistic model), a detailed investigation of this scenario is not conducted in the present study.

    \item \textbf{No Model Selection Procedures:} The simulation does not incorporate model selection techniques (e.g., stepwise selection, best subset selection, or information criteria-based selection). As a result, the performance of goodness-of-fit tests following automated model selection is not evaluated.

    \item \textbf{No Assessment of General Performance Metrics:} The study does not address general performance measures such as the Brier score, log-loss, overall accuracy, or $R^2$ statistics (including Cox \& Snell, Nagelkerke, or McFadden $R^2$), which are commonly used to quantify model fit and explained variation.

    \item \textbf{No Consideration of Missing Data:} All simulations assume complete data. Scenarios involving missing data, or the use of methods such as inverse probability weighting or multiple imputation for goodness-of-fit testing, are not considered.

    \item \textbf{No Evaluation of Informal Diagnostic Methods:} Informal diagnostic approaches, such as residual-based visualizations or graphical diagnostics for lack-of-fit, are not discussed or evaluated.

    \item \textbf{No Assessment of Variable Omission or Inclusion:} The simulation does not include scenarios where variables are intentionally omitted or included to assess the ability of tests to detect unimportant or missing predictors. Thus, the performance of likelihood ratio or Wald tests for variable significance is not examined in the simulation context.
    \item \textbf{Comparing Response to Predicted Probabilities from Other Models:} In addition to logistic regression, predicted probabilities can be obtained from alternative models such as probit regression, complementary log-log models, or even machine learning approaches like random forests, gradient boosting, or neural networks.
\end{itemize}

\section{Future Work}

The findings of this thesis, while comprehensive within its defined scope, naturally open up several important avenues for future research. The complex interplay between test performance, data structure, and model specification suggests that the task of assessing goodness-of-fit is far from solved. The following areas represent critical next steps for building upon the insights gained in this study.

\paragraph{Expanding the Simulation Scenarios}
A crucial extension of this work is to evaluate the top-performing tests under a wider array of misspecification scenarios.
\begin{itemize}
    \item[--] \textbf{Link Function Misspecification:} Our study focused on omitted terms, but misspecification of the logit link is a common and important problem. The Stukel test showed surprising power against omitted non-linearity, suggesting its utility might be broader than just link function assessment. A dedicated simulation study is needed to formally compare the power of the top-tier tests (e.g., McCullagh, Osius-Rojek, Stute-Zhu, and Stukel) in detecting various forms of link misspecification, such as asymmetry or non-standard tail behavior.

    \item[--] \textbf{Moderately Sparse and Grouped Data:} This thesis focused on the challenging case of extreme data sparsity. As noted by Kuss (2002), the relative performance of tests can change in moderately sparse or grouped data settings. Future work should investigate if tests like the Farrington test regain their high power in these contexts, potentially altering the final recommendations for practitioners who work with such data.
          \newpage
    \item[--] \textbf{Complex Covariate Structures:} The current simulation relied on independent covariates. Real-world data often contains moderate to high correlation between predictors. A future study should explore how multicollinearity affects the performance and stability of these goodness-of-fit tests.
\end{itemize}

\paragraph{Extending to Modern Modeling Paradigms}
The landscape of predictive modeling has evolved significantly, and goodness-of-fit research must evolve with it.
\begin{itemize}
    \item[--] \textbf{High-Dimensional Data and Regularization:} A major frontier is the assessment of goodness-of-fit for models in high-dimensional settings ($p > n$) or for models fitted with regularization techniques like LASSO and Ridge regression. The process of coefficient shrinkage fundamentally alters the model's properties, and the behavior of these GOF tests in such a context is largely unknown.

    \item[--] \textbf{Application to "Black-Box" Machine Learning Models:} While this thesis focused on logistic regression, the need for reliable calibration assessment is even more acute for complex machine learning models like Gradient Boosting Machines (GBM), Random Forests, and Neural Networks. Future research should investigate how these rigorous statistical tests can be adapted and validated for assessing the probabilistic output of such non-parametric, "black-box" models.
\end{itemize}

\paragraph{Methodological and Computational Enhancements}
The practical utility of a test is often determined by its computational feasibility.
\begin{itemize}
    \item[--] \textbf{Computationally Efficient Approximations:} The Projection-Based test and the le Cessie-van Houwelingen test were identified as powerful but computationally prohibitive. A significant methodological contribution would be the development of computationally efficient approximations or faster algorithms for these tests (e.g., through advanced GPU programming or theoretical shortcuts), making them more accessible for routine use.

    \item[--] \textbf{Improving Adaptive Tests:} The BAGofT test was found to be unreliable, in part because its adaptive partitioning requires prior knowledge of the omitted variable. Future research could focus on developing a more robust, "unsupervised" version of the test that can automatically discover the most relevant partitions without such prior knowledge, potentially creating a more powerful and practical tool.
\end{itemize}

By addressing these open questions, the statistical community can continue to develop and refine the tools necessary for building trustworthy and reliable predictive models in an increasingly complex data landscape.

\newpage
\section*{Summary}

Let us recall the words of the brilliant statistician and researcher Agresti: \newline

\emph{`In practice, there is no guarantee that a certain logistic regression model fits the data well.'}" \parencite{agresti2013categorical} \newline

At its core, this thesis was driven by a simple question: "Is this model a good fit for the data?" To answer this, we explored a wide range of statistical tests, from classic methods to modern machine learning techniques. This journey has made one thing clear: in the age of complex data, the answer to that simple question is anything but simple.
\vspace{0.5em}

\noindent The extensive simulation study in this work acted as a rigorous stress test for three dozen different methods. The results painted a clear picture: there is no single "best" test for every situation. We saw that classic tests like the Pearson and Deviance statistics fail in the sparse data settings they were not designed for. We also identified a core group of highly reliable and powerful tests—specifically those by McCullagh, Osius-Rojek, and Stute-Zhu—that should form the foundation of a modern toolkit for any researcher assessing a logistic regression model.
\vspace{0.5em}

\noindent However, the most critical lesson came from the real-world application to the low birth weight dataset. It revealed the dangerous gap that can exist between a statistical p-value and the practical reality of a model's performance. In our analysis, the majority of tests produced non-significant p-values, suggesting the model was adequate. At the same time, visual tools like the GIVITI calibration belt and the RMS reliability plot clearly showed that the model was systematically miscalibrated. This proves that relying on p-values alone is not enough.
\vspace{0.5em}

\noindent Therefore, the final message of this work is a call for a change in practice. We must move away from relying on a single, often-misunderstood p-value, and towards a more complete approach that combines multiple, robust statistical tests with careful visual diagnostics. The recommendations and the decision flowchart presented in this thesis are designed to provide a framework for this kind of critical thinking.
\vspace{0.5em}

\noindent This research is a contribution to an ongoing and important conversation. As predictive models become more deeply integrated into science, medicine, and our daily lives, the need for them to be trustworthy is more critical than ever. The principles of a thorough, transparent, and multi-faceted validation, as explored and recommended in this work, are essential for achieving the important goal of building models that are not just powerful, but also honest.

\newpage
\newpage

\newpage
\appendix
\chapter{Appendix: Derivations \& Codes}

{\small
\section{Ungrouping the data}
\label{sec:ungrouping_data}
\subsection{Data Expansion}
Ungrouping, also known as data expansion, is the process of transforming each row of grouped binary data—where each row summarizes $n_g$  observations with $y_g$  successes—into $n_g$  individual Bernoulli trials, each with $n_g = 1$.
This approach is widely used in fields such as biostatistics, epidemiology, econometrics, and the social sciences.
Ungrouping is the process of intentionally converting grouped data into a sparse format, where each row represents a single observation (i.e., each covariate pattern has $n_g = 1$), without losing or adding any information. This is exactly what was done with the \texttt{birthwgt} data by setting \texttt{birthwgt\_trials} to 1 for every row (see section \ref{sec:gof_tests_results}).

To illustrate, suppose we have a grouped dataset where each row contains a covariate pattern and a column \texttt{nobs} indicating the number of times that pattern occurs (i.e., the number of trials). For example:

\begin{center}
    \begin{verbatim}
age  lwt  race2  race3  smoke  y  nobs
18   100     0      0      1   0     2
20   120     0      1      0   1     2
17   119     0      1      0   2     3
\end{verbatim}
\end{center}

Here, the first row represents a covariate pattern that occurs 2 times, the second pattern also occurs 2 times, and the third pattern occurs 3 times, for a total of 7 observations.

To ungroup this data, we expand each row so that each occurrence becomes its own row, resulting in 7 unique rows (one for each trial), each with \texttt{nobs = 1}. For example, the ungrouped data would look like:

{
\centering
\begin{verbatim}
age  lwt  race2  race3  smoke  y  nobs
18   100     0      0      1   0     1
18   100     0      0      1   0     1
20   120     0      1      0   1     1
20   120     0      1      0   1     1
17   119     0      1      0   0     1
17   119     0      1      0   1     1
17   119     0      1      0   1     1
\end{verbatim}
}

In this way, the total number of rows equals the sum of the original \texttt{nobs} values, and each row now represents a single observation with its covariate values, preserving all the original information in a sparse (ungrouped) format.

The motivation for ungrouping the data stems from the fact that many goodness-of-fit tests are specifically designed for ungrouped (sparse) data, which is a central focus of this thesis. Only a few tests—namely, the Pearson test (Equation \ref{eq:pearson_group}), Deviance test, Farrington test (Equation \ref{eq:farrington}), and McCullagh test (Equation \ref{eq:mccullagh})—have explicit formulas for grouped data (i.e., can take number of \texttt{trials} ($n_g$) as input), largely because they are based on standardized Pearson residuals. In contrast, most other tests are intended for use with ungrouped data, making the ungrouped (sparse) format more broadly applicable for the analyses conducted here, they are built for the purpose that $Y \sim Binomial(1, \pi)$, hense y = 0 or 1 (i.e., Giviti-Test Section \ref{sec:giviti_calibration_test_and_belt_framework}).
\newpage
\paragraph{This idea was discussed by A. Agresti in (2007), An Introduction to Categorical Data Analysis, 2nd. edition \parencite{agresti2013categorical}, p.106 :}

\begin{quote}
    \textit{In Table 3.1 on snoring and heart disease in the previous chapter, 254 subjects reported snoring every night, of whom 30 had heart disease. If the data file has grouped binary data, a line in the data file reports these data as 30 cases of heart disease out of a sample size of 254. If the data file has ungrouped binary data, each line in the data file refers to a separate subject, so 30 lines contain a 1 for heart disease and 224 lines contain a 0 for heart disease. The ML estimates and SE values are the same for either type of data file.}
\end{quote}

\subsection{Frequency Weights (Retaining Grouped Data)}

One common approach to handling grouped binary data is to use frequency weights, where each row in the dataset represents a unique covariate pattern and an additional column (e.g., \texttt{nobs}) indicates how many times that pattern occurs. This method retains the grouped structure of the data and is supported in most statistical software: \textbf{Stata:} \texttt{[fw = nobs]}, \textbf{SAS:} \texttt{freq = nobs} \textbf{R:} \texttt{weights = nobs} in \texttt{glm()}
\\
\noindent\textbf{Example in R:}
\begin{verbatim}
glm(y ~ age + lwt + smoke, family = binomial, data = grouped_data, 
    weights = nobs)
\end{verbatim}

This approach yields identical coefficient estimates and deviance statistics as the ungrouped (expanded) data, but is computationally more efficient for large datasets. but in here \texttt{y} is the number of success not binary variable, which is not compatible many GOF tests.

\subsection{Binomial Responses (Successes/Trials Syntax)}

Another approach is to specify the response as a binomial variable, using the number of successes and the number of trials for each covariate pattern. This is particularly useful when applying classical goodness-of-fit (GOF) tests that are designed for grouped data.

\begin{itemize}
    \item \textbf{R:} \texttt{cbind(successes, failures)} in \texttt{glm()}
    \item \textbf{SAS:} \texttt{events/trials} syntax in \texttt{proc logistic} \parencite{kuss2002paper}
\end{itemize}

\textbf{Example in R:}
\begin{verbatim}
glm(cbind(y, nobs - y) ~ age + lwt + smoke, family = binomial,
     data = grouped_data)
\end{verbatim}

This method is preferred when classical GOF tests (such as Pearson or deviance tests) are required, as it directly models the grouped binomial outcomes. For more on structuring and tidying data for analysis, see the Tidy Data principles by Wickham \parencite{tidydata}.

\newpage
\section{Appendix: Code}

\subsection{$X^2, D$ and $HL$ simulation Code} \label{3-sim}
This Code were conducted to compare the Power of the test Pearson Chi Square $X^2$,
Deviance D and Hosmer Test, by adding quadratic term in true model , That is
neglected in the fited model.
Note : power in the probability of reject Null, while its really false.
%

\begin{lstlisting}[language=R,frame=single,basicstyle=\footnotesize\ttfamily,numbers=left,numbersep=5pt,backgroundcolor=\color{LightGray},breaklines=true,columns=fullflexible,keepspaces=true]
# Load necessary libraries
library(tidyverse)
library(broom)
library(ResourceSelection)  # For hoslem.test function
# Function to generate data
generate_data <- function(n, beta0, beta1, beta2 = 0) {
  x <- runif(n, -3, 3)
  logit_p <- beta0 + beta1 * x + beta2 * x^2
  p <- 1 / (1 + exp(-logit_p))
  y <- rbinom(n, 1, p)
  data.frame(x = x, y = y)}
# Function to perform Pearson chi-square test
perform_pearson_test <- function(data) {
    #false fitting, quadratic term neglected
    model <- glm(y ~ x, data = data, family = binomial) 
    pearson_residuals <- residuals(model, type = "pearson")
    chisq_stat <- sum(pearson_residuals^2)
    df <- nrow(data) - length(coef(model))
    1 - pchisq(chisq_stat, df)}
# Function to perform Hosmer-Lemeshow test
perform_hosmer_test <- function(data) {
  #false fitting, quadratic term neglected
  model <- glm(y ~ x, data = data, family = binomial) 
  hoslem.test(data$y, fitted(model))$p.value}
# Function to perform deviance test
perform_deviance_test <- function(data) {
  #false fitting, quadratic term neglected
  model <- glm(y ~ x, data = data, family = binomial) 
  1 - pchisq(model$deviance, model$df.residual)}
# Simulation function for all tests
simulate_power <- function(n_sim, n_sample, beta0, beta1, beta2 = 0, alpha = 0.05) {
  results <- replicate(n_sim, {
    data <- generate_data(n_sample, beta0, beta1, beta2)
    c(pearson = perform_pearson_test(data) < alpha,
      hosmer = perform_hosmer_test(data) < alpha,
      deviance = perform_deviance_test(data) < alpha)
  })
  rowMeans(results)}
# Function to calculate power for different sample sizes
power_curve <- function(sample_sizes, n_sim, beta0, beta1, beta2 = 0, alpha = 0.05) {
  powers <- sapply(sample_sizes, function(n) {
    simulate_power(n_sim, n, beta0, beta1, beta2, alpha)
  })
  data.frame(sample_size = rep(sample_sizes, 3),
    power = c(powers["pearson",], powers["hosmer",], powers["deviance",]),
    test = rep(c("Pearson", "Hosmer-Lemeshow", "Deviance"), 
    each = length(sample_sizes)))}
# Calculate power for different sample sizes
set.seed(123)  # for reproducibility
sample_sizes <- seq(50, 500, by = 50)
power_data <- power_curve(sample_sizes, n_sim = 500, beta0 = 0, beta1 = 1, 
beta2 = 0.5)
# Plot power curves
ggplot(power_data, aes(x = sample_size, y = power, color = test, linetype = test)) +
  geom_line() +
  geom_point() +
  scale_color_manual(values = c("Pearson" = "black", 
  "Hosmer-Lemeshow" = "black", "Deviance" = "black")) +
  scale_linetype_manual(values = c("Pearson" = "solid", 
  "Hosmer-Lemeshow" = "dotted", "Deviance" = "dashed")) +
  geom_hline(yintercept = 0.8, linetype = "dashed", color = "red" , size = 1) +
  scale_y_continuous(breaks = seq(0, 1, by = 0.2), limits = c(0, 1)) +
  labs(x = "Sample Size", y = "Power",
       title = "Power Curves for Goodness-of-Fit Tests in Logistic Regression",
       color = "Tests", linetype = "Tests") +
  theme_minimal(base_size = 14) +
  theme(legend.position = "bottom",
  text = element_text(family = "serif"),
  legend.text = element_text(size = 12),
  legend.title = element_text(size = 12),
  axis.title = element_text(size = 13),
  axis.text = element_text(size = 13),
  plot.title = element_text(size = 14, face = "bold"),
  panel.grid.minor = element_blank()
  )
# Print summary statistics
summary_stats <- power_data %>%
  group_by(test) %>%
  summarise(
    mean_power = mean(power),
    min_power = min(power),
    max_power = max(power)
  )
print(summary_stats)
\end{lstlisting}

\subsection{Average P-value Of Chi-Square Test As $n$ Increases, While Testing Mis-specified Fitted Model}
This code (with the same setting of Appendix \ref{3-sim}) was conducted to understand the behaviour of the Average P-values of the Chi-Square Test, where the fitted model was wrongly specified, expecting it to decrease so we approach the rejection region ($\alpha > p-value$). However, this was not the case.

\begin{lstlisting}[language=R,frame=single,basicstyle=\footnotesize\ttfamily,numbers=left,numbersep=5pt,backgroundcolor=\color{LightGray},breaklines=true,columns=fullflexible,keepspaces=true]
# Load necessary libraries
library(tidyverse)

# Function to generate data
generate_data <- function(n, beta0, beta1, beta2 = 0) {
  x <- runif(n, -3, 3)
  logit_p <- beta0 + beta1 * x + beta2 * x^2
  p <- 1 / (1 + exp(-logit_p))
  y <- rbinom(n, 1, p)
  data.frame(x = x, y = y)
}

# Function to perform Pearson chi-square test
perform_pearson_test <- function(data) {
  model <- glm(y ~ x, data = data, family = binomial)
  pearson_residuals <- residuals(model, type = "pearson")
  chisq_stat <- sum(pearson_residuals^2)
  df <- nrow(data) - length(coef(model))
  1 - pchisq(chisq_stat, df)
}

# Simulation function
simulate_pvalues <- function(n_sim, n_sample, beta0, beta1, beta2 = 0) {
  replicate(n_sim, {
    data <- generate_data(n_sample, beta0, beta1, beta2)
    perform_pearson_test(data)
  })
}

# Set parameters
set.seed(123)  # for reproducibility
sample_sizes <- seq(50, 2000, by = 50)
n_simulations <- 1000
beta0 <- 0
beta1 <- 1
beta2 <- 0.5  # Misspecification

# Run simulations for each sample size
results <- sapply(sample_sizes, function(n) {
  pvalues <- simulate_pvalues(n_simulations, n, beta0, beta1, beta2)
  c(mean_pvalue = mean(pvalues), 
    se_pvalue = sd(pvalues) / sqrt(n_simulations),
    power = mean(pvalues < 0.05))
})

# Create results table
results_table <- as.data.frame(t(results))
results_table$sample_size <- sample_sizes
results_table <- results_table %>%
  select(sample_size, everything()) %>%
  mutate(across(where(is.numeric), ~round(., 4)))

# Print results table
print(results_table)

# Save results to CSV
write.csv(results_table, "pearson_chisquare_results.csv", row.names = FALSE)

# Create a plot of mean p-values
ggplot(results_table, aes(x = sample_size, y = mean_pvalue)) +
  geom_line() +
  geom_point() +
  geom_errorbar(aes(ymin = mean_pvalue - se_pvalue, 
                    ymax = mean_pvalue + se_pvalue),
                width = 20) +
  labs(x = "Sample Size", y = "Mean P-value",
       title = "Mean P-value of Pearson Chi-Square Test vs Sample Size") +
  theme_minimal() +
  theme(text = element_text(family = "serif", size = 12),
        axis.title = element_text(size = 14),
        plot.title = element_text(size = 16, face = "bold"))

ggsave("pearson_chisquare_pvalues.png", width = 10, height = 6, dpi = 300)
\end{lstlisting}

\subsection{Conversion from Multinomial to Multivariate Normal for Grouped and Individual Data} \label{sec:multinomial_to_multivariate_normal}
\begin{lstlisting}[language=R,frame=single,basicstyle=\footnotesize\ttfamily,numbers=left,numbersep=5pt,backgroundcolor=\color{LightGray},breaklines=true,columns=fullflexible,keepspaces=true]
library(MASS)
library(mvnormtest)
library(ggplot2)
library(ggthemes)
library(extrafont)

simulate_multinomial <- function(n, J) {
  p <- rep(1/J, J)
  rmultinom(1, n, p) }

test_multivariate_normality <- function(data) {
  mshapiro.test(t(data))$p.value }

run_simulation <- function(sample_sizes, fixed_J = NULL, num_simulations = 100) {
  p_values <- sapply(sample_sizes, function(n) {
    J <- if(is.null(fixed_J)) n else fixed_J
    samples <- replicate(num_simulations, simulate_multinomial(n, J)
        , simplify = FALSE)
    mean(sapply(samples, test_multivariate_normality))     })
  return(p_values) }

sample_sizes <- c(20, 50, 100, 200, 500, 1000, 1500, 2000)

# Scenario 1: J increases with n
p_values_increasing_J <- run_simulation(sample_sizes)
# Scenario 2: J is fixed at 10
p_values_fixed_J <- run_simulation(sample_sizes, fixed_J = 10)

results <- data.frame(
  n = rep(sample_sizes, 2), p_value = c(p_values_increasing_J, p_values_fixed_J),
  scenario = rep(c("J increases with n", "J fixed at 10"), 
    each = length(sample_sizes)) )

ggplot(results, aes(x = n, y = p_value, color = scenario, group = scenario
    , linetype = scenario)) +
  geom_line() + geom_point() + scale_x_log10() +
  scale_linetype_manual(values = c("J increases with n" = 
    "dashed", "J fixed at 10" = "solid")) +
  labs(x = "Total Sample Size (n)", 
       y = "Average p-value",
       title = "Multivariate Normality Test: Increasing J vs Fixed J") +
  theme_base(base_family = "CM Roman") +  # Use Computer Modern Roman font
  theme(
    text = element_text(family = "CM Roman"),
    plot.title = element_text(size = 10, hjust = 0.5),
    axis.title = element_text(size = 10),
    axis.text = element_text(size = 10),
    legend.position = "right",
    legend.text = element_text(size = 10),
    legend.title = element_blank()   ) +
  scale_color_manual(values = c("J increases with n" = "blue", 
    "J fixed at 10" = "red"))

print("P-values for increasing J:")
print(p_values_increasing_J)
print("P-values for fixed J:")
print(p_values_fixed_J)
\end{lstlisting}

\newpage
\section{Outside Resources}
\subsection{Agresti (2007)}
\label{ap:agresti_2007}
\begin{figure}[h]
    \centering
    \includegraphics[width=0.4\textwidth]{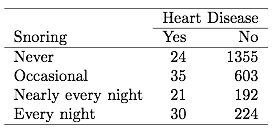}
    \label{fig:agresti_table1}
\end{figure}
Snoring and Heart Disease: Data from Agresti (2007) \parencite{agresti2013categorical}. The table shows the frequency of heart disease (Yes/No) by snoring category.
\subsection{Hosmer-Lemeshow Test (1997)}
\label{ap:hosmer_lemeshow_1997}
\begin{figure}[h]
    \centering
    \includegraphics[width=0.8\textwidth]{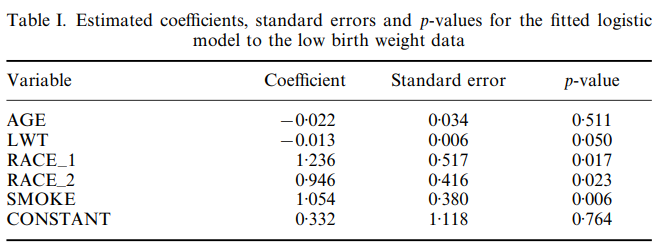}
    \label{fig:hosmer_table1}
\end{figure}
\begin{figure}[h]
    \centering
    \includegraphics[width=1\textwidth]{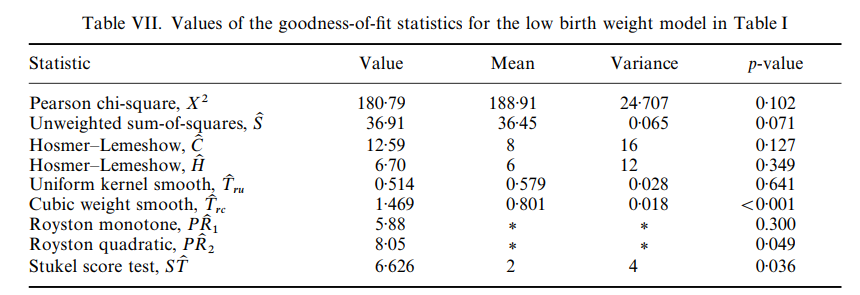}
    \label{fig:hosmer_table2}
\end{figure}

Both tables are reproduced from Hosmer and Lemeshow (1997) \parencite{Hosmer1997}. The results presented here closely align with those in this thesis (see section \ref{sec:real_application}), though some discrepancies remain. These differences are primarily attributable to the random noise introduced by Hosmer and Lemeshow in their original analysis, as well as variations in the algorithms implemented by different statistical software packages (see section \ref{sec:statistical_packages}).

\newpage
\subsection{Kuss (2002)}
\label{ap:kuss_2002}
\begin{figure}[h]
    \centering
    \includegraphics[width=0.8\textwidth]{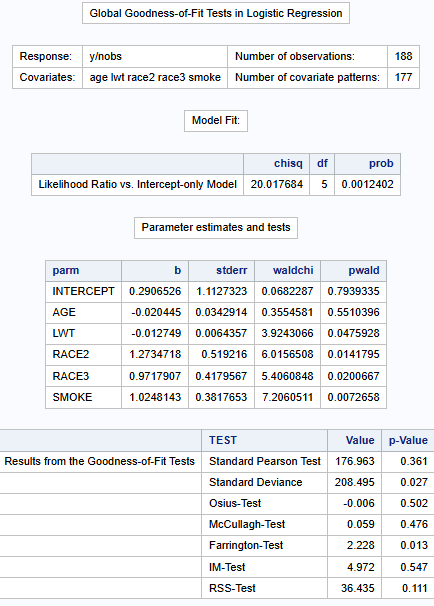}
    \label{fig:kuss_table1}
\end{figure}

It is evident that the results obtained by Kuss are highly consistent with those presented in this thesis (see section \ref{sec:real_application}) for nearly all tests (note that the RSS test corresponds to the Copas test in our context). In his analysis, Kuss applied the grouped data approach.

\newpage
\chapter{Algorithms}
\label{appendix:algorithms}

This appendix contains all the algorithms referenced in Chapter 3 (Selected Goodness-of-fit Tests) and Chapter 4 (Machine Learning and Bootstrap Algorithms). The algorithms are organized thematically, with traditional goodness-of-fit tests followed by modern calibration and bootstrap-based methods.

\section{Classical Chi-Square Type Tests}

\begin{algorithm}[H]
    \caption{Calculation of Pearson and Deviance Statistics for Logistic Regression}
    \label{alg:pearson_and_deviance_statistics}
    \begin{algorithmic}[1]
        \Procedure{CalculatePearsonAndDeviance}{data, fitted\_model}
        \State $\hat{\pi}_i \gets \textsc{GetFittedProbabilities}(fitted\_model, data)$
        \State $n \gets \textsc{NumberOfObservations}(data)$
        \State $Pearson (X^2) \gets 0$
        \State $Deviance (D) \gets 0$
        \For{$i = 1$ to $n$}
        \State $y_i \gets \textsc{ObservedOutcome}(data, i)$
        \State $\hat{\pi}_i \gets \textsc{FittedProbability}(fitted\_model, i)$
        \State $r_{i,pearson} \gets \frac{y_i - \hat{\pi}_i}{\sqrt{\hat{\pi}_i(1-\hat{\pi}_i)}}$ \Comment{Pearson residual}
        \State $Pearson \gets Pearson + r_{i,pearson}^2$
        \State $d_i \gets -2 \left[ y_i \log(\hat{\pi}_i) + (1-y_i)\log(1-\hat{\pi}_i) \right]$ \Comment{\small Dev. component}
        \State $Deviance \gets Deviance + d_i$
        \EndFor
        \State $p\_value_{Pearson} \gets 1 - \textsc{ChiSquareCDF}(Pearson, n-p^*)$
        \State $p\_value_{Deviance} \gets 1 - \textsc{ChiSquareCDF}(Deviance, n-p^*)$
        \If{$p\_value < 0.05$}
        \State \Return "Reject $H_0$: Evidence of poor fit"
        \Else
        \State \Return "Fail to reject $H_0$: No evidence of poor fit"
        \EndIf
        \EndProcedure
    \end{algorithmic}
\end{algorithm}

\begin{algorithm}[H]
    \caption{Hosmer-Lemeshow Test}
    \label{alg:hosmer_lemeshow_test}
    \begin{algorithmic}[1]
        \Procedure{HosmerLemeshowTest}{data, fitted\_model, G, test\_type}
        \State $\hat{\pi} \gets \textsc{GetFittedProbabilities}(fitted\_model, data)$
        \If{test\_type = "C"}
        \State $groups \gets \textsc{CreateEqualSizedGroups}(\hat{\pi}, G)$ \Comment{$G$ = number of groups}
        \ElsIf{test\_type = "H"}
        \State $groups \gets \textsc{CreateFixedCutpointGroups}(\hat{\pi}, G)$
        \EndIf
        \State $HL\_stat \gets 0$
        \For{$g = 1$ to $G$}
        \State $O_g \gets \textsc{ObservedEvents}(data, groups[g])$  \Comment{$O_g = \sum_{\{i: y_i \in groups[g]\}} y_i$}
        \State $n_g \gets \textsc{GroupSize}(groups[g])$
        \State $\overline{\hat{\pi}}_g \gets \textsc{MeanProbability}$  \Comment{ $\overline{\hat{\pi}}_g =  \frac{1}{n_g} \sum_{\{i: \hat \pi_i \in groups[g]\}} \hat \pi_i $}
        \State $e_g \gets \textsc{ExpectedEvents}(\hat{\pi}, groups[g])$
        \State \Comment{$e_g = n_g\overline{\hat{\pi}}_g  = \sum_{\{i: \hat \pi_i \in groups[g]\}} \hat \pi_i $}
        \State $HL\_stat \gets HL\_stat + \frac{(o_g - e_g)^2}{n_g\overline{\hat{\pi}}_g(1-\overline{\hat{\pi}}_g)}$
        \EndFor
        \State $p\_value \gets 1 - \textsc{ChiSquareCDF}(HL\_stat, G-2)$
        \If{$p\_value < 0.05$}
        \State \Return "Reject $H_0$: Evidence of poor fit"
        \Else
        \State \Return "Fail to reject $H_0$: No evidence of poor fit"
        \EndIf
        \EndProcedure
    \end{algorithmic}
\end{algorithm}

\begin{algorithm}
    \caption{Modified Hosmer-Lemeshow Test for Large Samples ($\text{mHL}_{\text{large}}$)}
    \label{alg:mHL_large}
    \begin{algorithmic}[1]
        \Procedure{mHLlarge}{$y, \hat{p}, G=10, \text{outsample}=\text{false}, \epsilon_0=\text{null}$}
        \State $n \gets \text{length}(y)$ \Comment{Sample size}
        \State $\text{dof} \gets \text{if outsample then } G \text{ else } G-2$ \Comment{Degrees of freedom}
        \If{$\epsilon_0 = \text{null}$}
        \State $\epsilon_0 \gets \sqrt{(\chi^2_{0.95,\text{dof}} - \text{dof})/10^6}$
        \EndIf
        \State $\hat{C} \gets \textsc{ComputeHLStatistic}(y, \hat{p}, G)$ \Comment{Hosmer $\hat{C}$ Test}
        \State $p\text{-value} \gets 1 - F_{\chi^2_{\text{dof},\epsilon_0^2n}}(\hat{C})$ \Comment{Compute p-value}
        \If{$p\text{-value} < 0.05$}
        \State \Return "Reject $H_0$: Model fit is not acceptable"
        \Else
        \State \Return "Fail to reject $H_0$: Insufficient evidence against acceptable fit"
        \EndIf
        \EndProcedure
    \end{algorithmic}
\end{algorithm}

The following algorithm uses these main inputs and variables:
\begin{itemize}
    \small
    \item $\boldsymbol{y}$: a vector containing the observed binary outcomes (each value is 0 or 1).
    \item $\hat{\boldsymbol{\pi}}$: a vector of predicted probabilities from the model (values between 0 and 1).
    \item $G$: the number of groups to divide the data into for the Hosmer-Lemeshow statistic (default is 10).$n$: the total number of observations (sample size).
    \item \texttt{outsample}: a logical value (true or false) indicating whether the test is being performed on out-of-sample data (default is false).
    \item $\epsilon_0$: the threshold value for what is considered an acceptable fit (if not provided, a default is calculated).
    \item \texttt{dof}: degrees of freedom for the test, which depends on $G$ and whether the data is out-of-sample.
    \item $F_{\chi^2_{\text{dof},\epsilon_0^2n}}$: the cumulative distribution function (CDF) of the non-central chi-square distribution, with degrees of freedom \texttt{dof} and non-centrality parameter $\epsilon_0^2 n$.
\end{itemize}

\begin{algorithm}[H]
    \caption{Pigeon-Heyse ($J^2$) Test (DOR)}
    \label{alg:pigeon_heyse_test}
    \begin{algorithmic}[1]
        \Procedure{PigeonHeyseTest}{data, fitted\_model, g}
        \State $\hat{\pi} \gets \textsc{GetFittedProbabilities}(fitted\_model, data)$
        \State $groups \gets \textsc{CreateEqualSizedGroups}(\hat{\pi}, G)$
        \State $J2\_stat \gets 0$
        \For{$g = 1$ to $G$}
        \State $o_g \gets \textsc{ObservedEvents}(data, groups[g])$
        \State $n_g \gets \textsc{GroupSize}(groups[g])$
        \State $\overline{\hat{\pi}}_g \gets \textsc{MeanProbability}(groups[g])$
        \State $e_g \gets n_g \overline{\hat{\pi}}_g$
        \State $\phi_g \gets \textsc{CorrectionFactor}(\hat{\pi}, groups[g])$
        \State $\phi_g \gets \sum_{\{i: \hat \pi_i \in groups[g]\}}\hat{\pi}_{ig}(1-\hat{\pi}_{ig})/n\overline{\hat \pi}_g(1-\overline{\hat \pi}_g)$
        \State $J^2\_stat \gets J^2\_stat + \frac{(o_g - e_g)^2}{n_g\overline{\hat{\pi}}_g(1-\overline{\hat{\pi}}_g)} \cdot \phi_g^{-1}$
        \EndFor
        \State $p\_value \gets 1 - \textsc{ChiSquareCDF}(J^2\_stat, G-1)$
        \If{$p\_value < 0.05$}
        \State \Return "Reject $H_0$: Evidence of poor fit"
        \Else
        \State \Return "Fail to reject $H_0$: No evidence of poor fit"
        \EndIf
        \EndProcedure
    \end{algorithmic}
\end{algorithm}

\section{Partitioning-Based Tests}

\begin{algorithm}[H]
    \caption{Pulkstenis and Robinson (PR) Test}
    \label{alg:pr_test}
    \begin{algorithmic}[1]
        \Procedure{PRTest}{data, fitted\_model}
        \State $X \gets \textsc{GetCategoricalCovariates}(data)$
        \State $\hat{\pi} \gets \textsc{GetFittedProbabilities}(fitted\_model)$
        \State $patterns \gets \textsc{CreateCovariatePatterns}(X)$ \Comment{Different from HL}
        \State $\chi^{*2} \gets 0$, $D^{*2} \gets 0$
        \State $C \gets \textsc{NumberOfUniqueCategoricalPatterns}(X)$
        \For{each $pattern$ in $C$} \Comment{Iterate over patterns, not deciles}
        \State $data\_subset \gets \textsc{GetDataSubset}(data, pattern)$
        \State $median\_prob \gets \textsc{Median}(\hat{\pi}[data\_subset])$
        \State $subgroups \gets \textsc{SplitByMedian}(data\_subset, \hat{\pi}, median\_prob)$
        \For{$h = 1$ to $2$}  \Comment{Two subgroups per pattern}
        \State $g \gets \textsc{GetGroupIndex}(pattern, h)$
        \State $o_g \gets \textsc{ObservedEvents}(subgroups[h])$ \Comment{$o_g = \sum_{\{i: y_i \in \text{group}_g\}} y_i$}
        \State $n_g \gets \textsc{GroupSize}(subgroups[h])$
        \State $\overline{\hat{\pi}}_g \gets \textsc{MeanProb}(subgroups[h], \hat{\pi})$ \Comment{$\overline{\hat{\pi}}_g = \frac{1}{n_g} \sum_{\{i: \hat{\pi}_i \in \text{group}_g\}} \hat{\pi}_i$}
        \State $e_k \gets n_k\overline{\hat{\pi}}_k$ \Comment{$e_k = \sum_{\{i: \hat{\pi}_i \in \text{group}_k\}} \hat{\pi}_i$}
        \State $\chi^{*2} \gets \chi^{*2} + \frac{(o_k - e_k)^2}{n_k\overline{\hat{\pi}}_k(1-\overline{\hat{\pi}}_k)}$
        \If{$o_k > 0$ and $o_k < n_k$}  \Comment{to avoid log(0)}
        \State $D^{*2} \gets D^{*2} - 2[o_k \log(\frac{o_k}{e_k}) + (n_k - o_k) \log(\frac{n_k - o_k}{n_k - e_k})]$
        \EndIf
        \EndFor
        \EndFor

        \State $K' \gets \textsc{NumberOfCategories}(X)$
        \State $k \gets \textsc{NumberOfCategoricalVariables}(X)$
        \State $df \gets (2C - 1)(K' - 1) - k - 1$ \Comment{Different df from HL}
        \State $p\_value\_chi \gets 1 - \textsc{ChiSquareCDF}(\chi^{*2}, df)$
        \State $p\_value\_D \gets 1 - \textsc{ChiSquareCDF}(D^{*2}, df)$
        \If{$p\_value\_chi < 0.05$ or $p\_value\_D < 0.05$}
        \State \Return "Reject $H_0$: Evidence of poor fit"
        \Else
        \State \Return "Fail to reject $H_0$: No evidence of poor fit"
        \EndIf
        \EndProcedure
    \end{algorithmic}
\end{algorithm}

\begin{algorithm}[H]
    \caption{Xie's Goodness-of-Fit Test}
    \label{alg:xie}
    \begin{algorithmic}[1]
        {\small
        \Procedure{XieGoodnessOfFitTest}{data, fitted\_model, G}
        \State $clusters \gets \textsc{ClusterCovariateSpace}(data, G)$
        \State $Xie\_stat \gets 0$
        \For{$g = 1$ to $G$} \Comment{Groups are made by clustering}
        \State $Xie\_stat \gets Xie\_stat + \frac{(O_g - E_g)^2}{n_g\overline{\hat{\pi}}_g(1-\overline{\hat{\pi}}_g)}$ \Comment{$O_g$,$E_g$ $n_g$,$\overline{\hat{\pi}}_g$ as in Hosmer-Lemeshow}
        \EndFor
        \State $p\_value \gets 1 - \textsc{ChiSquareCDF}(Xie\_stat, G - (p/2) - 1)$
        \If{$p\_value < 0.05$}  "Reject $H_0$: Evidence of poor fit"
        \Else "Fail to reject $H_0$: No evidence of poor fit"
        \EndIf
        \EndProcedure
        }
    \end{algorithmic}
\end{algorithm}

\section{Standardization-Based Tests}

\begin{algorithm}[H]
    \caption{Osius and Rojek Test for Individual Data}
    \label{alg:osius_rojek_test}
    \begin{algorithmic}[1]
        \Procedure{OsiusRojekTest}{fitted\_model, data}
        \State $\hat{\pi} \gets \textsc{GetFittedProbabilities}(fitted\_model)$
        \State $n \gets \textsc{GetSampleSize}(data)$
        \State $p^* \gets \textsc{GetNumberOfParameters}(fitted\_model)$ \Comment{Number of parameters in the model}
        \State $v_i \gets \hat{\pi}_i(1 - \hat{\pi}_i)$ \Comment{Variance for each observation, $\boldsymbol{v}$ is vector of $v_i$}
        \State $c_i \gets \frac{1 - 2\hat{\pi}_i}{v_i}$ \Comment{Correction term for each observation, $\boldsymbol{c}$ is vector of $c_i$}
        \State $X^2 \gets \sum_{i=1}^n \frac{(y_i - \hat{\pi}_i)^2}{v_i}$ \Comment{Pearson Chi-Square statistic}
        \State $\mathbf{X} \gets \textsc{GetModelMatrix}(data)$ \Comment{Design matrix of original covariates}
        \State $\mathbf{W} \gets \textsc{DiagonalMatrix}(v)$ \Comment{Diagonal matrix of weights}
        \State $\boldsymbol{\beta} \gets (\mathbf{X}^T\mathbf{W}\mathbf{X})^{-1}\mathbf{X}^T\mathbf{W}\mathbf{c}$ \Comment{WLS regression, c on X with weights $v$}
        \State $\hat{\mathbf{c}} \gets \mathbf{X}\boldsymbol{\beta}$ \Comment{Fitted values}
        \State $RSS \gets \sum_{i=1}^n v_i(c_i - \hat{c}_i)^2$ \Comment{Weighted residual sum of squares}
        \State $A \gets 0$ \Comment{Variance Correction factor, simplifies to 0 for individual data}
        \State $Z_{O\&R} \gets \frac{X^2 - (n-p^*)}{\sqrt{A + RSS}}$ \Comment{Osius and Rojek (Pearson Standardization)}
        \State $p\_value \gets 2 * (1 - \textsc{NormalCDF}(|Z_{O\&R}|))$ \Comment{Two-tailed p-value}
        \If{$p\_value < 0.05$}
        \State \Return "Reject $H_0$: Evidence of poor fit"
        \Else
        \State \Return "Fail to reject $H_0$: No evidence of poor fit"
        \EndIf
        \EndProcedure
    \end{algorithmic}
\end{algorithm}

\begin{algorithm}[H]
    \caption{The McCullagh Goodness-of-Fit Test Algorithm (Kuss, 2002)}
    \label{alg:mccullagh}
    \begin{algorithmic}[1]
        \State \textbf{Input:}
        \State Observed outcomes $y_i$ for each covariate pattern $i=1, \dots, n$.
        \State Number of trials $m_i$ for each pattern (for ungrouped data, $m_i=1$ for all $i$).
        \State Fitted probabilities $\hat{\pi}_i$ from the logistic regression model.
        \State Design matrix $\mathbf{X}$ of size $n \times p^*$, where $p^*$ is the number of parameters.

        \vspace{1em}
        \State \textbf{Step 1: Calculate the Pearson Chi-Square Statistic ($X^2$)}
        \State This is the uncorrected, standard Pearson statistic.
        \State $X^2 \gets \sum_{i=1}^{n} \frac{(y_i - m_i\hat{\pi}_i)^2}{m_i\hat{\pi}_i(1-\hat{\pi}_i)}$.

        \vspace{1em}
        \State \textbf{Step 2: Calculate Intermediate Components for Corrections}
        \State \textit{Weight Matrix:} $\mathbf{W} \gets \text{diag}(m_i\hat{\pi}_i(1-\hat{\pi}_i))$.
        \State \textit{Information Matrix Inverse:} $\mathbf{I}^{-1} \gets (\mathbf{X}^T \mathbf{W} \mathbf{X})^{-1}$.
        \State \textit{Leverages:} $h_i \gets (\mathbf{X} \mathbf{I}^{-1} \mathbf{X}^T)_{ii}$ (diagonal elements of the hat matrix).
        \State \textit{Skewness Correction Vector:} $u_i \gets \frac{1 - 2\hat{\pi}_i}{m_i\hat{\pi}_i(1-\hat{\pi}_i)}$. Let $\mathbf{u}$ be the vector of these values.
        \State \textit{Fitted Skewness Vector:} $\hat{\mathbf{u}} \gets \mathbf{X} \mathbf{I}^{-1} \mathbf{X}^T \mathbf{W} \mathbf{u}$.
        \State \textit{Binomial Moments for each observation $i$:}
        \State $k_{2i} \gets m_i\hat{\pi}_i(1-\hat{\pi}_i)$ (Variance)
        \State $k_{3i} \gets k_{2i}(1-2\hat{\pi}_i)$ (Third central moment)
        \State $k_{4i} \gets m_i(\hat{\pi}_i - 7\hat{\pi}_i^2 + 12\hat{\pi}_i^3 - 6\hat{\pi}_i^4)$ (Fourth central moment related term)

        \vspace{1em}
        \State \textbf{Step 3: Calculate the Corrected Mean ($E[T_{McC}]$)}
        \State The naive expectation is the degrees of freedom, $df = n-p^*$. McCullagh corrects this value.
        \State $E[T_{McC}] \gets (n-p^*) - \frac{1}{2}\sum_{i=1}^{n}\left(\frac{k_{4i} h_i}{k_{2i}}\right) + \frac{1}{2}\sum_{i=1}^{n}(\hat{u}_i k_{3i} h_i)$.

        \vspace{1em}
        \State \textbf{Step 4: Calculate the Corrected Variance ($Var(T_{McC})$)}
        \State \textit{Residual Sum of Squares of u:} $RSS_u \gets \mathbf{u}^T(\mathbf{W} - \mathbf{W}\mathbf{X}\mathbf{I}^{-1}\mathbf{X}^T\mathbf{W})\mathbf{u}$.
        \State $Var(T_{McC}) \gets \left(1 - \frac{p^*}{n}\right) \left(2\sum_{i=1}^{n}\frac{m_i-1}{m_i} + RSS_u\right)$.
        \State \Comment{For ungrouped data where all $m_i=1$, the first term inside the parenthesis is zero.}

        \vspace{1em}
        \State \textbf{Step 5: Compute the Final Statistic and p-value}
        \State Standardize the Pearson statistic using the corrected mean and variance:
        \State $Z_{McC} \gets \frac{X^2 - E[T_{McC}]}{\sqrt{Var(T_{McC})}}$.
        \State Calculate the one-sided p-value from the standard normal distribution:
        \State $p_{value} \gets 1 - \Phi(Z_{McC})$, where $\Phi$ is the standard normal CDF.
    \end{algorithmic}
\end{algorithm}

\begin{algorithm}[H]
    \caption{Farrington's Test for Goodness-of-Fit in Logistic Regression}
    \label{alg:farrington_test}
    \begin{algorithmic}[1]
        \Procedure{FarringtonTest}{data, fitted\_model}
        \State $\hat{\pi}_i \gets$ predicted probability for $i = 1, \ldots, N$ from fitted model
        \State $X^2_F \gets 0$
        \For{$i = 1$ to $N$}
        \State $A_i \gets \frac{(y_i - m_i\hat{\pi}_i)^2}{m_i\hat{\pi}_i(1 - \hat{\pi}_i)}$
        \State $B_i \gets \frac{-(1 - 2\hat{\pi}_i)}{m_i\hat{\pi}_i(1 - \hat{\pi}_i)}(y_i - m_i\hat{\pi}_i)$
        \State $X^2_F \gets X^2_F + A_i + B_i$
        \EndFor
        \State \Comment{The distribution of $X^2_F$ is approximately $\chi^2_{N-p^*}$, where $p^*$ is the number of model parameters}
        \State $df \gets N - p^*$
        \State $p\_value \gets 1 - \textsc{ChiSquareCDF}(X^2_F, df)$
        \If{$p\_value < 0.05$}
        \State \Return "Reject $H_0$: Evidence of poor fit"
        \Else
        \State \Return "Fail to reject $H_0$: No evidence of poor fit"
        \EndIf
        \EndProcedure
    \end{algorithmic}
\end{algorithm}

\section{Score and Likelihood-Based Tests}

\begin{algorithm}[H]
    \caption{Modified Tsiatis Test for Logistic Regression}
    \label{alg:tsiatis_test}
    \begin{algorithmic}[1]
        \Procedure{ModifiedTsiatisTest}{data, fittedmodel, g}
        \State Partition the covariate space into $g$ (any partitioning method can be used).
        \State Create indicator variables $\mathbf{I} = (I^{(1)}, ..., I^{(g-1)})$ for each region
        \State \Comment{Z statistics (component-wise standardised scores), easily computed with R}
        \State $\mathbf{Z} \gets \textsc{ScoreFunction}(\mathbf{x}'\hat{\boldsymbol{\beta}}, \mathbf{I}'\boldsymbol{\gamma})$
        \State $T \gets \mathbf{Z}'\mathbf{Z}$ \Comment{Compute test statistic}
        \State $df \gets g-1$ \Comment{Degrees of freedom, g - 1 is mentioned by Hosmer}
        \State $p_{value} \gets 1 - \chi^2_{\text{CDF}}(T, df)$
        \If{$p_{value} < \alpha$}
        \State \Return "Reject $H_0$: Evidence of poor fit"
        \Else
        \State \Return "Fail to reject $H_0$: No evidence of poor fit"
        \EndIf
        \EndProcedure
    \end{algorithmic}
\end{algorithm}

\begin{algorithm}[H]
    \caption{Tsiatis Score Test for Logistic Regression (Matrix Form)}
    \label{alg:tsiatis_score_test_matrix_form}
    \begin{algorithmic}[1]
        \Procedure{TsiatisScoreTest}{data, fittedmodel, $g$}
        {\small
        \State Partition the covariate space into $g$ groups (e.g., using $k$-means or quantiles)
        \State Create indicator variables $\mathbf{I} = (I^{(1)}, ..., I^{(g-1)})$ for $g-1$ groups (reference group omitted)
        \State Fit the null logistic regression model: $logit(\pi) = \mathbf{x}'\hat{\boldsymbol{\beta}}$
        \State Compute fitted probabilities: $\hat{\pi}_i = \frac{\exp(\mathbf{x}_i'\hat{\boldsymbol{\beta}})}{1 + \exp(\mathbf{x}_i'\hat{\boldsymbol{\beta}})}$ ,Compute residuals: $r_i = y_i - \hat{\pi}_i$
        \State Construct the cluster indicator matrix $\mathbf{X}_C$ (rows: observations, columns: $g-1$ indicators)
        \State Compute the score vector: $\mathbf{S} = \mathbf{X}_C^\top \mathbf{r}$
        \State Compute the main model matrix $\mathbf{X}$ (including intercept and covariates)
        \State Compute the diagonal weight matrix: $W = \mathrm{diag}(\hat{\pi}_i (1-\hat{\pi}_i))$
        \State Compute covariance blocks:
        \begin{itemize}
            \item $V_{11} = \mathbf{X}^\top W \mathbf{X}$
            \item $V_{12} = \mathbf{X}^\top W \mathbf{X}_C$
            \item $V_{21} = \mathbf{X}_C^\top W \mathbf{X}$
            \item $V_{22} = \mathbf{X}_C^\top W \mathbf{X}_C$
        \end{itemize}
        \State Compute $V = V_{22} - V_{21} V_{11}^{-} V_{12}$ \Comment{$V_{11}^{-}$ is a generalized inverse}
        \State Compute test statistic: $T = \mathbf{S}' V^{-} \mathbf{S}$
        \State Compute degrees of freedom: $df = \mathrm{rank}(V)$ \Comment{Can be approximated by $g-1$ in most cases}
        \State Compute $p$-value: $p = 1 - \chi^2_{\text{CDF}}(T, df)$
        \If{$p < \alpha$}
        \State \Return "Reject $H_0$: Evidence of poor fit"
        \Else
        \State \Return "Fail to reject $H_0$: No evidence of poor fit"
        \EndIf
        }
        \EndProcedure
    \end{algorithmic}
\end{algorithm}

\begin{algorithm}[H]
    \caption{Stukel Test for Logistic Regression Link Misspecification}
    \label{alg:stukel_test}
    \begin{algorithmic}[1]
        \State \textbf{Fit standard logistic regression:}
        Estimate $\hat{\beta}$ from $ logit(\pi_i) = x_i^\top \beta$

        \State \textbf{Compute linear predictors:}
        For each $i$, set $\hat{\eta}_i = x_i^\top \hat{\beta}$

        \State \textbf{Construct indicators and quadratic terms:}
        \For{$i = 1$ to $n$}
        \State $z_{1i} \gets 0.5 \cdot \hat{\eta}_i^2 \cdot I(\hat{\eta}_i > 0)$
        \State $z_{2i} \gets 0.5 \cdot \hat{\eta}_i^2 \cdot I(\hat{\eta}_i < 0)$
        \EndFor

        \State \textbf{Augment the logistic regression model:}
        Fit: $logit (\pi_i) = x_i^\top \beta + \alpha_1 z_{1i} + \alpha_2 z_{2i}$

        \State \textbf{Test for link misspecification:}\\
        \begin{itemize}
            \item \textbf{Score Test:}
                  \begin{itemize}
                      \item Compute the score vector $U$ for $(\alpha_1, \alpha_2)$ at $(0,0)$
                      \item Compute observed Fisher information matrix $\mathcal{I}$ at $(0,0)$
                      \item Calculate $TS = U^\top \mathcal{I}^{-1} U$ (compare to $\chi^2_2$)
                  \end{itemize}
            \item \textbf{Or, Likelihood Ratio Test:}
                  \begin{itemize}
                      \item Fit both standard and augmented models (Steps 1 and 4)
                      \item $LRT = 2 \times (\ell_{\text{augmented}} - \ell_{\text{standard}})$ (compare to $\chi^2_2$)
                  \end{itemize}
        \end{itemize}

        \State \textbf{Decision:}\\
        \begin{itemize}
            \item If $p$-value $<$ significance threshold (e.g., $0.05$), conclude evidence against logistic link fit.
            \item Otherwise, do not reject adequacy of logistic link.
        \end{itemize}

    \end{algorithmic}
\end{algorithm}

\begin{algorithm}[H]
    \caption{Deviance F-test}
    \label{alg:anova_f_test_for_deviances}
    \begin{algorithmic}[1]
        \Procedure{Anova F-Test for Deviances}{fitted\_model, data, num\_groups=10}
        \State $\hat{\pi} \gets \textsc{GetFittedProbabilities}(fitted\_model)$
        \State $r_d \gets \textsc{GetDevianceResiduals}(fitted\_model)$
        \State $cut\_points \gets \textsc{Quantiles}(1:n, \text{probs}=\{0.1, 0.2, ..., 1.0\})$ \Comment{Any Grouping Method can be used.}
        \State $G \gets \textsc{CutIntoGroups}(1:n, cut\_points, num\_groups)$
        \Comment{G might look like: [1,1,...,1,2,2,...,2,...,10,10,...,10]}
        \State $F \gets \textsc{OneWayANOVA}(response = r_d, factor = G)$
        \State $df_1 \gets num\_groups - 1$
        \State $df_2 \gets n - num\_groups$
        \State $p\_value \gets 1 - F_{CDF}(F, df_1, df_2)$
        \If{$p\_value < 0.05$}  "Reject $H_0$: Evidence of poor fit"
        \Else  "Fail to reject $H_0$: No evidence of poor fit"
        \EndIf
        \EndProcedure
    \end{algorithmic}
\end{algorithm}

\section{Information Matrix Tests}

\begin{algorithm}[H]
    \caption{The Information Matrix (IM) Test Algorithm (Kuss, 2002)}
    \label{alg:im_test}
    \begin{algorithmic}[1]
        \Procedure{IMTest}{y, $\hat{\pi}$, X}
        { \small
        \State \Comment{X: The design matrix of the model (with intercept)}
        \State \textbf{Step 1: Calculate Core Components}
        \State Calculate the Pearson residuals: $r_i = (y_i - \hat{\pi}_i) / \sqrt{\hat{\pi}_i(1-\hat{\pi}_i)}$
        \State \textbf{Step 2: Construct the Auxiliary Regressors ($W$)}
        \State Create the first set of regressors, $W_1 = D \cdot X$, where:
        \State \quad $D$ is a diagonal matrix with $D_{ii} = \sqrt{\hat{\pi}_i(1-\hat{\pi}_i)}$.
        \State Create the second set of regressors, $W_2 = F \cdot Z$, where:
        \State \quad $Z$ is a matrix containing the element-wise squares of the  matrix $X$ ($Z_{ij} = X_{ij}^2$).
        \State \quad $F$ is a diagonal matrix with $F_{ii} = \sqrt{\hat{\pi}_i(1-\hat{\pi}_i)}(1-2\hat{\pi}_i)$
        \State Combine these to form the full matrix of auxiliary regressors: $W = [W_1 | W_2]$.
        \State \textbf{Step 3: Perform the Auxiliary Regression}
        \State Perform an Ordinary Least Squares (OLS) regression of the Pearson residuals ($r$) on the auxiliary regressors ($W$).
        \State $\text{OLS fit} \gets \text{regress}(r \sim W)$
        \State \textbf{Step 4: Calculate the Test Statistic (ESS)}
        \State Explained Sum of Squares (ESS) from this auxiliary regression.
        \State $TSS \gets \sum r_i^2$ (Total Sum of Squares)
        \State $RSS \gets \sum (\text{residuals from OLS fit})_i^2$ (Residual Sum of Squares)
        \State $IM_{stat} \gets TSS - RSS$
        \State \textbf{Step 5: Compute the p-value}
        \State The test statistic follows a chi-square distribution under the null hypothesis. The degrees of freedom are equal to the number of regressors in the second part ($W_2$), which corresponds to the number of columns in the original design matrix $X$.
        \State $df \gets \text{number of the parameters in the model} = p+1$
        \State $p\_value \gets 1 - \text{P}(\chi^2(df) \leq IM_{stat})$
        \If{$p\_value < 0.05$}
        \State \textbf{Decision:} Reject the null hypothesis; model is likely misspecified.
        \Else
        \State \textbf{Decision:} Do not reject the null hypothesis; no evidence of misspecification.
        \EndIf
        \State \Return $IM_{stat}, p\_value$
        \EndProcedure
        }
    \end{algorithmic}
\end{algorithm}

\section{Smoothing-Based Methods}

\begin{algorithm}[H]
    \caption{Kernel-Based Smoothed Residual Test }
    \label{alg:kernel_based_smoothed_residual_test}
    \begin{algorithmic}[1]
        \Procedure{KernelSmoothedTest}{data, $h_n$} \Comment {$h_n$ is the bandwidth parameter of Kernel Smoothing}
        \State $\hat{\beta} \gets \textsc{FitLogisticModel(data)}$
        \State $\hat{\pi}_i  \gets \textsc{FittedValue(data,Model)}  $ \Comment{Estimated probabilities}
        \State $\hat{V} \gets \textsc{EstimateCovarianceMatrix = Diag($V_{ii}$)}$     \Comment{$V_{ii} = \hat{\pi}_i  (1- \hat{\pi}_i ) $}
        \State $ H \gets \textsc{HatMatrix = } VX[(X)^{\prime}VX]^{-1}(X)^{\prime}$
        \For{$i = 1$ to $n$}
        \State $W_i \gets \textsc{ComputeWeightsVector}(X_i, X, h_n)$ \Comment{Using kernel function}
        \EndFor
        \State $W \gets \textsc{WeightsMatrix}(W_i)$ \Comment{W is n by n, contain all weights for all residuals}
        \State ${A}_{r}=({I}-{M})^{\prime}{Q}_{r}({I}-{M})$ \Comment{ Where ${Q}_{r}= {V}^{-1/2}({W}^{\prime}{[Diag({W}{W}^\prime)}]^{-1}{W}){V}^{-1/2}$ }
            \State $\hat e \gets Y - \hat \pi$   \Comment{e is the residual Vector}
            \State $\hat r \gets V^{-1/2} \hat e$   \Comment{r is the Pearson residual Vector}
            \State $\hat r_s \gets W^\prime \hat r$ \Comment{$r_s$ is the smoothed Residual}
            \State $ \mathbf{ \hat{T_r}  \gets \hat r_s^\prime   D_r^{-1} \hat r_s =  \hat r^\prime W^\prime D_r^{-1} W \hat r} $       \Comment{Defined By Hosmer (1997)}
            \State $\widehat{\mathbf{Var}}(\hat{T}_{r}) \gets \cong2\left(\frac{2}{3}\right)^{p}\frac{\mathrm{trace}(\mathbf{W}\mathbf{W}^{\prime})}{n^{2}}$ \Comment{Approx. by le Cessie (1991)}
            \State $\mathbf{E(\hat{T}_r)}=\mathrm{trace}(\mathbf{A}_r\mathbf{V})$
            \State $p\_value \gets Pr[\chi^2(\nu)\geqslant b \mathbf{\hat{T}_r}] $ \Comment{Chi Square with $(\nu = 2\hat E^2/\hat {Var})$ degrees of freedom}
            \If{  $p\_value < 0.05$}   \Comment{ $(b = 2\hat E/\hat {Var})$}
            \State \Return "Reject $H_0$: Evidence of poor fit"
            \Else
            \State \Return "Fail to reject $H_0$: No evidence of poor fit"
        \EndIf
        \EndProcedure
    \end{algorithmic}
\end{algorithm}

\begin{algorithm}[H]
    \caption{Le Cessie and van Houwelingen (1995) Score Test Based on Random Effects Model}
    \label{alg:le_cessie_van_houwelingen_1995_score_test}
    \begin{algorithmic}[1]
        \Procedure{ScoreTest}{data, $R$}
        \State $\hat{\beta} \gets \textsc{FitLogisticModel}(data)$
        \State $\hat \pi \gets g(X, \hat{\beta})$ \Comment{Estimated probabilities}
        \State $\hat e \gets Y - \hat \pi$   \Comment{e is the residual Vector}
        \State $Q \gets e' R e$    \Comment{Q is the Smoothed Residual , $R = WW^\prime$ defined previously }
        \State $E(Q) \gets \text{trace}(RV)$ \Comment{V is the estimated covariance matrix of Y}
        \State $\text{Var}(Q) \gets \textsc{ComputeVariance}(Q, R, V)$ \Comment{Complex variance calculation}
        \State $T \gets (Q - E(Q)) / \sqrt{\text{Var}(Q)}$
        \State $p\_value \gets 1 - \Phi(T)$ \Comment{One-tailed test using standard normal distribution}
        \If{$p\_value < 0.05$}
        \State \Return "Reject $H_0$"
        \Else
        \State \Return "Fail to reject $H_0$"
        \EndIf
        \EndProcedure
    \end{algorithmic}
\end{algorithm}

\begin{algorithm}[H]
    \caption{Copas USS Test with Osius-Rojek Normal Approximation}
    \label{alg:copas_uss_test}
    \begin{algorithmic}[1]
        \Procedure{CopasOsiusRojekTest}{model} \Comment{X, y, $\hat{\pi}$ as defined in the previous Algorithms}
        \State $SSE \gets \sum_{i=1}^n (y_i - \hat{\pi}_i)^2$ \Comment{The unweighted sum of squared errors}
        \State $w_i \gets \hat{\pi}_i(1 - \hat{\pi}_i)$  \Comment{the weights based on fitted probabilities.}
        \State $d_i \gets 1 - 2\hat{\pi}_i$   \Comment{used as response in the weighted linear regression}
        \State $z \gets \textsc{lm.wfit}(X, d, w, \text{method="qr"})$
        \State $res \gets z.\text{residuals} \cdot \sqrt{z.\text{weights}}$
        \State $sd \gets \sqrt{\sum res^2}$
        \State $ev \gets \sum w_i$
        \State $Z \gets (SSE - ev) / sd$   \Comment{ normal approximation derived by Osius and Rojek}
        \State $p\_value \gets 2 \cdot (1 - \Phi(|Z|))$ \Comment{Two-sided p-value}
        \If{$p\_value < 0.05$}
        \State \Return "Reject $H_0$"
        \Else
        \State \Return "Fail to reject $H_0$"
        \EndIf
        \EndProcedure
    \end{algorithmic}
\end{algorithm}

\begin{algorithm}[H]
    \caption{GAM-based Goodness-of-Fit Tests (HL+GAM, PR+GAM, XIE+GAM)}
    \label{alg:gam_gof_test}
    \begin{algorithmic}[1]
        \Procedure{GAM\_GOF\_Test}{data, test\_type}
        \State \textbf{Step 1: Fit Overfitted GAM Model}
        \State Fit a Generalized Additive Model (GAM) to the binary response using all available covariates, including main effects, interactions, and smooth terms for continuous variables, with a logit link.
        \State $\hat{\pi} \gets$ Fitted probabilities from the GAM model for each observation.

        \State \textbf{Step 2: Form Groups for Goodness-of-Fit Test}
        \If{test\_type = HL+GAM}
        \State Partition data into $G=10$ groups (deciles) based on $\hat{\pi}$.
        \ElsIf{test\_type = PR+GAM}
        \For{each unique pattern of categorical covariates}
        \State Split data into two groups by the median of $\hat{\pi}$ within that pattern.
        \EndFor
        \ElsIf{test\_type = XIE+GAM}
        \State Encode categorical variables numerically.
        \If{$k < 5$}
        \State $G \gets 10$
        \Else
        \State $G \gets k+5$
        \EndIf
        \State Cluster the covariate space (all predictors) into $G$ groups using a clustering algorithm (e.g., k-means).
        \EndIf

        \State \textbf{Step 3: Compute Observed and Expected Counts}
        \For{each group $g$}
        \State $o_g \gets$ Number of observed positive outcomes in group $g$ \Comment{as Hosmer-Lemeshow}
        \State $e_g \gets$ Sum of fitted probabilities in group $g$ from logistic regression model.
        \EndFor

        \State \textbf{Step 4: Calculate Pearson Chi-Square Statistic}
        \State $X^2 \gets \sum_{g=1}^G \frac{(o_g - e_g)^2}{e_g(1 - \overline{\hat{\pi}}_g)}$ \Comment{$\overline{\hat{\pi}}_g$ is mean fitted probability in group $g$}
        \State Set degrees of freedom $df$ according to the literature (e.g., for HL+GAM, $df=G-2$; for XIE+GAM, $df=G-(k/2)-1$).

        \State \textbf{Step 5: Compute p-value}
        \State $p\_value \gets 1 - \textsc{ChiSquareCDF}(X^2, df)$

        \If{$p\_value < 0.05$}
        \State \Return "Reject $H_0$: Evidence of poor fit"
        \Else
        \State \Return "Fail to reject $H_0$: No evidence of poor fit"
        \EndIf
        \EndProcedure
    \end{algorithmic}
\end{algorithm}

\newpage

\section{Modern Calibration and Machine Learning Methods}

This section contains algorithms from Chapter 4 that represent modern approaches to calibration testing and machine learning-based goodness-of-fit assessment. These algorithms bridge traditional statistical inference with contemporary machine learning validation techniques.

\begin{algorithm}[H]
    \caption{Algorithm for the Unreliability (U) Test (LHR test)}
    \label{alg:unreliability_test}
    \begin{algorithmic}[1]
        \Procedure{UnreliabilityTest}{$\boldsymbol{y}$, $\boldsymbol{\hat{\pi}}$}
        {\small
        \State \Comment{$\boldsymbol{y}$: Vector of observed binary outcomes (0s and 1s) in the validation set}
        \State \Comment{$\boldsymbol{\hat{\pi}}$: Vector of predicted probabilities from the original model}

        \State \textbf{Step 1: Transform the predictor variable}
        \State Calculate the linear predictor ($logit$) from the original model's probabilities:
        \State $logit \gets \ln(\hat{\pi} / (1 - \hat{\pi}))$

        \State \textbf{Step 2: Fit the Cox validation model}
        \State Fit a new logistic regression model on the validation data with $logit$ as the sole covariate to estimate the calibration intercept ($\hat{\gamma}_0$) and slope ($\hat{\gamma}_1$):
        \State $\hat{fit}_{cal} \gets \textsc{LogisticRegression}(y \sim logit)$ \Comment{Equation \ref{eq:harrell_cal_model} Appendix \ref{app:calibration_model_proof}}

        \State \textbf{Step 3: Calculate the log-likelihood of the fitted calibration model ($\mathcal{L}_{cal}$)}
        \State Extract the maximized log-likelihood from the fitted calibration model object:
        \State $\mathcal{L}_{cal} \gets \textsc{logLikelihood}(\hat{fit}_{cal}) = \sum_{i=1}^{n} [y_{i} \ln(\hat{fit}_{cal,i}) + (1 - y_{i}) \ln(1 - \hat{fit}_{cal,i})]$

        \State \textbf{Step 4: Calculate the log-likelihood of the null model ($\mathcal{L}_{null}$)}
        \State This corresponds to a model with $\gamma_0=0$ and $\gamma_1=1$:

        \State $\mathcal{L}_{null} \gets  \textsc{logLikelihood}(\hat{\pi}_i) = \sum_{i=1}^{n} [y_{i} \ln(\hat{\pi}_{i}) + (1 - y_{i}) \ln(1 - \hat{\pi}_{i})]$

        \State \textbf{Step 5: Compute the Likelihood Ratio (LR) Test Statistic (recall \ref{sec:score_test_in_logistic_regression})}
        \State Calculate the statistic as per Equation \ref{eq:u_lr_statistic}. This value must be non-negative, as the fitted model's log-likelihood cannot be lower than the restricted null model's.
        \State $LR_{U} \gets 2 \times (\mathcal{L}_{cal} - \mathcal{L}_{null})$

        \State \textbf{Step 6: Compute the p-value}
        \State Compare the $LR_U$ statistic to a chi-square distribution with 2 degrees of freedom to test the significance of the miscalibration:
        \State $p\_value \gets 1 - \text{P}(\chi^2(2) \leq LR_{U})$

        \State \textbf{Step 7: Make the Decision Based on the p-value}
        \If{$p\_value < \alpha$}
        \State \Return "Reject $H_0$: Evidence of miscalibration".
        \Else
        \State \Return "Fail to reject $H_0$: No evidence of miscalibration".
        \EndIf
        }
        \EndProcedure
    \end{algorithmic}
\end{algorithm}

\begin{algorithm}[H]
    \caption{The GiViTI Calibration Test Framework}
    \label{alg:giviti}
    \begin{algorithmic}[1]
        \Procedure{GivitiCalibrationTest}{$\boldsymbol{y}$, $\boldsymbol{\hat{\pi}}$, devel}
        \State \Comment{devel: Context flag, either "external" or "internal"}
        \State $logit \gets ln(\frac{\hat{\pi}}{1-\hat{\pi}})$
        \State \textbf{Step 1: Set the starting degree for forward selection}
        \If{`devel` is "external"}
        \State $startDegree \gets 1$
        \Else[{if `devel` is "internal"}]
        \State $startDegree \gets 2$
        \EndIf

        \State \textbf{Step 2: Data-driven forward selection for polynomial degree}
        \State Fit a sequence of polynomial logistic models (Equation \ref{eq:giviti_poly_model}) starting from `startDegree`.
        \State At each step $k$, compute the LR statistic $D_k = 2(\mathcal{L}_{k+1} - \mathcal{L}_k)$.
        \State Stop and set the final degree $m=k$ if $P(\chi^2_1 > D_k) < (1 - \text{thres})$, \Comment{where \textbf{`thres`} is a significance threshold (e.g., 0.95).}
        \State \textbf{Step 3: Fit the final model} \Comment{Equation \ref{eq:giviti_poly_model}, Appendix \ref{app:calibration_model_proof}}

        \State Let $fit_m$ be the final selected polynomial model with degree $m$
        \State for instance, if the final model is a $m=3$, then it is a cubic model, and the model is $\text{logit}({\pi}) = \gamma_0 + \gamma_1 Logit + \gamma_2 Logit^2 + \gamma_3 Logit^3$ , where $Logit = \frac{\hat{\pi}}{1-\hat{\pi}}$.
        \State on R function can be used to fit this model:
        \State \texttt{glm(y $\sim$ Logit + I(Logit\^{}2) + I(Logit\^{}3), family = binomial)}
        \State \textbf{Step 4: Calculate the final test statistic}
        \State $\mathcal{L}_m \gets \textsc{logLikelihood}(fit_m)$ \Comment{log-likelihood of the final model}
        \State $\mathcal{L}_{\text{null}} \gets \sum_{i=1}^{n} [y{i} \ln(\hat{\pi}_{i}) + (1 - y_{i}) \ln(1 - \hat{\pi}_{i})]$ \Comment{same as U-test (Section \ref{sec:unreliability_test})}
        \State $T_m \gets 2 \times (\mathcal{L}_m - \mathcal{L}_{\text{null}})$

        \State \textbf{Step 5: Compute the p-value using the correct null distribution}
        \If{`devel` is "external"}
        \State $p\_value \gets 1 - \text{CDF}_{\text{external}}(T_m, m, \text{thres})$ \Comment{CDF details in Paper \parencite{nattino2014new}}
        \Else
        \State $p\_value \gets 1 - \text{CDF}_{\text{internal}}(T_m, m, \text{thres})$ \Comment{CDF details in Paper \parencite{nattino2015new}}
        \EndIf

        \State \Return Test statistic $T_m$ and $p\_value$
        \EndProcedure
    \end{algorithmic}
\end{algorithm}

\begin{algorithm}[H]
    \caption{eHL Test Algorithm}
    \label{alg:ehl}
    \begin{algorithmic}[1]
        {\small
            \Procedure{eHLTest}{$\mathbf{y}, \boldsymbol{\hat{\pi}}, B, s$}
            \State \Comment{$B$: number of splits (e.g., 10), $s$: split ratio (e.g., 0.5)}
            \State $e\_values \gets \emptyset$
            \For{$i = 1$ to $B$}
            \State \textbf{Step 1: Split the Data}
            \State Randomly partition the dataset $D$ into a training set $D_{train}$ (with fraction $s$ of the data) and a test set $D_{test}$.

            \State \textbf{Step 2: Non-parametric Calibration on Training Set}
            \State On $D_{train}$, fit an isotonic regression model of the outcomes $y_{train}$ on the predictions $P_{train}$.
            \State From the isotonic regression fit, derive a set of adaptive bins and their corresponding calibrated probabilities, $q_{train}$. This creates a data-driven calibration mapping.

            \State \textbf{Step 3: Calculate e-value on Test Set}
            \State Initialize the e-value for this split, $E_{split} \gets 1$.
            \For{each observation $j$ in $D_{test}$}
            \State Use the calibration mapping from Step 2 to find the calibrated probability $q_j$ for the prediction $P_j$.
            \State Calculate the individual e-variable (likelihood ratio) for observation $j$:
            \[
                E_j = \frac{q_j^{y_j}(1-q_j)^{1-y_j}}{P_j^{y_j}(1-P_j)^{1-y_j}}
            \]
            \State Update the total e-value for the split: $E_{split} \gets E_{split} \times E_j$.
            \EndFor
            \State Append $E_{split}$ to the list $e\_values$.
            \EndFor

            \State \textbf{Step 4: Aggregate and Convert to p-value}
            \State Calculate the final test statistic, $eHL$, by averaging the collected e-values:
            \[ eHL = \frac{1}{B} \sum_{i=1}^{B} E_{split, i} \]
            \State Convert the final $eHL$ value to a p-value:
            \[ p\_value = \min\left(1, \frac{1}{eHL}\right) \]
            \If{$p\_value < \alpha$}
            \State \Return "Reject $H_0$: Evidence of poor calibration/fit"
            \Else
            \State \Return "Fail to reject $H_0$: No evidence of poor calibration/fit"
            \EndIf
            \EndProcedure
        }
    \end{algorithmic}
\end{algorithm}

\begin{algorithm}[H]
    \caption{Spiegelhalter's z-test}
    \label{alg:spiegelhalter_z_test}
    \begin{algorithmic}[1]
        \Procedure{SpiegelhalterZTest}{$\mathbf{y}, \boldsymbol{\hat{\pi}}, \alpha$}
        {\small

            \State \textbf{Step 1: Calculate the z-statistic}
            \[
                Z = \frac{\sum_{i=1}^N (y_i - \hat{\pi}_i)(1 - 2\hat{\pi}_i)}{\sqrt{\sum_{i=1}^N (1 - 2\hat{\pi}_i)^2 \hat{\pi}_i (1 - \hat{\pi}_i)}}
            \]
            \State \textbf{Step 2: Make a decision}
            \State Find the critical value $z_{crit}$ from the standard normal distribution for significance level $\alpha$ (e.g., $z_{crit} \approx 1.96$ for $\alpha = 0.05$).
            \If{$|Z| > z_{crit}$}
            \State \Return "Reject $H_0$: Evidence of poor calibration"
            \Else
            \State \Return "Fail to reject $H_0$: No evidence of poor calibration"
            \EndIf
            \EndProcedure
        }
    \end{algorithmic}
\end{algorithm}

\section{Bootstrap and Adaptive Methods}

This section contains sophisticated bootstrap-based and adaptive algorithms that represent cutting-edge approaches to goodness-of-fit testing, particularly suited for complex data scenarios and machine learning model validation.

\begin{algorithm} [H]
    \caption{Binary Regression Adaptive Goodness-of-fit Test (BAGofT)}
    \label{alg:bagoft_test}
    \begin{algorithmic}[1]
        \Procedure{BAGofT2019}{data, MTA, $S$, $K$, $N_{min}$}
        \If{$S = 1$}
        \State $D_{n1}, D_{n2} \gets \textsc{SplitData}(data)$ \Comment{Split data once}
        \State $fitted\_model \gets \textsc{FitModel}(MTA, D_{n1})$
        \State $partitions \gets \textsc{GeneratePartitions}(D_{n1}, K, N_{min})$
        \State $best\_partition \gets \textsc{SelectBestPartition}(partitions, D_{n1}, fitted\_model)$
        \State $BAG\_stat \gets \textsc{ComputeBAGStatistic}(D_{n2}, best\_partition, fitted\_model)$
        \State $p\_value \gets 1 - \textsc{ChiSquareCDF}(BAG\_stat, K)$ \Comment{Compare to $\chi^2_K$ distribution}
        \If{$p\_value < 0.05$}
        \State \Return "Reject $H_0$"
        \Else
        \State \Return "Fail to reject $H_0$"
        \EndIf
        \Else
        \State $p\_values \gets \emptyset$
        \For{$s = 1$ to $S$}
        \State $D_{n1}, D_{n2} \gets \textsc{SplitData}(data)$
        \State $fitted\_model \gets \textsc{FitModel}(MTA, D_{n1})$
        \State $partitions \gets \textsc{GeneratePartitions}(D_{n1}, K, N_{min})$
        \State $best\_partition \gets \textsc{SelectBestPartition}(partitions, D_{n1}, fitted\_model)$
        \State $BAG\_stat \gets \textsc{ComputeBAGStatistic}(D_{n2}, best\_partition, fitted\_model)$
        \State $p\_value \gets 1 - \textsc{ChiSquareCDF}(BAG\_stat, K)$
        \State Append $p\_value$ to $p\_values$
        \EndFor
        \State $final\_p\_value \gets \textsc{Median}(p\_values)$
        \State $threshold \gets \textsc{Quantile}(\mathcal{N}(0.5, 1/(12S)), 0.05)$
        \If{$final\_p\_value < threshold$}
        \State \Return "Reject $H_0$: Evidence of poor fit"
        \Else
        \State \Return "Fail to reject $H_0$: No evidence of poor fit"
        \EndIf
        \EndIf
        \EndProcedure

    \end{algorithmic}
\end{algorithm}

\begin{algorithm}[H]
    \caption{Binary Regression Adaptive Goodness-of-fit Test (BAGofT) with Bootstrap p-value}
    \label{alg:bagoft_bootstrap_pvalue}
    \begin{algorithmic}[1]
        \Procedure{BAGofT}{data, model, $S$, partitioner, $N_{\text{sim}}$}
        \State Fit null model $M_0$ to \texttt{data}
        \State Compute observed test statistic $T_{\text{obs}}$ via \Call{SplitAGG}{data, $M_0$, $S$, partitioner}
        \For{$b = 1 \text{ to } N_{\text{sim}}$}
        \State Generate bootstrap data $D_b$ by simulating responses under $M_0$
        \State $T_b \gets$ \Call{SplitAGG}{$D_b$, $M_0$, $S$, partitioner}
        \EndFor
        \State Compute empirical p-value: $p = \frac{1}{N_{\text{sim}}} \sum_{b=1}^{N_{\text{sim}}} I(T_b \geq T_{\text{obs}})$
        \If{$p < \alpha$}
        \State \Return "Reject $H_0$"
        \Else
        \State \Return "Fail to Reject $H_0$"
        \EndIf
        \EndProcedure

        \Procedure{SplitAGG}{data, model, $S$, partitioner} \Comment{Same as Algorithm \ref{alg:bagoft_test}}
        \For{$s = 1 \text{ to } S$}
        \State Randomly split data into training set $D_{1,s}$ and test set $D_{2,s}$
        \State Fit $M_0$ on $D_{1,s}$
        \State Obtain partition $P_s$ using \texttt{partitioner} on $D_{1,s}$
        \State Compute BAG statistic $B_s$ for $D_{2,s}$ with partition $P_s$
        \State Compute $p$-value $p_s = 1 - F_{\chi^2}(B_s; K_s)$, where $K_s =$ number of cells in $P_s$
        \EndFor
        \State Return aggregate statistic (e.g., $T = \text{median}\{p_1, ..., p_S\}$ \textbf{or} $\min$, $\text{mean}$, etc.)
        \EndProcedure
    \end{algorithmic}
\end{algorithm}

\begin{algorithm}[H]
    \caption{Stute-Zhu Test Algorithm with Model-Based Bootstrap}
    \label{alg:tsz_bootstrap}
    \begin{algorithmic}[1]
        \State \textbf{Part 1: Calculate the Observed Statistic, $T_{SZ}^{obs}$}
        \State Fit the logistic regression model under the null hypothesis $H_0$ to the original data.
        \State Obtain the raw residuals, $\hat{r}_i = y_i - \hat{\pi}_i$, and the linear predictors, $logits$, for $i=1, \dots, n$.
        \State Sort the residuals $\hat{r}_i$ according to the values of their corresponding $logits$ to get $\hat{r}_{(i)}$.
        \State Compute the cumulative sums of the sorted residuals: $C_k = \sum_{i=1}^{k} \hat{r}_{(i)}$.
        \State Calculate the observed statistic: $T_{SZ}^{obs} \gets \frac{1}{n^2} \sum_{k=1}^{n} C_k^2$.

        \vspace{1em}
        \State \textbf{Part 2: The Bootstrap Procedure}
        \State Let $\hat{\pi}_i$ be the fitted probabilities from the initial model in Part 1.
        \For{$b = 1$ to $B$} \Comment{$B$ is the number of bootstrap replications (e.g., 200)}
        \State \textbf{Step 2a: Generate Bootstrap Sample}
        \State Generate a new vector of outcomes $Y^* = (y_1^*, \dots, y_n^*)$ where each $y_i^*$ is drawn from a Bernoulli distribution with success probability $\hat{\pi}_i$.
        \State Create the bootstrap dataset $(Y^*, X)$, using the original covariates.

        \State \textbf{Step 2b: Calculate Bootstrap Statistic}
        \State Fit the same logistic regression model to the bootstrap dataset.
        \State Using the new residuals and linear predictors from this bootstrap model, calculate the bootstrap statistic $T_{SZ}^{*(b)}$ following the same method as in Part 1.
        \EndFor

        \vspace{1em}
        \State \textbf{Part 3: Calculate the p-value}
        \State The bootstrap p-value is the proportion of bootstrap statistics that are greater than or equal to the observed statistic:
        \State $p_{value} \gets \frac{1}{B} \sum_{b=1}^{B} I(T_{SZ}^{*(b)} \geq T_{SZ}^{obs})$.
    \end{algorithmic}
\end{algorithm}

\begin{algorithm}[H]
    \caption{Projection-Based Test Algorithm with Model-Based Bootstrap}
    \label{alg:projection_bootstrap}
    \begin{algorithmic}[1]

        \State \textbf{Part 1: Calculate the Observed Statistic, $T_{proj}^{obs}$}
        \State Fit the logistic regression model under $H_0$ to the original data.
        \State Obtain the residual vector $\hat{\boldsymbol{\epsilon}}$ and the design matrix $\mathbf{X}$.
        \State Compute the $n \times n$ matrix $\mathbf{A}$ based on $\mathbf{X}$ as defined in Equation \ref{eq:A_ij}.
        \State Calculate the observed statistic: $T_{proj}^{obs} \gets \frac{1}{n^2} \hat{\boldsymbol{\epsilon}}^T \mathbf{A} \hat{\boldsymbol{\epsilon}}$.

        \vspace{1em}
        \State \textbf{Part 2: The Bootstrap Procedure}
        \State The bootstrap procedure is identical to the one described in Algorithm \ref{alg:tsz_bootstrap}. For each of the $B$ bootstrap replications:
        \State Generate a bootstrap sample $(Y^*, X)$.
        \State Fit the model to the bootstrap data.
        \State Calculate the bootstrap test statistic $T_{proj}^{*(b)}$ using the new residuals and the original design matrix $\mathbf{X}$ (or the new one, as they are identical).

        \vspace{1em}
        \State \textbf{Part 3: Calculate the p-value}
        \State The bootstrap p-value is the proportion of bootstrap statistics that are greater than or equal to the observed statistic:
        \State $p_{value} \gets \frac{1}{B} \sum_{b=1}^{B} I(T_{proj}^{*(b)} \geq T_{proj}^{obs})$.
    \end{algorithmic}
\end{algorithm}

\begin{algorithm}[H]
    \caption{The Lai and Liu Standardized H-L Test Algorithm}
    \label{alg:lailiu_bootstrap}
    \begin{algorithmic}[1]
        {\small
            \State \textbf{Input:} A large dataset of size $n$, a fitted logistic regression model, a target "standard" sample size $n_0$, and number of bootstrap replications $k$.

            \vspace{1em}
            \State \textbf{Part 1: Estimate the "Standard Power" via Bootstrap}
            \State Let the original large dataset have $m$ events (outcomes = 1) and $n-m$ non-events (outcomes = 0).
            \State Calculate the target number of events and non-events for the standard sample size, maintaining the event rate:
            \State $m_0 \gets \text{round}(n_0 \times (m/n))$
            \State $n_{0, non-event} \gets n_0 - m_0$
            \State Let $z_{\alpha}$ be the critical value for the H-L test, which is the $1-\alpha$ quantile of a $\chi^2$ distribution with $g-2$ degrees of freedom (typically $g=10$).
            \For{$i = 1$ to $k$} \Comment{$k$ is typically 1000 or more}
            \State \textbf{Step 1a: Create a Resampled "Standard" Dataset}
            \State Create a bootstrap dataset of size $n_0$ by randomly sampling (with replacement) $m_0$ observations from the original pool of $m$ events, and $n_{0, non-event}$ observations from the original pool of $n-m$ non-events.

            \State \textbf{Step 1b: Calculate the H-L Statistic}
            \State Fit the logistic regression model of interest to this smaller, resampled dataset.
            \State Compute the standard H-L test statistic, $T_{HL}^{(i)}$, using $g$ groups.
            \EndFor

            \State \textbf{Step 1c: Estimate the Standardized Power ($p_k$)}
            \State The standardized power, denoted $p_k$ in the paper, is the proportion of bootstrap statistics that exceed this critical value:
            \State $p_k \gets \frac{1}{k} \sum_{i=1}^{k} I(T_{HL}^{(i)} \geq z_{\alpha})$.

            \vspace{1em}
            \State \textbf{Part 2: The Probabilistic Decision Rule}
            \State The final decision is made probabilistically, with a rejection probability equal to the estimated power $p_k$.
            \State Generate a single random number, $r$, from a Uniform(0, 1) distribution.
            \If{$r < p_k$}
            \State Reject the null hypothesis, concluding there is evidence of lack of fit relative to the standard scenario.
            \Else
            \State Fail to reject the null hypothesis, concluding the model fit is acceptable.
            \EndIf
        }
    \end{algorithmic}
\end{algorithm}

\printbibliography[heading=bibintoc]

\end{document}